\documentstyle[pre,aps,multicol,eqsecnum,tighten,epsfig,array]{revtex}

\begin{document}

\title{Front propagation into unstable states: \\
Universal algebraic 
convergence towards uniformly translating pulled fronts}
\author{Ute Ebert$^{1,2} $ and Wim van Saarloos$^1$} 
\address{$^1$Instituut--Lorentz,
  Universiteit Leiden, Postbus 9506, 2300 RA Leiden, the Netherlands\\
$^2$CWI, Postbus 94079, 1090 GB Amsterdam, the Netherlands}
\date{submitted to Physica D on March 31, 1999 --- 
revised version from February 25, 2000} 
\maketitle


\begin{abstract}
  Fronts that start from a local perturbation and propagate 
  into a linearly unstable state come in two classes:
  pulled fronts and pushed fronts. The term ``pulled front'' expresses
  that these fronts are ``pulled along'' by the spreading of linear
  perturbations about the unstable state. Accordingly, their
  asymptotic speed $v^*$ equals the spreading speed of perturbations
  whose dynamics is governed by the equations linearized about the
  unstable state. The central result of this paper is the analysis of
  the convergence of asymptotically uniformly traveling pulled fronts
  towards $v^*$.  We show that when such fronts evolve from
  ``sufficiently steep''  
  initial conditions, which initially decay faster than $e^{-\lambda^*
    x}$ for $x\to \infty$, they have a {\em universal relaxation
    behavior} as time $t\to\infty$: The velocity of a pulled front
  always relaxes algebraically like 
  $v(t)=v^*-3/(2\lambda^*t)\;
  + (3\sqrt{\pi}/2)\;D\lambda^*/(D{\lambda^*}^2t)^{3/2}+O(1/t^2)$.
  The parameters $v^*$, $\lambda^*$, and $D$ are determined through a
  saddle point analysis from the equation of motion linearized about
  the unstable invaded state.  This front velocity is independent of
  the precise value of the front amplitude which one tracks to measure the
  front position.  The interior of the front is essentially slaved to
  the leading edge, and develops universally as
  $\phi(x,t)=\Phi_{v(t)}\left(x-\int^t dt'
    \;v(t')\right)+O(1/t^2)$, where $\Phi_{v}(x-vt)$ is a uniformly
  translating front solution with velocity $v<v^*$.  Our result, which can
  be viewed as a general center manifold result for pulled front
  propagation, is derived in detail for the well known nonlinear
  diffusion equation of type $\partial_t \phi
  =\partial_x^2\phi+\phi-\phi^3$, where the invaded unstable state is
  $\phi=0$. Even for this simple case, the subdominant $t^{-3/2}$ term
  extends an earlier result of Bramson. Our analysis is then
  generalized to more general (sets of) partial differential equations
  with higher spatial or temporal derivatives, to {\em p.d.e.}'s with
  memory kernels, and also to difference equations such as those that
  occur in numerical finite difference codes.  Our {\it universal}
  result for pulled fronts thus implies independence {\em (i)} of the
  level curve which is used to track the front position, {\em (ii)} of
  the precise nonlinearities, {\em (iii)} of the precise
  form of the linear operators in the
  dynamical equation, and {\em (iv)} of the precise initial
  conditions, as long as they are sufficiently steep.  The only
  remainders of the explicit form of the dynamical equation are the
  nonlinear solutions $\Phi_v$ and the three saddle point parameters
  $v^*$, $\lambda^*$, and $D$.  As our
  simulations confirm all our analytical predictions in every detail,
  it can be concluded that we have a complete analytical understanding
  of the propagation mechanism and relaxation behavior of pulled
  fronts, if they are uniformly translating for $t\to\infty$.  An
  immediate consequence of the slow algebraic relaxation is that the
  standard moving boundary approximation breaks down for weakly curved
  pulled fronts in two or three dimensions. In addition to our main
  result for pulled fronts, we also discuss the propagation and
  convergence of fronts emerging from initial conditions which are not
  steep, as well as of pushed fronts. The latter relax exponentially
  fast to their asymptotic speed.

\end{abstract}


\newpage

\begin{multicols}{2}

\tableofcontents

\end{multicols}

\newpage

\begin{multicols}{2}

\section{Introduction} \label{S1}

\subsection{Outline of the problem} \label{S11}

In this paper we address the rate of convergence or ``relaxation'' of
the velocity and profile of a front that propagates into an {\em
  unstable} state.  The particular fronts we analyze separate two
nonequilibrium homogeneous states, one of which is stable and one of
which is unstable, and are such that the asymptotic front solution is
a uniformly translating one. We assume that the unstable state is
initially completely unperturbed in a large part of space, and that
thermal and other noise are negligible.  Examples of such situations
arise in one form or another in
physics\cite{provansal,ahlers,babcock,fineberg,tsameret,ball,limat,langermk,goldstein,deissler1,deissler2,nb,salje,dorsey,ebert,luecke,ramses,eggers,cladis,martin,tucross,krug,torcini,martin2,chomaz,martin3,carpentier,ch,combustion},
chemistry\cite{combustion,fifebook,field,britton,chemfr1,chemfr2,chemfr3},
and biology\cite{fifebook,britton,murray}.  If the unstable state
domain is not perturbed by imperfect initial conditions or thermal
noise, it can only disappear through invasion by the stable state
domain.  We analyze the propagation of fronts formed in this process,
in particular the temporal convergence towards an asymptotic front
shape and velocity, and show that it is characterized by a universal
power law behavior in the so-called pulled regime.  We concentrate on
planar fronts, which thus can be represented in one spatial
dimension. However, our results for these and for the dynamical
mechanism also have 
important implications\cite{ebertmba} for the derivation of moving
boundary approximations\cite{fife,karma} for weakly curved fronts in
higher dimensions, as well as for the evaluation of the effects of
noise on fronts\cite{noise1,noise2,noise3,noise4,noise5,noise6,noise7},
especially the effect of multiplicative
noise\cite{multiplicativenoise,rocco}.

The problem of front propagation into an unstable state has a long
history, which dates back\cite{history} to the pioneering work by Kolmogoroff,
Petrovsky and Piscounoff (= KPP) \cite{KPP} and by Fisher
\cite{Fisher} on the nonlinear diffusion equation
\begin{equation}
\label{101}
\partial_t \phi = \partial_x^2 \phi + f(\phi)~,
\end{equation} 
where $f(\phi)$ is such that it has a homogeneous stable state
$\phi=1$ and a homogeneous unstable state $\phi=0$.  The early work on
this equation \cite{KPP,Fisher} was motivated by the biological
problem of 
gene spreading in a population. Since this work, the nonlinear
diffusion equation (\ref{101}), in particular the one with a simple
nonlinearity of the type
\begin{equation}
\label{102}
f=f_{\rm KPP}(\phi)=\phi-\phi^k ~,~k>1,~~{\rm e.g.,}~k=2~{\rm or}~3~,
\end{equation}
has become a standard problem in the mathematical literature
\cite{fifebook,britton,murray,aw,collet,mckean,depassier,textbook}.  
For the F-KPP equation defined by (\ref{101}) and
(\ref{102}), there exist dynamically stable uniformly translating
front solutions $\phi(x,t) \equiv \Phi_v(x-vt)$ for every velocity
$v\ge v^*=2\sqrt{f'(0)}$, and hence every one of these solutions is a
possible attractor of the dynamics for long times $t$. The resulting
dynamical behavior or ``velocity selection'' depends on the initial
conditions and has been investigated by a variety of methods
\cite{KPP,Fisher,aw,mckean};  essentially all its relevant properties 
have been derived rigorously \cite{aw}. E.g., following the lines of KPP
\cite{KPP}, Aronson and Weinberger proved rigorously \cite{aw}, that
every initial condition, that decays spatially at least 
as fast as $e^{-\lambda^* x}$ ($\lambda^*=v^*/2$) into the unstable
state for $x \rightarrow \infty$, approaches for large times the front 
solution
$\Phi_{v^*}(x-v^*t)$ with the smallest possible velocity $v^*$.  Most
of the rigorous mathematical methods can, however, not be extended to
higher order equations \cite{collet2}.

In physics, the interest in front propagation into unstable states
initially arose from a different angle. Since the late fifties, the
growth and advection of linear perturbations about a homogeneous
unstable state has been analyzed through an asymptotic large time
expansion of the Green's function of the linear equations
\cite{bers,huerre,landau}. Only ten to fifteen years ago did it become
fully clear in the physics community
\cite{stokes,dl,bj,dee2,vs1,dee,vs2,vs3,powell,oono,paquette}, that
there was actually an empirical but deep connection between the
rigorous results for the second order equations and some aspects of
the more general and exact but nonrigorous results for the growth of
linear perturbations. This has given rise to a number of
reformulations and intuitive scenarios aimed at understanding the
general front propagation problem into unstable states
\cite{dl,bj,vs1,vs2,vs3,powell,oono,paquette}.  

Although our results
bear on many of these approaches, our aim is {\em not} to introduce
another intuitive or speculative scenario. Rather, we will introduce
what we believe to be the first systematic analysis of the rate of
convergence or ``relaxation'' of the front velocity and profile in the
so-called ``linear marginal stability'' \cite{vs1,vs2} or ``pulled''
\cite{stokes,oono,paquette} regime. In this regime the asymptotic
front velocity is simply the linear spreading speed determined by the
Green's function of the linearized equations. Quite surprisingly, our
analysis even yields a number of {\em new} and {\em exact} results for
the celebrated nonlinear diffusion equation (\ref{101}), but it
applies equally well to (sets of) higher order partial differential equations
that admit uniformly translating fronts, to difference
equations, or to integro-differential equations. We will discuss such
equations in general, and then illustrate our results on the example
equations from Table I.  

For all such equations, our results have a remarkable degree of 
simplicity and universality: 
Pulled fronts always converge in time with {\em universal power laws
  and prefactors} that are independent of the precise form
of the equations {\em and} independent of the precise initial
conditions as long as they obey a certain steepness criterion. To be
precise, for equations such that the dynamically selected asymptotic
front is a uniformly translating 
pulled front, and for so-called {\em sufficiently steep} initial
conditions defined such that
$\lim_{x\to\infty}\phi(x,0)\;e^{\lambda x}=0$ for some $\lambda>\lambda^*$, 
we derive that the
asymptotic velocity convergence is given by the universal law
\begin{eqnarray}
\label{107}
\nonumber v(t)& =& v^* +\dot{X}~,~\\ \dot{X} & = &
-\frac{3}{2\lambda^*
  t}\left(1-\sqrt{\frac{\pi}{(\lambda^*)^2Dt}}\;\right)
+O\left(\frac{1}{t^2}\right)~.
\end{eqnarray}
The velocity $v^*$, the inverse length
$\lambda^*$ and the diffusion constant $D$ are in general obtained
from a saddle point expansion \cite{landau} for the equation of motion
linearized about the unstable state.
In a frame moving with velocity $v^*$ the quickest growing mode $k^*$
is identified by the complex saddle point equation
$\partial_k\left.\left[\omega(k)-v^*k\right]\right|_{k=k^*}=0$  where 
$\omega(k)$ is the 
dispersion relation of a Fourier mode $e^{-i \omega t+ikx} $.  In
the more usual decomposition into real functions this implies that
\cite{bers,bj,vs1,vs2}
\begin{equation}
\label{sadpoint}
\left. \frac{\partial \mbox{Im}~ \omega}{\partial \mbox{Im}~ k}
\right|_{k^*} = v^* ~, ~~ \left. \frac{\partial \mbox{Im}~
  \omega}{\partial \mbox{Re}~ k} \right|_{k^*} =0~.
\end{equation}
The speed of the frame is asymptotically the same as the speed of the
front if
\begin{equation}
\label{sadpoint2}
\frac{ \mbox{Im}~ \omega (k^*)}{ \mbox{Im}~ k^*} = v^*~.
\end{equation}
For the uniformly translating fronts that we will analyze here, we
have
\begin{equation}
  \mbox{Im}~k^* \equiv \lambda^*>0~,~~ \mbox{Re}~k^* =0~,~~
  \mbox{Re}~\omega(k^*)=0~,
\end{equation}
and a real positive diffusion coefficient $D$
\begin{equation}
\label{diffusioncoef}
D= \left.\frac{i\partial^2\omega}{2\partial k^2}\right|_{k^*} =\left.
\frac{\partial^2 \mbox{Im}~\omega}{2\partial \lambda^2}\right|_{k^*}~.
\end{equation}

While the velocity of a front is converging to its asymptotic value, so is the
profile shape. Note that $v(t)$ (\ref{107}) does not depend on the 
``height'' 
$\phi=h$, which is being tracked. In fact, if we define 
the velocity $v_\phi$ of the fixed amplitude $\phi=h$ through 
$\phi\left(x+\int^t d\tau\;v_\phi(\tau)\;,\;t\right)=h$, then up 
to order $1/t^2$ the velocity $v_{\phi}(t)= v^* +\dot{X}$ 
is {\em independent of the ``height''} $\phi=h$.  
Moreover, it is determined solely by properties of
the equation linearized about the unstable state, as Eqs.\ 
(\ref{sadpoint})--(\ref{diffusioncoef}) show. In this sense, we can
indeed speak of {\em pulling} of the front by the leading edge of the front.

The above expression for $v(t)$ contains {\em all} the universal
terms, since the next $1/t^2$ term in the long time expansion does
depend on initial conditions.  The above analytic results for the
universal velocity convergence as well as related ones for the
relaxation of the front profile which are summarized in Table II and
discussed in more detail below, are fully confirmed by extremely
precise numerical simulations.  Taken together, this study therefore
yields the understanding of the pulled front mechanism that so many
authors \cite{langermk,bj,vs1,dee,vs2,powell,oono,paquette,derrida}
have sought for.

In this paper, first the asymptotic long time behavior is worked out 
in detail and to high orders for the F-KPP equation (\ref{101}), 
(\ref{102}) in two matched asymptotic expansions in $1/\sqrt{t}$. 
Once we will have laid out the structure of this expansion, it is
clear that essentially the same matched expansions can be applied 
to other more
complicated types of equations, provided that they admit a family of
uniformly translating front solutions in the neighborhood of the
asymptotic ``pulled'' velocity $v^*$.  
Moreover, the two lowest order equations in the $1/\sqrt{t}$
expansion in the so-called leading edge region together with a 
boundary condition suffice to calculate the universal convergence.
The structure of these equations is virtually independent of the precise
form of the dynamical equation.  For more general equations, 

\begin{center}
\begin{tabular}{||p{8cm}||}
\hline \hline \\
\hspace{0.4cm} The EFK eq.:\vspace*{0.12cm} \\
$~~ \partial_t\phi = -\gamma\partial_x^4\phi+\partial_x^2\phi+f(\phi)
~~~\mbox{with } 0<\gamma<1/12~,$ \\ \\
\hspace*{0.4cm}The streamer eqs.: \vspace*{0.12cm} \\
$~\begin{array}{rl} 
\partial_t \sigma &=  D\partial_x^2\sigma+\partial_x(\sigma E)
+\sigma |E|e^{-1/|E|}~,
\nonumber\\ 
\partial_t E & = -D\partial_x\sigma-\sigma E~,
\end{array} $  \vspace*{0.12cm} \\ \\
\hspace*{0.4cm}A difference equation from kinetic theory: \vspace*{0.12cm} \\
$~~\partial_t C_i(t) = -C_i+C_{i-1}^2~,$\\ \\
\hspace*{0.4cm}A second order extension F-KPP eq.: \vspace*{0.12cm} \\ 
$~~\tau_2 \:\partial_t^2 \phi + \partial_t \phi=  \partial_x^2 \phi
+\phi - \phi^3 ~,$\\ \\
\hspace*{0.4cm}An equation with a memory kernel:\vspace*{0.12cm} \\
$~\begin{array}{rl}\partial_t \phi(x,t) & =  \partial^2_x \phi(x,t) 
+  \int_0^t dt'\; K(t-t')\;
\phi(x,t') \\ & ~~~ - \phi^k(x,t)~,\end{array}$\\ \\
\hspace*{0.4cm}Finite difference versions of the  F-KPP eq.\ like: 
\vspace*{0.12cm} \\
$~\begin{array}{rl}
\displaystyle \frac{u_j(t+\Delta t)-u_j(t)}{\Delta t} &= 
\displaystyle \frac{u_{j+1}(t)-2u_j(t)+u_{j-1}(t)}{(\Delta x)^2} \\ 
& \displaystyle ~~~~+u_j(t)-u^k_j(t)~.\end{array} $ \\ \\ 
\hline \hline 
\end{tabular}
\end{center}
\begin{minipage}{8.5cm}
{\bf Table I:} { Summary of the equations studied in detail in Section
\ref{S56} as examples of the general validity of our results for
higher order equations, coupled equations, difference equations, and
equations with a kernel. All these equations have pulled front
solutions whose asymptotic speed relaxes according to (\ref{107}).}   
\end{minipage} \vspace*{0.45cm} 

\noindent
we hence limit the discussion to the motivation and
analysis of these two equations.  Although we will give some
discussion of the assumptions that underly the expansion (like the one
that there is a nearby family of moving front solutions), a full
analysis of these as well as of the extension to nonuniformly
translating fronts, such as those arising in the EFK equation
of Table I for $\gamma>1/12$, in the Swift-Hohenberg
equation\cite{sh}, or in the complex Ginzburg-Landau
equation\cite{vs3}, will be left to future
publications\cite{willem1,spruijt,kees}.

For equations (\ref{101}), (\ref{102}) we simply have $v^*=2$,
$\lambda^*=D=1$. The first term in (\ref{107}) then reduces to a
wellknown
result of Bramson \cite{bramson}, who  rigorously proved that the
convergence to the asymptotic velocity $v^*$ is
$v(t)=v^*-3/(2\lambda^* t)$ uniformly, i.e., independent of the
amplitude $\phi$ whose position one tracks. The factor $3/2$ in this expression 
has often been considered puzzling, since the {\em linear} diffusion equation 
with localized initial conditions yields $v(t)=v^*-1/(2\lambda^* t)+\ldots~$.
In \cite{vs2}, it was argued that the
factor 3/2 in this result applies more generally to higher order
equations as well, but a systematic analysis or an argument for why
the convergence is uniform, was missing. Apart from this and a recent
rederivation \cite{derrida} of Bramson's result along lines similar in
spirit to ours\footnote{The main focus of the work by Brunet and
  Derrida \cite{derrida} is actually the correction to the asymptotic
  velocity if the function $f(\phi )$ has a cutoff $h$ such that
  $f_h(\phi) =0$ for $\phi <h$. The method the authors use to derive
  this, is actually closely related to the one they use to rederive
  Bramson's result, and to our approach. See in this connection also the recent paper by Kessler {\em et al.} \cite{kessler}. } and a few papers similar in
spirit to that of Bramson \cite{mckean,delft,gallay}, we are not aware
of systematic calculations of the velocity and profile
relaxation. Even for the convergence  
of the velocity in the celebrated nonlinear difusion equation, our
$1/t^{3/2}$ term appears to be new. 

From a different perspective, Powell {\em et al.} \cite{powell} also
considered the convergence properties of pulled fronts. These authors
studied the shapes of the front profiles in the nonlinear diffusion
equation and argued that they relax along the the family of unstable
uniformly translating front solutions. Although they realized that 
the velocity relaxation was algebraic and from below, 
they did not seem to realize
that the  dominant $-3/(2t)$ velocity correction was known from
earlier work \cite{bramson,vs2}. As we shall see below when we will
discuss the  
shape relaxation of fronts, our derivation is the first analytic
derivation and confirmation of the picture of Powell {\em et al.}, and
identifies the connection with the  velocity relaxation. 

Our results are not only of interest in their own right, but they also
have important implications. Since the asymptotic convergence towards
the attractor $\Phi^*$ is algebraic in time, the attractor alone might
not give sufficient information about the front after long but finite
times, since algebraic convergence has no characteristic time scale.
In particular, there is no time beyond which convergence can be
neglected. Such slow convergence means that in many cases,
experimentally as well as theoretically, one observes transients and
not the asymptotic behavior. In fact, in the very first explicit
experimental test of front propagation into unstable states in a
pattern forming system \cite{ahlers}, viz.\ Taylor-Couette flow, the
initial discrepancy between theory and experiment was later shown to
be related to the existence of slow transients \cite{luecke}.  The
slow convergence is important for theoretical studies as well: it is a
common experience (see, e.g., \cite{nb,dee,goldstein2}) that when
studying front propagation in the ``pulled'' regime numerically, the
measured transient front velocity is often below $v^*$. This is so
even though the {\em asymptotic} front speed can never be below $v^*$,
because no slower attractor of the dynamics exists.  This observation
finds a natural explanation in our finding that the rate of
convergence is always power law slow, and that the front speed is
always approached {\em from below}.

A second important implication of the absence of an intrinsic time
scale of the front convergence is the following. When we consider the
propagation of such fronts in more than one dimension in which there
is a coupling to another slow field (as, e.g., in the phase field
models\cite{karma,langer,bates}), the front dynamics does not
adiabatically decouple from the dynamics of the other field and from
the evolution of the curvature and shape of the front itself. This
implies that the standard moving boundary
approximation\cite{fife,karma,buckmaster} (which actually rests on the
assumption that the convergence on the ``inner scale'' is exponential)
can not be made.  Though this is intuitively quite obvious from the
power law behavior of the front convergence process, the connection
between the convergence and the breakdown of a moving boundary
approximation also emerges at a technical level: the divergence of the
solvability integrals that emerge when deriving a moving boundary
approximation turns out to be related to the continuity of the
stability spectrum of pulled fronts \cite{ebertmba}.  The break-down
of the solvability analysis for perturbations of the asymptotic front
in the pulled regime also has consequences for the evaluation of
multiplicative noise in such equations \cite{ebertmba,rocco}.

\subsection{Pushed versus pulled fronts, selection and convergence}
\label{S12}

Let us return to the well understood nonlinear diffusion equation
(\ref{101}) and discuss to which nonlinearities $f(\phi)$ our
prediction of algebraic convergence applies and why. If $f'(0)<0$, the
invaded state $\phi=0$ is linearly stable, and the construction of a
uniformly translating front $\phi(x,t)=\Phi_v(x-vt)$ poses a nonlinear
eigenvalue problem.  The solution with the largest eigenvalue $v$ is
the unique stable and dynamically relevant solution (unique up to
a translation, of course). As is well known and
discussed in Section \ref{S2}, any initial front that separates
the (meta)stable state $\phi=0$ at $x\to\infty$ from another stable state at
$x\to-\infty$ will converge exponentially in time to this unique attractor $\Phi_v$.  However,
whenever $f'(0)>0$, $\phi=0$ is unstable, there is not a unique
asymptotic attractor $\Phi_v$, but a continuous spectrum of nonlinear
eigenvalues $v$ which constitute the velocities of possible
attractors $\Phi_v$. The existence of a continuum of attractors of the
dynamics poses a so-called {\em selection problem}: From which initial
conditions  will the front dynamically approach which attractor?  The
attractor with the smallest velocity plays a special role, as its
basin of attraction are all ``sufficiently steep'' initial conditions,
as defined in Section \ref{S2}. It therefore will be refered to as
{\em the selected} front solution.

When we concentrate on these ``sufficiently steep'' initial conditions
and analyze the dependence on the nonlinearity $f$ in (\ref{101}),
the transition from exponential to algebraic convergence does {\em
  not} coincide with the transition from stability to instability
of the invaded state $\phi=0$, but with the transition between two
different mechanisms of front propagation into unstable states.
Indeed, it is known (see also Sect.\ \ref{S2}), that for $f'(0)>0$,
there are two different mechanisms for how the
selected front $\Phi_{sel}$ and its speed $v_{sel}$ are determined.
Either $\Phi_{sel}$ is found by constructing a so-called strongly
heteroclinic orbit for $\Phi_v$ from the full nonlinear equation. This
case is also known as case II \cite{bj} or nonlinear marginal stability
\cite{vs1,vs2}, or as pushing \cite{stokes,oono,paquette}. Or, the
selected velocity $v_{sel}$ is determined by linearizing about the
unstable state $\phi=0$, which case is known as case I or linear
marginal stability, or as pulling.  We henceforth will use the terms
``pushing'' and ``pulling'' for the two different propagation
mechanisms of a selected front evolving from steep initial conditions,
since they very literally express the 
different dynamical mechanisms.

In a pushed front just like in a front propagating into a (meta)stable
state, the dynamics is essentially determined in the nonlinear
``{\em interior part}'' of the front, where $\phi$ varies from close
to $\phi=0$ 
to close to the stable state. The construction of the selected
front as a strongly heteroclinic orbit in the pushed case continuously
extends into the construction of the heteroclinic orbit of the unique
attractor if the invaded state is (meta)stable ($f'(0)<0$).  For both
pushed fronts and fronts propagating into linearly stable states, the
spectrum of linear perturbations is bounded away from zero, so that
convergence towards the asymptotic front is exponential in time.

In a pulled front, the dynamics is quite different: As we shall see,
it is determined essentially in the region linearized about the
unstable state. We call this region the leading edge of the front.
Eq.\ (\ref{101}) is appropriate for analyzing the front interior.  We
will see in Section \ref{S24}, that a stability analysis performed in
this representation is not able to capture the convergence of a steep
initial condition towards a pulled front.  Rather the substitution
\begin{equation}
 ~\psi=\phi\; e^{\lambda^*\xi}~,~~ \xi=x-v^*t~,
\end{equation}
which we shall term  the {\em leading edge representation},
transforms  (\ref{101}) into
\begin{eqnarray}
\label{tip}
\partial_t\psi&=&\partial_\xi^2\psi+\bar{f}(\psi,\xi)~,\\
\bar{f}&\equiv &e^{\lambda^*\xi} \left[f\left(\psi
e^{-\lambda^*\xi}\right)-f'(0)\;\psi e^{-\lambda^*\xi}\right]\nonumber
\\ & = & O\left(\psi^2\;e^{-\lambda^*\xi}\right)~. \nonumber
\end{eqnarray}
This equation will turn out to be appropriate for analyzing a leading
edge dominated dynamics.  Note that $\bar{f}$ is at least of order
$\psi^2$ with an exponentially small coefficient as
$\xi\to\infty$. For large $\xi$, the dynamics is purely diffusive. If
the nonlinearity obeys $f(\phi)-f'(0)\phi<0$ for all $\phi>0$ ---
which is known as a sufficient criterion for pulling --- the
nonlinearity $\bar{f}$ is always negative. Then $\bar{f}$ purely damps
the dynamics in the region of smaller $\xi$. The dynamics evolving
under (\ref{tip}) is equivalent to simply linearizing (\ref{101})
about the unstable state in the large $\xi$ region --- there is only
one subtle but important ingredient from the requirement that the
dynamics in the linear region crosses over smoothly to the nonlinear
front behavior at smaller $\xi$, that actually enters our leading edge
analysis in the form of a boundary condition. In the leading edge
representation (\ref{tip}), this is brought out by the presence of the
sink-type term $\bar{f}$ which is nonzero  in a localized region
behind the leading edge. With this small caveat\footnote{Note though,
  that this subtle point is quite important --- as we shall see, the
  saddle point or pinch point analysis gives precisely the wrong
  prefactor for the leading $1/t$ convergence term because this
  boundary condition is not satisfied.}, we can conclude that the
leading edge of the front ``pulls the rest of the front along'', 
which is precisely the mechanism that gives rise to the universal
algebraic convergence behavior. In a pushed front, in contrast, the
nonlinearity ``pushes the leading edge forward'' and convergence is
exponential.

To illustrate this discussion by a concrete example, we note that when
the function $f(\phi)$ in the nonlinear diffusion equation is of the
form
\begin{equation}
\label{103}
f=f_\epsilon(\phi)=\epsilon\phi+\phi^{n+1}-\phi^{2n+1}~~, ~~n>0~.
\end{equation}
we  can rely on known analytic solutions for $\Phi_v$. In this case, 
the state $\phi=0$ is (meta)stable for $\epsilon<0$.  For
$0<\epsilon<(n+1)/n^2$, the selected front is pushed, and for 
$\epsilon>(n+1)/n^2$, it is pulled (see Section \ref{S2} and Appendix \ref{A1}).

At this point, a brief explanation of our use of the word ``metastable'' 
may be appropriate. For systems with a Lyapunov function, 
the word metastable is often used in physics to denote a linearly 
stable state, which does not correspond to the absolute minimum
of the Lyapunov function or ``free energy''.
A domain wall or front between the absolutely stable and a metastable 
state then moves into the metastable domain; one may therefore loosely
call a linearly stable state ``metastable'', if it is invaded
by another ``more stable'' state through the motion of a domain wall
or front.

The understanding of the two different dynamical mechanisms of pushing
and pulling in the nonlinear diffusion equation (\ref{101}) lays the
basis for the analysis of equations like those listed in Table I.
The essential step towards a generalization of the leading edge
representation (\ref{tip}) is done by a saddle point analysis, that
identifies which Fourier modes of linear perturbations of the unstable
state will dominate the long time dynamics. This analysis yields the
parameters $v^*$, $\lambda^*$, the diffusion constant $D$ and possible
higher order terms required for the leading edge representation.

\subsection{Sketch of method and results on front relaxation in the pulled regime} 
\label{S13}

Bramson's method \cite{bramson} to calculate algebraic
convergence is specifically adapted to equations of type (\ref{101}).
It is based on a representation of the diffusion equation by Brownian
processes, which are evaluated probabilistically.  Instead, we
construct the asymptotic convergence trajectory towards a known
asymptotic state by solving the differential equations in a systematic
asymptotic expansion which, though nonrigorous, extends immediately to
higher order equations. Our approach leads to {\em exact} results
since the expansion parameter are inverse powers of the time $t$, so
these terms become arbitrarily small in the asymptotic regime.

The idea of the method is that in a pulled front, the speed is
essentially set in the leading edge, where linearization of the
equation of motion about the unstable state is justified. This leading
edge has to be connected to what we will refer to as the interior part
of the front, defined to be the region where we have to work with the
full nonlinear equation. For the interior, we use the fact that for
large times the shape of the converging front will resemble the
asymptotic front, and thus can be expanded about it. We also
explicitly make use of the fact that the initial state $\phi(x,0)$ for
large $x$ is steeper than the asymptotic front profile
$\Phi^*=\Phi_{v^*}$ in the leading edge,  
i.e., $\phi(x,0)\;e^{\lambda^*x}\to0$ as $x\to\infty$. The structure
of the problem then dictates 
the expansion in $1/\sqrt{t}$. 

The structure of the expansion in $1/\sqrt{t}$ is the only real input
of the analysis; its selfconsistency becomes clear a posteriori and it
can be motivated from the earlier work on the long time expansion of the
Green's function of the linearized equations. Equivalently, the
self-consistency emerges from
the observation that the equation governing the convergence towards
the asymptotic front profile (\ref{tip}) reduces essentially to a
diffusion equation in the leading edge of the front.  The derivation
of the  exact results summarized in Table II is essentially based on
this ansatz.

\end{multicols}

\begin{figure} \label{fig1}
\vspace{0.5cm}
\epsfig{figure=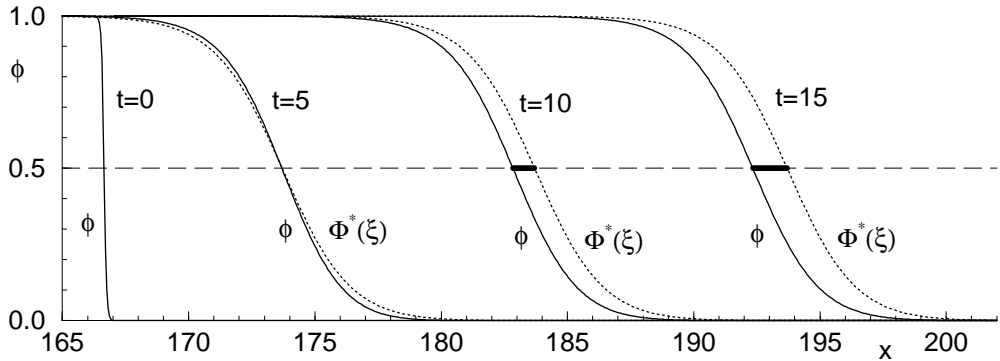,width=0.8\linewidth}
\vspace{0.5cm}
\caption{Illustration of the fact, that even though the shape of
a front profile is quite close to $\Phi^*$, the position of a front
is shifted logarithmically in time relative to the uniformly
translating profile $\Phi^*(\xi)$.\\
Solid lines: evolution of some initial condition $\phi(x,0)$
of the form (\ref{405}) 
under $\partial_t\phi=\partial_x^2\phi+\phi-\phi^3$ at times 
$t=5,10,15$. Dotted lines: evolution of $\phi(x,t)=\Phi^*(\xi)$,
$\xi=x-2t$, at times $t=5,10,15$. $\Phi^*$ is placed such, that the
amplitude $\phi=1/2$ coincides with that of $\phi(x,t)$ at time
$t=5$. The logarithmic temporal shift is indicated by the fat line.}
\end{figure}

\begin{multicols}{2}
 
The shape convergence is also obtained explicitly from our analysis.
The crucial input for the analysis is the right frame and structure to
linearize about. At first sight, a natural guess would be that for
large times, the actual shape of the front $\phi(x,t)$ should be
linearizable about the shape of the asymptotic front $\Phi^*(x-v^*t)$.
However, the algebraic velocity convergence (\ref{107}) implies, that
if a converging front profile $\phi$ is close to the asymptotic
uniformly translating front profile $\Phi^*(x-v^*t)$ at some time
$t_0$, the distance between the actual profile and $\Phi^*$ will
diverge at large times $t$ as $X(t)=-(3/2\lambda^*)\ln (t/t_0)+
\ldots$. This result which is illustrated in Fig.\ 1 implies that 
if we want to linearize $\phi$ about $\Phi^*$ at all times, 
{\em we have to move $\Phi^*$ along with the non-asymptotic 
velocity $v(t)$ (\ref{107}) of the converging front}.  A
crucial step for the analysis is thus to linearize about $\Phi^*(\xi_X)$
in a coordinate system
\begin{equation}
\label{108}
(\xi_X,t)~~,~~ \xi_X\equiv \xi-X(t)= x-v^*t -X(t)~,
\end{equation}
moving with the converging front. If we expand $\phi$ about
$\Phi^*(\xi_X)$ with $\xi_X$ from (\ref{108}) and then resum, we find that
the interior shape of the front is given by
\begin{equation}
\label{109}
\phi(x ,t)=\Phi_{v(t)}(\xi_X +x_0)+O\left(\frac{1}{t^2}\right)
\end{equation}
for $\xi_X\ll\sqrt{4Dt}$. $x_0$ expresses the translational degree of
freedom of the front. The uniformly translating front $\Phi_v(\xi)$ is
a solution of the ordinary differential equation for the uniformly
translating profile $\phi(x,t)=\Phi_v(x-vt)$ but with $v$ replaced by
the {\em instantaneous} value $v(t)$ of the velocity. E.g., for the
nonlinear diffusion equation (\ref{101}), $\Phi_v(\xi)$ is the
solution of
\begin{equation}
\label{1010}
-v \partial_\xi \Phi_v(\xi)  = \partial^2_{\xi} \Phi_v(\xi) + f(\Phi_v(\xi))~.
\end{equation}
Eq.\ (\ref{109}) also confirms that to leading order the interior is
slaved to the slow dynamics of the leading edge.  The transient
profiles $\Phi_{v(t)}$ in (\ref{109}) propagate with velocity $v(t)$
smaller than $v^*$ according to (\ref{107}).

For the special case of Eq.\ (\ref{101}) it is well known (see also
Section \ref{S2}), that when constructing a front $\Phi_v$ starting
from $\Phi_v=1$ at $\xi\to-\infty$, it eventually will become negative
for finite $\xi$ whenever $v<v^*$, and that globally such fronts
either do not exist or are dynamically unstable, depending on the
properties of $f$ for negative $\phi$.  However, only the positive
part of $\Phi_{v(t)}$ from $\xi\to-\infty$ up to $\xi\ll\sqrt{t}$
plays a role as a transient. That the convergence
trajectory is approximately given by $\Phi_{v(t)}$, was already
observed numerically in equations of type (\ref{101}) by Powell {\em
  et al.} \cite{powell}. Our analytical derivation of this result
actually holds for a larger class of equations, but at the same time
we find that it only holds up to a correction term of order $1/t^2$.
This non-universal correction is always non-vanishing.

For $\xi_X\gg\sqrt{4Dt}$, the transient crosses over to
\begin{eqnarray}
\label{1011}
\phi(x,t) = \alpha & \xi_X &\; e^{-\lambda^*\xi_X\;-\xi_X^2/(4Dt)} \;
\\
&\times  & \left( 1 + 
 O \left(\frac{1}{\sqrt{t}}\right)+O\left(\frac{1}{\xi_X}\right) \right).
\nonumber
\end{eqnarray}
The analytical expression for the universal correction of
order $1/\sqrt{t}$ in (\ref{1011}) is given by Eqs. (\ref{3066}) and 
(\ref{3068}) for the nonlinear diffusion equation and is generalized 
by  Eqs.\ (\ref{5047}) or (\ref{M14}), 
while the correction of order $1/t$ will depend on initial conditions, 
and is thus non-universal.

A crucial insight implemented above is that the front consists of 
different dynamical regions which have to be matched to each other.
The situation is sketched in Fig.\ 2. For a pulled front,
the Gaussian region (\ref{1011}) of the leading edge essentially
determines the velocity while the front interior (\ref{109})
is slaved to leading order. The Gaussian region might be preceeded
by a region of ``steepness'' $\lambda$ being conserved in time which for
sufficiently steep initial conditions $\lambda>\lambda^*$ has no
dynamical importance (where the steepness $\lambda$ is defined in Eq.\
(\ref{lambdadef}) below). Likewise, for flat initial conditions the dynamics
is dominated by the conserved $\lambda$ region, while pushed dynamics
is dominated by the front interior. In both of these cases, the intermediate
Gaussian region is absent. For the nonlinear diffusion equation (\ref{101}),
the different cases are discussed in Section \ref{S2} and summarized 
in Table IV.

Our results (\ref{107}) -- (\ref{1011}) are universal in four ways:\\ 
$\bullet$ They are independent of which ``height'' or level curve is
being tracked 
to define the front velocity.\\
$\bullet$ The predicted convergence behavior is independent of the
precise initial conditions, provided they decay quicker than
$e^{-\lambda^* |x|}$ far in the unstable regime.\\ $\bullet$ The
leading edge behavior (\ref{107}) and (\ref{1011}) is independent of
the precise nonlinearities. For Eq.\ (\ref{101}), the constants $v^*$,
$\lambda^*$ and $D$ depend on $f'(0)$ only. For the more general
equations, these constants are completely determined by the saddle
point expansion in the equation linearized about the unstable state.
\\ $\bullet$ If we analyze general equations like those listed in
Table I, our prediction for the interior part of
the front (\ref{109}) stays unchanged, as long as the front speed
stays determined by the linearization about the unstable state, i.e.,
the front stays pulled, and as long as the state behind the front
stays homogeneous.  The effect of the nonlinearities just gets
absorbed in appropriate functions $\Phi_{v}$.

The results summarized in this Subsection are the most central new results
of this paper. They are summarized, for easy reference in Table II.

\end{multicols}

\begin{figure} \label{fig2}
\vspace{0.5cm}
\epsfig{figure=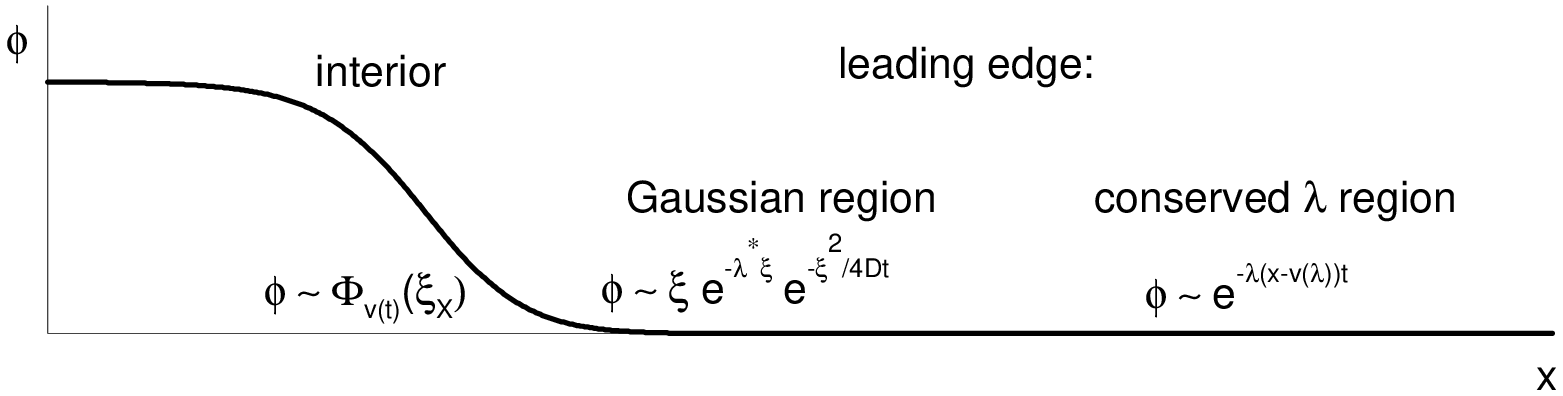,width=0.8\linewidth}
\vspace{0.5cm}
\caption{Sketch of a pulled front with the different dynamical regions:
interior = nonlinear region, leading edge = region linearized
about the unstable state. Depending on the initial conditions, 
the leading edge might still consist
of two different regions: a Gaussian region and a region of conserved
steepness $\lambda$. For $\lambda>\lambda^*$ (defining ``sufficiently steep''
initial conditions), the (intermediate) asymptotic Gaussian
region determines the velocity relaxation.
}
\end{figure}

\begin{center}
\begin{tabular}{||p{9cm}||}
\hline \hline
\begin{eqnarray} 
&&\mbox{Height independent velocity:}\nonumber\\
v(t) &=& v^* +\dot{X} \nonumber\\
      &=& v^* -\frac{3}{2\lambda^*
  t}\left(1-\sqrt{\frac{\pi}{(\lambda^*)^2Dt}}\;\right)
+O\left(\frac{1}{t^2}\right)~,~~ \nonumber\\
&&\mbox{where the saddle point analysis of the }\nonumber\\
&&\mbox{linearized equation yields }v^*, \lambda^*, D, 
\nonumber\\&&\mbox{cf.\ Table V  or Eqs. (\ref{sadpoint})-(\ref{diffusioncoef})}
\nonumber\\
&&\nonumber\\
&\Rightarrow&\mbox{Use the coordinate: } \xi_X= x-v^*t -X(t)~. \nonumber\\
&&\nonumber\\
&&\mbox{Front for }\xi_X\ll\sqrt{4Dt} \mbox{ (front interior)}:\nonumber\\
~~~~\phi(x,t)&=&\Phi_{v(t)}(\xi_X)+O\left(\frac{1}{t^2}\right)~,
\nonumber\\
&=&\Phi^*(\xi_X)+\dot{X}\;\eta_{\rm sh}(\xi_X)+O\left(\frac{1}{t^2}\right)~,~~
\nonumber\\
&&\mbox{where }\Phi_v(\xi) \mbox{ solves } \phi(x,t)=\Phi_v(x-vt)~,
\nonumber\\
&&\mbox{and }\eta_{\rm sh}(\xi)=\left.\delta\Phi_v(\xi)/\delta v\right|_{v^*} ~.~~
\nonumber\\
&&\nonumber\\
&&\mbox{Front for }\xi_X\gg\sqrt{4Dt} \mbox{ (leading edge)}:\nonumber\\
~~~~\phi(x,t)&=&\alpha\xi_X\;e^{-\lambda^*\xi_X}\;e^{-\xi_X^2/(4Dt)}
+\ldots~.
\nonumber
\end{eqnarray} \\
\hline \hline
\end{tabular}

~\\
~\\
~\\

\begin{minipage}{16.5cm}
{\bf Table II:} The central results on the universal algebraic
relaxation towards uniformly translating pulled fronts, see also Fig.\ 2.
These results apply to steep initial conditions in the nonlinear 
diffusion equation in the pulled regime (Case IV of Table IV, 
see Sections \ref{S3} and \ref{S4}) and to more general equations 
(see Section \ref{S5}).
\end{minipage}
\end{center}

\begin{multicols}{2}

\subsection{Organization of the paper} \label{S14}

Before embarking on our explicit calculation of the velocity and shape
convergence in the pulled regime, we review in Section \ref{S2} rather 
well-known results on the multiplicity, stability and convergence of
pushed fronts in the nonlinear diffusion equation, and discuss how far 
these results can be extended to pulled fronts or fronts emerging from 
``flat'' initial conditions. Since the convergence towards pulled
fronts cannot be derived by linear stability analysis, we set the
stage for  Section \ref{S3} by introducing the leading edge
transformation.
%
In the central Section \ref{S3} the detailed analysis of pulled
front relaxation in the the nonlinear diffusion equation (\ref{101})
is given. The detailed
numerical simulations that fully confirm our analytical predictions are
presented in Section \ref{S4}. In this Section we also pay attention
to the specific problems of spatial discretization and system size
arising in the numerical solution of pulled front propagation. In
Section \ref{S5} we  extend 
our analysis to more general equations, discuss the example
equations listed in Table I, and present numerical
results, again in excellent agreement with our analytical
predictions. Here the picture of a new center manifold theorem for
pulled front propagation emerges.
We then close the main body of the paper with a summary and outlook 
in Section \ref{S6}.

Since this is a long paper with  a large number of detailed
results of various types, and since we have made an attempt to make
our results accessible for readers from different fields, we introduce 
Table III as a ``helpdesk'' for the reader who wants to focus on  a
particular aspect of the front propagation problem only, or who wants
to get only an idea of the essential ingredients of our approach and
the main results.

We finally note that a brief sketch of our results can be found in
\cite{prl} and the lecture notes \cite{vs4}. Later extensions of the 
present work can be found in 
\cite{ebertmba,noise7,rocco,willem1,spruijt,kees}.

\end{multicols}

\begin{center}
\begin{tabular}{||p{0.5cm}p{6.75cm}p{0.5cm}|p{0.5cm}p{6.75cm}p{0.5cm}||}
 \hline \hline    \multicolumn{6}{||c||}{  } \\
\multicolumn{6}{||c||}{ THE READER'S HELPDESK} 
\\  \multicolumn{6}{||c||}{  } \\
  \hline   & & & & & \\
 \multicolumn{3}{||c}{{\em If......}}  &  
 \multicolumn{3}{|c||}{{\em Then our advice is......}} \\ 
&&&&& \\ \hline  &&&&&  \\ &
you want to focus right away on the relaxation calculation without
analyzing what a stability calculation tells and does not tell about
the relaxation of pulled fronts
 && & if you know what is meant with the ``pulled'' velocity $v^*$
 you can start with Section \ref{S3} immediately; if not, read Section
 \ref{S25}, \ref{S26} and possibly Section \ref{S531} first & \\ 
&&&&& \\ \hline & &&&& \\ &
you are already familiar with previous ideas concerning 
front selection in the physics literature, but want to get an idea of
our change of emphasis and of the new detailed results of this paper & &
&  to skim Section \ref{S21} for notation and a summary of the most 
important background material, to then check the appropriate Sections 
of \ref{S2} on points which are not clear from Section \ref{S21}, and then
to proceed to Section \ref{S3} & \\ 
& &&&& \\ \hline &&&&& \\ &
you (mainly) want to read about the connection between stability,
selection and relaxation & & &  
to read Section \ref{S2} with Table IV and for the generalization 
Sections \ref{S52} and \ref{S53} with their appendices & \\
&&&&& \\ \hline &&&&&\\ &
you only want to get an idea of the conceptual basis of the algebraic
convergence & & & to read Section \ref{S31} and possibly Sections 
\ref{S53}--\ref{S55} 
for the arguments concerning higher order partial differential
equations or other types of equations & \\
&&&&& \\ \hline &&&&& \\ &
you are unfamiliar with the concept of pulled velocity $v^*$ for
higher order equations and want to know how it is determined 
&&& to read Sections \ref{S31} and \ref{S531} (and possibly 
parts of Sections \ref{S54} and \ref{S55}) & \\
&&&&& \\ \hline &&&&& \\ &
you just want to see the numerical support for the algebraic
relaxation prediction from Tables II and V, or want to read about the
numerical intricacies of studying the pulled front convergence 
&&& read Section \ref{S4} on the nonlinear diffusion equation and 
Section \ref{S56} for higher order and coupled equations & \\ 
&&&&& \\ \hline &&&&& \\ &
you just want a toolkit for when to apply the predictions from Tables
II and V &&& to read Section \ref{S63} & \\ 
&&&&& \\ \hline \hline
\end{tabular}

~\\
~\\
~\\

\begin{minipage}{16.5cm}
{\bf Table III:} A guide through the paper for the efficient reader 
who wants to read about specific results only, or who already has 
some background knowledge on the problem of front propagation into 
unstable states.
\end{minipage}
\end{center}

\newpage

\begin{multicols}{2}

\section{Stability, selection and convergence in the nonlinear 
diffusion equation}  

\label{S2} 

In this Section, we provide the necessary background information on
fronts propagating into unstable states by reviewing a number of
results on the multiplicity and stability of uniformly translating
front solutions of the nonlinear diffusion equation
\cite{britton,murray,fife,depassier,KPP,Fisher,aw,bj,vs1,vs2,vs3,powell,oono,paquette,derrida,barenblatt,stability1,stability2,stability3,stability4,stability5,stability6,stability7}.
We also summarize to what extent the linear stability analysis of
these uniformly translating fronts allows us to solve the selection
problem, i.e., to determine the basins of attraction of these
solutions in the space of initial conditions and for different
nonlinearities $f$, and to what extent it allows us to
answer the related question of the
convergence rate and mechanism. It will turn out that the linear
stability analysis fails to explain how
pulled fronts emerging from sufficiently steep initial conditions
relax to their asymptotic speed and profile.  This sets 
the stage for a different approach to pulled fronts by introducing the
leading edge representation.

\subsection{Statement of problem and essential concepts} \label{S21}

In Sections \ref{S2} -- \ref{S4}, we analyze the nonlinear diffusion
equation
\begin{equation} \label{205} 
\partial_t \phi(x,t) = \partial_x^2 \phi + f(\phi)~,
\end{equation} 
where $f(\phi )$ is assumed to be continuous and differentiable. For
studying front propagation into unstable states, it is convenient to
take
\begin{eqnarray} \nonumber
  f(0)& = & 0=f(1)~,~f'(0)=1~,~\\ f(\phi) & > & 0 ~\mbox{ for all
    }~0<\phi<1~. \label{206}
\end{eqnarray} 
so that in the interval $[0,1]$ $f(\phi)$ has one unstable state at
$\phi=0$ and only one stable state at $\phi =1$.  Eq.\ (\ref{206})
implies that $f'(1)<0$.  Note, that we have specified the behavior of
$f(\phi)$ only on the interval $0\le\phi\le1$. This is all we need
since it can be shown by comparison arguments\cite{aw,note1} that an
initial state with $0\le\phi(x,0)\le1$ for all $x$ conserves this
property in time under the dynamics of (\ref{205}), (\ref{206}).

In passing we note that for a nonlinearity like (\ref{103}), a general
equation of the form
\begin{eqnarray} \label{201}
\partial_\tau\varphi&=& D \partial_y^2\varphi+F_\epsilon(\varphi)~~,
~~F_\epsilon(0)=0=F_\epsilon(\varphi_s)~,~ \nonumber\\ 
&&F_\epsilon'(0)=\epsilon~~,~~\varphi_s>0~,
\end{eqnarray}
results. It allows $\epsilon$ to take either sign. For $\epsilon<0$,
the state $\phi=0$ is linearly stable, for $\epsilon>0$, it is
unstable. Fronts propagating into metastable states ($\epsilon<0$)
will sometimes also be discussed briefly for comparison.  If
$\epsilon>0$, (\ref{201}) transforms to the normal form (\ref{205}) as
\begin{equation}
\label{202}
t=\epsilon\tau~,~x=\sqrt{\epsilon/D}\;y~,~\phi=\varphi/\varphi_s
~,~f(\phi)=\frac{F_\epsilon(\varphi)}{\epsilon\;\varphi_s}~.
\end{equation}
Hence velocities transform as
$dx/dt=\left[dy/d\tau\right]/\sqrt{D\epsilon}$.

{\em The front propagation problem} can now be stated as follows.
Consider some initial condition $0\le\phi(x,0)\le1$ with
\begin{equation} \label{207}
  \lim_{x\to\infty}\phi(x,0) = 0~~,~~\phi(x,0) > 0 \mbox{ for some
    }x~,
\end{equation}
that evolves under the equation of motion (\ref{205}) with (\ref{206})
into a front propagating to the right.
Which time-independent profile and which velocity will this front approach
asymptotically as time $t\to\infty$, if any? How quick will the
convergence to this asymptotic front be? Can we identify the
mechanisms that generate such dynamical behavior? Can we rephrase it
in such terms that we can generalize results to equations other than
(\ref{205})? These questions essentially concern the nature of the
front selection mechanism.

As is well known, the answers to these questions depend on more
specific properties of the initial condition as well as of the
nonlinearity $f(\phi)$. For the nonlinear diffusion equation, the
answer to the selection problem is known in full rigor, but we will
only review here those concepts which are important in a more general
context and which play a role in the subsequent relaxation analysis.
We now briefly outline these main concepts and results and
explain them in more detail in the rest of Section II.

{\em Existence of a family of front solutions.} For front propagation
into unstable states the selection problem is different and more
intricate than for bistable fronts (fronts between two linearly stable
states), since when one solves the {\em o.d.e.}\ for the uniformly
translating profile $\phi(x-vt)=\Phi_v(\xi)$ one finds that there is a
family of fronts solutions parametrized by the continuous variable $v$
that are possible attractors of the dynamics. 
This is in contrast to the situation for bistable fronts where the selected 
velocity $v$ is obtained simply as a nonlinear eigenvalue problem. 

{\em Steepness of a front.}
Most of our discussion  focusses more than earlier work on
the central and unifying role of the {\em steepness} $\lambda$ of the
leading edge of a front, defined as the asymptotic exponential decay
rate:
\begin{equation} \label{lambdadef}
  \phi(x,t) \stackrel{x\to\infty}{\sim} e^{-\lambda x}~~~
  \Leftrightarrow ~~~\lambda = -\lim_{x \to \infty} \left({{\partial
      \ln \phi}\over{\partial x}}\right) ~.
\end{equation}
When $\phi(x,t)$ decays faster than exponentially as $x\to \infty$, this
implies $\lambda= \infty$.

{\em Pulled and pushed fronts.} 
The family of uniformly translating and dynamically stable fronts 
$\Phi_v$ can be uniquely parametrized either by the velocity $v$
or by the spatial decay rate or steepness $\lambda$. The difference 
between pushed and pulled solutions is especially clear if we 
characterize them by $\lambda$. A given nonlinearity $f$ defines
two particular steepnesses: $\lambda_{sel}$ which characterizes
the pushed and pulled front solutions and $\lambda_{steep}$ which
characterizes the basin of attraction of these so-called selected fronts.
The front solution with $\lambda=\lambda_{sel}>1$ defines the pushed front,
while the pulled one has $\lambda=\lambda_{sel}=\lambda^*=1$. The continuous
family of dynamically stable front solutions that exists in addition
to these selected fronts, is parametrized by $\lambda<\lambda_{steep}\le1$.
The nature and construction of the fronts is discussed in more detail 
in Section \ref{S22}, together with a simple property of pulled fronts
which will play an important role in our later relaxation analysis,
namely the fact that the asymptotic large time profile of a pulled front 
is as $\Phi_{v^*}(\xi) \sim\xi \:e^{-\lambda^*\xi}$ for $\xi\gg1$.

We will characterize also an initial condition by its steepness 
$\lambda$ and call it {\em a sufficiently steep initial condition}, if
$\phi(x,t=0)$ decays to zero exponentially faster than
$e^{-\lambda_{steep}x}$ for some $\lambda_{steep}\le 1$, i.e.,
\begin{eqnarray} \label{208}
\mbox{sufficiently steep:}~~~ \phi(x,0) &  \stackrel{x\to\infty}{<} &
e^{-\lambda x}~,~~\\ \mbox{for some } \lambda & > & \lambda_{steep} ~,\nonumber
\end{eqnarray}
otherwise we call it {\em flat}:
\begin{equation}
\label{2008}
\mbox{flat:}~~~\phi(x,0) \stackrel{x\to\infty}{\sim} e^{-\lambda x}~~,~~
\lambda<\lambda_{steep}~,
\end{equation}
How $\lambda_{steep}$ is determined
by $f(\phi)$, will be discussed in Section \ref{S24}. We will see
that always $0<\lambda_{steep}\le 1$ for Eq.\ (\ref{205}), and in
particular that for pulled fronts 
\begin{equation}
\mbox{pulled fronts:} ~~~\lambda_{steep}=\lambda^*=1~,
\label{suffsteep}
\end{equation} 
while for pushed 
fronts $\lambda_{steep}<1$. The criterion (\ref{208}) for steepness
includes all initial conditions with bounded support or, e.g., the
initial condition $\phi(x,0)=\theta(-x)$ with $\theta$ the step
function.

Note that the intermediate case $\phi(x,0)\sim x^{-\nu}e^{-\lambda^* x}$
is neither sufficiently steep nor flat, according to our
definitions. In Section \ref{S3} 
we shall recover Bramson's \cite{bramson} observation that such
special initial conditions also lead to a $1/t$ relaxation of the
velocity profile, but with a $\nu$-dependent prefactor for $\nu <2$.

{\em Conservation of steepness.} In Section \ref{S25} we discuss what
we term conservation of steepness: if an initial condition is
characterized by a steepness $\lambda$, then at any finite time 
the steepness of $\phi(x,t)$ is the same as that of the initial condition
$\phi(x,t=0)$. (Note that the limits $t\to\infty$ and $x\to\infty$ do not 
commute.)

{\em The linear stability analysis} of
front solutions can be performed in detail for  the nonlinear
diffusion equation. As summarized in Section \ref{S23}, pushed fronts
have a gapped spectrum, while pulled fronts have a gapless spectrum
within their natural Hilbert space. 
In the selection analysis, we in general also need perturbations from 
outside this Hilbert space.

{\em Stability and selection.} In Section \ref{S24}, we discuss the
connection between the stability of front solutions and the selection
mechanism; this connection, which underlies much of the marginal
stability scenario \cite{bj,vs1,vs2}, hinges on the fact that the
conservation of steepness allows one to relate the steepness of the 
initial condition to the steepness of the late stage evolution of
the front that can be decomposed into an asymptotic front profile
plus a linear perturbation. The spectral decomposition of this 
perturbation is largely determined by the steepness of the initial
and the asymptotic state.

{\em Basins of attraction and rate of convergence} are also discussed
in Section \ref{S24}.
Flat initial conditions (\ref{2008}) approach a front characterized by
their initial $\lambda$. Sufficiently steep initial conditions (\ref{208})
in the pushed regime ($\lambda_{sel}>1$) evolve at late times into 
a pushed front corrected by linear perturbations that can be represented 
by eigenfunctions of the stability operator, whose spectrum has a gap.
Hence the convergence of a pushed front is exponential in time.
In contrast, the rate of convergence of pulled fronts ($\lambda_{sel}=1$) 
can not simply be obtained from the spectrum, as it is gapless, 
and generic perturbations are not spanned by the natural eigenfunctions.

{\em Leading edge and interior dominated dynamics.} Both the stability
analysis and our relaxation analysis bring out the importance of
distinguishing {\em
  leading edge dominated} from  {\em interior dominated} dynamics.
The most obvious form of leading edge dominated dynamics results from
flat exponential initial conditions (\ref{2008}) with finite steepness
$\lambda$. In this case, the asymptotic front speed is just the speed
\begin{equation}
\label{2044}
v(\lambda)=\lambda+\frac{1}{\lambda}
\end{equation}
with which the exponential tail
$e^{-\lambda x}$ propagates according to the linear dynamical equation
\begin{equation} \label{209}
\partial_t \phi = \partial_x^2 \phi + \phi + o(\phi^2)~.
\end{equation} 
This equation is obtained by linearizing about the unstable state
$\phi=0$, and is appropriate in the leading edge region. 
The more important leading edge dominated dynamics occurs, however,
for sufficiently steep initial conditions (\ref{208}) converging to a
{\em pulled} front. As already mentioned, for pulled fronts the
asymptotic front speed is just the linear spreading velocity $v^*$
determined in
the leading edge where the dynamics is essentially governed by the
linearized evolution equation. This type of leading edge dominated
pulled dynamics  occurs when the nonlinearities in $f(\phi) $ are 
mostly saturating so that they slow down the growth.  In passing we
note that we rederive  in Appendix \ref{bound} the well-known
sufficient criterion for pulling in the nonlinear diffusion equation, viz.
\begin{equation}
\label{2010}
f'(0)=\max_{0\le \phi\le 1} \frac{f(\phi)}{\phi}~,
\end{equation}
with the help of a transformation that we call the leading edge 
transformation \cite{prl,kees}. This form of a proof is generalizable 
to some other equations \cite{spruijt}.  Pulled fronts are
actually at the margin of leading edge domination: although the
linearized equation (\ref{209}) is sufficient to determine
$v_{sel}=v^*=2$, we will see in Section \ref{S3} that the {\em
  convergence} towards this velocity is governed by a nontrivial
interplay of the dynamics in the leading edge and the ``slaved''
interior.

Leading edge dominated dynamics contrasts with {\em interior
  dominated} dynamics, which occurs when the nonlinear function
$f(\phi)$ is such that steep initial conditions give rise to {\em
  pushed} fronts. For interior dominated or {\em pushed} dynamics,
$v_{sel}$ is associated with the existence of a strongly heteroclinic
orbit in the phase space associated with $\Phi_v(\xi)$ (Section
\ref{S22}). This means that the whole nonlinearity $f(\phi)$ is
needed for constructing $v_{sel}$, not only the linearization $f'(0)$
about the unstable state.  The linear stability analysis
of Section \ref{S23} implies that pushed fronts converge exponentially 
in time to their asymptotic speed (Section \ref{S24}).  
This type of dynamics extends smoothly towards fronts propagating 
into metastable states, i.e., towards $\epsilon<0$ in (\ref{201}).

While in this Section we consider the nonlinear diffusion equation
(\ref{205}), (\ref{206}) only, the straightforward extension to
generalized {\em p.d.e.}'s  of the form $F(\phi, \partial_x \phi,
\partial^2_x \phi, \partial_t \phi)=0$ can be found in Appendix
\ref{A3}. 

In the following subsections the above
assertions are further substantiated. Readers familiar with most of
the concepts and results listed above can proceed to Section \ref{S3}.

\subsection{Uniformly translating fronts: \\
  candidates for attractors and transients} \label{S22}

In this Section, we recall some well known properties
\cite{KPP,Fisher,aw,bj,vs1,vs2,paquette,arnold,guckenheimer,note2} 
of uniformly translating front
solutions of the nonlinear diffusion equation (\ref{205}),
(\ref{206}).  We transform to a coordinate
system moving with uniform velocity $v$: $(x,t) \to (\xi,t)$,
$\xi=x-vt$, so that the temporal derivative  transforms as
$\partial_t\big|_x = \partial_t\big|_\xi - v \partial_\xi\big|_t$.
For a front $\phi(x-vt)=\Phi_v(\xi)$ translating uniformly with
velocity $v$, the time derivative vanishes in the comoving frame
$\partial_t\big|_\xi\Phi_v=0$, and so $\Phi_v(\xi)$ obeys the ordinary
differential equation
\begin{equation} \label{2012} 
\partial_\xi^2 \Phi_v + v \partial_\xi \Phi_v + f(\Phi_v)=0~.
\end{equation}
In view of the initial condition (\ref{207}), we will throughout this
paper focus on the  
right-moving front and hence we impose the boundary conditions
\begin{eqnarray}  
  \Phi_v(\xi)  &\to&  1 ~~\mbox{for }\xi\to-\infty~,
\nonumber\\  
\Phi_v(\xi)  &\to&  0 ~~\mbox{for }\xi \to \infty~.
\label{2013}
\end{eqnarray}

Close to the stable state $\phi=1$, the differential equation can be
linearized about $\phi=1$ and solved explicitely. The general local
solution is a linear combination of $e^{-\tilde{\lambda}_\pm\xi}$ with
\begin{equation} \label{2014} 
  \tilde{\lambda}_\pm=\frac{v\pm\sqrt{v^2-4 f'(1)}}{2}~.
\end{equation} 
According to (\ref{206}), $f'(1)$ is negative. Thus for any real $v$,
$\tilde{\lambda}_+$ is positive and $\tilde{\lambda}_-$ is negative.
With the convention (\ref{2013}), only the negative root is
acceptable. So
\begin{equation} \label{2015} 
  \Phi_v(\xi)=1 \pm e^{-\tilde{\lambda}_-(\xi-\xi_0)} 
  + o(e^{-2\tilde{\lambda}_-\xi})
  \mbox{ for }\xi\to-\infty~.
\end{equation} 
The free integration constant multiplying $e^{-\tilde{\lambda}_-\xi}$
here has been decomposed into a sign
$\pm$ and a free parameter $\xi_0$ accounting for translation
invariance. Apart from translation invariance, there are two solution
for $\Phi_v$ close to $\phi=1$ distinguished by $\pm$.

A global view of the nature and multiplicity of solutions can be
obtained with a well known simple particle-in-a-potential analogy. This
analogy has of course been exploited quite often in various types of 
approaches\cite{bj,gunton,langer3,bray}, and only works for the nonlinear
diffusion equation, not for equations with higher spatial derivatives;
for these, we have to rely on a construction of solutions as
trajectories in phase space as sketched around Eq.\ (\ref{2022}).  

The particle-in-a-potential analogy is based on the identification of
equation (\ref{2012}) with the equation of motion of a classical
particle with friction in a potential. One identifies
$\Phi_v$ with a spatial coordinate, $\xi$ with time, $v$ with a
friction coefficient, and $f$ with the negative force,
$f=-\mbox{force}=\partial_{\phi} V(\phi)$ derived from the potential
$V(\phi)=\int^\phi d\phi'\;f(\phi')$. The potential has a maximum at
$\phi=1$ and a minimum at $\phi=0$. The construction of $\Phi_v$ is
equivalent to the motion of a classical particle with ``friction'' $v$
in this potential, where at ``time'' $-\infty$ the particle is at rest
at the maximum of $V$.  Obviously for any positive ``friction'' $v>0$,
the particle will never reach the minimum at $\phi=0$, if it takes off
from the maximum at $\phi=1$ towards $\phi>1$.  It will always reach
$\phi=0$ if it takes off towards $\phi<1$. Thus for every
$v>0$, there is a unique uniformly translating front (unique up to
a translation), that starts as (\ref{2013}) and reaches
$\phi=0$ monotonically. Close to $\phi=1$ it is given by the $-$
branch in Eq.\ (\ref{2015}).

Let us be more specific on how $\phi=0$ is approached.  If the
``friction'' $v$ is sufficiently large, the motion of the particle
will be overdamped when it first approaches $\phi=0$, it will
reach $\phi=0$ only for ``time'' $\xi\to\infty$, and form a monotonic 
front over the whole $\xi$ axis. This behavior continues down to
a critical value of the ``friction'' $v_c$. It defines the critical 
velocity $v_c$ as the smallest velocity at which
$\Phi_v(\xi)$ monotonically reaches $\Phi_v(\xi)\to0$ at
$\xi\to\infty$. (As we will discuss in Sect.\ \ref{S23}, 
a uniformly translating front $\Phi_v$ is dynamically stable 
if and only if $v\ge v_c$.) If $v<v_c$, the particle will reach $\phi=0$ 
at a finite ``time'' $\xi$ and cross it.  What then happens, depends on
$f(\phi)$ for negative arguments.  If $f'(0)=1$ for both positive and
negative arguments $\phi$ as in the case of the nonlinearities
(\ref{102}) or (\ref{103}),
the particle might oscillate a finite or an infinite number of times 
through $\phi=0$ and reach $\phi=0$ asymptotically for $\xi\to\infty$ as
\begin{equation} \label{2018} 
  \Phi_v(\xi)=\left\{\begin{array}{ll}
A_v \;e^{-\lambda_-\xi} + B_v \;e^{-\lambda_+\xi} & \mbox{ for }v>2~,\\
(\alpha\xi+\beta) \;e^{-\lambda^*\xi} & \mbox{ for }v=v^*=2 ~,\\ 
 C_v\;e^{-\lambda_0 \xi} \;\cos k (\xi-\xi_2) & \mbox{ for }|v|<2~, \\ 
\end{array}\right.  \label{phiasymp}
\end{equation} 
where
\begin{eqnarray} \label{lambda1}
  \lambda_\pm(v) & = & \lambda_0(v) \pm \mu(v) ~~(v>2)~,~~ \lambda_0(v) =
  \frac{v}{2} ~~(\mbox{all}~v) ~,\nonumber\\
  && \\ \mu(v) & = & \frac{ \sqrt{v^2-4} }{2}
  ~~(v>2)~,~~ k(v)= \frac{\sqrt{4-v^2}}{2} ~~(v<2)~,
  \nonumber\\ && \\ \lambda^* & = &
  \lambda_0 (v^*) ~=~\lambda_\pm (v^*) ~=~1 ~~(v=v^*=2)~.
   \label{2018a}
\end{eqnarray}
The solution (\ref{2018}) of the equation linearized about $\phi=0$
contains two free parameters for every $v$.  These parameters are 
determined by the unique approach of the front $\Phi_v$ from $\phi=1$ 
and will in general both be non-vanishing.
The special value $v^*=2$ is determined by linearization about
the unstable state. As can be seen from (\ref{2018}), 
it is a lower bound on the critical velocity $v_c$.
At this value of the velocity the two roots
$\lambda_+$ and $\lambda_-$ coincide. As a result, the
asymptotic profile is not the sum of two exponentials, but an
exponential times a first order polynomial in $\xi$. 

Depending on the nonlinearity $f$, the critical $v_c$ can be 
determined by two different mechanisms that turn out to
distinguish pushed ($v_c>v^*$) or pulled ($v_c=v^*$) fronts.
Suppose first that upon lowering $v$ the front solutions $\Phi_v$ 
remain monotonic till $v=v^*$. In this case $v_c=v^*$ is determined by 
the equation linearized about the unstable state, and we will see,
that sufficiently steep initial conditions (\ref{208}) evolve 
into pulled fronts. A second possibility is the
following. At very large $v$, the front solution is certainly
monotonic, since in the particle-on-the-hill analogy the particle slowly
creeps to the minimum of the potential for large ``friction'' $v$.
Hence $A_v$ in (\ref{2018}) is positive for large $v$.  Now,  
depending on the nonlinearities, it may happen upon lowering $v$ 
that at some velocity $v=v^\dagger$, $A_{v^\dagger}=0$. The
front is nonmonotonic for $v<v^\dagger$ as $A_v$ will be negative for
$v<v^\dagger$. Hence in this case $v_c=v^\dagger$ and pushed fronts
result. The pushed velocity $v^\dag$ thus emerges from the global 
analysis of the whole nonlinear front, and not only from linearization 
about the unstable state.

For uniformly translating pulled fronts we will use the short hand
notation $\Phi^* \equiv \Phi_{v^*}$. For large $\xi$ they are
asymptotically
\begin{equation} \label{2019}
  \Phi^*(\xi) \equiv \Phi_{v^*}(\xi)\stackrel{\xi\to\infty}{\sim}
  \xi\;e^{-\xi}~,
\end{equation}
since in general the coefficient $\alpha$ in (\ref{phiasymp}) is
nonzero. This particular form will  in Section \ref{S3} turn out 
to have important consequences for the convergence of pulled fronts: 
it determines the prefactor of the $1/t $ relaxation term. 

For fronts with velocity $v>v^*$, the smaller $\lambda$ will
dominate the large $\xi$ asymptotics, so generically
\begin{equation} \label{2020}
  \Phi_v(\xi) \stackrel{\xi\to\infty}{\sim} e^{-\lambda_- \xi}~.
\end{equation}
However, for a front solution with velocity $v^\dag$, we have
$A_{v^\dag}=0$, and so
\begin{equation} \label{2021}
  \Phi^\dag(\xi) \equiv \Phi_{v^\dag}(\xi)
  \stackrel{\xi\to\infty}{\sim} e^{-\lambda_+ \xi}~.
\end{equation}

An alternative formulation that can be generalized to higher order
equations is the following. A construction of front solutions of Eq.\ 
(\ref{2012}) is equivalent to a construction of trajectories in a
phase space ($\Phi_v , \Psi_v \equiv \partial_\xi \Phi_v ) $ in which
the flow is given by
\begin{equation} \label{2022} 
\partial_\xi {\Phi_v\choose\Psi_v}={\Psi_v\choose -v\Psi_v-f(\Phi_v)}~.
\end{equation} 
Front solutions correspond to trajectories between the fixed points
$(\Phi_v,\Psi_v)=(1,0)$ and $(0,0)$. These are thus heteroclinic
orbits in phase space. Out of the $(1,0)$ fixed point come
two trajectories in opposite directions along one eigenvector
according to (\ref{2015}). When we follow the direction for
which $\Phi_v$ decreases for increasing $\xi$, its behavior near the
$(0,0)$ fixed point is given by (\ref{2018}). Now, since the flow
depends continuously on $v$, so will $A_v$ and $B_v$ in (\ref{2018}).
For large $v$ $A_v$ is positive, and from the construction of the
flow in phase space one sees that $A_v$ may change sign on lowering
$v$. The largest $v$ with $A_v=0$ determines the change from monotonic
to non-monotonic fronts.  At this $v=v^\dag$, the trajectory flows
into the stable $(0,0)$ fixed point along the most strongly
contracting eigendirection --- this is precisely what is expressed in
(\ref{2021}). For this reason, the solution $\Phi^\dag$ is
referred to by Powell {\em et al.} \cite{powell} as a strongly
heteroclinic orbit. 
In \cite{vs3}, this solution was referred to as ``the nonlinear front
solution''.

In summary, the main results of the preceding analysis are:
\\
$\bullet$ For every $v\ge v_c$, there is a uniformly translating front
$\Phi_v$ with velocity $v$, which monotonically connects $\phi=1$ at
$\xi\to-\infty$ to $\phi=0$ at $\xi\to\infty$.  All $\Phi_v$ with
these properties are uniquely determined by $v$ up to translation
invariance.
\\
$\bullet$ For every $0<v<v_c$, there is a unique front solution
$\Phi_v$, that translates uniformly with velocity $v$, and that
monotonically connects $\phi=1$ at $\xi\to-\infty$ to $\phi=0$ at some
finite $\xi=\bar{\xi}$. 
\\
$\bullet$ Depending on the nonlinearities, the change from monotonic
to nonmonotonic behavior can either occur at the velocity $v^*$, with
$v^*=2$ for (\ref{205}) and (\ref{206}), or at a larger velocity
$v^{\dagger}$: $v_c=\mbox{max}[v^*,v^\dagger$]. If $v^\dagger$ exists,
it is the largest velocity at which there is a strongly heteroclinic
orbit. 

The results for invasion into either metastable ($f'(0)<0$) or 
unstable states ($f'(0)>0$) and for $v_c=v^\dag>v^*$
and $v_c=v^*$ are summarized in $v(\lambda)$ plots in Fig.\ 3,
which show the multiplicity of stable uniformly translating fronts
$\Phi_v$ parametrized by either $v$ or $\lambda$.

The results of this subsection play a role in the subsequent analysis:
\\ 
{\em (i)} There are important connections \cite{bj} between the
properties of the uniformly translating front solutions and the
stability of these fronts (see Section \ref{S23}).  In particular
front solutions with velocity $v\geq v_c$ are dynamically stable and
possible attractors of the long time dynamics. Fronts with velocity
$v<v_c$ either do not exist or are unstable.  \\ 
{\em (ii)} The results for front selection can be 
easily formulated in terms of the properties of these 
uniformly translating solutions\cite{bj,vs1,vs2}: for
sufficiently steep initial conditions the dynamically selected
velocity coincides with $v_c$: $v_{sel}=v_c$. If $v_{sel}=v^\dagger$,
we speak of the {\em pushed} regime, while if $v_{sel}=v^*$ we speak of
{\em pulled} fronts.  \\ 
{\em (iii)} We will see in Section \ref{S3}, that the positive 
monotonic part of the front solutions $\Phi_v(\xi)$ with velocity 
$v<v^*$ plays a role in the convergence behavior in the interior region 
of pulled fronts. Note, however, that while a solution $\Phi_v(\xi)$ 
of the {\em o.d.e.}\ has according to (\ref{phiasymp}) an oscillatory 
leading edge for large $\xi$, that causes the dynamic instability 
of these solutions, the relaxing front is approximated by $\Phi_{v(t)}$
only in the interior front region and crosses over to a different
functional form in the leading edge. This behavior is in agreement with 
the conservation of positivity of the solution in a nonlinear diffusion 
equation, if the initial condition was positive.\\
All arguments essentially also apply to higher order equations, 
though then the positivity and monotonicity properties of the 
solutions loose their distinguished role.

\end{multicols}

\begin{figure}\label{fig3}
\epsfig{figure=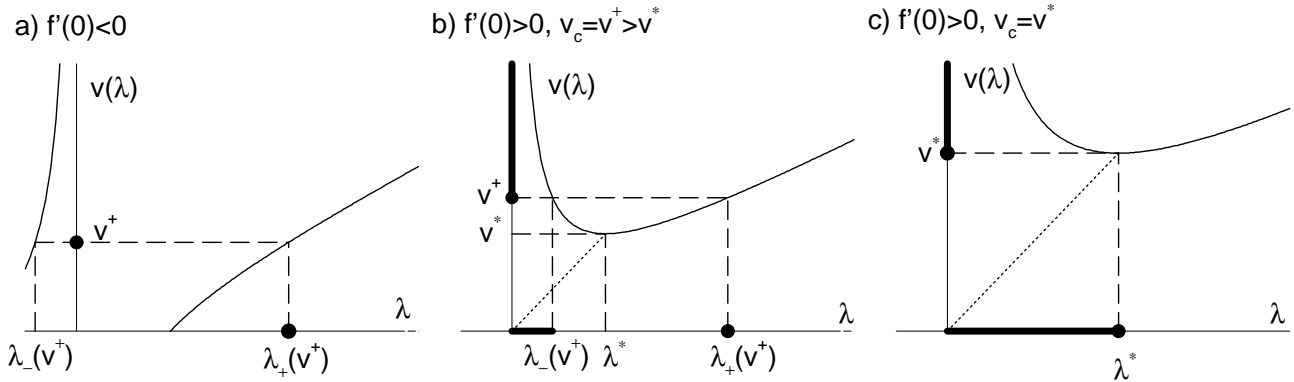,width=0.9\linewidth}
\caption{Steepness $\lambda$ (\ref{lambdadef}) versus velocity 
$v(\lambda)=\lambda+f'(0)/\lambda$ with solid line for real $\lambda$ and
dotted line for real part of complex $\lambda$.
$v^\dag$ is the pushed velocity derived from global analysis, $v^*$
the linear spreading velocity. A fat line or point on the axes denotes 
the possible attractors $\Phi_v(x-vt)$ of the dynamics, parametrized
either by velocity $v$ or by steepness $\lambda$.
$a)$ The case $f'(0)<0$ corresponding to front propagation into 
a (meta)stable state. In this case, there is a unique
attractor with velocity $v_{sel}=v^\dag$ and steepness 
$\lambda_{sel}=\lambda_+(v^\dag)$.
$b)$ and $c)$  The case $f'(0)>0$ corresponding to front propagation into 
an unstable state. In this case there is a continuum of attractors 
parametrized by $v\ge v_c$.
$b)$ The pushed regime: $v_{sel}=v_c=v^\dag>v^*$. The steepness 
$\lambda_{sel}=\lambda_+(v^\dag)$
of the steepest attractor is isolated just as in case $a$.
There is a continuous family of fronts parametrized by 
$0<\lambda<\lambda_{steep}=\lambda_-(v^\dag)$.
$c)$ The pulled regime: $v_{sel}=v_c=v^*$. The steepness $\lambda^*=
\lambda_{steep}=\lambda_{sel}$ of the steepest attractor is at 
the margin of the $\lambda$-continuum of attractors.}
\end{figure} 

\begin{multicols}{2}

\subsection{Linear stability analysis of moving front solutions}

\label{S23}

To study the linear stability of a uniformly translating front
$\Phi_v$, we linearize about it in the frame $\xi=x-vt$ moving with
the constant velocity $v$, by writing
\begin{equation} \label{2023}
\phi(\xi,t)=\Phi_v(\xi)+\eta(\xi,t)~.
\end{equation}
Inserting (\ref{2023}) into (\ref{205}), we find to linear order the
equation of motion for $\eta(\xi,t)$
\begin{equation} \label{2024} 
\partial_t\eta={\cal L}_v \eta+O(\eta^2)
\end{equation} 
with the linear operator
\begin{equation} \label{2025} 
{\cal L}_v = \partial_\xi^2 + v\partial_\xi +f'\Big(\Phi_v(\xi)\Big)~.
\end{equation} 
${\cal L}_v$ is not self-adjoint, so left and right eigenfunctions
will differ.  The trouble is caused by the linear derivative
$v\partial_\xi$.  It can be removed by the following transformation
\cite{stability1,bj}:
\begin{eqnarray} \label{2026} 
\psi&=&e^{v\xi/2}\;\eta~,\\
\label{2027} 
{\cal H}_v&=&-\;e^{v\xi/2}\;{\cal L}_v\;e^{-v\xi/2}~.
\end{eqnarray}
${\cal H}_v$ is the linear Schr\"odinger operator
\begin{equation} \label{2029} 
{\cal H}_v=-\partial_\xi^2+V(\xi)~~,~~
V(\xi)={\textstyle\frac{v^2}{4}}-f'\Big(\Phi_v(\xi)\Big)~,
\end{equation}
and the equation of motion (\ref{2024}) transforms to
\begin{equation} \label{2028} 
-\partial_t\psi={\cal H}_v \psi+O\left(\psi^2\;e^{-v\xi/2}\right)~.
\end{equation} 
This Schr\"odinger problem is, of course, well known to physicists
\cite{messiah} (see also
\cite{stability1,stability2,stability3,stability4,stability5,stability6,textbook}),
as long as $\psi$ lies in the natural Hilbert space of ${\cal H}_v$.

However, the transformation (\ref{2026}), (\ref{2027}) increases the weight
of the leading edge ($\xi\to\infty$) by a factor $e^{v\xi/2}$, while
it enhances convergence at $\xi\to-\infty$. 
Therefore, only perturbations with
\begin{equation}
\label{hilsp}
\lim_{\xi\to\infty} |\eta|\;e^{\lambda_0(v)\;\xi} <\infty
~~~\mbox{with }~\lambda_0(v)=\textstyle\frac{v}{2}
\end{equation}
are spanned by the eigenfunctions within the conventional Hilbert space 
of ${\cal H}_v$. For the selection analysis in Section \ref{S24} below,
this function space in general is not sufficient.
As is discussed in detail in Appendix D, 
one can construct eigenmodes of ${\cal L}_v$ outside the Hilbert
space defined by (\ref{hilsp}). With this extension of the function space,
scalar products of arbitrary eigenfunctions might be divergent,
so one looses the efficient tool of projection onto eigenfunctions
by taking inner products. 
Nevertheless, in most cases generic perturbations still can be 
decomposed into these eigenfunctions, except in the case of pulled 
fronts: the linear perturbation $\eta$ of a 
sufficiently steep front $\phi\sim e^{-\lambda\xi}$ with 
$\lambda>1$ (\ref{208}) about the asymptotic pulled front 
$\Phi^*\sim(\alpha\xi+\beta)\;e^{-\xi}$ (\ref{2018}), (\ref{2019})
will decay asymptotically as 
\begin{equation}
\eta=\phi-\Phi^*\stackrel{\xi\gg1}{\sim}-(\alpha\xi+\beta)\;e^{-\xi}~.
\end{equation}
Since there is only one zero mode of translation
with a slightly different asymptotic behavior
\begin{equation}
\label{zero}
{\cal L}_v\;\eta_0=0~~,~~
\eta_0=\partial_\xi\Phi^*\stackrel{\xi\gg1}{\sim}
-(\alpha\xi+\beta-\alpha)\;e^{-\xi}~, 
\end{equation}
the asymptotics of $\eta$ cannot fully be decomposed into eigenfunctions. 
Here the double root structure of the leading edge with $\alpha\ne0$
plays a crucial role, as it later will do again.

The most important conclusions from the present discussions and
the detailed Appendix \ref{stabilityapp} are:

1) Non-monotonic fronts are intrinsically unstable, and generically
will not be approached by any initial condition.

2) Monotonic fronts propagating with velocity $v$ are stable against
perturbations steeper than $e^{-\lambda_-(v)\xi}$. 

3) Perturbations $\eta$ about pushed fronts $\Phi^\dag$, that decay 
more rapidly than $\lambda_-(v^\dag)$ have a gapped spectrum --- see
Eq.\ (\ref{hoiute}). The same holds for perturbations about fronts 
$\Phi_v$ with a velocity $v>v^\dag$, if their steepness is
larger than $\lambda_-(v)$.

4) The spectrum of pulled fronts is gapless and cannot be decomposed
into eigenfunctions of ${\cal H}_v$ even outside the conventional
Hilbert space.

Before closing this subsection, we  note that although the
particle-on-a-hill analogy for $\Phi_v$ or the 
mapping onto the Schr\"odinger equation for $\eta_\sigma$ are
insightful and very efficient ways to arrive at our results for
existence and stability of uniformly translating front solutions, the
analysis by no means relies on these. In fact, much of the phase space
analysis can easily be generalized to higher order equations as 
those shown in Table I.
E.g., in the stability analysis of non-monotonic fronts, the
discrete set of solutions with $A_v=0$ plays a
particular role.  For equations like the EFK equation from Table I,  
monotonicity ceases to be a criterium, but conditions like $A_v=0$ 
defining so-called strongly heteroclinic solutions continue 
to play a central role in the stability analysis, as is discussed 
in Appendix \ref{A6} to Sect.\ \ref{S5}.

\subsection{Consequences of the stability analysis for selection and 
  rate of convergence; marginal stability} \label{S24}

Suppose now, that we start with an initial condition $\phi(x,0)$ in the
nonlinear diffusion equation (\ref{205}) with a given nonlinearity
$f(\phi)$, and then study the ensuing dynamics. What will the linear
stability analysis tell us about 
the asymptotic ($t\to\infty$) state and the rate of convergence? It
turns out that  the issue of selection is more closely related to that 
of stability than one might expect at first sight. The reason is the
conservation of steepness discussed in more detail in  Section
\ref{S25} below: {\em If
  initially at $t=0$ the steepness $\lambda$ defined in
  (\ref{lambdadef}) is nonzero (finite or infinite), then at any
  finite time $t <\infty$ the steepness is conserved:}
\begin{equation} \label{lambdacons}
  \phi(x,0) \stackrel{x\to\infty}{\sim} e^{-\lambda x}~
  \Longrightarrow
 \phi(x,t) \stackrel{x\to\infty}{\sim} e^{-\lambda x}
 \mbox{ for all }t<\infty~.
\end{equation} 
Note that the limits $x\to\infty$ and $t\to\infty$ do not commute.
We characterize the initial condition by its steepness
$\lambda_{init}$ defined by
\begin{equation}
\label{2042}
\phi(x,t=0) \stackrel{x\to\infty}{\sim} e^{-\lambda_{init} x}~.
\end{equation}
As a consequence of (\ref{lambdacons}), we can use $\lambda_{init}$ to
characterize not only the initial conditions but also the profile at
any later time $0\le t<\infty$, when the front velocity might be
already close to its asymptotic value.

The conservation of steepness (\ref{lambdacons}) entails
that a front characterized by an initial steepness $\lambda_{init}$,
will be characterized by the same steepness after
any finite time, so also at a late stage when the
velocity and shape of a front are close to their asymptotic limits. 
At such a late stage, the front $\phi$ can be decomposed into
a possible attractor $\Phi_v(x-vt)$ of the dynamics
plus a linear perturbation $\eta$ 
as in (\ref{2023}). Characterize the attractor $\Phi_v$ that we investigate 
by its steepness $\lambda_{asympt}$. The resulting 
perturbation $\eta(x,t)=\phi(x,t)-\Phi_v(x-vt)$ then will have steepness
\begin{equation}
\label{lameta} \label{2043}
\lambda_\eta=\min\left[\lambda_{init},\lambda_{asympt}\right]~.
\end{equation}
Whether the perturbation $\eta$ will grow or decay, that means, whether
$\Phi_v$ with a particular velocity $v$ is the attractor of the evolution
of $\phi$ or not, is determined by the decomposition of the perturbation
$\eta$ into eigenmodes of the linear operator. Whether this spectrum
has growing eigenmodes, depends on the operator and the functions space
defined by the steepness $\lambda_\eta$.
With the tools of the stability analysis from Appendix \ref{stabilityapp},
the selection question can therefore be rephrased purely in terms of 
$\lambda_\eta$, $\lambda_{asympt}$ and the two steepnesses 
$\lambda_{steep}$ and $\lambda_{sel}$ characterizing the nonlinearity $f$:
for pushed fronts
\begin{equation}
\label{2045}
\lambda_{steep}=\lambda_-(v^\dag)~~,~~
\lambda_{sel}=\lambda_+(v^\dag)~~,~~
v_{sel}=v^\dag~,
\end{equation}
and for pulled fronts
\begin{equation}
\label{2045a}
\lambda_{steep}=\lambda_{sel}=\lambda^*~~,~~
v_{sel}=v^*~,
\end{equation}
in the notation of (\ref{lambda1}) -- (\ref{2018a}).

A detailed discussion of the question to what
extent one can understand 
the selection and rate of convergence of fronts
following this line of analysis, is given in Appendix
\ref{consequencesapp} and summarized in Table IV below. The two most
important conclusions for our purposes concern fronts evolving from
sufficiently steep initial conditions:

{\em (i)} The gapped spectrum of the conventional Hilbert space
for pushed fronts implies that the relaxation towards pushed front 
solutions is exponential in time. 

{\em (ii)} Even after extending linear stability analysis
  beyond the Hilbert space, {\em it is not possible to derive the rate
  of convergence of pulled fronts from the stability spectrum},
  since it is gapless, and generic perturbations cannot be decomposed
  into eigenmodes of the linear stability operator even in an
  enlarged functions space.

We finally note that the usual marginal stability viewpoint is to 
characterize the family of stable front solutions $\Phi_v$ 
by the velocity $v$; from this perspective, the front velocity
$v_{sel}$ selected by the sufficiently steep initial conditions is at
the edge of a continuous spectrum of {\em stable} solutions with $v\ge
v_{sel}$. In this sense, both the pushed and the pulled attractors are
  marginally stable \cite{bj,vs1,vs2}. The picture changes, however,
  when the attractors are not characterized by the velocity $v$, but
  by their asymptotic steepness $\lambda$, see Fig.\ 3. 
  The pulled front then still is at the margin 
  of a continuous spectrum, while the pushed front is isolated just
  like the bistable front.

\subsection{The dynamics of the leading edge of a front}

\label{S25}
In this Section, we reconsider the dynamics in the leading edge in more
detail, first to demonstrate the conservation of steepness expressed by
(\ref{lambdacons}), second to clarify the dynamics that ensues
from flat initial conditions, and third to lay the basis for the
quantitative analysis of the relaxation of pulled fronts in Section \ref{S3}. 

\subsubsection{Equation linearized about $\phi=0$}

\label{S251}

When we analyze the leading edge region of the front, where $|\phi|\ll1$, 
we to lowest order can neglect $o(\phi^2)$ in (\ref{209}) and analyze
\begin{equation}
\label{2049}
\partial_t\phi=\partial_x^2\phi+\phi~.
\end{equation}
We first explore the predictions of this equation, before exploring
the corrections due to the nonlinearity $f$ in Section \ref{S252}.

Eq.\ (\ref{2049}) is a linear equation, so the superposition of solutions
again is a solution. A generic solution is, e.g., an exponential
$e^{-\lambda x}$.  It will conserve shape and propagate with velocity
$v(\lambda)=\lambda+1/\lambda$ (\ref{2044}):
\begin{equation}
\label{2050}
\phi(x,t)\sim e^{-\lambda\;\left[x-v(\lambda)t\right]}~.
\end{equation}
The minimum of $v(\lambda)$ is given by $v^*=v(\lambda^*=1)=2$.

Consider now a superposition of two exponentials $c_1 \;e^{-\lambda_1
  x} +c_2 \;e^{-\lambda_2 x}$. Without loss of generality, we can
assume the maximum velocity to be
$v_{max}=\max[v(\lambda_1),v(\lambda_2)]=v(\lambda_1)$. In the
coordinate system $\xi_1=x-v(\lambda_1)t$ the temporal evolution then
becomes
\begin{eqnarray}
\label{2051}
\phi(x,t) &=& c_1 \;e^{-\lambda_1\xi_1} + c_2 \;e^{-\lambda_2\xi_1}
\;e^{-\sigma t}~,
\\
\label{2052}
\sigma&=&\lambda_2\;\Big(v(\lambda_1)-v(\lambda_2)\Big)>0~.
\end{eqnarray}
Clearly, the contribution of $\lambda_2$ decays on the time scale
$1/\sigma$, and so for large times $\gg 1/\sigma$, the velocity of 
a so-called level curve of $\phi=const.>0$ in an $x,t$ diagram  
will approach $v(\lambda_1 )$ and the profile will converge to 
$e^{-\lambda_1\xi_1}$ (see \cite{kessler} for a similar type of analysis). 
The steepness of the 
leading edge at $\xi\to\infty$, on the other hand, will be given by
$\lambda_{min}=\min[\lambda_1,\lambda_2]$ for all times $t<\infty$.

This simple example already backs up much of our discussion of
perturbations outside the Hilbert space in Appendix \ref{consequencesapp},
that apply to the Cases II and III in Table IV:\\ 
1) The limits $\xi\to\infty$ and $t\to\infty$ in general do not commute.\\
2) The steepness $\lambda=\min_i[\lambda_i]$ is a conserved quantity 
at $x\to\infty$ and $t<\infty$. As the explicit example of \cite{evsp}
shows, for equations for which one can derive a comparison theorem,
the conservation of steepness can easily be derived rigorously.
\\ 
3) The velocity of a constant amplitude $\phi=const.>0$ will be 
governed by the quickest mode present $v=\max_i[v(\lambda_i)]$ at 
large times $t\gg1$.

Let us now analyze initial conditions steeper than any exponential.
Quite generally, an initial condition $\phi(x,0)$ evolves under
(\ref{2049}) as\footnote{Eliminate the linear growth term in (\ref{2049})
by the transformation $\phi=e^t\bar{\phi}$, solve the diffusion equation
  $\partial_t\bar{\phi}=\partial_x^2\bar{\phi}$, and transform back.}
\begin{equation} \label{2053}
  \phi(x,t)=\int_{-\infty}^\infty dy\;\phi(y,0)\;\;
  \frac{e^{-\big[(x-y)^2-4t^2\big]/(4t)}}{\sqrt{4\pi\;t}}~.
\end{equation}
Assume for simplicity, that the initial condition $\phi(y,0)$ is
strongly peaked about $y=0$, so that for large times, we can neglect
the spatial extent of the region where $\phi(y,0)\neq 0 $ initially.
Upon introducing the coordinate $\xi=x- 2t$ we get
\begin{equation} \label{2054}
  \phi(x,t) \propto \frac{e^{-\xi-\xi^2/(4t)}}{\sqrt{t}}~~ 
  \mbox{for }t\gg1~.
\end{equation}
This general expression leads to three important observations:\\ 
1) The steepness of the leading edge characterized by $\lambda=\infty$ 
at $\xi\to\infty$ indeed is conserved for all finite times $t<\infty$.\\ 
2) At finite amplitudes $\phi=const.>0$ and large times $t$, the
steepness of the front propagating towards $\xi\to\infty$
approaches $\lambda^*=1$ and the velocity approaches $v^*=2$.\\ 
3) Eq.\ (\ref{2054}) furthermore implies, that a steep initial condition 
like $\phi(y,0)$ approaches the asymptotic velocity $v^*$ as
\begin{equation} \label{2055}
  v(t)_{lin} = v^* +\dot{\xi}_h = 2 - \frac{1}{2t}
  +O\left(\frac{1}{t^{2}}\right)~,
\end{equation}
where we defined the position $\xi_h(t)$ of the amplitude $h$ in the
comoving frame $\xi=x-2t$ as $\phi(\xi_h(t),t)=h$. Eq.\ (\ref{2055}) 
is then obtained simply by solving 
$\ln \phi =-\xi_h -\xi_h^2/4t -(\ln t)/2=const.$

This algebraic convergence  is consistent with the gapless spectrum
of linear perturbations, and as such it identifies the missing link in
the analysis of the relaxation of pulled fronts.  However,
Bramson's work \cite{bramson} shows, that the qualitative prediction
of convergence as $1/t$ is right, but the coefficient of $1/t$ is
wrong. In fact, the mathematical literature \cite{mckean} has established
(\ref{2055}) as an upper bound for the velocity of a pulled front in a
nonlinear diffusion equation. The algebraic convergence clearly comes
from the $1/\sqrt{t}$ prefactor characteristic of the fundamental
Gaussian solution of the diffusion equation (\ref{2054}) --- 
this qualitative mechanism will be found to be right in Section \ref{S3}.

We finish our discussion of solutions of the linearized equation
(\ref{2049}) with another illustrative example.  After
the discussion of the solution (\ref{2050}) one might be worried about
initial conditions with $\lambda\gg1$. Such an initial condition is
steep according to our definition, so it should approach the velocity
$v^*$. But according to (\ref{2050}), it approaches the larger
velocity $v(\lambda)$. However, even in the framework of the
linearized equation, this paradox can be resolved: An initial
condition $e^{-\lambda x}$ on the whole real axis is, of course,
unphysical, and we in fact only want this behavior at $x\gg1$, where
$\phi$ is small.  Let us therefore truncate the exponential for small
$x$ by writing, e.g., $\phi(x,0)=\theta(x) \;e^{-\lambda x}$, with
$\theta$ the step function. Insertion into (\ref{2053}) yields the
evolution
\begin{equation}
\label{2056}
\phi(x,t) = e^{-\lambda[x-v(\lambda)t]}\; \frac{1+\mbox{erf
    }\frac{x-2\lambda t}{\sqrt{4t}}}{2}~,
\end{equation}
where erf $x = 2\pi^{-1/2} \int_0^x dt\;e^{-t^2}$ is the
errorfunction.  For $t\gg 1$ the crossover region where $x \approx
2\lambda t$ separates two different asymptotic types of behavior:
\begin{equation}
\label{2057}
\phi(x,t) \approx \left\{\begin{array}{ll} \normalsize
e^{-\lambda[x-v(\lambda)t]} ~~&~\mbox{for }x\gg2\lambda t\\ 
\normalsize \frac{e^{-(x-2t)-(x-2t)^2/4t}}{\sqrt{4\pi t}\;\lambda
  (1-x/(2\lambda t))} ~~&~\mbox{for }x\ll2\lambda t
\end{array}\right.
\end{equation}
In the region of $x\gg2\lambda t$ we find our previous solution
(\ref{2050}) with conserved leading edge steepness and velocity $v(\lambda)$,
while in the region of $x\ll2\lambda t$ we essentially recover
(\ref{2054}), with $\xi=x-2t$. 

Considering the three different velocities --- $v(\lambda)$ for the
region of conserved $\lambda$, $v^*=2$ for the ``Gaussian'' region behind, 
and $2\lambda$ for the crossover region between the two asymptotes --- the
distinction between flat and steep initial conditions now comes about
quite naturally:\\ $a)$ For flat initial conditions, we have
$\lambda<1$, and an ordening of velocities as
$2\lambda<v^*<v(\lambda)$. The crossover region then moves slower than
both asymptotic regions, so for large times the region of finite
$\phi$ will be dominated by $e^{-\lambda[x-v(\lambda)t]}$.\\ $b)$ For
steep initial conditions, we have $\lambda>1$, and the velocities
order as $v^*<v(\lambda)<2\lambda$. The crossover region then will
move quicker than both asymptotic regions, and the region of finite
$\phi$ will be dominated by $e^{-\xi-\xi^2/4t}/\sqrt{t}$, where $\xi=x-2t$.

We finally note that the above results can also be reinterpreted in
terms of the intuitive picture advocated in \cite{vs1,vs2}: The group
velocity $v_{gr}(\lambda)= dv(\lambda)/d\lambda$ of a near exponential
profile in the leading edge is, according to (\ref{2044}), negative 
for $\lambda < 1$ and positive for $\lambda >1$. In this way of
thinking, the region with steepness $\lambda$ in the case considered
above expands when $\lambda <1 $ since the crossover region moves back
in the comoving frame [case ($a$)], and it moves out of sight 
towards $\xi\to\infty$ for $\lambda >1$ [case ($b$)], 
since the crossover region moves faster than the local comoving frame.

\subsubsection{Leading edge representation of the full equation} \label{S252}

Just as the linear stability analysis of the front was insufficient
to cover the full dynamical behavior of the nonlinear diffusion
equation (\ref{205}) and in particular the dynamics of the leading edge,
so is the linearized equation (\ref{2049}).
In Section \ref{S3}
we will see that only through joining these complementary approaches,
we can gain a quantitative understanding of the convergence of steep 
initial conditions towards a pulled front $\Phi^*$. 

The shortcomings of the linearized equation (\ref{2049}) become quite 
clear by confronting it with what we will call the leading edge 
representation of the full equation (\ref{205}):
\begin{equation}
\label{2058}
\partial_t\psi=\partial_\xi^2\psi+\bar{f}(\psi,\xi)~,
\end{equation}
where we transformed with
\begin{equation}
\label{20058}
\psi=\phi\;e^{\lambda^*\xi}~~,~~\xi=x-v^*t~.
\end{equation}
The parameters are $\lambda^*=v^*/2=\sqrt{f'(0)}$.
This transformation eliminates the terms of order $\psi$
and $\partial_\xi\psi$ from the linear part of the equation.
The nonlinearity is
\begin{eqnarray}
\label{200058}
\bar{f}(\psi,\xi)&=&e^{\lambda^*\xi}\;
\left(f\left(\psi\;e^{-\lambda^*\xi}\right)
             -f(0)-f'(0)\;\psi\;e^{-\lambda^*\xi}\right)
             \nonumber\\
             &=& O\left(\psi^2\;e^{-\lambda^*\xi}\right)~.
\end{eqnarray}
This transformation is quite comparable to the transformation of a 
linear perturbation $\eta$ into the Schr\"odinger picture
as in (\ref{2025}), (\ref{2026}). However, we here transform
the full nonlinear equation, and not only the linearization
about some asymptotic solution.

For, e.g., $f(\phi)=\phi-\phi^3$ we have $\bar{f}=-\psi^3\;e^{-2\xi}$.
When we neglect $\bar{f}$ in (\ref{2058}), the equation is equivalent
to the linearization about $\phi=0$ (\ref{2049}). The linearization is correct
for $\xi\gg1$, but the presence of the crossover towards a different 
behavior for smaller $\xi$ has important consequences for the solutions
of the full nonlinear diffusion equation.

In particular, for the leading edge of a pulled front 
$\Phi^*\sim(\alpha\xi+\beta)\;e^{-\xi}$ (\ref{2018}), we generically
find $\alpha\ne0$ and accordingly the leading edge behavior (\ref{2019}). 
This leading edge behavior will play a central role in Section \ref{S3}. 
In Section \ref{S22} we derived $\alpha\ne 0$ from the uniqueness 
of the trajectory in phase space, i.e., from the construction
of the whole front from $\phi=0$ up to $\phi=1$. We now will give
a different argument for $\alpha\ne0$ from the analysis of (\ref{2058}),
that does not rely on constructing the whole solution up to $\phi=1$.

The front $\Phi^*$ propagates uniformly with velocity $v^*=2$,
so in the frame $\xi=x-2t$ it is stationary. $\Psi^*=\Phi^* e^\xi$
then solves
\begin{equation}
\label{2059}
\partial_\xi^2\Psi^*+\bar{f}(\Psi^*,\xi)=0~.
\end{equation}
The boundary conditions (\ref{2013}) for $\Phi^*$ imply for $\Psi^*$:
\begin{equation}
\label{2060}
\Psi^*(\xi)\sim\left\{\begin{array}{ll}
\alpha\xi+\beta ~~&\mbox{for }\xi\to\infty\\
O(e^{\:\lambda^*\xi}) &\mbox{ for }\xi\to-\infty\end{array}\right.
\end{equation}
The solution $\Psi^*=\alpha\xi+\beta$ for $\xi\to\infty$ can 
directly be derived from (\ref{2059}) and the condition, that $\Phi^*$
vanishes at $\xi\to\infty$. Now integrate
(\ref{2059}) over the real $\xi$ axis, and find
\begin{equation}
\label{2061}
\alpha = - \int_{-\infty}^\infty d\xi\;\bar{f}\left(\Psi^*,\xi\right)~.
\end{equation}
The integral on the right hand side is well-defined, since $\bar{f} $
vanishes exponentially, both for $\xi\to -\infty$ and for $\xi \to \infty$.
Clearly, a nonlinearity $\bar{f}\ne 0$ generically implies $\alpha\ne0$
and hence the leading edge behavior (\ref{2019}) for a pulled $\Phi^*$ front. 
Only for particular nonlinearities $f$, we occasionally 
find $\alpha=0$ (see (\ref{3068}) and Appendix \ref{A1}).  
Having $\alpha=0$  is obviously only possible if $\bar{f}$ has
terms of opposite sign, so that its spatial average vanishes. For the
nonlinearity of form $f=\phi-\phi^k$ with $k>1$ (\ref{102}), we find
$\alpha \neq 0$ always, and in this case the term
$\bar{f}$ acts like a localized sink term in the diffusion equation
(\ref{2059}) for $\psi$. This interpretation is especially useful for
the discussion of the non-uniformly translating fronts
\cite{willem1,kees}.

\subsection{Concluding remarks --- interior and edge dominated dynamics} 
\label{S26}

Table IV summarizes the results of this Section for various
nonlinearities and initial conditions. For pushed fronts, the
stability analysis gives essentially all the ingredients to determine
the rate of convergence for pushed fronts. For pulled fronts emerging
from sufficiently steep initial conditions one finds however (see Case
IV of Appendix \ref{consequencesapp}) that the linear stability analysis
is not the appropriate tool: the spectrum is gapless and the rate of
convergence can not be determined from the spectrum.

The crucial insight for the further analysis is that a relaxing front
can be decomposed  into different dynamical regions. Linear stability
analysis is the appropriate tool for the interior dominated dynamics
of a pushed front. Pulled  fronts and fronts evolving from ``flat''
initial conditions are leading edge dominated. This calls for
different methods of analysis. The relaxation of pulled fronts emerging
from sufficiently steep initial conditions will be addressed in the
next section.

\end{multicols}
\newpage
\begin{center}
\begin{tabular}{||c||c|c|c||} \hline \hline 
& \multicolumn{3}{c||}{ } \\
& \multicolumn{3}{c||}{{\em NONLINEARITY $f(\phi)$}}\\  
& \multicolumn{3}{c||}{ } \\ 
{\em INITIAL CONDITION} &&& \\
$\phi(x,0) \sim e^{-\lambda_{init} x}$ & $~~~$metastable 
($f'(0)<0$):$~~~$ & $~$unstable ($f'(0)>0$),$~$ 
& $~$unstable ($f'(0)>0$),$~$\\
as $x\to\infty~~$ & & {\em pushed} regime: & {\em pulled} regime: \\
& $v_{sel}=v^\dag>0$ & $v_{sel}=v^\dag>v^*$      & $v_{sel}=v^*$ \\
&         & $v^*=2\sqrt{f'(0)}>0$ & $v^*=2\sqrt{f'(0)}>0$ \\
 \hfill &&&\\
\hline\hline
&&&\\
& Case I:        & Case I:               & Case IV:\\
{\em steep}:
& pushed dynamics, & pushed dynamics, & pulled dynamics, \\
$\lambda_{init}>\lambda_0(v_{sel})$
& $\lambda  \longrightarrow  \lambda_+(v^\dag)$ 
& $\lambda  \longrightarrow  \lambda_+(v^\dag)$ 
& $\lambda  \longrightarrow  \lambda^*$ \\
 (including $\lambda=\infty$)
& $v(t)=v^\dag+O(e^{-\sigma t})$ & $v(t)=v^\dag+O(e^{-\sigma t})$ 
& $v(t)=v^*+O(1/t)$ \\
&&&\\
\hline
&&&\\
& Case II:        & Case II:                      &\\
{\em steep}:
& pushed dynamics, & pushed dynamics,             &  not applicable, 
since \\
$~~\lambda_-(v_{sel})<\lambda_{init}<\lambda_0(v_{sel})~~$
& $\lambda  \longrightarrow  \lambda_+(v^\dag)$ 
& $\lambda  \longrightarrow  \lambda_+(v^\dag)$ 
& $~~\lambda_\pm(v^*)=\lambda_0(v^*)=\lambda^*~~$ \\
& $v  \longrightarrow  v^\dag$ & $v  \longrightarrow  v^\dag$ & \\
& generically: & generically: & \\
& $v(t)=v^\dag+O(e^{-\sigma t})$ & $v(t)=v^\dag+O(e^{-\sigma t})$ 
&  \\
&&&\\
\hline
&&&\\
&  & Case III:                      & Case III: \\
&  & leading edge & leading edge \\
{\em flat}:
& not applicable, since & dominated dynamics, & dominated dynamics, \\
$0<\lambda_{init}<\lambda_-(v_{sel})$
& $\lambda_-(v^\dagger) < 0$
& $~\lambda  \longrightarrow  \lambda<\lambda_-(v^\dag)~$ 
& $~\lambda  \longrightarrow  \lambda<\lambda^*~$ \\
& & $v \longrightarrow v(\lambda)>v^\dag$ & $v \longrightarrow v(\lambda)>v^*$ \\
& & generically: & generically: \\
&  & $~v(t)=v(\lambda)+O(e^{-\sigma t})~$ 
& $~v(t)=v(\lambda)+O(e^{-\sigma t})~$ \\
&&&\\
\hline \hline
\end{tabular}

~\\
~\\
~\\

\begin{minipage}{16.5cm}
{\bf Table IV:} Table of initial conditions and nonlinearities,
resulting in relaxation cases I -- IV from Appendix
\ref{consequencesapp}.
Fronts at all times $t$ are characterized by their steepness 
$\lambda$ (\ref{lambdadef}) in the leading edge, and an arrow
$\longrightarrow$ indicates the evaluation of the quantity for $t \to \infty$.
The nonlinearity only enters through the existence of a 
strongly heteroclinic orbit $\Phi_v(x-vt)$ with $v_{sel}=v^\dag>v^*$
(see Section \ref{S22}) or its non-existence (then $v_{sel}=v^*$). 
$v_{sel}$ determines $\lambda_{\pm,0}(v_{sel})$ as in (\ref{lambda1}), 
which in turn classifies the initial conditions. Pushed or pulled 
dynamics are special cases of 
interior or leading edge dominated dynamics for steep initial conditions.
Cases I -- III are treated in Appendix \ref{consequencesapp}  with
stability analysis methods 
and generically show exponential relaxation.
Case IV is not amenable to stability analysis methods.
It shows algebraic relaxation and is treated from Section \ref{S3} on.
\end{minipage}
\end{center}

\newpage

\begin{multicols}{2}

\section{Universal pulled convergence of steep
  initial conditions in the nonlinear diffusion equation}
\label{S3}

In the present Section, we will combine our understanding of the
dynamics of the leading edge and of the interior of a front into one
consistent analytical frame, that allows us to calculate the long time 
convergence of 
steep initial conditions towards a pulled front --- as we discussed in
the previous section, the relaxation can in this case not be obtained
from the linear stability analysis of the asymptotic solution.  
The different 
dynamical regions of such a front are sketched in Fig.\ 2. We match an
expansion in the interior, that resembles features of the linear
stability analysis, to an expansion of the leading edge. Both
expansions are asymptotic expansions in $1/\sqrt{t}$. This approach
allows us to derive the power law convergence of the front velocity
and the front profile towards $\Phi^*$. This convergence is
universal in leading and subleading order and we calculate all
universal convergence terms analytically. For clarity we present the
detailed calculation for the nonlinear diffusion equation in this
Section first, and then discuss the generalization in Section \ref{S5}.

\subsection{Observations which motivate our approach} \label{S31}

\subsubsection{Asymptotic steepness of leading edge
  determines rate of convergence}
\label{S311} 

Our calculation of the spreading of the leading edge under the
linearized equation in Section \ref{S25} gave qualitatively the right
results, but failed to reproduce the quantitative results for the
nonlinear equation: Inserting sufficiently steep initial conditions
(\ref{208}) into the {\em linearized} equation (\ref{209}), we found
that the asymptotic shape (\ref{2054}) approaches $e^{-\xi}$ times a
Gaussian for $t\to\infty$ and $\xi\gg1$ and that this implies for the
asymptotic convergence that $v(t)_{lin}=2-1/(2t)+\ldots$ (\ref{2055}).
For the {\em nonlinear} equation we know that the asymptotic front
profile $\Phi^*$ behaves as $\Phi^*\sim\xi\;e^{-\xi}$ for $\xi\gg 1$
(\ref{2019}) and Bramson has derived with probabilistic methods, that
$v(t)=2-3/(2t)+\ldots$ independent of the height at which the velocity
is measured \cite{bramson}. 

How the exact result of Bramson \cite{bramson} comes out naturally and
generally is brought out quite clearly by rephrasing an argument of
\cite{vs2} as follows (see also \cite{derrida,prl}).

Let us work in the leading edge representation (\ref{2058}),
(\ref{20058}), and let us from here on use the co-moving variable
$\xi$ specifically for the frame moving with the pulled velocity 
$v^*=2$,
\begin{equation}
\xi= x-v^*t = x-2t~.
\end{equation}
The fundamental similarity solution of the diffusion-type equation
(\ref{2058}) for the leading edge variable
$\psi=\phi e^{\lambda^*\xi}$ in
the region where the nonlinearity can be neglected, is of course
the Gaussian
\begin{equation} \label{302} 
  \psi_0(\xi,t)=\frac{e^{-\xi^{2}/(4t)}}{\sqrt{4\pi t}}~.
\end{equation} 
It reproduces our solution (\ref{2054}) for $\phi$. But also any
derivative of the Gaussian $\psi_n=\partial_{\xi}^n \psi_0$ solves
(\ref{2058}) for $\xi\gg1$. The $\psi_n/\psi_0$ are simply Hermite
polynomials \cite{messiah2,abra}. In particular, the dipole solution
\begin{equation} \label{303} 
  \psi_1(\xi,t)=\partial_{\xi}\psi_0 \propto
  \frac{\xi \;e^{-\xi^{2}/(4t)}}{t^{3/2}}
\end{equation}  
also solves the diffusion equation (\ref{2058}) for $\xi\gg1$ and
has the proper asymptotics $\Phi\;e^{\xi}\propto\xi$ for
$t\to\infty$. Transforming (\ref{303}) back to $\phi$, we find
\begin{equation} \label{304} 
  \phi(x,t) \propto (x-2t)\;e^{-[x-2t+(3/2)\ln
    t]}\;e^{-(x-2t)^{2}/(4t)}~.
\end{equation} 
If we now trace the position $2t+X_h(t)$ of  the
point where $\phi$ 
reaches the amplitude $h$ in the original $x$ frame, we find by
solving $\phi(2t+X_h,t)=h $ from 
(\ref{304}) for $X_h(t)\ll\sqrt{4t}$
\begin{equation} \label{305} 
  v(t)=2+\dot{X}_h = 2-\frac{3}{2t}+\ldots~,
\end{equation} 
in agreement with Bramson's result. This indicates that for large
times $t\gg 1$ and far in the leading edge $\xi\gg1$, the converging
front is approximately given by (\ref{303}) if $\alpha\ne0$ in
(\ref{2018}) --- remember that $\alpha \neq 0$ implies that
$\Phi^*(\xi) \sim \xi e^{-\xi}$ for large $\xi$.  We will see indeed
that (\ref{303}) does emerge as the 
dominant term in a systematic asymptotic expansion in the leading edge
region. For reasons explained below, it is, however, more convenient
to formulate this expansion in a slightly different frame.

\subsubsection
{Interior follows leading edge:
  uniform convergence}

The above argument shows that the leading $-3/(2t)$ velocity
correction is due to the diffusion-type dynamics in the leading edge. 
Why would the convergence to the asymptotic profile be {\em uniform}, 
i.e., be independent of the level curve whose position is tracked (the
level curve is the curve in an $x$-$t$ plot that indicates the values
where $\phi$ reaches a particular level or ``height'')? 
The answer to this question is intuitively
quite simple. As $\dot{X}_h \simeq -3/(2t)$, $X_h \simeq -3/2 \ln t$.
If we compare the position $X_{h_1}$ of a height $h_1$ in the 
leading edge ($h_1\ll1$) with a position $X_{h_2}$ of a height $h_2$ 
in the interior ($h_2=O(1)$) where the dynamics of $\phi$
is described by the nonlinear equation, we will have
$X_{h_2}=X_{h_1}-W(h_1,h_2)$, where $W$ is the width of the front
between these two heights.  Clearly, if $W$ approaches a finite value 
for long times, we need to have also $X_{h_2} \simeq 3/2 \ln t$ 
in dominant order as $t\to \infty$, and hence also 
$\dot{X}_{h_2}=-3/(2t) +\cdots$. 
But an equation of motion like (\ref{101}) has front solutions whose
width is finite, so we expect indeed that $W=O(1)$ for large times. Our
analysis will confirm this expectation. 
In other words, the leading order velocity correction as $-3/(2t)$ is 
set by the dynamics of the leading edge, and because of the finite 
asymptotic width of the front, the convergence is {\em uniform}, 
i.e., independent of $h$.

\subsubsection{Choose proper frame and subtraction for the interior} 
\label{S312}

The above observations have another important consequence.  After the
front has evolved for some time, we will find it selfconsistent to
assume, that its shape will resemble the asymptotic shape $\Phi^*$. If
we want to understand the interior part of the front, it might at
first sight seem appropriate to linearize the converging front $\phi$
about the asymptotic front $\Phi^*$.  However, the profile $\Phi^*$
propagates uniformly with velocity 2, while as we saw above, the
transient profile $\phi$ propagates with velocity $v^*-3/(2t)$. Thus,
if the interior regions of the $\phi$- and the $\Phi^*$-fronts are at
about the same part of space at time $t_0$, their distance will {\em
  diverge} as $(3/2)\ln(t/t_0)$ as $t$ grows! This was already illustrated
in Fig.\ 1. Hence, linearization of $\phi$
about the asymptotic profile $\Phi^*$ during the whole time evolution
requires to move $\Phi^*$ along with the velocity $2-3/(2t)+\ldots$ of
$\phi$ and {\em not} with its proper velocity 2. Our expansion is
therefore based on writing $\phi$ as
\begin{equation} \label{307}
  \phi(\xi,t)=\Phi^*(\xi_X)+\eta(\xi_X,t)~,
\end{equation}
where
\begin{equation} \label{306} \label{pre1}
  \xi_X = \xi -X(t)= x-2t-X(t)~.
\end{equation}
This Ansatz anticipates that we need to shift  the profile
$\Phi^*$ an appropriate distance  $X(t)\propto \ln t$,
and that with 
a proper choice of $X(t)$, $\eta$ becomes a small and decaying
perturbation.

\subsubsection{Choose proper expansions and match leading edge to interior}
\label{S313}

We will need two different expansions for the leading edge and for the
interior. The expansions have
to be chosen such that they can be matched in overlapping intervals
through resummation of the expansions.

Since we use the coordinate system (\ref{306}) in the interior, we
also should use it in the leading edge. The leading $1/t$ contribution
from the leading edge suggests to expand $\eta$ in the interior as
$\eta_1(\xi_X)/t + \ldots$, and we shall see indeed that such a
form emerges automatically from the ansatz (\ref{307}). The
appropriate variable for $\xi_X\gg\sqrt{t}$ in the leading edge, on the
other hand, is the similarity variable of the diffusion equation
\begin{equation} \label{308}
z=\frac{\xi_X^2}{4t}~,
\end{equation}
as suggested by (\ref{302}) -- (\ref{304}). Expressing $\xi_X$ by $z$
and $t$ introduces a dependence on $1/\sqrt{t}$. We find, that it is
actually consistent to expand the interior in powers of $1/\sqrt{t}$
(instead of $1/t$) times functions of $\xi_X$, and the leading edge also
in powers of $1/\sqrt{t}$ times functions of $z$.

The structure of these expansions is essentially our only input. Given
this structure, the leading and subleading order universal terms of
the expansions  are uniquely determined.

\subsection{Expansion in the interior region} 
\label{S32}

We first analyze the interior part of the front where $\phi$ varies
from close to 0 to close to 1. We work in the comoving frame
$\xi_X=x-v^*t -X(t)$ of (\ref{306}), where $X$ will have to be determined.
We expand $\phi$ about $\Phi^*(\xi_X)$ as in (\ref{307}). Because of
translation invariance, we have the freedom to fix the position of
$\Phi^*$ and the zero of the coordinate system by imposing
\begin{equation} \label{309} 
\phi(0,t) = \frac{1}{2}~\mbox{ and }~\Phi^*(0)=\frac{1}{2}
~\Rightarrow~\eta(0,t)=0~.
\end{equation} 
For $\Phi^*$, one has
$\Phi^*(-\infty )=1$, and we also assume that $\phi$
approaches 1 for $\xi_X\to-\infty$.  This results in the second
condition on $\eta$
\begin{equation} \label{3010} 
\lim_{\xi_X \to-\infty} \eta(\xi_X,t)=0~.
\end{equation} 

We insert $\phi$ into the equation of motion (\ref{101}), transform
$x$ to the coordinate $\xi_X$ (\ref{306}) and find for $\eta$ the 
equation\footnote{\label{footnoteref} Throughout this paper, we shall  
suppress the index ${\tiny X}$ on partial derivatives with respect 
to $\xi_X$ for notational convenience. Since 
$\left.\partial_{\xi_X}\right|_t = \left.\partial_\xi\right|_t$, 
this does not lead to  any ambiguities.}
\begin{equation} \label{3011} 
\partial_t \eta = \partial_\xi^2 \eta + v^* \partial_\xi \eta
+ \dot{X}\partial_\xi\Phi^*+f(\Phi^*+\eta)-f(\Phi^*)~.
\end{equation}

Once $\eta$ is small enough because time has evolved sufficiently long,
$f(\Phi^*+\eta)$ can be expanded in $\eta$ and we find
\begin{equation} \label{3012} 
\partial_t \eta = {\cal L}^* \eta + \dot{X}\;\partial_\xi\Phi^* + 
\dot{X}\;\partial_\xi\eta + \frac{f''(\Phi^*)}{2}\;\eta^2 + 
O(\eta^3)~,
\end{equation} 
where
\begin{equation} \label{3013} 
{\cal L}^* = \partial_\xi^2 + v^* \partial_\xi +f'(\Phi^*(\xi_X)) 
\end{equation} 
is the linearization operator (\ref{2025}) for $v=v^*$.

In Sections \ref{S251} and \ref{S311} we have argued, that one expects
$\dot{X}(t)=O(t^{-1})$. Asymptotic balancing in
(\ref{3012}) then requires, that the leading order term of $\eta$ is
of the same order $\eta=O(t^{-1})$.  We therefore try to expand
as $\eta=\eta_1(\xi_X)/t+\ldots$.  We have argued, that connecting the
interior expansion to the leading edge expansion requires an ordening
in powers of $1/\sqrt{t}$.  So we choose the ansatz
\begin{eqnarray} \label{3014}
\label{7}
\dot{X} & = &\frac{c_1}{t}+\frac{c_{3/2}}{t^{3/2}}+\frac{c_2}{t^2}+\ldots~,
\\
\label{8} \label{3015}
\eta(\xi_X,t)&  = & \frac{\eta_{1}(\xi_X)}{t}+\frac{\eta_{3/2}(\xi_X)}{t^{3/2}}
+\ldots~.
\end{eqnarray}
Substitution of the above expansions into (\ref{3013}) and ordering in
powers of $1/\sqrt{t}$ yields a hierarchy of {\em o.d.e.}'s of second
order:
\begin{eqnarray} \label{3016}
{\cal L}^* \eta_1     &=& - c_1\partial_\xi \Phi^* ~,
\\ \label{3017}
{\cal L}^* \eta_{\frac{3}{2}} &=& - c_{\frac{3}{2}}\partial_\xi \Phi^* ~,
\\ \label{3018}
{\cal L}^* \eta_2     &=& - \eta_1 - c_1\partial_\xi \eta_1 
- c_2\partial_\xi \Phi^* - f''(\Phi^*) \eta_1^2/2 ~,
\\ \label{3019}
{\cal L}^* \eta_{\frac{5}{2}} &=& - {\scriptstyle \frac{3}{2}}
\eta_{\frac{3}{2}} - c_1\partial_\xi \eta_{\frac{3}{2}} 
- c_{\frac{3}{2}}\partial_\xi \eta_1 - c_{\frac{5}{2}}\partial_\xi \Phi^* 
\nonumber\\
&& - f''(\Phi^*) \eta_1 \eta_{\frac{3}{2}}\qquad\mbox{ etc., generally:}
\\ \label{3020}
{\cal L}^* \eta_{\frac{n}{2}} &=& - \;\frac{n-2}{2}\;
\eta_{\frac{n-2}{2}} - \sum_{m=2}^{n-2} c_{\frac{m}{2}}
\partial_\xi \eta_{\frac{n-m}{2}}
 - c_{\frac{n}{2}}\partial_\xi \Phi^*
\nonumber\\
&& - \left. \sum_{k=2}^{\infty} \; \frac{f^{(k)}(\Phi^*)}{k!} 
\left( \sum_{m_k} \eta_{m_k} \right)^k  \right|_{\sum_k {m_k} =\frac{n}{2}}~.
\end{eqnarray}
It is important to realize that we do not need to drop nonlinear
terms, but that the 
expansion of $f(\Phi^*+\eta)$ in powers of $\eta$ is also ordered in
powers of $1/\sqrt{t}$. So the higher order terms $\eta^n$ find their
natural place as inhomogeneities in the equations for $\eta_i$ for
$i\ge 2$. The hierarchy of {\em o.d.e.}'s is such that the
differential equation for $\eta_i$ contains inhomogeneities that
depend only on $\eta_j$ with $j<i$. The equations therefore can be
solved successively. Each $\eta_i$ solves
a second order differential equation, and the two constants of
integration are fixed by the two conditions (\ref{309}) and
(\ref{3010})\footnote{Had we introduced an $\eta_{1/2}$, we would have  found
  the equation ${\cal L}^*\eta_{1/2}=0$ with the unique solution
  $\eta_{1/2}=0$.}.

Note also, that the time dependent collective coordinate $X(t)$ in
$\xi_X=\xi-X(t)$ only enters Eqs. (\ref{3016})-(\ref{3020}) in the
form of the constants $c_{n/2}$, which at this point are still
undetermined, and that the functions $\eta_{n/2}$ obey  {\em o.d.e.}'s.

Let us now compare $\eta=\phi-\Phi^*$ to the variations of the profile
shape with velocity $v$,
\begin{equation} \label{3021}
\delta=\Phi_{v^*+\dot{X}}-\Phi^*=\dot{X}\;\eta_{\rm sh} 
+ \frac{\dot{X}^2}{2}\;\eta_{\rm sh}^{(2)} +\ldots~,
\end{equation}
where $\eta_{\rm sh} \equiv \delta \Phi_v/ \delta v |_{v^*} $ is a
``shape mode'', which gives the change in the profile under a change
in $v$. By considering variations of $v$ in the {\em o.d.e.}\ for the
profile $\Phi_v$, we find that $\eta_{\rm sh}$ and $\eta_{\rm sh}^{(2)}$ obey
\begin{eqnarray} \label{3023} \label{pre2}
{\cal L}^*\eta_{\rm sh}&+&\partial_\xi\Phi^*=0~,
\\ \label{3024}
{\cal L}^*\eta_{\rm sh}^{(2)}&+&2\partial_\xi\eta_{\rm sh}
+f''(\Phi^*)\;\left(\eta_{\rm sh}\right)^2=0~.
\end{eqnarray}

Upon comparing (\ref{3023}) -- (\ref{3024}) with (\ref{3016}) --
(\ref{3018}), we can identify
\begin{eqnarray} \label{3025}
\eta_1     &=& c_1\eta_{\rm sh} ~,\\ \label{3026}
\eta_{3/2} &=& c_{3/2}\eta_{\rm sh}~,\\ \label{3027}
\eta_2     &=& c_2\eta_{\rm sh} +\frac{c_1^2}{2} \eta_{\rm sh}^{(2)} + c_1\rho~,
\end{eqnarray} with $\rho$ a  correction term, that solves
the equation
\begin{equation} \label{3030} 
{\cal L}^*\rho+\eta_{\rm sh}=0~.
\end{equation}
In these differential equations, $\eta_{\rm sh}$, $\eta_{\rm sh}^{(2)}$ and
$\rho$ obey the conditions
\begin{eqnarray} \label{3028} \label{pre3}
\eta_{\rm sh}(0)=0 ~,&~\eta_{\rm sh}^{(2)}(0)=0~,& ~\rho(0)=0~,\\
\label{3029}
\eta_{\rm sh}(-\infty)=0 ~,&~\eta_{\rm sh}^{(2)}(-\infty)=0~,& ~\rho(-\infty)=0~,
\end{eqnarray}
cf.\ (\ref{309}) and (\ref{3010}).

$\rho$ is the first nonvanishing term that indicates the difference
between the transient profile $\phi(x,t)$ and the uniformly
translating front solution with the instantaneous velocity
\begin{equation} \label{3032} \label{pre5}
v(t)=v^*+\dot{X}~.
\end{equation} as resummation of $\phi$  yields
\begin{eqnarray} \label{3031} \label{pre4}
\phi(\xi_X,t)=\Phi_{v(t)}(\xi_X)& + &\frac{c_1}{t^2}\;\rho(\xi_X) +\\ \nonumber 
& + &\frac{t_0}{t^2}\;\eta_{\rm sh}(\xi_X)
+O\left(\frac{1}{t^{5/2}}\right)~.
\end{eqnarray} 
This equation confirms that up to order $1/t^2$, the profile shape is
given by the solution $\Phi_{v(t)}$ of the {\em o.d.e.}\ with the
instantaneous velocity $v(t)$.

Some remarks on these results are in place:

1) We see, that the dynamics in the front interior is slaved to the
evolution of $v(t)$ imposed by the leading edge, as we anticipated in
Section \ref{S312}.

2) The fact that the profile $\phi$ is up to order $t^{-2}$ given by
$\Phi_{v(t)}(\xi_X)$ can be traced back to the fact that since $v(t)$
varies as $t^{-1}$, the time derivative term $\partial_t \phi$ in the
dynamical equation generates terms of order $t^{-2}$. This is why the
first two equations in the hierarchy, (\ref{3016}) and (\ref{3017}),
coincide with the {\em o.d.e.} (\ref{3023}) for the shape mode. 

3) Based on numerical data, Powell et al.\ \cite{powell} have
conjectured, that $\phi$ converges along the trajectory in function
space formed by the $\Phi_v$'s with $v<v^*$.  We here have derived
this result analytically, and identify the velocity $v$ of the
transients $\Phi_v$ with the actual instantaneous velocity
$v=v^*+\dot{X}$ of the front. We find a non-vanishing correction of
order $1/t^2$ to $\phi\approx\Phi_{v^*+\dot{X}}$.

4) The transients $\Phi_v$ have always $v<v^*$ at late times, since we
will find that $c_1=-3/2$, in accord with the discussion of Section
\ref{S311}. Note that as discussed in Section \ref{S22}, such $\Phi_v$
are positive from $\xi_X\to-\infty$ up to a finite value of $\xi_X$ only. 
For the
transient (\ref{3031}) we need only the positive part of $\Phi_v$. The
transient (\ref{3031}) crosses over to a different functional form,
before $\Phi_v$ becomes negative.

5) There is a non-universal contribution of order $1/t^2$ to
(\ref{3031}).  It is non-universal, because it depends on initial
conditions: The structure of our expansion (\ref{3014}), (\ref{3015})
is an asymptotic expansion about $t\to\infty$, that does not fix
$t=0$.  We thus can expand in $1/(t-t_0)=1/t+t_0/t^2+O(1/t^3)$
just as well as in $1/t$. This allows us to add an arbitrary multiple
of $\eta_{\rm sh}/t^2$ to $\phi$ in Eq.\ (\ref{3031}). The order
$1/t^2$ term in (\ref{3031})  is thus always non-zero,  because
the functions $\rho(\xi_X)$ in (\ref{3033}) is non-vanishing and
not a multiple of $\eta_{sh}(\xi_X)$, but its precise value will
depend on initial conditions.

6) The expansion is an asymptotic expansion \cite{bender}. Thus, when
we will have determined the coefficients $c_1$ and $c_{3/2}$ in
(\ref{3014}) later, these are the {\em exact} prefactors if we expand
the velocity and shape in inverse powers of $t$ in the limit $t\to
\infty$. However, the expansion will not have a finite radius of
convergence in $1/\sqrt{t}$.

\subsection{Interior shape expanded towards the leading edge} \label{S33}

We now will see, that for $\xi_X\ge O(\sqrt{t})$ the structure of our
expansion (\ref{3015}) breaks down. We then have to resum the terms
and use a different expansion\footnote{Actually, the interior
  expansion also breaks down for $\xi_X \to -\infty$. There too, a
  different expansion can be used, and this expansion can be matched
  to the one we introduced for the interior region. We will not
  discuss this further here, as it is of no further consequence.}.

Let us calculate the contributions $\eta_i$ from (\ref{3016}) --
(\ref{3020}) explicitly in the leading edge region, where $\xi_X\gg1$
and $\phi,\Phi^*\ll 1$. In this region ${\cal L}^*$ (\ref{3013}) and
$\Phi^*$ (\ref{2018}) are
\begin{equation} \label{3033} 
{\cal L}^*=\partial_\xi^2+2\partial_\xi+1 
~~,~~\Phi^*=(\alpha\xi_X+\beta)\;e^{-\xi_X}~.
\end{equation} 
We remove the exponential through the transformation
\begin{equation} \label{3034} 
{\cal L}^*= e^{-\xi_X}\;\partial_\xi^2\;e^{\xi_X}~~,~~
\eta_{\frac{n}{2}}=e^{-\xi_X}\;\psi_{\frac{n}{2}}
~~,~~\phi=e^{-\xi_X}\;\psi~.
\end{equation} 
The differential equations determining the $\psi_{\frac{n}{2}}$ are
explicitly
\begin{eqnarray} \label{3035}
\partial_\xi^2 \psi_1     &=& c_1 (\alpha\xi_X+\gamma) 
~~~,~~~\gamma=\beta-\alpha~,
\\
\partial_\xi^2 \psi_{\frac{3}{2}} &=& c_{\frac{3}{2}}(\alpha\xi_X+\gamma) ~,
\nonumber\\
\partial_\xi^2 \psi_2     &=& [-1+c_1(1-\partial_\xi)] \psi_1 
+ c_2(\alpha\xi_X+\gamma) ~,
\nonumber\\
\partial_\xi^2 \psi_{\frac{5}{2}} &=& 
[-{\scriptstyle\frac{3}{2}}+c_1(1-\partial_\xi)] \psi_{\frac{3}{2}} 
+ c_{\frac{3}{2}}(1-\partial_\xi) \psi_1 + \nonumber\\ 
& & \hspace*{3.5cm} +  c_{\frac{5}{2}}(\alpha\xi_X+\gamma) ~,
\nonumber\\
&& \qquad\mbox{ etc., generally:}
\nonumber\\
\partial_\xi^2 \psi_{\frac{n}{2}} &=& 
[-{\scriptstyle\frac{n-2}{2}}+c_1(1-\partial_\xi)]\psi_{\frac{n-2}{2}} 
\nonumber\\
&& + \sum_{m=3}^{n-2} c_{\frac{m}{2}}
(1-\partial_\xi) \psi_{\frac{n-m}{2}}
+ c_{\frac{n}{2}}(\alpha\xi_X+\gamma) ~,
\nonumber
\end{eqnarray}
where we have omitted exponentially small corrections of order
$e^{-\xi_X}$ in the inhomogeneities on the r.h.s.\ of the equations.
The conditions (\ref{309}) and (\ref{3010}) on $\eta$ do not influence
the solution in the leading edge.

The equations (\ref{3035}) are easily solved. For $\psi=e^{\xi_X}\phi$
we find in the region $\xi_X \gg1$
\begin{eqnarray} \label{3036}
\psi = e^{\xi_X}\Phi^*  + \sum_{n=2}^\infty
\frac{\psi_{\frac{n}{2}}}{t^{n/2}}&=&
\\
 \alpha\;\xi_X &+& \beta + \nonumber\\
+ \frac{c_1\alpha\;\xi_X^3}{3!\;t} &+& \frac{c_1\gamma\;\xi_X^2}{2!\;t}
+O \left(\frac{\xi_X}{t}\right)
\nonumber\\
&+&\frac{c_{\frac{3}{2}}\alpha\;\xi_X^3}{3!\;t^{3/2}}
+O \left(\frac{\xi_X^2}{t^{3/2}}\right)
\nonumber\\
+ \frac{c_1(c_1-1)\alpha\;\xi_X^5}{5!\;t^2} &+&
\frac{c_1[(c_1-1)\gamma-c_1\alpha]\;\xi_X^4}{4!\;t^2} +\ldots
\nonumber\\
&+&\frac{c_{\frac{3}{2}}(2c_1-{\scriptstyle \frac{3}{2}})\alpha\;\xi_X^5}
{5!\;t^{5/2}} +\ldots
\nonumber\\
+ \frac{c_1(c_1-1)(c_1-2)\alpha\;\xi_X^7}{7!\;t^3} &+& \ldots
\nonumber
\end{eqnarray}
Obviously, for $\xi_X\ge\sqrt{t}$, the expansion is not properly
ordered in powers of $1/\sqrt{t}$ anymore, since, e.g., $\xi_X^3/t$
eventually will become larger than $\xi_X$. A quick inspection of
(\ref{3036}) shows that we can continue to work in an $1/\sqrt{t}$
expansion if we use the variable $z=\frac{\xi_X^2}{4t}$ (\ref{308})
instead of $\xi_X$.  The expression (\ref{3036}) can be identified with
\begin{eqnarray} \label{3037}
\psi &=& \sqrt{t}\;
\alpha\left((4z)^{1/2}+\frac{c_1(4z)^{3/2}}{3!}
+\frac{c_1(c_1-1)(4z)^{5/2}}{5!}\right.
\nonumber\\
&&\qquad\qquad\quad\left.+\;\frac{c_1(c_1-1)(c_1-2)(4z)^{7/2}}{7!}+\ldots\right)
\nonumber\\
&&+\;t^0\;\left(\beta+\frac{c_1(\beta-\alpha)(4z)}{2!}
+\frac{c_{\frac{3}{2}}\alpha(4z)^{3/2}}{3!} + O(z^2)\right)
\nonumber\\
&&+\;O(1/\sqrt{t})
\end{eqnarray}
This resummed expansion anticipates the crossover to the expansion in
$z$ and $1/\sqrt{t}$ below  for $\xi_X, t \gg 1$,
$z=\xi_X^2/4t =O (1)$, which will fix the coefficients $c_1$,
etc. Note that for  $c_1=-3/2$,  terms of order $\sqrt{t}$  sum up  to
$\alpha \sqrt{4zt}\; e^{-z}= \xi_X 
e^{-\xi_X^2/4t}$, which is, in dominant order, the behavior already
anticipated in Section \ref{S311}.

Instead of resumming the interior expansion explicitly, it is  much
more transparent to write an expansion directly in terms of 
powers of $1/\sqrt{t}$ and the similarity variable $z$ of the diffusion
equation. This approach, which amounts to a matching procedure,
is the subject of the next subsection.

\subsection{Analysis of the leading edge} \label{S34}

We now take up the analysis of the leading edge region $\xi_X \ge
O(\sqrt{t})$ in the case
that the initial conditions are sufficiently steep, so that for
$\psi=\phi\;e^{\xi_X}$
\begin{equation} \label{3038}
  \lim_{\xi_X \to\infty} \psi(\xi_X,t)< e^{-\delta \xi_X}~,~~~\delta >0~.
\end{equation} 
Note that according to the discussion of Section \ref{S25}, this
condition holds at any finite time $t<\infty$ if it is obeyed initially at
$t=0$.

We have already argued in Sections \ref{S25} and \ref{S311} that the
asymptotic profile of the leading edge might be expected to be
somewhat like a Gaussian in $\xi_X$ and $t$ times a Hermite polynomial.
Also the resummation of the interior front solution suggests such a
form for large $\xi_X$. We now investigate this expansion more
systematically, and will show that it actually takes the form of a
Gaussian times a generalization of Hermite polynomials, namely
confluent hypergeometric functions \cite{abra}.

In passing, we stress that the arguments from
\ref{S311} can be compared directly to our calculation here  only to lowest order, because
we now work with the coordinate $z=(x-2t-X(t))^2/(4t)$, while we
presented our earlier intuitive arguments in the coordinate
$z^*=(x-2t)^2/(4t)$. Of course, one can also set up a systematic
expansion in the latter coordinate $z^*$, but this requires the
introduction of logarithmic terms for a proper matching to the
interior part of the front. Working throughout in the shifted frames
$\xi_X=x-2t-X(t)$ or $z$  avoids this altogether.

In the coordinates $\xi_X$ and $t$, the equation of motion for $\psi$ in
the leading edge region is (Recall that the earlier leading edge representation in
(\ref{2058}) was in the frame $\xi=x-2t$)
\begin{equation} \label{3039} 
\partial_t\psi=\partial_\xi^2\psi+\dot{X}(\partial_\xi-1)\psi+o(e^{-\xi_X})~.
\end{equation} 

The differential operators transform under change of coordinates to
$z=\xi_X^2/(4t)$ and $t$ as
\begin{equation} \label{3040} 
\partial_t\big|_\xi = \partial_t\big|_z - \frac{z}{t}\partial_z\big|_t
~,~\partial_\xi\big|_t=\sqrt{\frac{z}{t}}\;\partial_z\big|_t~.
\end{equation} 
Motivated by the form (\ref{2054}) and the discussion of Section
\ref{S311}, we extract the Gaussian $e^{-\xi_X^2/(4t)}=e^{-z}$ from
$\psi$ by writing:
\begin{equation} \label{3041}
  \psi(\xi,t) = e^{-z}\;G(z,t)~~,~~z=\frac{\xi_X^2}{4t}~.
\end{equation}
This extraction also allows us to make contact later with functions
tabulated in \cite{abra}. The dynamical equation (\ref{3039}) is
equivalent to the equation for $G$:
\begin{eqnarray} \label{3042}
  \lefteqn{\left[z\partial_z^2+\left(\frac{1}{2}-z\right)\partial_z
    -\frac{1}{2}-t\partial_t-c_1\right]\;G=} \nonumber\\ && ~~~~~~~~ =
  \left[(\dot{X}t-c_1)+\dot{X}\sqrt{t}\sqrt{z}(1-\partial_z)\right]\;G
    ~.
\end{eqnarray}
The equation is organized such, that the differential operators of
order $t^0$ are on the l.h.s.\ of the equation, while the r.h.s.\ has
the operators of order $t^{-1/2}$ and smaller.

In analogy to our earlier expansion (\ref{3015}), we now make an
ansatz for $G$ in powers of $1/\sqrt{t}$ times functions of $z$.  A
glimpse at the form of the interior shape expanded towards the leading
edge (\ref{3037}) tells us, that the expansion should start with the
order $\sqrt{t}$. We write
\begin{equation} \label{3043} 
  G(z,t) = \sqrt{t}\;g_{\frac{-1}{2}}(z) + g_0(z)
  +\frac{g_{\frac{1}{2}}(z)}{\sqrt{t}} + \ldots ~.
\end{equation} 
Insertion of this ansatz into (\ref{3042}) again results in a
hierarchy of ordinary differential equations, that can be solved
successively:
\begin{eqnarray} \label{3044}
  \lefteqn{ \left[z\partial_z^2+\left(\frac{1}{2}-z\right)\partial_z
    -1-c_1\right]\;g_{\frac{-1}{2}} =0~, \hspace*{1cm} }
\\ \label{3045}
&& \nonumber\\ \lefteqn{
  \left[z\partial_z^2+\left(\frac{1}{2}-z\right)\partial_z
    -\frac{1}{2}-c_1\right]\;g_0 =}
\\ \label{3046}
&& ~~~ = \left[c_{\frac{3}{2}} +
c_1\sqrt{z}(1-\partial_z)\right]\;g_{\frac{-1}{2}}~, \nonumber\\ &&
\nonumber\\ \lefteqn{
  \left[z\partial_z^2+\left(\frac{1}{2}-z\right)\partial_z
    -c_1\right]\;g_{\frac{1}{2}} =}
\\ 
&& ~~~ = \left[c_2 +
c_{\frac{3}{2}}\sqrt{z}(1-\partial_z)\right]\;g_{\frac{-1}{2}}
\nonumber \\&& ~~~~~~~~~~~~~~~~
+\left[c_{\frac{3}{2}} + c_1\sqrt{z}(1-\partial_z)\right]\;g_0~,
\nonumber
\end{eqnarray}
etc. The general solution of the homogeneous equations with two
constants of integration $k_{\frac{n}{2}}$ and $l_{\frac{n}{2}}$ can
be found in \cite{abra}, they are confluent hypergeometric functions. 
These special solutions  $g_{\frac{n}{2}}^{sp}$ of 
the inhomogeneous can also generally be expressed in terms of double
integrals over known functions, as is  discussed in Appendix 
\ref{g0app}. Below we will, however, just guess the series expansion
of the special function $g_0^{sp}$ we need.  We write the general solution as
\begin{eqnarray} \label{3047}
  g_{\frac{n}{2}}(z) &=& g_{\frac{n}{2}}^{sp}(z) +
  k_{\frac{n}{2}}\;M\left(c_1+\frac{1-n}{2}, \frac{1}{2}, z \right)
  \nonumber\\ && ~~~~+\; l_{\frac{n}{2}}\;\sqrt{z}\;
  M\left(c_1+\frac{2-n}{2}, \frac{3}{2}, z \right)~,
\end{eqnarray}
where the functions $M(a,b,z)$ can be expressed by the Kummer series
\cite{abra}
\begin{eqnarray} \label{3048}
M(a,b,z)&=&1+\frac{a\;z}{b}+\frac{a(a+1)\;z^2}{b(b+1)\;2!} +
\ldots+\frac{(a)_nz^n}{(b)_n n!}+\ldots~, \nonumber\\ \mbox{with}&&
(a)_n=\prod_{k=1}^n (a+k-1)=\frac{\Gamma(a+n)}{\Gamma(a)}~.
\end{eqnarray}
Just as in the integration of the interior shape in Section \ref{S32},
there are two constants of integration to be determined in every
solution $g_{\frac{n}{2}}$. In addition, however, the $c_i$ are not
just parameters of the equations as in Section \ref{S32}, but they now
have to be determined also. The conditions we use to determine these
three constants per equation, are now (\ref{3037}) and (\ref{3038}) in
analogy to the two conditions (\ref{309}) and (\ref{3010}) for the
$\eta_{\frac{n}{2}}$: $(i)$ The solution $g_{\frac{n}{2}}$ has to
agree with the expansion of the interior towards the leading edge
(\ref{3037}) for $z \ll 1$. Then the coefficients of $z^0$ and
$z^{1/2}$ in (\ref{3037}) determine the constants of integration
$k_{\frac{n}{2}}$ and $l_{\frac{n}{2}}$.  $(ii)$ The transients at any
finite time have to be sufficiently steep in the sense that they obey
(\ref{3038}) at any finite time $t$. Because of the form the
expansions (\ref{3042}) and (\ref{3044}), we require that each term
$g$ in the expansion diverges for  $z\gg 1$ at most as a power law of
$z$, {\em not} exponentially as $e^z$. In addition, $\psi = e^{z} G$
should not diverge as $t \to \infty$, but approach a time independent
limit. This gives another condition on the constants of integration,
that can be obeyed only for a particular choice of
$c_{\frac{n+3}{2}}$.  With these choices of the constants, the small
$z$ expansion of $\psi=e^{-z}G$ from (\ref{3041}) and (\ref{3043})
becomes identical with the interior shape expanded towards the leading
edge (\ref{3037}).

We will solve the first two equations (\ref{3044}) and (\ref{3045}) 
explicitly, since they determine the
universal terms of the velocity correction $\dot{X}$. In particular,
the solution for $g_{\frac{-1}{2}}$ (\ref{3044}) will connect to our
qualitative discussion of the leading $1/t$ velocity convergence term
(\ref{305}) in Section \ref{S311}. Eq.\ (\ref{3045}) will give the
universal subleading term\footnote{Don't 
confuse the expansion in $1/\sqrt{t}$ of the velocity in
  (\ref{305}) or (\ref{3014}) with the denominators in (\ref{302}) and
  (\ref{303}).  These powers of $1/\sqrt{t}$ in the
  $\xi^*$-representation are absorbed into the $X(t)$ of $\xi$ in the
  $\xi$-representation, as is sketched in (\ref{304}).}
of order $1/t^{-3/2}$.

Let us now start with the solution of the homogeneous leading order 
equation (\ref{3044}), where
$g_{\frac{-1}{2}}^{sp}(z)=0$. The constants of integration are fixed
by (\ref{3037}) as $k_{\frac{-1}{2}} = 0$ and $l_{\frac{-1}{2}}=
2\alpha$.  Therefore $g_{\frac{-1}{2}}(z)$ is after matching to the
interior
\begin{equation} \label{3049}
  g_{\frac{-1}{2}}(z)=
  2\alpha\;\sqrt{z}\;M\left(c_1+\frac{3}{2},\frac{3}{2},z\right)~.
\end{equation}
In order to analyze, how $c_1$ is determined by the matching and the
requirement that all transients are exponentially
steeper for $\xi_X \to \infty$ than
the asymptotic profile, we first recall the large $z$ behavior of
Kummer functions $M(a,b,z)$ \cite{abra}: For positive $b$ 
each term of the series
(\ref{3048}) is finite.  For $a$ not zero nor a negative integer, the
series is infinite. For $a$ zero or a negative integer $a=-n$, the
series is finite, since all terms from order $z^{n+1}$ on contain the
factor $(a+n)=0$, and for $b=1/2$ or 3/2, these finite polynomials are
Hermite polynomials. The large $z$ asymptotics of $M(a,b,z)$ for
positive $b$ is
\begin{equation} \label{3050} 
  M(a,b,z) \stackrel{z\to\infty}{\sim}\left\{
\begin{array}{ll}\frac{\Gamma(b)}{\Gamma(a)}\; z^{a-b}\;e^z 
  & \mbox{ for } -a\notin{\cal N}_0~,\\ 
  \frac{(a)_{|a|}z^{|a|}}{(b)_{|a|} {(|a|)}!} & \mbox{ for }
  -a\in{\cal N}_0~,
\end{array}\right.\end{equation} 
where ${\cal N}_0 $ denotes zero plus the positive integers.
If one inserts  (\ref{3050}) into (\ref{3049}) ones find for
$\xi_X \gg\sqrt{t}$
\begin{equation} \label{3051} 
  \phi\propto \xi_X\;e^{-\xi_X} 
\left\{\begin{array}{ll} 
  (\xi_X^2/t)^{\:c_1}                   &  -c_1-3/2\notin {\cal N}_0\\ 
  (\xi_X^2/t)^{-c_1-3/2}\;e^{-\xi_X^2/(4t)}&  -c_1-3/2\in {\cal N}_0
\end{array}\right. .
\end{equation} 
For $-c_1-3/2$ not a positive integer, we see from (\ref{3051}) that
$\psi(\xi_X,t)$ does not converge exponentially fast to zero, in
violation of the condition (\ref{3038}). Accordingly, for
so-called  sufficiently steep initial conditions that obey
(\ref{3038}) we
conclude that  $c_1+3/2$ has to be zero or a negative
integer.  Possible solutions are
\begin{equation} \label{3052} 
\begin{array}{ll}
  c_1=\frac{-3}{2}~~,~~&g_{\frac{-1}{2}}(z)=2\alpha\;\sqrt{z}~,\\ 
  c_1=\frac{-5}{2}~~,~~&g_{\frac{-1}{2}}(z)=2\alpha\;\sqrt{z}\;
  \left(1-\frac{2z}{3}\right)~,\\ 
    c_1=\frac{-7}{2}~~,~~&g_{\frac{-1}{2}}(z)=2\alpha\;\sqrt{z}\;
    \left(1-\frac{4z}{3}+\frac{4z^2}{15}\right)~,
\end{array}
\end{equation} 
etc., with the $g_{\frac{-1}{2}}$ given by Hermite polynomials. 

There are two ways to argue, why generically $c_1=-3/2$ is the
appropriate solution. $(a)$ If the initial condition is always
non-negative, e.g., because $\phi$ is a density, the transient may not
have nodes, so $c_1=-3/2$ is the only possible solution.  
$(b)$ If one can create a front with nodes whose leading edge after
some evolution is the superposition of the solutions in (\ref{3052}), 
the solution with $c_1=-3/2$ propagates quickest, so the other 
contributions will be convected to the back, and the $c_1=-3/2$ 
solution will dominate at large times \cite{willem2}. This argument 
coincides with the argument from Section \ref{S23}, that fronts with 
nodes generically are not attractors for the long time dynamics for 
the nonlinear diffusion equation (\ref{101}). A similar reasoning 
for the leading edge region can be developed from the arguments
in Section \ref{S64}.
Furthermore we have checked various initial conditions
with nodes numerically and we have found, that either the node gets
stuck behind the evolving front or moves away to $\xi\to\infty$ with
velocity larger than $v^*$, leaving in both cases a leading edge of
the front behind that develops with $c_1=-3/2$. We thus find for
initial conditions (\ref{3038}) steeper than $\Phi^*$ generically
\begin{equation} \label{3053} 
  c_1=\frac{-3}{2}~~,~~g_{\frac{-1}{2}}(z)=2\alpha\;\sqrt{z}~.
\end{equation} 
This solution is identical with the order $\sqrt{t}$ of $\psi e^z$
with $\psi$ from (\ref{3037}). For $\phi$ we find in the region
$\xi_X \gg1$ linearizable about the unstable state in leading order
\begin{eqnarray} \label{3054} 
\phi&=&\alpha\xi_X\;e^{-\xi_X-\xi_X^2/(4t)}\left(1+O(1/ \xi_X) +O(1/
\sqrt{t})\right) 
\\ \label{3055}
\xi_X&=&x-v^*t+\frac{3}{2}\;\ln t+O(1/\sqrt{t})~,
\end{eqnarray} 
consistent with the arguments from Section \ref{S311}.

Integration of $g_0$ now gives the subleading universal terms, which
are $O(1/\sqrt{t})$ in (\ref{3054}) and (\ref{3055}).
Insertion of (\ref{3053}) into (\ref{3045}) results in
\begin{equation} \label{3056}
  \left[z\partial_z^2+\left(\frac{1}{2}-z\right)\partial_z +1\right]
    g_0 = 2\alpha \left(\frac{3}{4}+c_{\frac{3}{2}}\sqrt{z}
    -\frac{3}{2}\;z\right)~.
\end{equation} 
We now can follow Appendix \ref{g0app} for the general solution of the
inhomogeneous equation, or we rather can guess a special solution
of the inhomogeneous equation by noting that the function
\begin{equation} \label{3058} 
  F_N(z)=\sum_{n=N}^\infty \frac{(1)_{n-2}\;
    z^n}{\left(\frac{1}{2}\right)_n\;n!}
\end{equation} 
is proportional to a truncated Kummer series $M(-1,\frac{1}{2},z)$
(\ref{3048}) and solves
\begin{equation} \label{3059} 
  \left[z\partial_z^2+\left(\frac{1}{2}-z\right)\partial_z +1\right]
    F_N(z) = \frac{z^{N-1}}{\left(\frac{1}{2}\right)_{N-1}\;(N-1)}~,
\end{equation} 
The special solution of the inhomogeneous
equation (\ref{3056}) is then easily seen to be
\begin{equation} \label{3057} 
  g_0^{sp}(z)=2\alpha\;\left(\frac{3}{4}+2 c_{\frac{3}{2}}\sqrt{z}
  -\frac{3}{4}\;F_2(z)\right)~.
\end{equation} 
Upon comparing (\ref{3058}) to (\ref{3048}) and (\ref{3050}), one
finds
\begin{equation} \label{3060} 
  g_0^{sp}(z) \stackrel{z\to\infty}{\sim} -\;\frac{3}{2}\;\alpha
  \;\sqrt{\pi}\;z^{-3/2}\;e^z~.
\end{equation} 
The general solution (\ref{3047}) of (\ref{3056}) is thus
\begin{eqnarray} \label{3061} 
  g_0(z)&=&g_0^{sp}(z) + k_0 \;(1-2z) +
  l_0\;\sqrt{z}\;M\left(\frac{-1}{2},\frac{3}{2},z\right) \nonumber\\ 
  &\stackrel{z\ll1}{=}& \left(\frac{3\alpha}{2}+k_0\right)
  +\left(4\alpha c_{\frac{3}{2}}+l_0\right) \sqrt{z}+O(z)
\\ \label{3062}
&\stackrel{z\to\infty}{\sim}& -\left(\frac{3}{2}\;\alpha \;\sqrt{\pi}+
\frac{l_0}{4}\right)\; z^{-3/2}\;e^z ~,
\end{eqnarray} 
where we have used (\ref{3050}) and (\ref{3060}) for the large $z$
asymptotics.  Compare now the small $z$ expansion (\ref{3061}) to
(\ref{3037}).  One obviously has to identify
\begin{equation} \label{3063}
  \frac{3\alpha}{2}+k_0=\beta~~,~~ 4\alpha c_{\frac{3}{2}}+l_0=0~.
\end{equation}
If $g_0$ would decay asymptotically as $z^{-3/2}\;e^z$ for large $z$
(\ref{3062}), the subleading contribution of order $1/\sqrt{t}$ in
$\phi$ (\ref{3054}) would not decay like a Gaussian $e^{-\xi_X^2/(4t)}$
as the leading order term does, but it would decay algebraically like
$\xi_X^{-3}(4t)^{3/2}$. This would destroy the ordering of our expansion
(\ref{3043}) and lead to a divergence of $\psi$ for $t\to \infty$.
Thus the coefficient of the leading order term
$z^{-3/2}\;e^z$ in $g_0$ (\ref{3062}) has to vanish:
\begin{equation} \label{3064}
  \frac{3}{2}\;\alpha \;\sqrt{\pi}+ \frac{l_0}{4}=0~.
\end{equation}
Eqs.\ (\ref{3063}) and (\ref{3064}) fix all constants $k_0$, $l_0$ and
$c_{\frac{3}{2}}$. The velocity correction of order $1/t^{3/2}$ is
\begin{equation} \label{3065} 
  c_{\frac{3}{2}}=\frac{3\sqrt{\pi}}{2}~,
\end{equation} 
and the analytic solution for $g_0(z)$ is
\begin{eqnarray} \label{3066} \label{pre6}
  g_0(z)=\beta\;(1-2z) &+&3\alpha\;\left(z-\frac{F_2(z)}{2}\right)\\ 
  &+&6\alpha\;\sqrt{\pi\:z}\;
  \left(1-M\left(\frac{-1}{2},\frac{3}{2},z\right)\right)~, \nonumber
\end{eqnarray}
with $\alpha$ and $\beta$ the coefficients of the asymptotic leading
edge shape $\Phi^*(\xi)=(\alpha\xi+\beta)\;e^{-\xi}$ for $\Phi^*\ll1$.
Note, that the subleading term $\beta$ contributes only the rather
trivial $(1-2z)$ term, while the coefficient of the leading $\alpha$
contains all nontrivial terms. The result (\ref{3065}) and
(\ref{3066}) reproduces the order $t^0$ in (\ref{3037}) identically.

We summarize the results obtained from the analysis of the leading
edge: The appropriate coordinate system is $\xi_X =x-v^*t-X(t)$, and the
universal velocity correction is given by
\begin{equation} \label{3067}
  \dot{X}=-\:\frac{3}{2t}\left(1-\sqrt{\frac{\pi}{t}}\:\right)
  +O\left(\frac{1}{t^2}\right)~.
\end{equation} 
The shape in the leading edge, where $\phi\ll1$, is given in terms of
the variables $\xi_X$ and $t$ by
\begin{eqnarray} \label{3068}  \label{pre7}
  \phi(\xi_X,t)&=&e^{-\xi_X-\xi_X^2/(4t)}\;G\left(\frac{\xi_X^2}{4t},t\right)
\\
&=&e^{-\xi_X-\xi_X^2/(4t)}\;\left(\alpha\xi_X+g_0\left(\frac{\xi_X^2}{4t}\right)+  \right. \\ \nonumber & & \hspace*{2cm} 
\left. + \frac{1}{\sqrt{t}} \; g_{\frac{1}{2}}\left(\frac{\xi_X^2}{4t}\right) +\ldots \right)~,
\nonumber
\end{eqnarray} 
with $g_0(z)$ from (\ref{3066}).

Eqs.\ (\ref{pre6}) -- (\ref{pre7}) is the second part of our final
result, valid in the leading edge of the front, where $\phi\ll 1$. It
complements our earlier result (\ref{pre4}), valid in the interior of
the front, with functions $\kappa_v$ from (\ref{pre2}) and $\rho_v$
from (\ref{3030}).

\subsection{Summarizing remarks} \label{S35}
Let us end this section by putting these analytical results into
perspective:

1)  The requirement that the leading edge remains steeper than the
asymptotic profile $\Phi^*$ at any finite time together with the
requirement that it converges to $\Phi^*$ as $t\to \infty$ determines
the 
velocity convergence constants $c_{\frac{n}{2}}$. These constants are
thus determined in the leading edge by the initial conditions. They
are just parameters in the equations for the interior (\ref{3016}) --
(\ref{3020}).

2) The leading order velocity correction $c_1$ reproduces Bramson's
result \cite{bramson}, which he derived through solving the
(nonlinear) diffusion equation with probabilistic methods. The
universal subdominant $1/t^{3/2}$ is new.

3) According to our discussion in connection with the interior
expansion, $g_{\frac{1}{2}}$ and $c_2$ should be termed non-universal,
because the change from $1/\sqrt{t}$ to $1/\sqrt{t-t_0}$ in the
asymptotic expansion about $t\to\infty$ changes these terms. As for
(\ref{3031}), we conclude that at least parts of these terms depend
on initial conditions and are therefore non-universal.

4) We stress once more that the full expansion is only asymptotic in 
$1/\sqrt{t}$, but that the prefactors of the $1/t$ and $1/t^{3/2}$
terms are exact.

5) The leading edge expansion is an intermediate asymptotics
in $z$ valid for $1\ll z \ll \sqrt{t}$ or $\sqrt{t} \ll \xi_X \ll t$,
resp. Above, we extensively made use of the cross-over to the interior 
expansion for $z\ll1$. Let us now look into the break-down for 
$z\ge O(\sqrt{t})$, i.e., for $\xi_X\ge O(t)$.
This second breakdown immediately follows from inserting into (\ref{3043})
our results $g_{\frac{-1}{2}}(z)=O(\sqrt{z})$ and $g_0(z)\ge O(z)$ 
(in fact $g_0(z)=O(z\ln z)$ according to Appendix \ref{g0app}).
This new crossover actually needs to exist in view of our
discussion in Section \ref{S251}: The steepness $\lambda$ is conserved
for $x\to\infty$ for all times $t<\infty$. It will retain the information 
about the precise initial condition. This region of conserved steepness
at $\xi_X>O(t)$ crosses over to the universal Gaussian leading edge region
for $\xi_X<O(t)$, which determines the universal relaxation behavior 
as discussed above. The region of conserved stepness $\lambda$ at 
$\xi_X>O(t)$ has no further consequence for the dynamics, if the initial
steepness is only $\lambda>\lambda^*$. It will disappear towards
$\xi_X\to\infty$ by outrunning the leading edge region with an 
approximately constant speed. This scenario is sketched in Fig.\ 2.

6) Our result is valid in the pulled regime but it does not apply at
the bifurcation point from the pulled to the pushed regime.  For
nonlinearity (\ref{103}) this means, that the analysis applies for
$\epsilon >3/4$\cite{vs2}.  Only then $\Phi^*(\xi)\propto\xi \;e^{-\xi}$,
which is one of the essential ingredients of our asymptotic analysis.
For $\epsilon <3/4$, the front is pushed, and convergence is
exponential, as discussed in Sections \ref{S23} and \ref{S24}. For
$\epsilon=3/4$, precisely at the pushed/pulled transition,
$\Phi^*(\xi)\propto e^{-\xi}$. In this case, convergence is still
algebraic, but the analysis of this chapter does not apply exactly.
The convergence analysis, however, can be set up along the same lines.
As shown in Appendix \ref{pu/pu} we then get instead of (\ref{3067})
\begin{equation} \label{3067b}
  \dot{X}=-\:\frac{1}{2t}\left(1- \frac{1}{2} \sqrt{\frac{\pi}{t}}\:\right)
  +O\left(\frac{1}{t^2}\right)~.
\end{equation} 
Note that the factor 3/2 of the $1/t$ term is replaced by 1/2 at the
bifurcation point. Along the lines of the arguments of Section
\ref{S311} this can be understood simply from the fact that at the
bifurcation point the asymptotic behavior of $\Phi^* $ is as
$\Phi^*(\xi) \sim e^{-\xi}$, {\em not} as $\xi e^{-\xi}$, and hence
that the simple Gaussian leading edge solution $ e^{-\xi
  -\xi^2/4t}/\sqrt{t}$ matches to the asymptotic front profile in leading
order. However, the velocity relaxation (\ref{3067b}) at
the pushed/pulled transition does contain a universal subleading term of
order $1/t^{3/2}$ that is absent in the relaxation of the linear
equation (\ref{2055}).

7) Up to now we excluded the particular initial conditions $\phi(x,0)
\simeq x^{-\nu} e^{-x}$ from our discussion, since they are neither
sufficiently steep nor flat according to our definition. It is amusing 
to see that also such initial conditions can be treated with our
approach. For sufficiently steep initial conditions, we discarded the
case that $-c_1-3/2$ would be different from an integer or zero after 
Eq.\ (\ref{3051}), because it would violate the exponential bound
(\ref{3038}) for large $\xi_X$. However, for the above particular 
initial conditions, the asymptotic behavior (\ref{3038}) 
is replaced by $\psi\approx\xi_X^{-\nu}$ for $\xi_X\gg1$. 
For any $\nu<2$ one concludes immediately from Eq.\ (\ref{3051}) that 
\begin{equation}
\phi(x,0)  \simeq x^{-\nu}e^{-x}~\Longrightarrow ~ v(t) = 2-
\frac{\nu+1}{2t} + \ldots ~,
\end{equation}
a result also derived by Bramson \cite{bramson}. In other words, in
the case in which the initial conditions are intermediate between
sufficiently steep and flat, the prefactor $c_1$ {\em does} depend on
the initial conditions and may even change sign, but the relaxation 
is still power law like. To get  the next order term in the expansion 
for these special initial conditions, our expansion will probably have 
to be generalized. We will comment on this in Section \ref{S6}.

\end{multicols}

\newpage

\begin{multicols}{2}

\section{Simulations of pulled fronts in the nonlinear diffusion
equation} 
\label{S4}

In this section, we present simulation data for fronts 
in the nonlinear diffusion equation 
$\partial_t\phi=\partial_x^2\phi+f(\phi)$ (\ref{101})
propagating into the unstable state $\phi=0$, 
and compare these with our analytical predictions. 
In particular, we thoroughly investigate fronts 
with the nonlinearity $f(\phi)=\phi-\phi^3$, 
so that the equation becomes
\begin{equation} \label{403}\label{402}
\partial_t\phi=\partial_x^2\phi+\phi-\phi^3.
\end{equation}
This equation forms pulled fronts with $v^*=2$ and $\lambda^*=1=D$, 
if the initial conditions are sufficiently steep.
As an example of a nonlinearity allowing for both pushed and pulled 
fronts, we also present data for $f(\phi)=\epsilon\phi+\phi^3-\phi^5$
for $\epsilon = 0.56$ and 0.96.

As an initial condition, we here always choose 
\begin{equation}
\label{405}
\phi(x,0)=\frac{1}{1 + e^{\lambda_{init} (x-x_0)}} \to\left\{
\begin{array}{ll} e^{-\lambda_{init} (x-x_0)} & x\to\infty\\ 
                  1 & x\to-\infty\end{array}
\right.~.
\end{equation}
According to our analytical results, all initial conditions with 
initial steepness $\lambda^*<\lambda_{init}\le\infty$ exhibit the same
universal relaxation behavior asymptotically as $t\to\infty$,
if the front is pulled. We indeed do find this in our simulations. 
Below we only present simulations for $\lambda_{init}=10$. 

The section is organized into a discussion of the specific numerical 
features of pulled fronts (Section \ref{S41}), the presentation of 
the raw simulation data for nonlinearities $f(\phi)=\phi-\phi^3$ and 
$f(\phi)=\epsilon\phi+\phi^3-\phi^5$ (Section \ref{S42}),
and a detailed comparison of the simulations for
(\ref{403}) with the analytical predictions (Section \ref{S43}).

\subsection{Numerical features specific to pulled fronts} \label{S41}

To integrate a given initial condition $\phi(x,0)$ forward 
in time $t$ for a nonlinear diffusion equation, we use a semi-implicit 
algorithm which is explained in detail in Section \ref{S566},
Eq.\ (\ref{5098}). When running the program, we have to choose a 
spatial and temporal discretization $\Delta x$ and $\Delta t$, 
a system size $0 \le x \le L$, 
and a position $x_0$ of the initial condition within the system.
Comparing results for different parameters $\Delta x$, $\Delta t$, 
$L$, and $x_0$ to each other and to the analytical predictions
in the extreme precision of often better than 6 significant figures, 
we find two features specific to the particular dynamic mechanism of 
pulled fronts:

\subsubsection{Effect of finite difference code} \label{S411}

The numerical results of the simulation depend of course on the step 
sizes $\Delta x$ and $\Delta t$ of the finite difference code. 
In fact, in Section \ref{S566} we will have collected all analytical 
tools to calculate the corrections to $v^*=2$, $\lambda^*=1$ and $D=1$, 
that depend on the numerical
integration scheme and on the parameters $\Delta x$ and $\Delta t$.
All data presented here are derived for $\Delta x = 0.01 = \Delta t$.
For a pulled front in a nonlinear diffusion equation solved with
a semi-implicit scheme, our analytical prediction (\ref{semi-im})
yields $v^*=2.000075$, $\lambda^*=0.999954$, and $D=1.00035$.

\subsubsection{Effect of finite system size} \label{S412}

In contrast to a pushed front, the final $t\to\infty$ relaxation of 
a pulled front very sensitively depends on system size $L$ and
front position $x_0$. This effect is closely related to the
pulled mode of propagation and the breakdown of the linear stability 
analysis. Because the half-infinite space $x\gg1$ of the leading edge
dominates the dynamics, the very long time dynamics of the front 
is sensitive to the region at
$x\gg1$, even if though there $\phi\ll1$. More precisely,
the diffusive spreading of the linear perturbation as in Eq.\
(\ref{303}) or (\ref{pre7}) that determines the speed, 
strongly depends on the boundary
conditions at $\xi_X=x-v^*t-X(t)=O(\sqrt{4Dt}\;)$.

For this reason, we shift the front back
to its original position $x_0$ within the system after every time step 
$t_2-t_1=1$. This eliminates the $x$-interval
$0\le x\le x_{shift}\approx v^*$ on the back side of the front 
from our data, while a new $x$-interval $L-x_{shift}\le x\le L$ 
has to be created. 
One might assume, that this procedure yields good results
for integration times $T$ up $T=O\left((L-x_0)^2/(4D)\right)$
because of the diffusive nature of the spreading.
However, the precision noticeably breaks down earlier because of
the arbitrariness of the newly created $x$-interval 
$L-x_{shift}\le x\le L$ in the shift process. Filling this region with
the constant $\phi(x)=\phi(L-x_{shift})$ creates a
flat initial condition, and the front accelerates
beyond $v^*$ for sufficiently long times.
We therefore use $\phi(x)\equiv0$ in this region.
The observed velocities $v_\phi(t)$ for $L<\infty$
then always will stay below those in the infinite system 
$L\to\infty$. The simulations in the finite system are close to
those in the infinite system up to times $T=O\Big((L-x_0)/v^*\Big)$.

\subsection{Simulation data} \label{S42}

\subsubsection{$f(\phi)=\phi-\phi^3$: pulled fronts} \label{S421}

As an example, we will extensively discuss simulations of
Eq.\ (\ref{403}). We present data with initial
conditions (\ref{405}) and $\lambda_{init}=10$, where the initial
condition is located at $x_0=100$ in a system of size $L=1000$.
According to our estimate above, the simulations then should
be reliable up to times $t$ of order $(L-x_0)/2=450$.
We present data up to $t=400$. The data from this simulation is 
evaluated in a sequence of figures showing increasing detail and 
precision.

Fig.\ 1 already showed the temporal evolution of a sufficiently steep
initial condition under the equation of motion (\ref{402}). It shows both, 
the total displacement of the front, and the evolution of the front shape.
We now choose different presentations that show these
two different aspects of the dynamics separately and in higher precision.

Let us first study the evolution of the front shape:
In Fig.\ 4, we present $\phi(\xi_X,t)$ as a function of $\xi_X$, 
where $\xi_X=x-v^*t-X(t)$, Eq.\ (\ref{pre1}), is adjusted such 
that $\phi(0,t)=1/2$ (\ref{309}) for all times $t$. 
The remaining dynamics in this frame is then the
pure evolution of the shape from its steep initial profile
$\phi(\xi_X,0)$ towards its flatter asymptotic profile
$\phi \to \Phi^*(\xi_X)$ as $t\to\infty$. Fig.\ 4($a$) shows
$\phi$ as a function of $\xi_X$ on the interval $-5<\xi_X<5$.
One sees the interior or nonlinear part of the front.
Fig.\ 4($b$) shows $\log\phi$ in the range $10^{-90}<\phi<1$.
This plot is appropriate to show the development of the leading edge,
which here essentially determines the dynamics.
Accordingly, a very different range of $\xi_X$ has to be plotted,
namely $0<\xi_X<190$. As is sketched already in Fig.\ 2,
the leading edge consists of two regions, namely
the ``Gaussian'' region, through which the asymptotic steepness
$\lambda^*$ spreads in time towards larger $\xi_X$, and the region
of conserved steeness $\lambda=\lambda_{init}$ in front of it.
In fact, Fig.\ 4($b$) shows that the initial $\lambda_{init}=10$
on the level $\phi=10^{-90}$ is still fully present for times $t=1$ and 2,
while at later times it gradually approaches $\lambda^*=1$.
At higher levels, $\phi=10^{-10}$ say, this process of replacement
of one steepness by the other is essentially completed at time $t=70$,
while at level $10^{-90}$, it is not completed even at time $t=400$,
where the simulation stops.

In Fig.\ 5, we focus on the second feature, namely the displacement 
of the front. We plot the velocity $v_\phi(t)$ of various amplitudes 
$\phi$ as a function of $t$. According to our previous definition,
we identify $v_{1/2}(t)=v^*+\dot{X}(t)$. For comparison, the predicted
asymptotic value $v^*$ is plotted as a dashed line.
In Fig.\ 5($a$), the non-universal initial
transients up to time $t=20$ are shown on the range $0<v<3$. 
In Fig.\ 5($b$), the velocities are plotted up to time $t=400$ 
on the velocity interval $1.97<v<2$. One observes, \\
--- that for fixed $t$, the velocity $v_\phi(t)$ is the smaller,
the larger $\phi$ is. This is an immediate consequence of the
fronts becoming flatter in time, cf.\ Fig.\ 4.\\
--- that the $v_\phi(t)$ for large $t$ approach a value largely
independent of $\phi$, that is still far from the asymptotic value $v^*$.
We will see below, that this is the signature of the shape relaxing
like $v_{\phi_1}(t) - v_{\phi_2}(t) \propto 1/t^2$ as $t\to\infty$,
while the overall relaxation is $v_\phi(t)-v^*\propto 1/t$.

\end{multicols}

\begin{figure} \label{fig5}
\vspace{0.5cm}
\begin{center}
\epsfig{figure=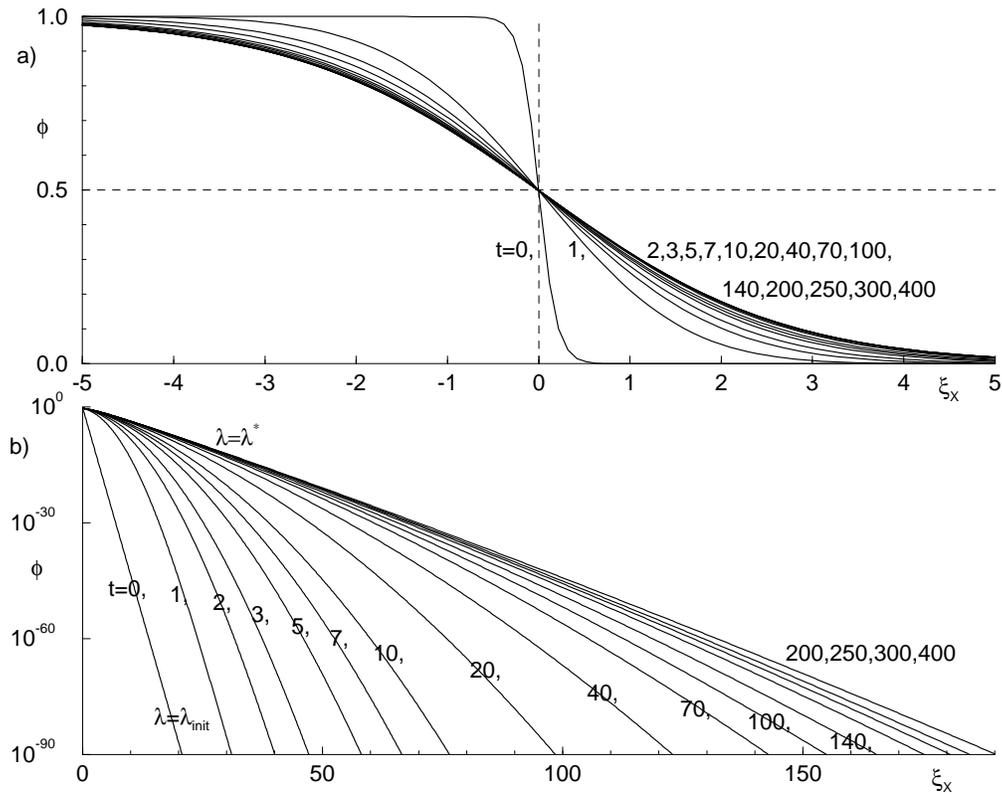,width=0.65\linewidth}
\end{center}
\vspace{0.5cm}
\caption{Simulation of the evolution of the shape of a front under 
(\ref{402}) at the times denoted in the figure. 
The initial condition is (\ref{405}) with $\lambda=\lambda_{init}=10$.
The comoving frame $\xi_X$ is chosen in such a way that $\phi(\xi_X=0,t)=1/2$ 
for all $t$. $a)$ A plot of $\phi$ versus $\xi_X$ shows mainly the interior 
of the front. 
$b)$ A plot of $\log \phi$ versus $\xi_X$ for sufficiently large $\xi_X$
shows mainly the leading edge of the front. 
Note the different scales of $\xi_X$.}
\end{figure}

\begin{figure} \label{fig6}
\vspace{-1cm}
\begin{center}
\epsfig{figure=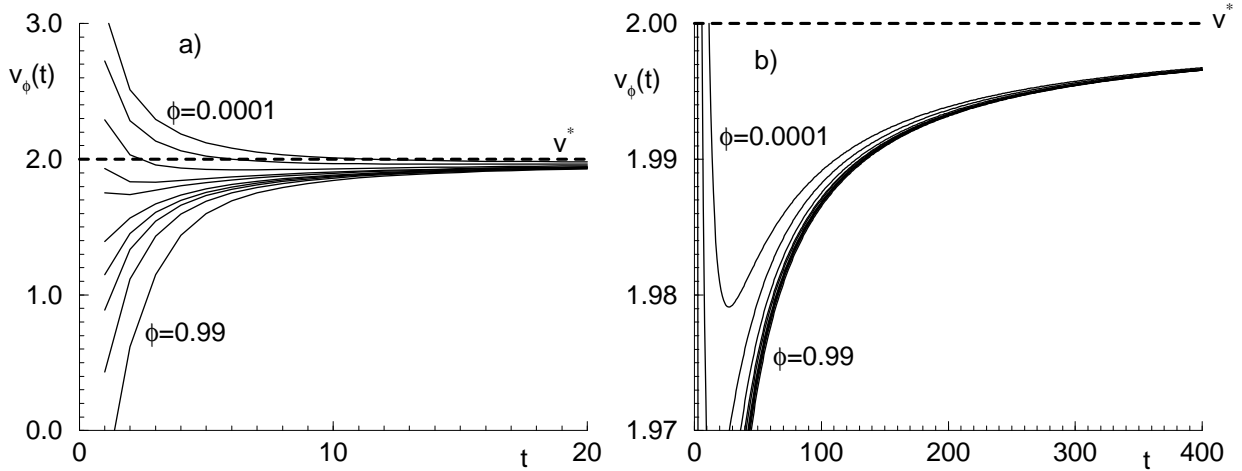,width=0.8\linewidth}
\end{center}
\vspace{0.5cm}
\caption{The same simulation as in Fig.\ 4.
Now the velocities $v_\phi(t)$ of amplitudes $\phi=0.99$,
0.9, 0.7, 0.5, 0.3, 0.1, 0.05, 0.01, 0.001, and 0.0001 (solid lines) 
are shown as a function of time $t$. The asymptotic velocity $v^*$
is marked by the dashed line. 
$a)$ Initial transients for times $0\le t\le 20$.
$b)$ The same data plotted for longer times $0\le t \le 400$ on 
an enlarged scale of $v$. The velocities $v_\phi(t)$ become largely
independent of the ``height'' $\phi$, and together slowly approach $v^*$
in agreement with the predicted universal algebraic relaxation.}
\end{figure}

\begin{multicols}{2}

\subsubsection{$f(\phi)=\epsilon\phi+\phi^3-\phi^5$:
pushed versus pulled fronts} \label{S422}

A well-known example of a nonlinear diffusion equation (\ref{101})
exhibiting both pushed and pulled fronts is given by the nonlinearity
(\ref{103}) with $n=2$:
\begin{equation}
\label{406}
\partial_\tau\varphi=\partial_y^2\varphi+\epsilon\varphi+\varphi^3-
\varphi^5~.
\end{equation}
This equation for $\epsilon<0$ is often used as a phenomenological 
(Ginzburg-Landau type) mean field model for a first order transition.
Likewise, its extension to a complex field is often used to model
a subcritical bifurcation in pattern forming systems.
According to arguments recalled 
in Appendix \ref{A1}, fronts of (\ref{406}) are pushed for
$\epsilon<3/4$, and pulled for $\epsilon>3/4$.

The rescaling necessary to bring (\ref{406}) to our standard form 
(\ref{206}) is discussed in (\ref{201}) and (\ref{202}), and yields
\begin{eqnarray}
\label{407}
\partial_t\phi&=&\partial_x^2\phi+\phi+\frac{1}{\bar{\epsilon}}\;\phi^3-
\left(1+\frac{1}{\bar{\epsilon}}\right)\;\phi^5~,\\
\mbox{where}&&
\label{408}
\bar{\epsilon}=\frac{\sqrt{1+4\epsilon}-1}{2} ~,~
\varphi_s^2=1+\bar{\epsilon}~.
\end{eqnarray}
The critical $\bar{\epsilon}$, where the pushed/pulled transition
occurs, is $\bar{\epsilon}_c=0.5$. 

We present data for the pushed front with $\bar{\epsilon}=0.4$ 
($\epsilon=0.56$) and the pulled front with $\bar{\epsilon}=0.6$
($\epsilon=0.96$). The initial condition is the one given before 
in (\ref{405}). The system size is $L=250$
and the front is located at $x_0=50$. The data therefore should
be reliable up to time of order 100, so the data presented extend
over $0\le t\le100$. 

In Fig.\ 6, we plot $v_\phi(t)$ as a function of $t$ for both 
values of $\bar{\epsilon}$ as solid lines, in the same way as 
the plot of Fig.\ 6 for the other nonlinearity. The dashed lines denote 
the asymptotic pulled velocity $v^*=2$ predicted for $\bar{\epsilon}=0.6$,
and the asymptotic pushed velocity (cf.\ Appendix \ref{A1})
\begin{equation}
\label{409}
v^\dag=\frac{1+4\bar{\epsilon}}{\sqrt{3\bar{\epsilon}(1+\bar{\epsilon})}}
=2.00594~~\mbox{ for }\bar{\epsilon}=0.4~.
\end{equation}
Fig.\ 6 shows $(i)$ that the simulated fronts in fact do approach
the predicted asymptotic velocities,
$(ii)$ that up to time $t\le10$ both fronts show quite similar
initial transients, $(iii)$ that for time $t\gg10$, however, the relaxation
towards the asymptotic velocity $v^\dag$ for $\bar{\epsilon}=0.4$
is much more rapid than that towards $v^*$ for $\bar{\epsilon}=0.6$.
This very clearly illustrates the difference between pushed
exponential and pulled algebraic relaxation, despite the tiny difference 
between $v^*$ and $v^\dag$.

We do not plot the figures of shape relaxation equivalent to Fig.\ 4, 
the only difference being that in the pushed case 
the region of conserved steepness $\lambda=\lambda_{init}$ at 
$\xi_X\gg1$ is invaded by a region of steepness 
$\lambda^\dag$ rather than by the pulled steepness $\lambda^*$.
$\lambda^\dag$ is determined by the front interior, while $\lambda^*$
is determined by the Gaussian region of the leading edge.

\end{multicols}

\begin{figure} \label{fig7}
\vspace{0.5cm}
\begin{center}
\epsfig{figure=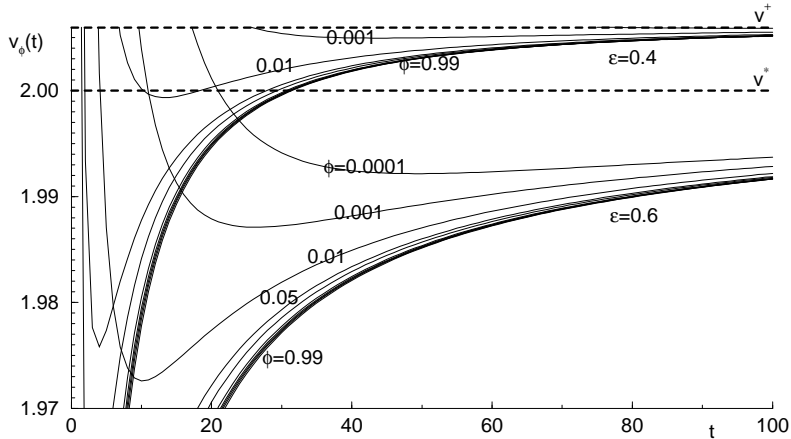,width=0.5\linewidth}
\end{center}
\vspace{-1cm}
\caption{Plot of $v_\phi(t)$ as a function of $t$ as in Fig.\ 5($b$),
but now for Eq.\ (\ref{406}). Simulations for $\bar{\epsilon}=0.4$
($\epsilon=0.56$) and $\bar{\epsilon}=0.6$ ($\epsilon=0.96$) are shown.
The dashed lines denote the asymptotic pulled velocity $v^*=2$ of the 
front with $\bar{\epsilon}=0.6$, and the asymptotic pushed velocity 
$v^\dag=2.00594$ of the front with $\bar{\epsilon}=0.4$. 
Note the quick exponential relaxation towards $v^\dag$ in contrast 
to the slow algebraic relaxation towards $v^*$. Further away from 
the transition $\bar{\epsilon}=0.5$ ($\epsilon=0.75$) from pulled
to pushed front propagation, the relaxation in the pushed regime
is even faster and the difference $v^\dag-v^*$ is larger.}
\end{figure}

\begin{multicols}{2}

\subsection{Comparison of simulations and analytical predictions} \label{S43}

We now return to our extensive simulation of the pulled front formed 
by the F-KPP equation $\partial_t\phi=\partial_x^2\phi+\phi-\phi^3$,
and compare the simulation data to our analytical predictions
from Table II with $v^*=2$, $\lambda^*=1=D$.

\subsubsection{Analysis of the velocity data} \label{S432}

We first concentrate on the analysis of the velocity data
$v_\phi(t)$ from Fig.\ 5. The prediction for the 
velocities $v_\phi(t)$ of the amplitudes $\phi$ is derived from
the expressions in Table II through
$v_\phi(t)=-\partial_t\phi/\partial_x\phi|_{\phi\;\rm fixed}$.
The result is
\begin{equation}
\label{4014}
v_\phi(t)=v^*+\dot{X}-\ddot{X}\;\left.\frac{\eta_{sh}}{\partial_\xi\Phi^*}
\right|_{\phi\;\rm fixed}+ O\left(\frac{1}{t^3}\right)~.
\end{equation}
Remember, that $\dot{X}$ is universal only till order $1/t^{3/2}$
and will exhibit contributions in order $1/t^2$, that depend on
initial conditions. The {\it difference} $v_{\phi_1}(t)-v_{\phi_2}(t)$,
however, will turn out to be independent of initial conditions 
up to order $1/t^{5/2}$.
Let us now test these predictions on the simulations in a series
of plots with growing precision in Figs.\ 7 -- 9.

As the velocity correction $\dot{X}$ is $1/t$ in leading order,
we plot $v_\phi(t)$ as a function of $1/t$ in Fig.\ 7, for the time
range $5<t<400$ in Fig.\ 7($a$), and for $100<t<400$ in Fig.\ 7($b$).
The dashed lines present the predicted asymptotes $v^*+\dot{X}=
2+\dot{X}_1(t)$ (the lower dashed line), and 
$v^*+\dot{X}=2+\dot{X}_{3/2}(t)$ (the upper dashed line), where we define
\begin{equation}
\label{4015}
\dot{X}_1(t)=-\frac{3}{2t}~~,~~
\dot{X}_{3/2}(t)=-\frac{3}{2t}\;\left(1-\sqrt{\frac{\pi}{t}}\right)~. 
\end{equation}

First of all, in comparing Figs.\ 7($a$) and 7($b$), we recognize
the asymptotic nature of the $1/\sqrt{t}$ expansion:
whether the $\dot{X}_1$ or the $\dot{X}_{3/2}$ asymptote gives the better
prediction, depends on the time scale: If we neglect the upper
three solid lines with velocities $v_\phi(t)$ for the very small
amplitudes $\phi=0.01$, 0.001, and 0.0001, the asymptote $2+\dot{X}_1$
clearly fits much better in Fig.\ 7($a$) for times $5<t<400$
--- while the asymptote $2+\dot{X}_{3/2}$ essentially 
coincides with $v_{0.001}(t)$, an observation we have no analytical
explanation for. For times $100<t<400$ in Fig.\ 7($b$),
however, the coincidence with $2+\dot{X}_{3/2}$ is excellent
for all $\phi$, and $2+\dot{X}_1$ very clearly is 
``far off'' on this very detailed scale.
Hence we will work below with the asymptote $2+\dot{X}_{3/2}(t)$,
and we present data for the time regime $20<t<400$ in Figs.\ 8 and 9.

Let us now further zoom in on the $\phi$ dependent velocity corrections 
(\ref{4014}) to $\dot{X}$. 
Fig.\ 8($a$) shows $v_\phi(t)-2-\dot{X}_{3/2}$ as a function of
$\ddot{X}_{3/2}=3/(2t^2)\;\big(1-(3/2)\sqrt{\pi/t}\big)$.
According to the prediction (\ref{4014}), the plot for small values
of $\ddot{X}_{3/2}\to0$ should show 
essentially straight $\phi$-dependent lines, all approaching
$v_\phi(t)-2-\dot{X}_{3/2}\to0$ as $\ddot{X}_{3/2}\to0$. Clearly,
that is what they do. 

Fig.\ 8($b$) shows one further step of precision aiming now at the
precise value of $v^*$: (\ref{4014}) predicts
\begin{equation}
\label{4016}
\frac{v_\phi(t)-v^*-\dot{X}}{\ddot{X}}=
-\left.\frac{\eta_{sh}}{\partial_\xi\Phi^*}\right|_{\phi\;fixed}
+ O\left(\frac{g(\phi)}{t}\right)~.
\end{equation}
However, the evaluation of this expression with $\dot{X}_{3/2}$
(\ref{4015}) yields $\phi$-independent corrections of order
$1/\sqrt{t}$:
\begin{eqnarray}
\label{4017}
\lefteqn{\frac{v_\phi(t)-v^*-\dot{X}_{3/2}}{\ddot{X}_{3/2}}=}
\\
&=&-\left.\frac{\eta_{sh}}{\partial_\xi\Phi^*}\right|_{\phi}
+\frac{2c_2}{3}+\frac{3c_2\sqrt{\pi}+2c_{5/2}}{3\sqrt{t}}
+ O\left(\frac{\bar g(\phi)}{t}\right)~.\nonumber
\end{eqnarray}
Remember, that the constants $c_2$, $c_{5/2}$ etc.\ depend on the
initial conditions. According to (\ref{4017}),
if we plot $\big(v_\phi(t)-v^*-\dot{X}_{3/2}\big)/\ddot{X}_{3/2}$
as a function of $1/\sqrt{t}$, we expect these functions to 
approach a $\phi$-dependent constant as $1/\sqrt{t}\to0$.

Fig.\ 8($b$) shows, that they in fact do so --- but only if
we choose the correct value of $v^*$! The dotted lines
show the function for $v^*=2$, the fat solid lines for
$v^*=2.000075$. The latter value is the analytical prediction of $v^*$
taking the finite gridsize corrections of the numerical code
into account, as explained in Sections \ref{S411} and \ref{S566}. 
The two values of $v^*$ differ in the 6th significant figure.
Fig.\ 8($b$) thus is an extremely precise demonstration of the correctness of 
our analytical arguments from both Sections \ref{S3} and \ref{S5},
since it clearly confirms our predictions to more than 6 significant
figures! 

Our test in Fig.\ 8($b$) is so sensitive, because we divide in Fig.\ 8($b$)
by the small quantities $\ddot{X}_{3/2}$, which are of order $10^{-5}$.
Without this division
the difference of the $v^*$'s in Fig.\ 8($a$) is not yet visible. 
The plot in Fig.\ 8($b$) shows that we fully understand the specific
numerical features of pulled front solutions, both the effect of the finite 
difference code and of the finite system size, cf.\ Sect.\ \ref{S41}.

We can eliminate $v^*$ and the nonuniversal corrections
$-c_2/t^2$ etc.\ by plotting $(v_\phi(t)-v_{0.5}(t))/\ddot{X}_{3/2}(t)$
as a function of $1/t$. Now (\ref{4014}) predicts
\begin{equation}
\label{4018}
\frac{v_\phi(t)-v_{0.5}(t)}{\ddot{X}_{3/2}}=
-\left.\frac{\eta_{sh}}{\partial_\xi\Phi^*}\right|_{\phi}
+ O\left(\frac{1}{t}\right)~.
\end{equation}
Fig.\ 9 shows this plot with the solid lines for $\phi=0.99$, 
0.5, 0.01, and 0.0001. The crosses on the axis are not(!)
extrapolated from the curves, but they mark the predicted
asymptotes $-\left.\eta_{sh}/\partial_\xi\Phi^*\right|_{\phi}$
for $\phi=0.99$, 0.5, 0.01, and 0.0001. The necessary data on 
$\eta_{sh}(\xi)$ and $\Phi^*(\xi)$ are derived from the numerical solution
of the appropriate {\em o.d.e.}'s, and completely independent from the
numerical integration of the {\em p.d.e.}\ for the initial value problem.
The coincidence of the extrapolated {\em p.d.e.}\ data with
the analytically predicted {\em o.d.e.}\ asymptote is most convincing.

\end{multicols}

\begin{figure} \label{fig8}
\vspace{0.5cm}
\begin{center}
\epsfig{figure=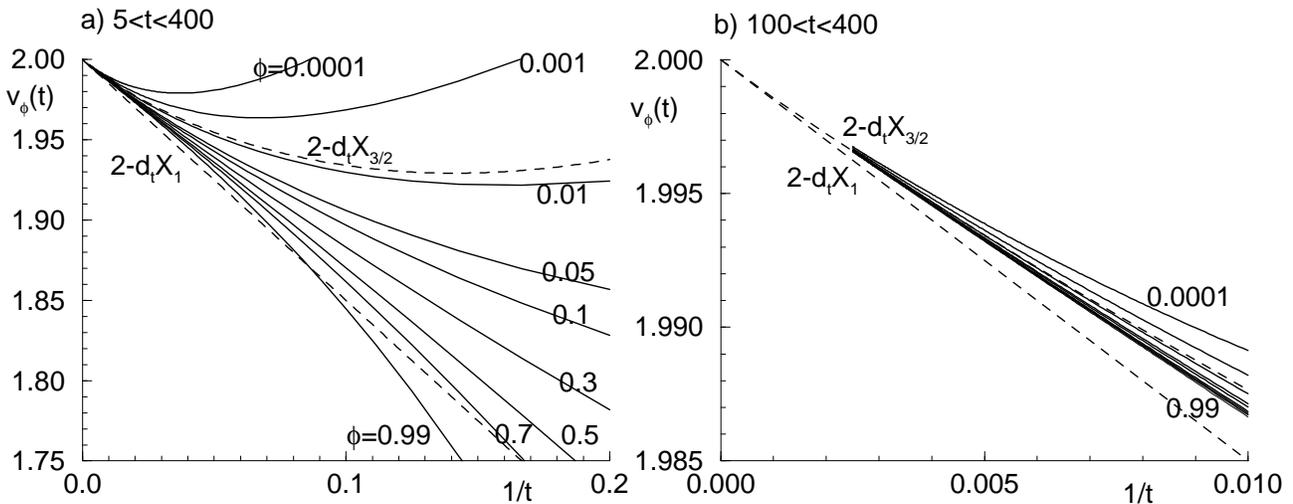,width=0.9\linewidth}
\end{center}
\vspace{0.5cm}
\caption{The data $v_\phi(t)$ from Fig.\ 5, but now plotted over $1/t$. 
The lower straight dashed line is the asymptote $v(t)=v^*+\dot{X}_1(t)$
(\ref{4015}), the upper curved dashed line is the asymptote
$v(t)=v^*+\dot{X}_{3/2}(t)$ with $v^*=2$. $a)$ time regime $5<t<400$, 
$b)$ time regime $100<t<400$. Note that due to the $t^{-3/2}$ correction
term, the effective slope in this plot is less than $3/2$, 
even at these long times.}
\end{figure}

\begin{figure} \label{fig9}
\vspace{0.5cm}
\begin{center}
\epsfig{figure=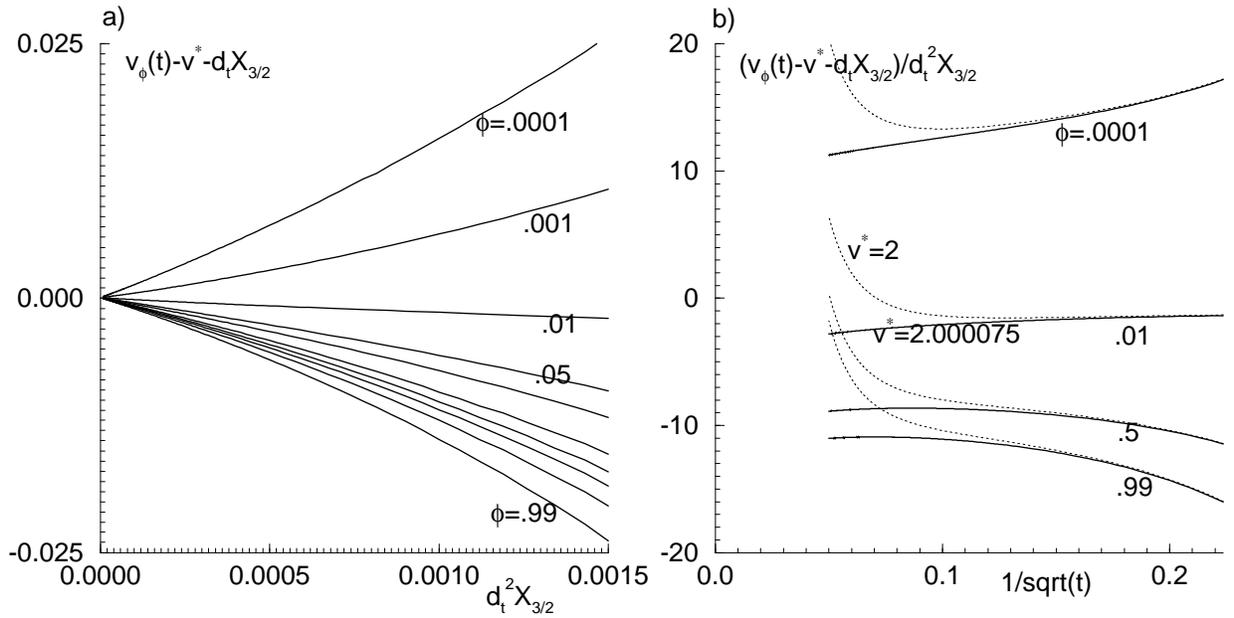,width=0.8\linewidth}
\end{center}
\vspace{1cm}
\caption[]{The data $v_\phi(t)$ from Figs.\ 5 and 7 for times 
$20 \le t \le 400$ in different representations. 
$a)$ $v_\phi(t)-2-\dot{X}_{3/2}$ as a function of $\ddot{X}_{3/2}$. 
See Eq.\  (\ref{4015}) for the definition of $\dot{X}_{3/2}$.
$b)$ $\big(v_\phi(t)-v^*-\dot{X}_{3/2}\big)/\ddot{X}_{3/2}$ as a function
of $1/\sqrt{t}$ for $\phi=0.99$, 0.5, 0.01, and 0.0001.
Dotted lines: $v^*=2$, solid lines: with the corrected value
$v^*=2.000075$ for our numerical scheme and gridsize according to 
Section \ref{S566}.}
\end{figure}

\begin{figure} \label{fig10}
\vspace{0.5cm}
\begin{center}
\epsfig{figure=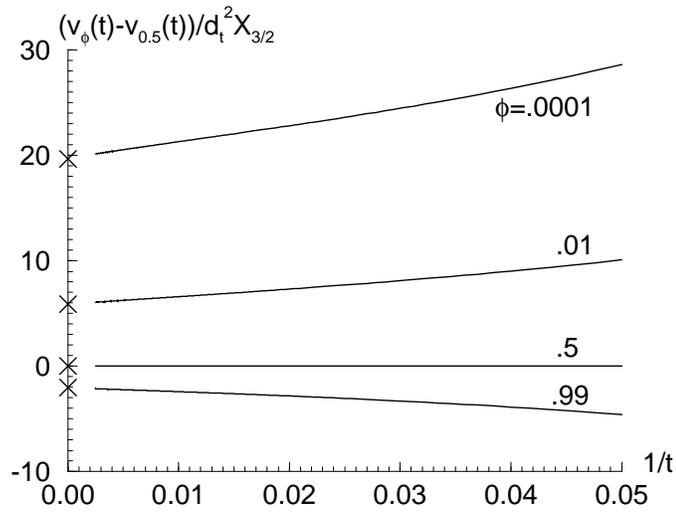,width=0.4\linewidth}
\end{center}
\vspace{0.8cm}
\caption[]{The solid lines are again the data from Figs.\ 5, 7, 8 
for times $20\le t\le400$,
now plotted as $(v_\phi(t)-v_{0.5}(t))/\ddot{X}_{3/2}$ over $1/t$
for $\phi=0.99$, 0.5, 0.01, and 0.0001. As explained in the text, 
this eliminates a nonuniversal $1/\sqrt{t}$ term, that depends on 
the initial conditions. The crosses result
from solving the {\em o.d.e.}'s for $\Phi^*$ and $\eta_{sh}$ numerically
and plotting $-\eta_{sh}/\partial_\xi\Phi^*\big|_\phi$ for 
$\phi=0.99$, 0.5, 0.01, and 0.0001. Eq.\ (\ref{4018}) predicts
that the lines should extrapolate to the crosses. Since they do,
and since $\ddot{X}(t)$ is of order $10^{-5}$ at the latest times,
these data confirm our predictions with extreme precision.}
\end{figure}

\newpage

\begin{multicols}{2}

\subsubsection{Analysis of the shape data} \label{S433}

We now leave the analysis of the velocity data, and come back to the shape 
data from Fig.\ 4. Table II immediately yields
\begin{equation}
\label{4019}
\frac{\phi(\xi_X,t) - \Phi^*(\xi_X)}{\dot{X}\;\eta_{sh}(\xi_X)}=
1+O\left(\frac{1}{t}\right).
\end{equation}
This gives the clue on how to rewrite the shape data $\phi(\xi_X,t)$
at different times as a function of $\xi_X$. The solutions of the 
{\em o.d.e.}'s
for $\eta_{sh}$ and $\Phi^*$ that are needed for evaluating (\ref{4019}) 
are derived numerically. They have been used for generating the crosses 
in Fig.\ 9 and are now also used in Fig.\ 10. 

Plotting the l.h.s.\ of Eq.\ (\ref{4019}) allows us to combine
the information about the interior from Fig.\ 4($a$) and the information
about the leading edge from Fig.\ 4($b$) into one plot.
In Fig.\ 10($a$), we do not divide by $\dot{X}$, but present 
the data at the small times $t=1$, 2, 3, 5, 7, 10, and 20
as $-(\phi-\Phi^*)/\eta_{sh}$ over $\xi_X$.
In Fig.\ 10($b$), the data at the large times $t=20$, 40, 70, 100,
140, 200, 300, and 400 are shown as $(\phi-\Phi^*)/(\dot{X}\eta_{sh})$ 
over $\xi_X$, where we use again the approximation $\dot{X}=\dot{X}_{3/2}$
(\ref{4015}). For comparison, both plots also show $\Phi^*(\xi_X)$
and $\xi_X=0$ as dashed lines. Also Fig.\ 10($b$) has the large time
prediction $(\phi-\Phi^*)/\dot{X}\eta_{sh}\to1$ as $t\to\infty$ 
as a dotted line.

Fig.\ 10($a$) shows how the interior of the front rapidly relaxes. 
Fig.\ 10($b$) demonstrates $(i)$ that with
$\dot{X}=\dot{X}_{3/2}$ we indeed have chosen the
correct asymptote, and $(ii)$ how the predicted asymptotic value
$(\phi-\Phi^*)/(\dot{X}\eta_{sh})\to 1$ as $t\to\infty$ is approached 
from above in the interior of the front and from below in the leading 
edge.

Note, that in Fig.\ 10($b$) all lines approximately cross one
point of height unity far in the leading edge. We have no intuitive
or analytical understanding of this observation.

\end{multicols}

\begin{figure} \label{fig11}
\vspace{0.5cm}
\epsfig{figure=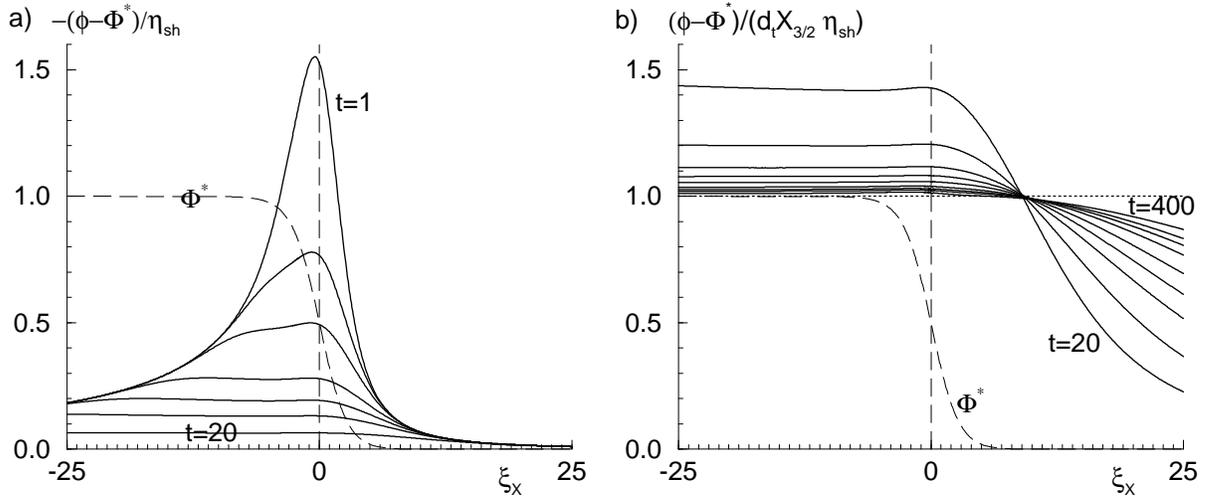,width=0.9\linewidth}
\vspace{0.8cm}
\caption{In this figure the shape data from Fig.\ 4 are represented 
differently using data from the numerical solution of the {\em o.d.e.}'s 
for $\Phi^*$ and $\eta_{sh}$.
$a)$ $-(\phi-\Phi^*)/\eta_{sh}$ (solid) as a function of $\xi_X$ for times
$t=1$, 2, 3, 5, 7, 10, and 20. $b)$ $(\phi-\Phi^*)/(\dot{X}_{3/2}\eta_{sh})$
(solid) as a function of $\xi_X$ for times $t=20$, 40, 70, 100, 140,
200, 250, 300, 400. Dotted line: predicted asymptote
$(\phi-\Phi^*)/(\dot{X}\eta_{sh})\to 1$ as $t\to\infty$.
Dashed lines in ($a$) and ($b$) give $\Phi^*(\xi_X)$ and $\xi_X=0$
for orientation.}
\end{figure}

\newpage

\begin{multicols}{2}

\section{Generalization of pulling to higher order (sets of) equations} 
\label{S5}

\subsection{Introduction}\label{S51}

In the last fifteen years, it has become clear that many of the
observations and intuitive notions concerning the behavior of front
solutions of the nonlinear diffusion equation (\ref{101}) generalize
to higher order equations or systems of coupled {\em p.d.e.}'s. First
of all, taking the spreading velocity $v^*$ of a linear perturbation
of the unstable state [Eqs.\ (\ref{sadpoint}) and (\ref{sadpoint2})]
as the generalization of $v^*=2\sqrt{f'(0)}$ for (\ref{101}), we
observe that there are numerous examples
\cite{goldstein,nb,ramses,krug,bj,dee,vs2,vs3} of fronts whose
asymptotic velocity approaches the pulled value $v^*$ given by
(\ref{sadpoint2}). So there is no doubt that the mechanism of fronts
``being pulled along'' by the leading edge generalizes to a large
class of equations.  Second, there is also quite a bit of evidence
for the existence of a pushed regime in more complicated equations. In
a number of cases, the pushed regime was again found to be related to
the existence of a strongly heteroclinic solution with velocity
$v^\dag>v^*$.  An example of a non-monotonic but still uniformly
translating pushed front solution in the EFK equation is shown in
Fig.\ 7 of \cite{vs2}.  In the quintic complex Ginzburg Landau
equation, it has turned out to be possible to solve for a strongly
heteroclinic front profile exactly, and in numerical simulations it
was empirically found that this solution does play the same role in
the front selection process as the pushed front $\Phi^\dag$ in the
nonlinear diffusion equation \cite{vs3}.  Pushed fronts also emerge in
coupled amplitude equations for chaotic domain boundary motion
\cite{tucross}.  For extensions of the Swift-Hohenberg equation there
are numerical and perturbative indications that both pulled and pushed
 regimes occur, and that one can tune the front velocity from one
regime to the other with one of the nonlinear terms in the equation
\cite{vs2}.

Much of our understanding of the above general findings has been
intuitive and empirical, or based on conjectures. We shall now show
that many of our results for the second order nonlinear diffusion
equation generalize to other equations, not only to (sets of) partial
differential equations of higher order, but also to other types of
equations like difference-differential equations \cite{krug,evsp}, or
differential equations with memory kernels \cite{odo}. We will
concentrate here on equations whose relevant front solutions are
uniformly translating.
For {\em p.d.e.}'s in this class, essentially the whole classification
of nonlinearities and initial conditions $\phi(x,0)$ in Table IV
applies, provided the uniformly translating fronts $\Phi_v$, and in
particular the fronts $\Phi^*$ and $\Phi^\dag$ exist. Many 
aspects of the stability analysis can be generalized, while 
the relaxation of pulled fronts requires the generalization of
the calculation in Section \ref{S3}.
This generalization, that we will develop below, leads to {\em new}
and {\em explicit} predictions for the front convergence in the pulled
regime, as summarized in Table II. The
fact that these predictions for various examples are fully corroborated
numerically in Section \ref{S56} makes us conclude that the velocity
selection and relaxation of uniformly translating fronts is now
essentially understood even for general sets of equations.

While this paper was nearing completion, it was becoming increasingly
clear that even though pattern forming fronts --- both fronts leading
to regular periodic patterns, as in the Swift-Hohenberg equation
\cite{sh,dl,vs2}, and fronts leading to chaotic patterns as in some
parameter ranges of the complex Ginzburg-Landau equation --- present
additional complications, our most central result for the universal
algebraic velocity relaxation carries over even to these. We will
leave this discussion to the future \cite{spruijt,kees}, and focus 
here on {\em p.d.e.}'s whose asymptotic pulled fronts are uniformly 
translating front solutions of the type $\Phi^*(x-v^*t)$, just as in 
the nonlinear diffusion equation.

In writing this section, we face the following dilemmas:
\begin{itemize}
\item[{\em (i)}] The extension of both the stability considerations of
  uniformly translating front solutions of section \ref{S2} and of the
  relaxation analysis of pulled fronts of section \ref{S3} depends
  quite crucially on two ingredients: First, that the front
  propagation into unstable states is 
  in the pulled regime, and, second, that there is a family of uniformly
  translating front solutions around $\Phi^*(x-v^*t)$: Only then can the
  relaxation in the front interior be along the manifold of front
  solutions according to $\phi(x,t) = \Phi_{v(t)} (\xi_X) + O(1/t^2)$.
  However, to our knowledge there is no general theory concerning the
  conditions under which fronts are pulled and concerning the
  multiplicity of front solutions: For particular equations under
  study or for some restricted classes of equations, one can often
  convince oneself that the front should be pulled and that $\Phi^*$
  should be a member of a family of front solutions, but a general
  theory is lacking.
  
\item[{\em (ii)}] An immediate jump to the most general (but abstract)
  case is pedagogically not justified and moreover would assume
  knowledge of the derivation of the pulled velocity $v^*$ that most
  readers probably do not have.
\end{itemize}

We have chosen to deal with this dilemma by simply summarizing our
main assumptions and our results concerning the extensions of Section
\ref{S2} to more general equations below, relegating the details of
the analysis to Appendices. Then, we proceed with the relaxation
analysis of pulled fronts in two steps. We first consider in Section
\ref{S53} the analysis of a single {\em p.d.e.}\ which is of first
order in time but of arbitrary order in space. After that, the
extension to {\em p.d.e.}'s that are of higher order in time is
discussed in Section \ref{S54}. The  
extension to even more general classes of equations, including
difference equations or integro-differential equations, e.g., 
with memory kernels, is then immediate, 
as we discuss in Section \ref{S55}. We there also discuss
coupled equations. Section \ref{S56} contains the
explicit analytical and numerical results for several of the
equations listed in Table I.

\subsection{Basic assumptions underlying the relaxation analysis of
  pulled fronts; generalization of Table IV} \label{S52}

Most of the results discussed in Section \ref{S2} for the nonlinear 
diffusion equation can be generalized to higher order nonlinear 
partial differential equations, as well as to difference or 
integro-differential equations and to coupled equations:\\
$\bullet$ The family of solutions can be parametrized as well by the steepness
$\lambda$ which gives the rate of exponential decay
of $\Phi_v(\xi)$ as $\xi \to \infty$.\\
$\bullet$ If there are one or more strongly heteroclinic solutions, then
at each velocity where such a solution exists, there is a strongly
heteroclinic mode of the linear stability operator which changes
stability, i.e., which is such that the mode is stabilizing for
fronts with velocities larger than this value and destabilizing for velocities
less than this value. This implies in particular that the pushed velocity
$v^\dagger$ is the largest velocity at which there is a strongly
heteroclinic front solution $\Phi^\dagger_v$, and that front solutions
with $v<v^\dagger$ are
unstable (see Appendices \ref{A6} and \ref{A7}).\\
$\bullet$ The linear spreading velocity $v^*$, given by Eqs.\ 
(\ref{5a24}) and (\ref{5a24a}) below, is the pulled front speed and
coincides with the minimum of the
velocities  of uniformly translating fronts $v(\lambda )$ (see
Section \ref{S532}).\\
$\bullet$ If there are no strongly heteroclinic solutions with
$v>v^*$, all front solutions with $v>v^*$ are stable to perturbations
which are steeper than $\lambda^*$, while front solutions with $v<v^*$
are unstable: the pulled front solution is then the slowest and
steepest solution which is stable. \\
$\bullet$ The fronts that dynamically emerge from steep initial
conditions (falling off faster than $e^{-\lambda^* x}$) converge to
pulled fronts propagating with speed $v^*$.

In this Section, we will investigate the front relaxation 
under the assumptions:
\begin{itemize}
\item[{\em A}] {\em The front solutions are pulled, i.e., starting
    from a steep initial condition the asymptotic front speed
    $v_{sel}$ equals the linear spreading speed $v^*$ given by Eqs.\ 
    (\ref{5a24}) and (\ref{5a24a}) below.}
\item[{\em B}] {\em The asymptotic front is uniformly
    translating , i.e., of the form
    $\Phi^*(x-v^*t)$, and it is a member of a continuous family of
    uniformly translating solutions $\Phi_v(x-vt)$, parametrized by $v$. }
\end{itemize}

To put our general assumptions {\em A} and {\em B} into perspective,
we  note that 
for a given equation the existence of a family of front solutions can
often be demonstrated by counting arguments. This is shown in Appendix
\ref{A6} for {\em p.d.e.}'s of first order in time that are invariant
under space reflection. Such counting arguments also lead one to
expect that generically either $\Phi^*(x-v^*t)$ is a member of a
continuous family of front solutions, or there is no uniformly
translating front solution $\Phi^*$ at all. For, if there is a
discrete set of front solutions (solutions $\Phi_v$ exist at isolated
values of the velocity), there is no particular symmetry reason to
have one at $v=v^*$, since the existence of an isolated solution
depends on the full nonlinear behavior of the ordinary differential
equation, not just on the properties near one of the asymptotic fixed
points. We comment in Section \ref{S6} on what might happen when there
is no uniformly translating front solution, even though the front
dynamics is pulled.

\subsection{Pulled front relaxation in single {\em p.d.e.}'s 
of first order in time}

\label{S53}

In the present Section C, we discuss an arbitrary {\em p.d.e.}
\begin{equation}
\label{5a2}
F\Big(\phi,\;\partial_x\phi,\;\ldots,\;\partial_x^N\phi,\;
\partial_t\phi\Big)=0~.
\end{equation}
for a single field $\phi(x,t)$. We assume that $F$ is analytic in all
its arguments, and that the equation admits homogeneous steady state
solutions $\phi=0$ and $\phi=1$. Moreover, we assume $\phi=0$ to be
linearly unstable and $\phi=1$ to be linearly stable, and we consider
fronts connecting these two asymptotic states as in (\ref{2013}).
Also, according to our assumption {\em B}, Eq.\ (\ref{5a2}) admits 
a continuous
  family of uniformly translating fronts $\phi(x,t)=\Phi_v(x-vt)$.  The
  linearization of some front $\phi(x,t)$ about some $\Phi_v$
  generalizes from (\ref{2023}) -- (\ref{2025}) to
\begin{equation} \label{5a5}
\phi(x,t) = \Phi_v(\xi)+\eta(\xi,t)~~,~~
\partial_t\eta = {\cal L}_v(\xi) \eta+O(\eta^2)~,
\end{equation}
where the linear operator is now
\begin{equation} \label{5a6}
{\cal L}_v(\xi)=\sum_{n=0}^N f_n(\xi)\;\partial_\xi^n 
+v\;\partial_\xi ~~,~~
f_n(\xi)=\frac{- F_n(\xi)}{F_{N+1}(\xi)} ~.
\end{equation}
Here $F_n(\xi)$ denote the functional derivatives of $F$:
\begin{equation} \label{5a7}
F_n(\xi) = \left.\frac{\delta F(\phi^{(0)},\ldots,\phi^{(N+1)})}
{\delta \phi^{(n)}}
\right|_{\begin{array}{l} \scriptstyle 
          \phi^{(m)}=\partial_\xi^m\Phi_v(\xi), \;m<N+1\\
          \scriptstyle 
          \phi^{(N+1)}=-v\partial_\xi\Phi_v(\xi)\end{array}} ~,
\end{equation} 
In order that the $F_{n}$ have no singularities, $F_{N+1}$ should be
of one sign; for convenience, we take $ F_{N+1}(\xi)<0$ for all $\xi$
and rescale time $t$ such that $ F_{N+1}(\infty)=-1$.  We also assume, that
$F_N(\xi)$ neither vanishes nor changes sign for any $\xi$.

\subsubsection{The pulled velocity $v^*$} \label{S531}

In the pulled regime and with steep initial conditions, the asymptotic
front velocity equals the linear spreading velocity $v^*$, i.e., the
velocity with which a localized perturbation spreads according to the
linearized equations. Since the calculation of $v^*$ forms the basis 
of our subsequent analysis, we summarize its derivation in the context 
of our first order {\em p.d.e.}\ (\ref{5a2}).
The general formulation in Section \ref{S55}, which is necessary
to treat difference equations or integro-differential equations,
is closest to the original ``pinch point'' analysis \cite{bers,landau},
from which many of these ideas originally emerged.

In the rest frame $(x,t)$, the equation linearized about $\phi=0$ is
\begin{equation}
\label{5a14}
\partial_t\phi={\cal L}_0(\infty)\phi=\sum_{n=0}^N a_n \partial_x^n \phi~,
\end{equation} which is the generalization of (\ref{2049}), and
where we introduced the short hand notation $a_n=f_n(\infty)$.  The
dispersion relation $\omega(k)$ of a Fourier mode
$e^{\:ikx-i\omega(k)t}$ is given by
\begin{equation}
\label{5a15}
 - i\omega(k) = \sum_{n=0}^N a_n\;(ik)^n~.
\end{equation}
Since we later will again characterize fronts by
their exponential spatial decay rate $\lambda=-ik$,
we already define the growth rate $s(\lambda)$ of the
steepness $\lambda$ as
\begin{equation}
\label{defs}
s(\lambda)=\mbox{Re }\left(-i\omega(i\lambda)\right) = 
\mbox{Re }\sum_{n=0}^N a_n\;(-\lambda)^n
\end{equation}
for later use.
We restrict the analysis to equations where the temporal growth
rate $\mbox{Re }(-i\omega(k))$ in (\ref{5a15}) will be negative
for short wave length Fourier modes $k$, i.e., where 
\begin{equation}
\label{restr}
\mbox{Re }a_N\;(\pm i)^N < 0~,
\end{equation}
since otherwise all smooth solutions will be 
unstable against perturbations of arbitrarily short wave lengths.

An arbitrary initial condition $\phi(y,0)$ will develop under
(\ref{5a14}) as
\begin{eqnarray} \label{5a16}
\phi(x,t)&=&\int_{-\infty}^\infty dy\;G(x-y,t)\;\phi(y,0)~,
\\ \label{5a17}
G(x,t)&=&\int_{-\infty}^\infty
\frac{dk}{2\pi}\;e^{ikx-i\omega(k)t}~.
\end{eqnarray}
in generalization of (\ref{2053}).

For sufficiently steep initial conditions $\phi(y,0)$, the
asymptotic behavior of $\phi(x,t)$ can be obtained from the large-time
asymptotics of the Green's function $G$ (\ref{5a17}) that can be evaluated 
by a saddle point integration \cite{bender} (also known as ``steepest
decent approximation'').
The result will depend on the frame of reference.  In an arbitrary
coordinate system $\xi=x-vt$ with $v$ fixed, a saddle point $k_n$
is a saddle of $-i\omega(k)+ivk$,
\begin{equation}
\label{5a18}
\left.{{d}\over{dk}} (-i\omega(k)+ivk) \right|_{k_n} = 0
~~\Longrightarrow~~
\left. { {d\omega(k)}\over{dk}} \right|_{k_n} = v~.
\end{equation}
A polynomial of degree $N$ (\ref{5a15}) generically has
$N-1$ saddle points $k_n$, $n=1,\ldots,N-1$, (\ref{5a18})
in the complex $k$ plane.  
The integral  
(\ref{5a17}) is therefore  dominated by the saddle point with the
largest growth rate through which we can lead the $k$-contour by
continuously deforming it off the real axis. If the contour can be
deformed to go through several, the relevant one is thus  that particular saddle
point $k^*(v)$ of the ones we can reach that has 
the maximal growth rate:
\begin{equation}
\label{defvstar}
\mbox{Re }(-i\omega(k^*)+ivk^*) =
\max_n \mbox{ Re }(-i\omega(k_n)+ivk_n)~.
\end{equation}
It will have
\begin{equation} \label{5a19}
D(v)=\frac{1}{2}\;\left.\frac{d^2i\omega(k)}{dk^2}\right|_{k^*(v)}
~~,~~\mbox{Re }D>0~.
\end{equation}
We stress that Eq. (\ref{5a18}) only expresses the  
 condition for the {\em existence} of a saddle point. Which saddle
 point the $k$-contour can be made to go through by contour
 deformation is a condition that depends on the {\em global}
 properties of $\omega(k)$  that can
 only by analyzed for a {\em given} dispersion relation. It is {\em not a
   local condition}. For a further discussion of this point we refer
 to Appendix \ref{utesfavorite}: we proceed  by assuming that the saddle point
 indicated by a star $*$ is the one that obeys (\ref{defvstar}) and
 this condition, without distinguishing this underlying condition with our
 notation.

The expansion of the integral (\ref{5a17})
about the saddle point $k^*(v)$ can be performed 
in a frame with arbitrary velocity $v$ and yields
\begin{equation} \label{5a20}
G(x,t)=e^{ik^*\xi+\:\big(-i\omega(k^*)+ivk^*\big)\;t} \;\;{\cal I}_v(\xi,t)
~~,~~\xi=x-vt~.
\end{equation} 
The integral ${\cal I}_v(\xi,t)$ is expressed after substitution of
$(k-k^*)=\kappa/\sqrt{t}$ as
\begin{eqnarray} \label{5a21} \nonumber
{\cal I}_v &=& \int_{-\infty}^\infty \frac{d\kappa}{2\pi\sqrt{t}}\;
e^{i\kappa\xi/\sqrt{t}-D\kappa^2+O(D_3\kappa^3/\sqrt{t})}
\\ \label{5a22}
&=&\frac{e^{-{\xi}^2/(4Dt)}}{\sqrt{4\pi Dt}} \;
\left(1+O\left(\frac{D_3 \xi}{D^2 {t}}\right) \right)
\end{eqnarray} 
for large $t$ and arbitrary $\xi$. Obviously $D$ plays the
role of a diffusion coefficient. $D_3$ is defined
below in (\ref{diffdef}). 

Generically, the growth (or decay) rate of the saddle point mode 
$\mbox{Re }(-i\omega(k^*(v))+ivk^*(v))$ will be nonvanishing.
We now define the particular linear spreading or pulled velocity 
$v^*$ through $\mbox{Re }(-i\omega(k^*)+iv^*k^*) = 0$, or
\begin{equation}
\label{5a24}
v^* = {{\mbox{Im } \omega(k^*)}\over{\mbox{Im } k^*}}
={{s(-ik^*)}\over{\mbox{Im } k^*}}
~~,~~k^*=k^*(v^*)~.
\end{equation}
This means, that in the frame moving with velocity $v^*$, 
the absolute value of the Green's function (\ref{5a20}) neither
grows nor decays in leading order.
$v^*$, $k^*$ and $\omega(k^*)$ are determined by (\ref{defvstar}),
(\ref{5a24}), and by Eq.\ (\ref{5a18}) evaluated at $v^*$:
\begin{equation} \label{5a24a}
\left. { {d\omega(k)}\over{dk}} \right|_{k^*} = v^*~.
\end{equation}
In addition, the solution determines $D=D(v^*)$ (\ref{5a20}).

Note that the leading order large $t$ result (\ref{5a20}), 
(\ref{5a22}) for the Green's function $G$ in (\ref{5a16})
is diffusive just like in (\ref{2054}),
despite the fact, that we are dealing here with an equation with
higher spatial derivatives. We shall see in Sects.\ \ref{S54} 
and \ref{S55} that this even remains true for much more general 
types of equations.

Note also that in our discussion of {\em p.d.e.}'s in this
and the next Section, we only take the spatial Fourier transform
of $G(x,t)$, as in (\ref{5a17}) above. However, the most general
formulation, which also applies to difference equations or 
integro-differential equations, is most
conveniently done by taking a Fourier transform in space and a
Laplace transform in time. In the present context, the Green's
function $\hat G(k,\omega)$ is then defined as 
\begin{eqnarray}
\hat G(k,\omega) &=& \int_0^{\infty} dt \int_{-\infty}^{\infty} dx\;
e^{-ik x + i\omega t} \;G(x,t) = \frac{1}{S(k,\omega)}~, \label{5a17b}
\nonumber\\
&&\mbox{where }S(k,\omega)=i\omega(k)-i\omega~,
\end{eqnarray} 
and the long time asymptotics is determined by the double roots
of the characteristic equation $S(k,\omega(k))=0$ (\ref{5a15}). 
We defer this type of formulation, which closer follows 
the ``pinch point'' analysis of \cite{bers,landau}, to Section \ref{S55}.

In practice, one first will drop condition
(\ref{defvstar}) and generically derive $N$ solutions 
$(k^*,v^*)$ from (\ref{5a24}), (\ref{5a24a}) for a given dispersion
relation. But as we already pointed out above, not all of these may be 
appropriate saddle points for the dynamics. 
Typically there are solutions with $\lambda^* \equiv
\mbox{Im}k^* >0$ and $v^*>0$, which describe a profile spreading
to the right and solutions with $\lambda^*<0$ and $v^*<0$
describing the spreading to the left, and the $k$-contour will have to 
be deformed through the appropriate one for the left- and right-moving 
front.
These solutions are related by symmetry, if the original {\em
  p.d.e.}\ is symmetric under space reflection: if (\ref{5a15}) only
contains even powers of $k$ and if the $a_n$ are real, then for every 
solution $(k^*,v^*)$ there is a solution $(-k^*,-v^*)$. Moreover, 
as mentioned already above, there might be various nontrivial saddle
point solutions which are  
not related by symmetry, if the degree $N$ of spatial derivatives 
is sufficiently large. The saddle point analysis as well as the arguments 
of Section \ref{S251} for the competition between different solutions of
the linearized equations clearly show that then the dynamically relevant
solution is the one with the largest velocity $v^*$ through which the
$k$-contour can be led.  

However, choosing the saddle point with the largest $v^*$ might
according to counting arguments (as in Appendix \ref{A6}) be
inconsistent with assumption {\em B} from Sect.\ \ref{S52}
of the existence of a family of uniformly translating fronts, 
since one expects the multiplicity of front solutions to be 
different for every saddle point $(v^*,k^*)$. The discussion 
of this issue we defer to Section \ref{S62}, as for the applications 
discussed in Sect.\ \ref{S56}, this problem does not rise.

\subsubsection{Uniformly translating solutions  $\Phi_v$}
\label{S532}

In the analysis of the nonlinear diffusion equation in Sect.\ 
\ref{S2}, we saw that the uniformly
translating solution $\Phi_v$ decayed as $e^{-\lambda \xi}$ with
$\lambda$ real for $v \geq v^*$. Here, $\lambda=\lambda_-(v)$ 
(\ref{lambda1}) is the smallest root of $v=s(\lambda)/\lambda$,
where $s(\lambda)$ (\ref{defs}) here equals $s(\lambda)=\lambda^2+1$. 
$v\ge v^*$ implied $\mbox{Re
  }k=0$, $\lambda = \mbox{Im }k >0$. These front solutions were found
to be stable to perturbations which are steeper than the front
solution $\Phi_v$ itself provided there is no pushed front 
solution $v_c=v^\dag$. The solutions with $v<v^*$ had $\mbox{Re }k
\neq 0$, $\mbox{Re }\omega \neq 0$, and were unstable.

We will focus here on the immediate generalization of these 
results, i.e., assume that fronts with $v \geq v^*$ have $\mbox{Re
  }k=0$ , so that their asymptotic spatial decay is as $e^{-\lambda
  \xi}$.  In particular, this gives for the pulled fronts
\begin{eqnarray} 
\label{realdef1} 
\mbox{Re } k^* & = &0~, ~~\hspace*{1.3cm} \mbox{Re } \omega(k^*)=0~,
\nonumber\\  \label{5a25} \lambda^* & \equiv & \mbox{Im } k^* >0~, ~~
s(\lambda^*)  \equiv
\mbox{Im } \omega(k^*) >0 ~,
  \\  D & = &  \left. \frac{1}{2} \frac{d^2 s }{d
    \lambda^2}\right|_{\lambda^*}  >0, ~~\mbox{Im }
  D=0 ~.\nonumber
\end{eqnarray}
With this assumption we consider only the generic case, that
dynamically accessible uniformly translating solutions of 
{\em real} equations will be characterized by a real spatial decay 
rate $\lambda$ and a real growth rate $s$, and that they will leave 
a homogeneous state $\phi=1$ behind. This
might exclude some pathological cases of uniformly translating 
front solutions, that are not characterized by a real 
$\lambda$\footnote{Elsewhere \cite{spruijt}, we will discuss
an extension of the notion of uniformly translating fronts
that allows to write pattern forming fronts 
in the Swift-Hohenberg equation as uniformly
translating solutions of a suitable set of {\em complex}
amplitude-like modes. For these $\mbox{Re }k\neq 0$. Similar
considerations hold for fronts in the complex Ginzburg-Landau equation
itself \cite{vs3}.}.

If the saddle point obeys (\ref{realdef1}), the expression for $t\gg1$
(\ref{5a20}) and $v=v^*$ for the Green's function $G$ reduces to
\begin{eqnarray}
\label{5a26}
G(\xi,t) &=& e^{-\lambda^*\xi}\;\frac{e^{-{\xi}^2/(4Dt)}}{\sqrt{4\pi
    Dt}} \; \left(1+O\left(\frac{D_3 \xi }{D^2 {t}}\right)\right)~,
\nonumber\\ &&~~\xi=x-v^*t~.
\end{eqnarray}
Except for a rescaling of time and length scales with the real
constants $\lambda^*$ and $D$, this is precisely the functional form
of (\ref{2054}).

If we consider the velocity $v(\lambda)$ of the family of front
solutions whose asymptotic spatial decay is as $e^{-\lambda \xi}$ with
real $\lambda$, then it is straight forward to see {\em that $\Phi^*$
  is the slowest of all these uniformly translating fronts:} According
to the linearized equation (\ref{5a14}) the solution in the leading
edge is as $e^{-\lambda x-i\omega(i\lambda)t}$.  The resulting
velocity $v$ is
\begin{equation} \label{50033}
  v(\lambda)=\frac{-i\omega(i\lambda)}{\lambda}= \frac{s
    (\lambda)}{\lambda} ~\mbox{ for all }\lambda~.
\end{equation}
The minimum of this curve is given by
\begin{eqnarray} \label{50034}
  0=\left.\frac{\partial
    v(\lambda)}{\partial\lambda}\right|_{\lambda^*}
    &=& \left.\frac{1}{\lambda}\;\left( \frac{\partial s
      (\lambda)}{\partial\lambda} -\frac{s
      (\lambda)}{\lambda}\right)\right|_{\lambda^*}~.
\\ \left. \frac{d^2 v(\lambda)}{d\lambda^2}\right|_{\lambda^*} & = &  
\frac{ 2D }{\lambda^*} > 0~.
\end{eqnarray}
 
Taking into account that $\omega(k)$ is analytic, Eqs.\ (\ref{50033})
and (\ref{50034}) are equivalent to Eqs.\ (\ref{5a24}) and
(\ref{5a24a}), because at a saddle of an analytic function, the maximum
as a function of real $k$ coincides with a minimum as a function of
imaginary $k$.

The analysis of the stability of the uniformly translating solutions
proceeds largely as in Section \ref{S23}: the existence of a family of
front solutions implies, according to counting arguments as
given in Appendix \ref{A6}, that there is at least a continuous
spectrum of eigenmodes of the stability operator.
Indeed, if we again write the temporal behavior of the stability
eigenmodes as $e^{-\sigma t}$ and the steepness of the modes as
$\Lambda$, and if we first focus on the spectrum of perturbations
that is also continuous in $\Lambda$,
then we have for the front solutions with 
$v(\lambda) \geq v^*$: $\sigma=-(s(\Lambda)-v(\lambda )\Lambda)$. 
Expanding the $\Lambda$ of the perturbation about the
$\lambda$ of the front, we then get
\begin{eqnarray}
  \sigma(\Lambda) & \approx& -\left( \frac{\partial
    s(\lambda)}{\partial \lambda} - v(\lambda)\right) (\Lambda
    -\lambda)~,\nonumber\\ & = & - \lambda \frac{\partial
      v(\lambda)}{\partial \lambda}~
    (\Lambda-\lambda)~,\label{sigmastab}
\end{eqnarray}
using (\ref{50033}) in the second line.
Since we showed above that ${\partial v}/{\partial \lambda }<0$
for $\lambda<\lambda^*$ ($v>v^*$), $\sigma(\Lambda)>0$ for $\Lambda >
\lambda$. This generalizes the result (\ref{hoiute}) for the 
nonlinear diffusion
equation that the front solutions $\Phi_v$ are stable to modes from
the continuous spectrum which are steeper than the front itself.
In addition to the continuous $\Lambda$ spectrum there again
may be discrete perturbation modes associated with the existence
of pushed front solutions.

We show in Appendix \ref{A6} that the existence of a strongly
heteroclinic front solution $\Phi^\dagger$ implies the existence of
unstable strongly heteroclinic stability modes for $v<v^\dagger$,
again in parallel to the results for the nonlinear diffusion equation. 
The central assumption of our further analysis is of course that we 
are in the pulled regime, and hence that such solutions are absent.

We finally note that the fact that $v(\lambda)$ has a minimum for
$\lambda=\lambda^*, v=v^*$, implies that for $v<v^*$ front solutions
decay to zero in an oscillatory manner for $\xi\to \infty$ as they
have $\mbox{Re} k \neq 0$. 
By expanding the function $v(\lambda)$ about the bifurcation point at
$v^*, \lambda^*$, it is easy to show that for small $|v-v^*|$, this
branch of solutions has $\mbox{Im}(k-k^*)=\lambda-\lambda^* \approx
(\lambda^*)^2 v'''/(12 D) |v-v^*|$, $\mbox{Re}k \approx
\sqrt{\lambda^* |v-v^*|/D} $, where $v'''= \frac{\partial^3 v(\lambda
  )}{\partial \lambda^3}|_{\lambda^*}$. One usually has $v'''<0$ and
then such solutions are unstable according to a slight generalization
of (\ref{sigmastab}).

\subsubsection{The leading edge representation}
\label{S533}

As in our analysis of the pulled dynamics of the nonlinear diffusion 
equation, we will find it expedient to study the large time 
asymptotics in the leading edge by using the leading edge representation 
$\psi$. For uniformly translating fronts, the immediate generalization 
of the transformation (\ref{20058}) from Section \ref{S2} is
\begin{equation}
\label{5a27}
\psi(\xi,t) = \phi(x,t)\;e^{\lambda^*\xi}~~,~~\xi=x-v^*t~.
\end{equation}
The linearized dynamical evolution equation for the leading edge
representation now generalizes   (\ref{2058}) to
\begin{equation} \label{50037a} \label{5a28}
\partial_t\psi={\cal D}\psi + o\left(\psi^2\;e^{-\lambda^*\xi}\right)~,
\end{equation}
where
\begin{equation} \label{50036a} \label{5a29}
{\cal D} = e^{\:\lambda^*\xi}\;{\cal L}_{v^*}(\infty)\;e^{-\lambda^*\xi}
= \sum_{n=2}^N D_n\;\partial_{\xi}^n~.
\end{equation}
A short calculation (Appendix \ref{A8}) reveals that the constants
$D_n$ can be expressed  
in terms of the dispersion relation $\omega(k)$ (\ref{5a15}) as
\begin{eqnarray}
\label{5a30}
D_n& = &
  \left.\frac{1}{n!}\;\frac{\partial^n}{\partial(-\lambda)^n}\;
    \Big(-i\omega(i\lambda)-v^*\lambda\Big)\right|_{\lambda=\lambda^*}
\label{diffdef}\\
 & = &  \left.\frac{1}{n!}\;\frac{\partial^n}{\partial(-\lambda)^n}\;
    \Big(s(\lambda)-v^* \lambda\Big)\right|_{\lambda=\lambda^*}~. 
    \nonumber
\end{eqnarray}
Note, that in this generalized leading edge representation (\ref{5a28})
the coefficients of $\psi$ and $\partial_{\xi}\psi$ again
are vanishing. This is an immediate consequence of the proper
choice of $v^*$ and $\lambda^*$. In fact, for uniformly translating 
fronts (\ref{5a25}) $D_0=0$ is equivalent to the proper choice
of the velocity $v^*$ (\ref{5a24}) and $D_1=0$ is equivalent
to the saddle point equation (\ref{5a24a}) fixing $\lambda^*$
for given $v^*$. $D_2$ is obviously identical to $D$ from (\ref{5a19}).
We will see below, that in the leading edge, the contribution 
proportional to $D_2= D$ gives the dominant contribution,
while $D_3$ appears only in the subdominant term, similar to
what we already observed in (\ref{5a22}). We therefore will 
essentially recover the results of the nonlinear diffusion equation
(\ref{101}), which had the particular property of $D_n=0$ for $n>2$.

\subsubsection{The relaxation analysis} \label{S534}

We have now laid the ground work for the extension of
the analysis of the relaxation of
pulled fronts for our more general equation (\ref{5a2}) in the case 
of sufficiently steep initial conditions which as before are
characterized by the requirement that 
\begin{equation}
\label{5a32} \label{5035}
\lim_{x\to\infty} \phi(x,0)\;e^{\lambda x}=0~
\end{equation}
for some $\lambda>\lambda^*$.
The analysis in Section \ref{S3} for the nonlinear diffusion equation
(\ref{101}) was based on the following steps:
\begin{itemize}
\item[]{\em Step 1.}\\ The proper choice of the comoving coordinate system
\begin{equation}
\label{5036}
\xi_X=x-v^*t-X(t)~~,~~\dot{X}=\frac{c_1}{t}+\frac{c_{3/2}}{t^{3/2}}+\ldots~,
\end{equation}
allowing for a logarithmic shift $X(t)\propto\ln t$ in comparison with to
the asymptotic coordinate system $\xi=x-v^*t$.
\item[]{\em Step 2.}\\ An expansion of $\phi$ in the nonlinear interior
part of the front about the asymptotic front profile $\Phi^*(\xi_X)$,
taken, however, not in the frame moving with velocity $v^*$, but in the
frame $\xi_X=x-\int^t dt' v(t')$ with velocity $v(t)=v^*+\dot{X}(t)$.
\item[]{\em Step 3.}\\ A resummation of this expansion of $\phi$ in the 
cross-over
region towards the leading edge, where the new variable $z=\xi_X^2/(4t)$
is introduced for the region with $\xi_X\ge O(\sqrt{t})$.
\item[]{\em Step 4.}\\ An analysis of the leading edge in variables $z$ and $t$,
where $\phi$ now is linearized about the unstable state $\phi=0$, and not
about $\Phi^*$.
The two boundary conditions that $\phi$ crosses over to the functional form 
of Step 3 for $z\ll1$, and that $\phi$ is steeper than $\Phi^*$ for
$z\gg1$, now determine both the functional form of $\phi$ and
the constants $c_{n/2}$ in $\dot{X}$. (We can think of this as a
matching procedure.) In this analysis, the fact that
the  parameter $\alpha\ne0$ in the asymptotics $\Phi^*(\xi_X) =
(\alpha \xi_X+\beta) e^{-\lambda^* \xi_X}$ is nonzero (see Section \ref{S252})
 plays a central role.
\end{itemize}

The generalization of these steps to our equation (\ref{5a2}) which is
 of  higher order in space, is actually quite straightforward. 
We again use the general coordinate $\xi_X$ (\ref{5036}) with $\dot{X}(t)$
to be determined. The interior expansion $\eta(\xi_X,t)=\phi-\Phi^*(\xi_X)$
from Section \ref{S32} applies literally, except that we now need to
use the linear operator ${\cal L}^*={\cal L}_{v^*}(\xi_X)$ from
(\ref{5a6}). Accordingly, also the resummation (\ref{3031}) again is
valid, and we again have 
\begin{equation}
\label{5037}
  \phi = \Phi_{v(t)} (\xi_X) + O\left(\frac{1}{t^2}\right)~,
\end{equation}
with $\Phi_v$ a uniformly translating solution  of (\ref{5a2}) with
velocity $v$. The correction
$O(1/t^2)$ is again non-vanishing and non-universal, as it
depends on the precise initial conditions.

The expansion of the interior shape towards the leading edge (\ref{3037})
depends on both the differential operator ${\cal L}^*$ for $\xi\to\infty$
and on the shape of the asymptotic front $\Phi^*$ (\ref{3033}).
Eq.\ (\ref{3033}) is generalized to
\begin{equation}
\label{5038}
\Phi^*(\xi)\stackrel{\xi \gg 1}{\sim}(\alpha\xi+\beta)\;e^{-\lambda^*\xi}
+\ldots~,
\end{equation}
since the saddle point expansion in Section \ref{S532} implies that
for a pulled front $\Phi^*$, two roots of the dispersion relation
coincide. Generally,
\begin{equation}
\label{5039}
\alpha\ne0 ~,
\end{equation} 
since a calculation resulting in a generalization of (\ref{2061}) can be set
up along similar lines: If there is a bounded uniformly translating solution
$\Phi^*(\xi)$, then upon going to  the leading edge representation and 
integrating the equation for $\Psi^*(\xi)$ once over $\xi$, we
find that $\alpha$ can be expressed in terms of the spatial integral
over the nonlinear terms. 

How does the leading edge develop under inclusion of
the higher spatial derivatives? First of all
we observe that the large-$t$-solutions (\ref{5a26}) and
(\ref{2054}) of the linearized equation (\ref{5a14})
are in leading order identical up to rescaling. In other words, 
the saddle point approximation renders the spreading around 
the asymptotic exponential solution diffusive.
This suggests that the leading edge can be analyzed by the same type 
of similarity variables $(z,t)$ as in (\ref{3041}). In fact, in our 
shifted coordinate frame $\xi_X$ (\ref{5036}) the leading edge 
representation is
\begin{eqnarray} \label{5040a}
\phi(x,t)&=&e^{-\lambda^*\xi_X}\psi(\xi_X,t)~,
\\ \label{5040}
\partial_t\psi&=&{\cal D}\psi+\dot{X}\left(\partial_\xi-\lambda^*\right)\psi
+o(e^{-\lambda^*\xi_X})~,
\end{eqnarray}
with the differential operator ${\cal D}$ from Eq.\ (\ref{5a29}).
After a rescaling with
\begin{eqnarray} \label{5041}
  \zeta_Y&=& \lambda^*\xi_X~~,~~\tau=D_2{\lambda^*}^2t~~,~~
  d_n=\frac{D_n{\lambda^*}^n}{D_2{\lambda^*}^2}~, 
  \nonumber\\ 
  \dot{Y}&=&\frac{\dot{X}\lambda^*}{D_2{\lambda^*}^2}=
  \frac{C_1}{\tau}+\frac{C_{\frac{3}{2}}}{\tau^{3/2}}+\ldots~,
  \nonumber\\
  C_n&=&c_n\lambda^*\;\left(D_2{\lambda^*}^2\right)^{n-1}~,
\end{eqnarray}
this equation takes the form
\begin{equation} \label{5042}
\partial_\tau\psi=
\left(\partial_\zeta^2+\sum_{n=3}^Nd_n\partial_\zeta^n\right)\psi
  +\dot{Y}\left(\partial_\zeta-1\right)\psi~,
\end{equation}
that is the same as Eq.\ (\ref{3039}) except that there are now
the higher derivatives $\partial_\zeta^n$. As we show explicitly in Appendix
\ref{hot}, the leading edge can be analyzed with the same ansatz as in
(\ref{3041}) and (\ref{3043}),
\begin{eqnarray} \label{5044}
\psi(\zeta_Y,\tau)&=&e^{-z}\;G(z,\tau)~~,~~z=\frac{\zeta_Y^2}{4\tau}~,\\
G(z,\tau)&=&\sqrt{\tau}\;g_{\frac{-1}{2}}(z)+g_0(z)
+\frac{g_{\frac{1}{2}}(z)}{\sqrt{\tau}}+\ldots ~,\nonumber
\end{eqnarray}
and in rescaled variables, one gets
\begin{eqnarray} \label{5047}
C_1&=&\frac{-3}{2}~~,~~C_{\frac{3}{2}}=\frac{3\sqrt{\pi}}{2}~~,
~~g_{\frac{-1}{2}}(z)=2\alpha\sqrt{z}~,\\
g_0(z)&=&\beta\;(1-2z) +3\alpha(1+d_3)z-2\alpha d_3z^2-\frac{3\alpha}{2}\;F_2(z)
\nonumber\\
&&~~+6\alpha\;\sqrt{\pi\:z}\;
\left(1-M\left(\frac{-1}{2},\frac{3}{2},z\right)\right)~.
\nonumber
\end{eqnarray}
In these variables the result is identical with that for the nonlinear
diffusion equation in Section \ref{S3}, except for the
additional terms proportional to $d_3$ in $g_0(z)$. 
In particular, the velocity parameters $C_1$ and $C_{3/2}$ and the leading
order contribution $g_{\frac{-1}{2}}(z)$ are independent of the value
of $d_3$, just like the subdominant term $\beta$ from (\ref{5038})
 enters $g_0(z)$ but not the other quantities. 
That for the problem written in variables 
$z$ and $\tau$, $d_3$ can only contribute in subleading order, 
is in fact immediately obvious after the transformation. It is surprising, 
however, that the subleading velocity coefficient $C_{3/2}$
is independent of the value of $d_3$. We will find it to be unchanged
even for much more general equations.

In terms of the unscaled variables, the universal algebraic convergence
of the velocity is given by
\begin{equation}
\label{5051}
v(t) = v^* -\frac{3}{2\lambda^* t}
\left(1-\sqrt{\frac{\pi}{(\lambda^*)^2Dt}}\;\right) + \cdots
\end{equation}
where $v^*$ and $\lambda^*$ are determined by the saddle point 
equations (\ref{5a24}) and (\ref{5a24a}) together with 
(\ref{defvstar}), and where the diffusion coefficient 
$D$ (\ref{5a19}) equals $D_2$ from (\ref{diffdef}).
The central results of this analysis are summarized in Table II.

\subsection{Generalization to single {\em p.d.e.}'s  of higher order in time} 
\label{S54} 
 
We now proceed in two further steps of generalization.  In the present
Section we first discuss partial differential equations for a single
field $\phi(x,t)$, which include higher order temporal derivatives as
well as mixed temporal and spatial derivatives. These are of the form
\begin{eqnarray} \label{506} \label{5a1}
  F\Big(\phi,&\;&\partial_x\phi,\;\ldots,\;\partial_x^N\phi,\;\partial_t\phi,
  \;\ldots,\;\partial_t^M\phi, \nonumber\\ &&
  \partial_t\partial_x\phi,\;\ldots,\;\partial_t^M\partial_x^N\phi\Big)=0~,
\end{eqnarray} 
generalizing (\ref{5a2}) to $M \ge 1$. In Section \ref{S55}, we then
also deal with difference or integro-differential equations and
coupled equations.

The extension to equations of type (\ref{5a1}) presents no conceptual
difficulty --- we will follow here a route that is the immediate
generalization of the discussion in the previous Section. The new
elements in the discussion will be the fact that higher order temporal
derivatives and mixed spatial and temporal derivatives are generated
in the dynamical equation for the leading edge representation $\psi$,
but as we shall see these turn out not to affect the expression for
the velocity relaxation and for the relaxation of the shape in the
interior front region. The notation in (\ref{matrixS}) -- (\ref{M10}),
which may strike the reader at first sight as unnecessarily heavy,
prepares for the discussion of even more general equations and sets of
equations in \ref{S55}, where finding a proper scalar leading edge 
representation is less straight forward than here.

If we linearize (\ref{5a1}) about $\phi=0$, we get an equation of the
form
\begin{equation}
\label{M1}
\sum_{m=0}^M \sum_{n=0}^N
a_{mn}\;\partial_t^m\;\partial_x^n\;\phi(x,t) + o(\phi^2)=0~.
\end{equation}
For solving the initial value problem in time, it is convenient to
Fourier-transform in space
\begin{equation}
\label{M2}
\phi(x,t)=\int_{-\infty}^\infty
\frac{dk}{2\pi}\;e^{ikx}\;\tilde{\phi}(k,t)~.
\end{equation}
Below we will use the superscript $\tilde{\;}$ to denote a quantity
Fourier transformed in space.

The Fourier transformation of (\ref{M1}) results in an {\em o.d.e.}\ of
order $M$ for every Fourier mode $\tilde{\phi}(k,t)$:
\begin{equation}
\label{M3}
\sum_{m=0}^M A_m(k)\;\partial_t^m\;\tilde{\phi}(k,t)=0
~~,~~A_m(k)=\sum_{n=0}^N a_{mn} (ik)^n~.
\end{equation}
Obviously, we need $M$ functions to specify the initial conditions.
We write these as an $M$-dimensional vector:
\begin{equation}
\label{M4}
\underline{\tilde{\phi}} =
(\tilde{\phi},\partial_t\tilde{\phi},\ldots,\partial_t^{M-1}\tilde{\phi})~.
\end{equation}
The equation of motion (\ref{M3}) can now be written in Fourier space
as
\begin{equation}
\label{M5}
\partial_t\;\underline{\tilde{\phi}}(k,t)=
-\underline{\underline{\tilde{T}}}(k)\cdot\underline{\tilde{\phi}}(k,t)~,
\end{equation}
with the $M \times M$ matrix
\begin{equation}
\label{M6}
\underline{\underline{\tilde{T}}}(k) = \left(\begin{array}{ccccc} 0 &
-1 & 0 & \cdots & 0 \\ 0 & 0 & -1 & & 0 \\ &\vdots & & \ddots & \\ 0 &
0 & 0 & \cdots & -1 \\ \frac{A_0}{A_M} & \frac{A_1}{A_M} &
\frac{A_2}{A_M} & \cdots & \frac{A_{M-1}}{A_M}
\end{array}\right)~.
\end{equation}
For later use, we here already define the matrix
\begin{equation}
\label{matrixS}
\underline{\underline{\hat{S}}}(k,\omega)
=A_M(k)\;\left(\underline{\underline{\tilde{T}}}(k) -
i\omega\;\underline{\underline{1}}\right)~,
\end{equation}
which later will result from a Fourier-Laplace transformation as in
(\ref{5a17b}). Here and below, we use the superscript $\hat{\;}$ to
denote a Fourier-Laplace transformed quantity to distinguish it from
spatially Fourier transformed quantities, which are indicated with a
tilde.

The $M$ eigenvalues $\omega_m(k)$ ($m=1,\ldots,M$) of the matrix
$\underline{\underline{\tilde{T}}}(k)$ are determined by the
characteristic equation $S\big(k,\omega_m(k)\big)=0$, where 
$S(k,\omega)$ is the characteristic polynomial
\begin{eqnarray}
\label{M8}
S(k,\omega) &=& \mbox{det}\;\underline{\underline{\hat{S}}}(k,\omega)
= \sum_{m=0}^M A_m(k)\;(-i\omega)^m \nonumber\\ &=&
\sum_{m=0}^M\sum_{n=0}^N a_{mn}\;(-i\omega)^m\;(ik)^n~.
\end{eqnarray}
Defining the eigenvectors $\underline{\tilde{U}}_{\:m}(k)$ of the
matrix $\underline{\underline{\tilde{T}}}(k)$ through
\begin{equation}
\label{5a51}
\underline{\underline{\tilde{T}}}(k)\cdot\underline{\tilde{U}}_{\:m}(k)=
i\omega_m(k)\;\underline{\tilde{U}}_{\:m}(k)~,
\end{equation}
and their adjoints through
\begin{equation}
\label{5a52}
\underline{\tilde{U}}_{\:m}^\dag(k)\cdot
\underline{\tilde{U}}_{\:n}(k) =\delta_{mn}~,
\end{equation}
the matrix $\underline{\underline{\tilde{T}}}(k)$ can be written as
\begin{equation}
\label{M7}
\underline{\underline{\tilde{T}}}(k) = \sum_{m=1}^M i\omega_m(k)\;
\underline{\tilde{U}}_{\:m}(k) \times
\underline{\tilde{U}}_{\:m}^\dag(k)~,
\end{equation}
where $\times$ denotes the outer product.

Now (\ref{M5}) is easily integrated in time and the Fourier
transformation inverted. We find in generalization of (\ref{5a16}) and
(\ref{5a17}):
\begin{eqnarray}
\label{M9}
\underline{\phi}(x,t)&=&\int_y\underline{\underline{G}}(x-y,t)\cdot
\underline{\phi}(y,0)~, \\ 
\label{M10}
\underline{\underline{G}}(x,t)&=& \sum_{m=1}^M \int
\frac{dk}{2\pi}\;e^{ikx-i\omega_m(k)t}\;
\underline{\tilde{U}}_{\:m}(k) \times
\underline{\tilde{U}}_{\:m}^\dag(k)~.  \nonumber\\ 
\end{eqnarray}
Obviously, the quickest growing mode $\underline{\tilde{U}}_{\:m}(k)$
--- characterized now by Fourier mode $k$ and branch of solutions $m$
--- again will be determined by a saddle point (\ref{5a24}),
(\ref{5a24a}).  Even more so than in the case of a first order
equation, we can in general have more than one saddle point, as each
branch of the disperion relation can in principle have one or more
saddle points (a
trivial example for two coupled equations is discussed in Appendix
\ref{Adoublekpp}). Again, the relevant saddle point is the one through 
which the $k$-contour can be deformed and which has
the largest 
velocity $v^*$ in the comoving frame. The associated saddle point
values are denoted as 
$k^*=i\lambda^*$, $D$ etc. As before, we assume uniform translation 
as in (\ref{5a25}), so that $k^*$ and $\omega(k^*)$ are purely imaginary.
Suppose, that $v^*$ lies on the branch $\omega_1(k)$. We
then find in the comoving frame $\xi=x-v^*t$ for long times $t$:
\begin{equation}
\label{GVD}
\underline{\underline{G}}(\xi,t)=
e^{-\lambda^*\xi}\;\frac{e^{-\xi^2/(4Dt)}}{\sqrt{4\pi Dt}}\;
\underline{\tilde{U}}_{\:1}(k^*) \times
\underline{\tilde{U}}_{\:1}^\dag(k^*) +\ldots ~,
\end{equation}
in generalization of (\ref{5a26}).

This result shows that in the long time limit the Green's function
$\underline{\underline{G}}$ projects onto the eigendirection
$\underline{\tilde{U}}_{\:1}(k^*)$. The result (\ref{GVD}) is not
restricted to the explicit form (\ref{M6}) of the matrix
$\underline{\underline{T}}$, so it applies to sets of coupled {\em
  p.d.e.}'s just as well as they also can be written in the form
(\ref{M5}).  Projection onto the eigendirection
$\underline{\tilde{U}}_{\:1}(k^*)$ then defines the scalar leading
edge equation resulting from coupled {\em p.d.e.}'s. We will further
exploit this property in the following section.

In the present section we just use (\ref{GVD}) to calculate $v^*$ and
$\lambda^*$, and to demonstrate why the leading edge transformation
catches the relevant dynamics.  Proceeding as in earlier Sections, the
scalar equation (\ref{M1}) now transforms under the leading edge
transformation with $v^*$ and $\lambda^*$ to
\begin{eqnarray}
\label{M11}
\phi(x,t)&=&e^{-\lambda^*\xi}\; \psi(\xi,t)~~,~~\xi=x-v^*t~,
\\
\label{M12}
0&=&\sum_{m=0}^M \sum_{n=0}^N a_{mn}
\left(\partial_t-v^*\partial_\xi+v^*\lambda^*\right)^m
  \left(\partial_\xi-\lambda^*\right)^n \psi \nonumber\\ 
    &=&\sum_{m=0}^M \sum_{n=0}^{M+N}
    b_{mn}\;\partial_t^m\;\partial_\xi^n\;\psi(\xi,t)~.
\label{Mfinal}
\end{eqnarray}

Just as the $a_{mn}$ from Eq.\ (\ref{M1}) can be written in terms of
derivatives of the characteristic polynomial $S(k,\omega)$ (\ref{M8})
as
\begin{equation}
  a_{mn}=\left.  \frac{(i\partial_\omega)^m}{m!} \;
  \frac{(-i\partial_k)^n}{n!}\; S(k,\omega)\right|_{(k=\omega=0)}~,
\end{equation} 
so  the $b_{mn}$ can be written as derivatives as well,
similar to (\ref{5a30}). In showing this, it simplifies the notation
to use coordinates  
expanded about the saddle point by introducing the variables
\begin{equation}
\label{omOmdef}
\Omega = \omega -v^* k~,~~~ q=k-k^* =k -i\lambda^*~,
\end{equation}
and by defining
\begin{eqnarray}
\label{5b61}
  S^*(q,\Omega) &=& S(k^*+q,\omega^*+v^*q + \Omega) ~,
  \\
  &&~~~~ k^*=i\lambda^*
  ~~,~~\omega^*=v^*k^*~.
  \label{ss*neu}
  \nonumber
\end{eqnarray}
When we will later consider the Fourier-Laplace transform of
$\psi(\xi,t)$ in the frame $\xi$, the frequency in this frame will
turn out to be $\Omega$ and the wavenumer will turn out to be $q$,
since $e^{-i\omega t+ikx}= e^{-\lambda^* \xi} (e^{-i\Omega t+iq
  \xi})$. Accordingly, the long time--small gradient expansion of
$\psi(\xi,t)$ will correspond to a small $\Omega$--small $q$
expansion. Indeed, in line with this interpretation, inspection of
(\ref{Mfinal}) shows that the $b_{mn}$ are simply
\begin{eqnarray}
\label{bmnS}
b_{mn}&=&\left.  \frac{(i\partial_\omega)^m}{m!} \;
\frac{(-i)^n(\partial_k+v^*\partial_\omega)^n}{n!}\;
S(k,\omega)\right|_{(k^*,v^*k^*)} \nonumber\\ &=& \left.
\frac{(i\partial_\Omega)^m}{m!} \; \frac{(-i\partial_q)^n}{n!}\;
S^*(q,\Omega)\right|_{(q=\Omega=0)}~.
\end{eqnarray}
We will discuss the precise correspondence between the formulation in
terms of $S$ and the dispersion relation $\omega_1(k)$ below, and just note
here that the saddle point equations that determine $\lambda^*$
and $v^*$ are expressed by
\begin{equation}
\label{saddle}
b_{00}=S^*(0,0) =0 ~, ~~~
b_{01}=-i\partial_q S^*(q,\Omega)|_{q=\Omega=0}=0~.
\end{equation} 
After dividing the whole equation (\ref{M12}) by $b_{10}$ and
introducing the notations
\begin{equation}
\label{ratios}
D_n=-\frac{b_{0n}}{b_{10}}~~,~~w=\frac{b_{11}}{b_{10}}~~,~~
\tau_1=\frac{b_{20}}{b_{10}}~~,~~ \mbox{etc.}~,
\end{equation}
the terms with the lowest derivatives are
\begin{eqnarray}
\label{M13}
\Big(\partial_t+\tau_1\;\partial_t^2+\ldots&-&
D_2\;\partial_\xi^2-D_3\;\partial_\xi^3+\ldots
\\&+& w\;\partial_t\partial_\xi+\ldots\Big)\;\psi
+o(\psi^2\;e^{-\lambda^*\xi})=0~.  \nonumber
\end{eqnarray}
This is the leading edge equation in its most general form.  Note
that after the leading edge transformation, the coefficient $w$ may be
nonzero even if the coefficient $a_{11}=0$ of
$\partial_t\partial_x\phi$ in the original equation of motion
(\ref{M3}) vanishes.

To show the connection with our discussion of first order equations in
earlier Sections, it is instructive to analyze the relation between
$S$ and the dispersion relation.  The various branches $\omega_m(k)$
or $\Omega_m(q)$ of the dispersion relation are defined implicitly
through the roots of
\begin{equation}
  S(k,\omega_m(k)) =0 ~~~\Longleftrightarrow ~~~S^*(q,\Omega_m(q))=0~,
  \label{dispdefnew}
\end{equation}
As before, let $\omega_1(k)$ ($\Omega_1(q)$) be the branch on which
the saddle point determining $v^*$ lies. Upon differentiating
(\ref{dispdefnew}) once with respect to $k$ or $q$ and using Eqs.\
(\ref{bmnS}) and (\ref{saddle}), we get our familiar result
\begin{equation}
  \left. \frac{d\omega_1(k)}{dk} \right|_{k^*} = v^* ~~~
    \Longleftrightarrow~~~ \left.
    \frac{d\Omega_1(q)}{dq}\right|_{q=0}=0~.
\end{equation}
Likewise, by differentiating (\ref{dispdefnew}) twice we get
\begin{equation}
  \left. \frac{ d^2 \Omega_1(q)}{dq^2}\right|_{q=0} = \left. \frac{d^2
      \omega_1(k)}{dk^2}\right|_{k^*} = - \left. \frac{\partial_q^2
      S(q,\Omega)}{\partial_\Omega S(q,\Omega)}
  \right|_{q=\Omega_1(0)=0}~. \label{derivativesofom}
\end{equation}
If we combine this with the expression $D=-b_{02}/b_{10}$, we
recover our familiar expression
\begin{eqnarray}
  D & = & \left. \frac{\partial_q^2 S(q,\Omega)}{2i \partial_\Omega
    S(q,\Omega)} \right|_{q=\Omega_1(0)=0}~ = \left. \frac{i d^2
    \Omega_1(q)}{2 dq^2}\right|_{q=0}~, \nonumber \\ & = & \left.
  \frac{i d^2 \omega_1(k)}{2 dk^2}\right|_{k^*} ~. \label{Dexprfinal}
\end{eqnarray}
For the case of an equation which is of first order in time, one can
easily check that our general expression for $D_n$ reduces to the one
given before in (\ref{5a30}), $D_n= (-i/n!) d^n\omega/d(ik)^n
|_{k^*}$.

Before we discuss the consequences of (\ref{M13}), we note in passing
that formally we could have proceeded directly from the
linearized equation of motion (\ref{M1}) to the leading edge
representation (\ref{Mfinal}) and hence to (\ref{M13}), by
choosing the two parameters $v^*$ and $\lambda^*$ such that the two
conditions $b_{00}=0=b_{01}$ are obeyed. The detour from this
straightforward transformation via the saddle point analysis was taken
to bring out the physical origin of the transformation in this context
and to show why one has to use the saddle point $(v^*,\lambda^*)$ with the
largest $v^*$ through which the contour can be deformed. In addition,
it explicitly shows how 
a particular ``direction'' $\underline{U}_1(k^*)$ of the vector
field $\underline{\phi}$ corresponds to the slow leading edge
dynamics. We will see in the next section that for coupled equations
there is some freedom in choosing the projection onto a scalar 
leading edge variable.

Let us now analyze the implications of the leading edge representation
(\ref{M13}). First of all, we observe, that a uniformly translating
pulled front $\Phi^*(\xi)=e^{-\lambda^*\xi} \Psi^*(\xi)$ still will
have the form (\ref{5038}) $\Psi^*(\xi)=(\alpha\xi+\beta)$, and that
the argument for $\alpha\ne0$ from Section \ref{S252} still does
apply.

Can the extra terms $\tau_1\:\partial_t^2\psi$,
$w\:\partial_t\partial_\xi\psi$ etc.\ change our relaxation prediction
from Section \ref{S53}? A short inspection shows, that after rewriting
the equation in variables $z$ and $t$, cf.\ (\ref{5041}) --
(\ref{5044}) and (\ref{3040}), $w\:\partial_t\partial_\xi\psi$ will be
of the same subleading order in $1/\sqrt{t}$ as
$D_3\;\partial_\xi^3\psi$, while both the terms
$\tau_1\:\partial_t^2\psi$ and $D_4\;\partial_\xi^4\psi$ will be one
order lower. Also, when rewriting the equation in the variable
$\xi_X=x-v^*t-X(t)$, higher temporal derivatives will create terms
like $\ddot{X}$ and $\dot{X}^2$ from the exponential factor in the
leading edge transformation
$\phi(\xi_X,t)=e^{-\lambda^*\xi_X}\psi(\xi_X,t)$. Since these are of
order $1/t^2$, they do not influence the leading and subleading terms.

We do not repeat the detailed calculation here, because it completely
follows the lines of the earlier one.  One finds that the result again
is given by (\ref{5047}), except that the subleading $g_0(z)$ picks up
another polynomial contribution from $w$ besides the one from $D_3$,
namely
\begin{equation}
\label{M14}
g_0(z)=g_0(z)\Big((\ref{5047})\Big)+2\alpha\;w\lambda^*\;
\left(z^2-\frac{3}{4}\right)~.
\end{equation}
The uniform velocity relaxation is invariably
\begin{equation}
\label{M15}
v(t) = v^* -\frac{3}{2\lambda^* t}
\left(1-\sqrt{\frac{\pi}{(\lambda^*)^2Dt}}\;\right) + \cdots~,
\end{equation}
and the interior part of the front is again slaved to the tip like
\begin{equation}
\label{M16}
\phi(x,t) = \Phi_{v(t)} (\xi_X) + O\left(\frac{1}{t^2}\right)~.
\end{equation}
So the predictions from Table II also apply to {\em p.d.e.}'s with
higher temporal derivatives like (\ref{5051}), if the front is pulled.

Thus we reach the important conclusion that {\em the universal power
  law convergence is not an artefact of the diffusion-type character
  of the nonlinear diffusion equation: it holds generally in the
  pulled regime of uniformly translating fronts, because the expansion
  about the saddle point, which governs the dynamics of the leading
  edge representation $\psi$, is essentially diffusive}.

\subsection{Further generalizations}\label{S55}    

We now complete the last step in our discussion, and show that our
results hold much more generally: even if the original dynamical
equation is not a {\em p.d.e.}, the dynamical equation for the
appropriate  leading
edge variable $\psi$ is {\em still} the same diffusion type equation
(\ref{M13}), and consequently, our results for the velocity and shape
relaxation from Table II do apply.

When we have a set of coupled equations, we can view them as components 
of a vector field, using a notation as in (\ref{M5}) with a different 
matrix $\underline{\underline{\tilde{T}}}(k)$.
The main complication we are facing  in this case  
is that the leading edge dynamics then not only ``selects'' a velocity
$v^*$ in the pulled regime, but also  an associated  eigendirection 
$\underline{\tilde{U}}_m(k)$ in
this vector space --- this eigendirection determines the relative
values of the various fields in the leading edge of the front. 
The long time dynamics  
in the frame moving with the pulled velocity $v^*$ is then associated
with a slow dynamics along this eigendirection, while the dynamics
along the other eigendirections is exponentially damped. The appropriate
{\em scalar} leading edge variable $\psi$ will then turn out to be
nothing but the projection of the dynamics along this slow direction.

The second complication is that we now consider
equations whose temporal dependence is not
necessarily of differential type $\partial_t^N$: they may just as
well be of difference type or contain
memory kernels.  To treat such equations, we  also perform 
a Laplace transformation in time besides the Fourier transformation
in space just as in (\ref{5a17b}) by defining
\begin{equation}
\hat{\phi}_m(k,\omega) = 
\int_0^\infty dt\;e^{i\omega t}\;\tilde{\phi}_m(k,t)~.
\end{equation}
We thus consider dynamical systems that after the
Fourier-Laplace-transformation of the equations, linearized about the
unstable state,  are of the form
\begin{eqnarray}
\label{gen1}
&&\sum_{m=1}^M \hat{S}_{nm}(k,\omega) \;\hat{\phi}_m(k,\omega) = 
\sum_{m=1}^M \tilde{H}_{nm}(k)\; \tilde{\phi}_m(k,t=0)~,
\nonumber\\
&&~~~~~n=1,\ldots,M
\end{eqnarray}
The terms on the right hand side generally arise upon partial
integration of temporal derivative terms, when we take the Laplace
transform. They contain the initial conditions.
%
Before exploring the implications of (\ref{gen1}), we first discuss in
more detail the type of systems whose linear dynamical equations can
be written in the above form.

{\em Sets of p.d.e.'s:} 
Single or coupled {\em p.d.e.}'s can generally be written in the
matrix notation 
$\left(\partial_t+\underline{\underline{\tilde{T}}}(k)\right)\cdot
\underline{\tilde{\phi}}(k,t)=0$, Eq.\ (\ref{M3}), and
after Laplace transformation immediately yield (\ref{gen1}),
with the matrices
$\underline{\underline{\hat{S}}}(k,\omega)=A_M(k)\;\left(
\underline{\underline{\tilde{T}}}(k)-i\omega\:
\underline{\underline{\tilde{1}}}\right)$ as before in (\ref{matrixS}),
and $\underline{\underline{\tilde{H}}}(k)=A_M(k)\;
\underline{\underline{1}}$. 
The leading edge behavior of single {\em p.d.e.}'s, where the
matrix $\underline{\underline{\tilde{T}}}(k)$ has the
explicit form (\ref{M6}),
was discussed in the previous section.
For coupled {\em p.d.e.}'s, the derivation of a scalar
leading edge equation is not as straight forward, and also leaves
some freedom, as we discuss below and for an example in Appendix
\ref{wim}.  
Nevertheless, we will see that the results summarized in Table II are
robust, in that they do not depend on the particular choice made.
We discuss examples of single {\em p.d.e.}'s 
in Sects.\ \ref{S561} and \ref{S562}, and an example of sets of 
{\em p.d.e.}'s in \ref{S563}.
Of course, if one has a {\em p.d.e.}\ for a single scalar field $\phi$,
one can directly take the Fourier-Laplace transform without writing
$\phi$ as a vector field. This yields a slight generalization of
(\ref{gen1}), the most important difference being that $H$ then also
depends on $\omega$. Our results can obviously also be obtained via
this route -- see Section \ref{S552} for further details.

{\em Difference-differential equations:} When we have difference
equations in space, the equations can also be reduced to  the above
form --- the only difference is that upon Fourier transformation in
space, the $k$-values can be restricted to lie in a finite interval
(the ``Brillouin zone'', in physics terminology). An example will be
discussed in Section \ref{S564}.   Likewise, when we analyze
a dynamical equation with finite time difference, the Laplace integral
can be replaced by a sum over integer times but the ``frequency''
remains a continuous variable. The only difference is that upon
Laplace inversion, the  integral is over a finite interval of $\omega$
values. Examples of  difference equations in both space and time, arising
from numerical schemes, can be found in \ref{S566}.

{\em Equations with memory or spatial kernels:} If the equation has
memory and/or spatial kernels of the type $\int dx' \int_0^t dt'
K(x-x',t-t')\phi(x',t')$ \cite{odo,pik}, then upon Fourier-Laplace 
transformation these just give rise to terms of the form
$\hat{K}(k,\omega)\hat{\phi}(k,\omega)$ 
in (\ref{gen1}), as will be illustrated 
with a simple example in Section \ref{S565}. The only
difference with the case of {\em p.d.e.}'s from this point of view
then is that the elements $\hat{S}_{mn}$  are not polynomials in
$\omega$ and $k$, but more general functions of these arguments.

\subsubsection{Long time asymptotics of the Green's function via
a  Fourier-Laplace transformation} \label{S551}

We now return to the problem of extracting the long time behavior of
the dynamical equation (\ref{gen1}) in Laplace-Fourier
representation. In analogy with our earlier analysis of {\em p.d.e.}'s,
and following \cite{bers,landau}, we introduce the Green's 
function\footnote{A different choice for the definition of
the Green's function is $\underline{\underline{\hat{G}}}(k,\omega) = 
\underline{\underline{\hat{S}}}(k,\omega)^{-1}\cdot
\underline{\underline{\hat{H}}}(k)$, which avoids the convolution 
of the initial condition with $\underline{\underline{H}}(z)$ in
(\ref{green}), and also for equations of 
the form (\ref{M3}) leads to the easier expression
$\underline{\underline{\hat{G}}}(k,\omega) =\left(
\underline{\underline{\tilde{T}}}(k)-i\omega\:
\underline{\underline{\tilde{1}}}\right)^{-1}$. The advantage of the
choice (\ref{defg}) is that we consistently work with
derivatives of $S = \det \underline{\underline{\hat{S}}}$.}
$\underline{\underline{G}}(k,\omega)$ of the linear equations, 
defined by
\begin{equation}
\label{defg}
  \underline{\underline{\hat{G}}}(k,\omega) = 
  \underline{\underline{\hat{S}}}(k,\omega)^{-1} ~.
\end{equation}
$\underline{\underline{\hat{S}}}^{-1}$ is the 
inverse of the matrix $\underline{\underline{\hat{S}}}$.
Eq.\ (\ref{gen1}) now immediately can be solved as
\begin{equation}
\underline{\hat{\phi}}(k,\omega) =
\underline{\underline{\hat{G}}}(k,\omega) 
\cdot \underline{\underline{\tilde{H}}}(k) 
\cdot \tilde{\underline{\phi}}(k,t=0)~.
\label{gen3}
\end{equation}
We write the eigenvectors and eigenvalues of 
$\underline{\underline{\hat{S}}}$ in analogy to (\ref{5a51}) -- (\ref{M7}) as
\begin{equation}
\underline{\underline{\hat{S}}}(k,\omega)\cdot
\underline{\hat{U}}_{\:m}(k,\omega)
=u_m(k,\omega)\;\underline{\hat{U}}_{\:m}(k,\omega)~.
\end{equation}
The determinant of $\hat{S}$ can now be written as
\begin{equation}
\label{Sfactor}
S(k,\omega)=\mbox{det }\underline{\underline{\hat{S}}}(k,\omega)=
\prod_{m=1}^M u_m(k,\omega)~.
\end{equation}
and the characteristic equation 
\begin{equation}
\label{charac}
u_m(k,\omega_m(k))=0
\end{equation}
determines the dispersion relation $\omega_m(k)$ of the mode with  
eigendirection $\underline{\hat{U}}_{\:m}(k,\omega)$. 
Note that each eigenvalue $u_m(k,\omega)$ may be a nonlinear function
of $k$ and $\omega$. Therefore it can happen that the equation
$u_m(k,\omega) =0$ specifies more than one branch 
$\omega(k)$ of the dispersion relation. For simplicity, we will not
distinguish this possibility with our notation, but we stress, that our 
results are generally valid.
For equations of the form (\ref{M5}), we can identify
$u_m(k,\omega)=A_M(k) \left(i\omega_m(k)-i\omega\right)$ and 
$\underline{\hat{U}}_{\:m}(k,\omega)=\underline{\tilde{U}}_{\:m}(k)$. 

Upon inverting the Fourier and the Laplace transformation, 
where the Laplace inversion requires a sufficiently large real
$\gamma$,  
we now find for the Green's function in the comoving frame $\xi=x-vt$:
\begin{eqnarray}
\underline{\phi}(\xi,t) &=&
\int dy\;\underline{\underline{G}}(\xi-y,t) 
\cdot \int dy'\;\underline{\underline{H}}(y-y') 
\cdot \underline{\phi}(y',0)~,
\nonumber\\
\label{green}
\underline{\underline{G}}(\xi,t)&=&\int_{-i\gamma-\infty}^{-i\gamma+\infty}
\frac{d\omega}{2\pi}
\int_{-\infty}^\infty \frac{dk}{2\pi}\;e^{ik\xi-i(\omega-vk) t}\;
\underline{\underline{\hat{G}}}(k,\omega)~,
\nonumber\\
\underline{\underline{\hat{G}}}(k,\omega) &=&\sum_{m=1}^M \;
\frac{\underline{\hat{U}}_{\:m}(k,\omega) \times
\underline{\hat{U}}_{\:m}^\dag(k,\omega)}{u_m(k,\omega)}~.
\label{gkomex}
\end{eqnarray}
The expression for $\underline{\underline{G}}(\xi,t)$ is the immediate
generalization of (\ref{M10}). When we evaluate the Fourier-Laplace
inversion of $\underline{\underline{G}}(\xi,t)$ in the long time limit, 
each term in the sum (\ref{gkomex}) can be evaluated by a so-called
``pinch point'' analysis \cite{bers,landau} making use 
of expansions about zeroes of $u_m(k,\omega)$. We then need to deform 
not only the contour of $k$-integration, as in the saddle point analysis 
in the previous sections, but also the contour of $\omega$ integration.
The pinch point analysis is based on first evaluating the $k$-integral,
and then the resulting $\omega$-integral. Alternatively, we can
extract the long time dynamics by first closing the $\omega$-contour,
and then performing the $k$-integral. This last route is closer 
to the one of Sect.\ \ref{S54}. For a further discussion of both
approaches and of the global conditions that determine  which of the
saddle points or pinch 
points is dynamically relevant, we refer to Appendix
\ref{utesfavorite}.  As before, we use the $*$ to denote the
appropriate solution that satisfies these conditions.

As always, there can in principle be several
saddle point or pinch point solutions through which the integration contour
can be deformed, and if this happens, the relevant 
one is the one corresponding to the largest velocity $v^*$. If we again 
write $u_1(k,\omega)$
for the eigenvalue  on which this solution lies and as before use a
superscript ${}^*$ for functions which are written in terms of the
transformed variable $\Omega$ and $q$ as in (\ref{omOmdef}) and 
(\ref{5b61}), the saddle or pinch point equations assume their familiar form
\begin{eqnarray}
& &  u_1  (k^*,\omega^*) =  0  \Longleftrightarrow  u^*_1(0,0)=0
~,\\ \nonumber
 (\partial_k & + &  v^* \partial_\omega ) u_1(k,\omega)|_{k^*,\omega^*}=0 
\Longleftrightarrow  \partial_q u^*_1(q,\Omega)|_{0,0} =0 ~.
\end{eqnarray}
Note that since $S$ is the product of all eigenvalues, cf. Eq.\
(\ref{Sfactor}), these equations are equivalent to those given before
in terms of $S$, Eq.\ (\ref{saddle}). Likewise, we get for the
long time asymptotics of the Green's function the immediate
generalization of (\ref{GVD}),
\begin{eqnarray}
\label{GVD2} & &
\underline{\underline{G}}(\xi,t) = \\ \nonumber & & ~~~
e^{-\lambda^*\xi}\;\frac{e^{-\xi^2/(4Dt)}}{\sqrt{4\pi Dt}}\;
\frac{\underline{\tilde{U}}_{\:1}(k^*,\omega^*) \times
\underline{\tilde{U}}_{\:1}^\dag(k^*,\omega^*)}
{i\partial_\omega u_1(k,\omega)|_{(k^*,\Omega^*)}} +\ldots ~,
\end{eqnarray}
which is our usual Gaussian expression again, with $D$ given by its
familiar expression
(\ref{Dexprfinal}).

Our strategy in deriving the long--time front dynamics is always to use
the long--time evaluation of the Green's function just to show how the
pulled velocity $v^*$ and the dominant exponential behavior
$e^{-\lambda^* \xi}$ emerge, and to motivate why the leading edge
variables $\psi(\xi,t)$ have essentially slow diffusive dynamics. The
analysis of the slow $\psi$ dynamics and the matching to the front
interior is most properly done by going back to the {\em p.d.e.}(s)
for the spatio-temporal evolution of $\psi$. Switching back to the
space--time formulation for $\psi$ comes out most directly from
Fourier-Laplace inversion of the small-$q$ and small-$\Omega$
expansion of the $\psi$-equation. Indeed, for $\underline{\psi}(\xi,t)$ the
appropriate  Green's function is $e^{\lambda^*\xi}
\underline{\underline{G}}(\xi,t)$ and according to (\ref{green}) we have  
\begin{equation}
e^{\lambda^*\xi}\;\underline{\underline{G}}(\xi,t)
 \label{gmatrix2}
  = \int \frac{d\Omega}{2\pi}
\int \frac{dq}{2\pi}\;e^{iq\xi-i\Omega t}\; 
\underline{\underline{\hat{G}}}^*(q,\Omega)~,
\end{equation}
which confirms that $\Omega$ and $q$ are the proper Fourier-Laplace
variables of the leading edge variables $\underline{\psi}$.

\subsubsection{The case of a single field}\label{S552}

In contrast to our earlier matrix notation,
a single equation for a single field $\phi(x,t)$ after Fourier-Laplace 
transformation can also be written in a scalar form:
\begin{eqnarray}
\label{ScalEq}
& & S(k,\omega)\;\hat{ \phi} (k,\omega ) =\label{eqagain} \\ & &
\nonumber ~~~\mbox{function}
\left\{ k, \;\tilde{\phi}(k,0), \;\partial_t \tilde{\phi}(k,t)|_{t=0},\;
   \cdots \right\}~.
\end{eqnarray}
 The most common and direct way to
arrive at the above equation is  by performing  a
Fourier-Laplace transformation on the original dynamical equation. In
this case one immediately gets the characteristic
function $S(k,\omega) $ on the left hand side, while the partial
integrations (or partial summations in the case of difference equations,
where also the derivatives in the initial condition terms are
replaced by finite difference versions)
of higher order temporal derivatives yield $\omega$-dependent initial
condition terms on the right in (\ref{eqagain}). Of course, we can also
arrive at this equation via the route of Section \ref{S54}, where we
introduced  a vector notation for a scalar {\em p.d.e.}\ of higher
order in time, so that  the dynamical equation is of matrix form
(\ref{gen1}). Indeed, when we then calculate $\det
\underline{\underline{S}}(k,\omega)$ with
$\underline{\underline{S}}(k,\omega)= A_M(k)\left(
\underline{\underline{\hat{T}}}(k)-i\omega
\underline{\underline{\hat{1}}} \right) $  by developing the
determinant along the last row of the matrix, one easily sees that one
just retrieves the above result. 

Of course, the asymptotic analysis of $\phi(\xi,t)$ parallels the
earlier discussion of Section \ref{S54}, irrespective of whether or
not the equation is written in vector form. Again, the asymptotic
spreading speed is given by a saddle point of $S(k,\omega)$. However,
as we have seen, for analyzing the proper front dynamics we want to
return to the dynamical equation for the leading edge variable $\psi$.
For the case of a {\em p.d.e.}, this can be done simply by
transforming the original equation for $\phi$ to the leading edge
representation $\psi(\xi,t)=e^{\lambda^* \xi} \phi(\xi,t)$, but for
difference equations or equations with memory terms, additional steps
are clearly necessary. The general analysis is based on the
observation that in the leading edge representation, the dynamical
equation is of the form
\begin{equation}
\label{tipdyn}
S^*(q,\Omega) \; \hat{\psi}(q,\Omega)= \mbox{initial condition terms}. 
\label{homeq}
\end{equation}
If we expand $S^*$ in $q$ and $\Omega_m$ and perform an inverse
Fourier-Laplace transform, we  immediately arrive at the {\em p.d.e.}\
(\ref{Mfinal}) for $\psi(\xi,t)$ with coefficients $b_{mn}$ given 
in terms of the derivatives of
$S^*$ according to (\ref{bmnS})! From there on, the analysis
completely follows 
the one in the last part of Section \ref{S54}, and we recover again
all our familiar expressions for the relaxation of the front velocity
and the profile.

We stress that for a {\em given} equation, the transformation to the
leading edge variable can be done {\em exactly}. If this is done for a
{\em p.d.e.}, we again get a {\em p.d.e.}\ of finite order. As no
approximations are made, the resulting equation still allows one to study 
the fast or small scale dynamics in the linear region as well. For
finite difference equations or for integro-differential equations, the
transformation to the leading edge variable $\psi$ still results in a
finite difference equation or an integro-differential equation:  the
usual {\em p.d.e.}\ for $\psi$ then only emerges if {\em in addition} a
gradient expansion is made for $\psi$. Such an expansion will
obviously contain an 
infinite number of terms. (We will see explicit
examples of this in Sections\ \ref{S564}, \ref{S565}, and \ref{S566}).
Normally, such an expansion is not of much use.
However, when we turn to the long time relaxation towards pulled
fronts in the leading edge, $\psi$ becomes 
{\em arbitrarily smooth and slow} and hence {\em the
derivatives become  nicely ordered}. Moreover,  the long-time
large-scale relaxation of $\psi$ corresponds precisely to the low-frequency 
small-wavenumber behavior of the Fourier-Laplace transform and this is
why the expansion of $S^*$ gives the proper evolution equation to
analyze the front relaxation: As
(\ref{bmnS}) shows, the coefficients $b_{mn}$ in this equation are
then nothing but the expansion coefficients of the characteristic
equation $S^*(q,\Omega)$ for small $q$ and $\Omega$.  In other words,
{\em independently of whether we started from a differential, a difference
or an integro-differential equation, we find at this point always the same
  p.d.e.\ for the leading edge variable $\psi$, and hence the same
  expression for the velocity relaxation!}
  
Let us finally remark, that instead of the leading edge transformation,
we   could also have performed a leading edge projection
onto the slow dynamics, as discussed in the following section. 
We will show, that the results of Table II do not depend
on this choice.

\subsubsection{The case of a set of fields and possible projections}\label{S553}

For dynamical equations which inherently consist of sets of equations
for more than one field, one obviously can only arrive at an equation
for a scalar variable $\psi$ by some kind of projection onto the slow
direction. The way in which one projects out the slow dynamics clearly 
entails a certain freedom of choice. For a given equation the
``best'' choice may be obvious, but in general there is some
ambiguity. We  illustrate this explicitly in Appendix \ref{Adoublekpp}.

We note first, that a vector field $\underline{\hat{\psi}}(q,\Omega)$
can be decomposed into its dynamical components $\hat{\pi}_m(q,\Omega)$ as
\begin{eqnarray}
\underline{\hat{\psi}}(q,\Omega)&=&\sum_{m=1}^M
\hat{\pi}_m(q,\Omega)\;\underline{\hat{U}}_{\:m}^*(q,\Omega)~,
\\
&&~\hat{\pi}_m(q,\Omega)=
{\underline{\hat{U}}_{\:m}^*}^\dag(q,\Omega)\cdot
\underline{\hat{\psi}}(q,\Omega)~,
\label{funnychoice}
\end{eqnarray}
where the superscript ${}^*$ on the eigenvectors $\underline{U}_{\:m}$ and
eigenvalues $u_m$ is to remind us that these are written in terms
of the variables $q$ and $\Omega$. 
 
Each $\hat{\pi}_m(q,\Omega)$ has its own dynamics, cf.\ (\ref{tipdyn}),
\begin{equation}
u_m^*(q,\Omega)\;\hat{\pi}_m(q,\Omega)=
\mbox{initial condition terms}_{\;m}~.
\end{equation}
The natural projection onto a scalar leading edge variable is thus 
onto the eigendirection with the largest $v^*$, which we denote
with $\underline{\hat{U}}_{\:1}^*(q,\Omega)$. We then identify
the scalar leading edge variable with $\hat{\pi}_1(q,\Omega)$.
Inverting now the Fourier-Laplace transformation, we find
a {\em p.d.e.}\ for $\pi_1(\xi,t)$ of the form (\ref{M12})
with the coefficients
\begin{equation}
\label{bmn1}
  b_{mn}^{(1)} = \left.  
 \frac{(i\partial_\Omega)^m}{m!} \;
 \frac{(-i\partial_q)^n}{n!}\;
 u^{*}_1(q,\Omega)\right|_{(q=\Omega=0)}~.
\end{equation}
Defining the saddle point parameters just as in (\ref{saddle}) and 
(\ref{ratios}) for Eq.\ (\ref{M12}), they in general will depend
on whether we derived the coefficients from $S$ or from $u_1$.
However, we will argue below, that the saddle point parameters
$v^*$, $\lambda^*$ and $D$ do not depend on this choice.

Though the projection onto $\underline{\hat{U}}_{\:1}^*(q,\Omega)$
is formally the simplest one, the direction of projection is actually
not very practical, as it depends on $q$ and $\Omega$.
In practice, one will want to project along a fixed direction.
Our previous analysis, summarized by Eq.\ (\ref{GVD2}), indeed
suggested to project the long time dynamics of the Green's function onto
$\underline{U}_1(k^*,\omega^*)=\underline{\hat{U}}_{\:1}^*(0,0)$.
Projection of $\underline{\hat{\psi}}(q,\Omega)$ onto this
eigendirection yields
\begin{eqnarray}
\hat{\psi}^p(q,\Omega)&=&{\underline{\hat{U}}_{\:1}^*}^\dag(0,0)
\cdot\underline{\hat{\psi}}(q,\Omega)\\
&=&\sum_{m=1}^M
\hat{\pi}_m(q,\Omega)\;{\underline{\hat{U}}_{\:1}^*}^\dag(0,0)
\cdot\underline{\hat{U}}_{\:m}^*(q,\Omega)~.
\nonumber
\end{eqnarray}
Now only for $q\approx 0\approx\Omega$, we have 
$\hat{\psi}^p(q,\Omega)\approx\hat{\pi}_1(q,\Omega)$, while 
for finite $q$ and $\Omega$, also $\hat{\pi}_m(q,\Omega)$ with
$m>1$ will contribute. Inverting the Fourier-Laplace transform
and working in the frame $\xi=x-v^*t$, we find the contributions
from $\hat{\pi}_{m>1}$ to decay exponentially in time. 
Such contributions we encountered already a number
of times before, for the first time in Sect.\ \ref{S25}.
The more important contribution comes from the coefficient of
$\hat{\pi}_1$, which is ${\underline{\hat{U}}_{\:1}^*}^\dag(0,0)
\cdot\underline{\hat{U}}_{\:1}^*(q,\Omega)=1-O(q,\Omega)$.
These algebraic corrections in $q$ and $\Omega$ actually modify
the $b_{mn}$ for the projection $\hat{\psi}^p(q,\Omega)$
in comparison to (\ref{bmn1}), except for the diffusion coefficient
$D$, as we will see below.

Still other projections might be physically useful as is illustrated 
on the explicit example of Appendix \ref{Adoublekpp}.
We now turn to the consequences of all these different choices.

\subsubsection{The freedom of projection and the universality of
Tables II and V} \label{S554}

At first sight, the leading edge transformation or the different
leading edge projections each determine their own saddle or
pinch point equations or expansion parameters $b_{mn}$, compare, e.g.,
(\ref{ss*neu}) with (\ref{bmn1}).

{\em Nevertheless, the definition of the saddle or pinch 
point parameters $v^*$, $\lambda^*$ and $D$ in Table V 
does not depend on the choice of the leading edge
transformation or projection, and hence the universal
relaxation results for the velocity $v(t)$ and the shape 
$\Phi_{v(t)}$ in Table II are independent of these as well.}

For the saddle/pinch point equations of Table V this conclusion 
is based on two observations: $(i)$ $S(k,\omega)$ contains $u_1(k,\omega)$ 
as a factor (\ref{Sfactor}). The saddle point is determined by
a double root in $k$ of $u_1(k,\omega)$, which can be written as
\begin{eqnarray}
\label{root}
u_1^*(q,\Omega)&=&
b_{10}^{(1)}\;\left(-i\Omega+Dq^2+\ldots\right)
\\
=u_1(k,\omega)&=&
b_{10}^{(1)}\;\left(-i(\omega-v^*k)+D(k-k^*)^2+\ldots\right)~.
\nonumber
\end{eqnarray}
$D$ here obviously is defined as $D=-b_{02}^{(1)}/b_{10}^{(1)}$
with $b_{mn}^{(1)}$ from (\ref{bmn1}).
The root (\ref{root}) fully determines the lowest derivatives 
of $S=\prod_m u_m$ at the saddle point $q=0=\Omega$ --- 
up to a constant prefactor, resulting from the other factors in $S$. 
$(ii)$ The saddle point parameters
are defined by homogeneous equations (\ref{saddle}) or
ratios of derivatives (\ref{ratios}). So the prefactors depending
on differentiation of either $u_1$ or $S$ will cancel in the
equations that determine $v^*$, $\lambda^*$ and $D$. In particular,
$D$ defined by $D=-b_{02}/b_{10}$ in (\ref{ratios}) is identical
with $D=-b_{02}^{(1)}/b_{10}^{(1)}$ here and with
other $D$'s resulting from different projections.

The subleading terms $D_3$ and $w$ for the scalar leading edge
variable in (\ref{M13}), in contrast, do depend on the choice
of projection. Hence, as there always will be a leading
edge equation of form (\ref{M13}), and as the universal results
summarized in Table II do not depend on the values of $D_3$ or $w$,
Table II is a universal result, {\em independent of the particular 
projection chosen}. The subleading contribution
$g_0(z)$ in the leading edge will always be solved as in (\ref{M14}), 
so it will not depend on initial conditions, {\em but it will depend 
on the direction of projection through the parameters $D_3$ and $w$}.

In conclusion, we reiterate  that  the relaxation results
also apply to dynamical equations other than {\em p.d.e.}'s, because
the dynamics of the leading edge representation $\psi$ becomes
arbitrarily slow
and diffusive for long times. This allows one to do a gradient
expansion in time and space for $\psi$, {\em even if the original
  equations are not p.d.e.'s!} In this case the path of analysis
via the Fourier-Laplace transformation and pinch point analysis 
is necessary. For equations, that are of differential form in time,
Fourier transformation in space and saddle point analysis is 
sufficient.

\end{multicols}

\begin{center}
\begin{tabular}{||p{10.5cm}||}
\hline\hline
\hspace{0.5cm} \begin{eqnarray} 
\mbox{Definition of }\omega_m(k):&&S(k,\omega_m(k))=0\nonumber\\
&&\nonumber\\
\mbox{Saddle point equations:}&&
(\mbox{definition: }\lambda^*=ik^*)\nonumber\\
S(k^*,\omega^*)=0~~&\Longleftrightarrow&~~\omega^*=\omega_m(k^*)\nonumber\\
\left.\left(\partial_k+v^*\partial_\omega\right)S\right|_{(k^*,\omega^*)}=0
~~&\Longleftrightarrow&~~
v^*=\left.\frac{\partial\omega_m(k)}{\partial k}\right|_{(k^*,\omega^*)}
\nonumber\\
&&\nonumber\\
\mbox{Comoving frame:}&&\nonumber\\
\mbox{Im }\left(\omega^*-v^*k^*\right)=0
~~&\Longleftrightarrow&~~
v^*=\frac{\mbox{Im }\omega_m(k^*)}{\mbox{Im }k^*}\nonumber\\
&&\nonumber\\
\mbox{Diffusion constant:}&&\nonumber\\
D=\left.\frac{-i\left(\partial_k+v^*\partial_\omega\right)^2S}
{2\;\partial_\omega S}\right|_{(k^*,\omega^*)}
~~&\Longleftrightarrow&~~
D=\left.\frac{i\partial^2\omega_m(k)}{2\;\partial
    k^2}\right|_{(k^*,\omega^*)}
\nonumber
\end{eqnarray} \\
\hspace*{0.5cm} {\em In general, only saddle points with } Re $D>0$ {\em are
relevant}
\\ \hspace*{0.5cm} {\em In this paper only saddle points with D real
  are considered.}\\ \\
\hline\hline
\end{tabular}

~\\
~\\
~\\

\begin{minipage}{16.5cm}
{\bf Table V:} The saddle or pinch point equations, determining
$v^*$, $k^*=i\lambda^*$ and $D$ for a given characteristic
function $S(k,\omega)=\mbox{det }\underline{\underline{\hat{S}}}(k,\omega)$.
If there are several saddle point solutions  that satisfy the global
conditions that determine which saddle point solution is dynamically
relevant (see Appendix \ref{utesfavorite}),
take the one with the largest $v^*$.
\end{minipage}
\end{center}

\begin{multicols}{2}

\subsection{Applications} 
\label{S56}

In this subsection we support the above arguments by summarizing the
results of numerical simulations of three equations --- a spatially
fourth order 
{\em p.d.e.}, a set of two coupled {\em p.d.e.}'s and a
difference-differential equation --- which are all in complete
agreement with our predicted universal relaxation trajectory
as in Table II,
consisting of the velocity convergence (\ref{M15}), the slaved
interior (\ref{M16}), and the cross-over to a diffusive type of
dynamics in the leading ege for $\xi \gtrsim \sqrt{t}$. We also
briefly consider a {\em p.d.e.}\ with second order temporal derivatives,
an extension of the nonlinear diffusion equation with a memory
kernel, and the discretization corrections in the Euler and in the
semi-implicit numerical integration method for a nonlinear diffusion
equation. The last results were used
already in Section \ref{S4} in our numerical study of the nonlinear
diffusion equation.

\subsubsection{The EFK equation}
\label{S561}

The EFK (``extended Fisher Kolmogoroff'') equation is an extension 
of the nonlinear diffusion equation\cite{vs2,dee}, which has
been investigated quite intensely in the mathematical literature
\cite{bert}. It reads
\begin{equation} \label{502}
\partial_t \phi=\partial_x^2\phi -\gamma\partial_x^4\phi + \phi- \phi^3~.
\end{equation}
A straightforward calculation \cite{vs1} shows that the saddle point 
equations (\ref{5a24}), (\ref{5a24a}) and (\ref{5a19}) yield
\begin{eqnarray}
\label{v*efk}
v^* & = & 2\lambda^*\;(1-2\gamma{\lambda^*}^2)~, \nonumber \\
\lambda^* & = & \left(\frac{1-\sqrt{1-12\gamma}}{6\gamma}\right)^{1/2}~,
\nonumber\\
D & = & \sqrt{1-12\gamma}~~ ~~ ~~ ~~ \mbox{ for } \gamma<{{1}\over{12}}~.
\end{eqnarray}
For $\gamma > 1/12$, the saddle point solution has $\mbox{Re }k^* \neq
0$, and in agreement with this, the pulled fronts in this equation are
then found to be non-uniformly translating and to generate periodic
patterns\cite{dee}. We will therefore focus here on the regime $\gamma
< 1/12$. The arguments of the Appendix of \cite{vs2} for the multiplicity 
of front solutions (summarized in Appendix \ref{A6}) give evidence that this
equation indeed admits a family of uniformly translating fronts in
this regime. One also can prove that the front cannot propagate
with a larger velocity than $v^*$ if the initial conditions are
sufficiently steep \cite{willem1,spruijt}. 
The convergence towards the pulled front solution should
therefore be given by Eq.\ (\ref{5051}) for $v(t)$ and Eq.\ (\ref{5037})
or Eqs.\ (\ref{5044})--(\ref{5047}) for the interior of the leading edge
of the front profile. Fig.\ 11 shows some of the results
of our numerical simulations for $v(t)$ at $\gamma=0.08$. 
This value of $\gamma$ is closely below the bifurcation value 
$\gamma_c=1/12=0.083$. The plot is of the same type as in Fig.\ 8($a$) 
for the nonlinear diffusion equation. 

The numerical grid sizes of the simulation are $\Delta x = 0.01 = \Delta t$.
The system size is $L=200$, the initial condition is characterized by
$\lambda_{init}=20$ and $x_0=25$.
The analytical prediction for $\gamma=0.08$ is according to 
(\ref{v*efk}) in the limit $\Delta x\rightarrow 0, \Delta t
\rightarrow 0$: $D=0.2$, $\lambda^*=\sqrt{5/3}=1.29$, and
$v^*=4.4\cdot\lambda^*/3=1.89$. The ratio between the $1/t$- and
the $1/t^{3/2}$-contribution in $v(t)$ according to (\ref{5051}) is measured
on the time scale 
\begin{equation}
\label{T}
T=1/({\lambda^*}^2D)~,
\end{equation} 
as in the dimensional analysis (\ref{5041}).
For $\gamma=0.08$, we have $T=3$. 
The plot of Fig.\ 8($a$) gave good results from time $t=20$ on, 
where $T=1$. It is therefore consistent, that the plot of 
Fig.\ 11 with $T=3$ is good from times $t=60$ on. We thus plot
here the time interval $60 \le t \le 200$.
One  can already anticipate from the plot that again a correction
of $v^*$ for the numerical finite difference code will be required
if we proceed to even higher precision. In conclusion, we find 
the results to be in full accord with our analytical predictions.

\end{multicols}

\begin{figure} \label{fig12}
\vspace{0.5cm}
\begin{center}
\epsfig{figure=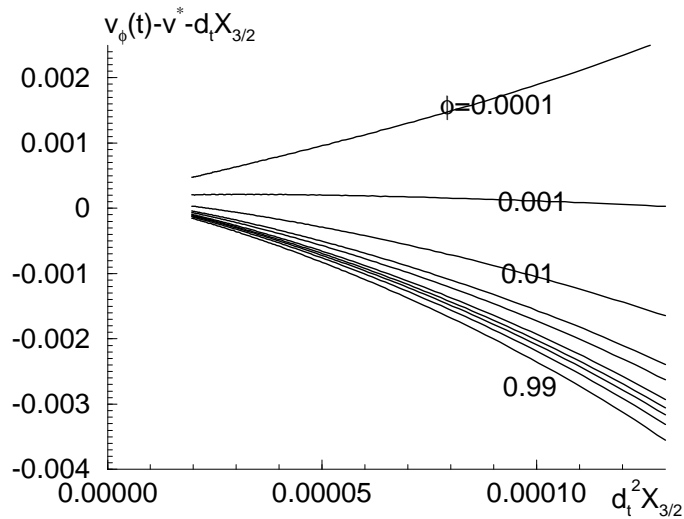,width=0.4\linewidth}
\end{center}
\vspace{0.5cm}
\caption{Velocity relaxation in the EFK equation (\ref{502}) 
for $\gamma=0.08$:
Plot of $v_\phi(t)-v^*-\dot{X}_{3/2}$ as a function of $\ddot{X}_{3/2}$
as in Fig.\ 8($a$) for times $60 \le t \le 200$. 
System size $L=200$, front position $x_0=25$, 
initial steepness $\lambda_{init}=20$ in (\ref{405}). 
Grid sizes $\Delta x = 0.01 = \Delta t$.}
\end{figure}

\begin{multicols}{2}

\subsubsection{The streamer equations}
\label{S563}

Streamers are discharge patterns which result from the competition
between an electron avalanche formation due to impact ionization, and
the screening of the electric field by charges. For planar streamer
fronts, the equation for the electron density
$\sigma$ and electric field $E$ are \cite{ebert}
\begin{eqnarray} \label{508}
\partial_t\sigma &=& D_\sigma\partial_x^2\sigma+\partial_x(\sigma E)+
\sigma f_{str}(E)~,
\nonumber\\
\partial_t E &=& -D_\sigma \partial_x\sigma-\sigma E~,
\end{eqnarray}
where we have assumed that in the region $x\gg1$, where the
electron density vanishes $\sigma^+=\sigma(x\to\infty,t)=0$,
the electric field $E^+=E(x\to\infty,t)$ does not change
in time: $\partial_t E^+=0$. The field dependent ionization
rate has a functional form like $f_{str}(E)=|E|\;e^{-1/|E|}$.
This is the functional form we use in our simulations.
The state $(\sigma,E)=(0,E^+)$ is unstable,
and also for these equations it is known \cite{ebert}, 
that they admit a one parameter family of uniformly translating 
front solutions. The dispersion relation for linear 
perturbations about the unstable state $\sigma=0$, $E=E^+<0$ reads
$-i \omega(k)=ikE^+ + f_{str}(E^+) -D_\sigma k^2$, where we choose 
to analyze the leading edge in a projection onto the $\sigma$-axis.
The saddle point equations (\ref{5a24}), (\ref{5a24a}) and (\ref{5a19})
then yield
\begin{eqnarray}
v^* & = & -E^+ + 2 \sqrt{D_\sigma f_{str}(E^+)}~,\nonumber\\
\lambda^* & = & \sqrt{f_{str}(E^+) /D_\sigma}~,\\
D & = & D_\sigma~.
\end{eqnarray}
Again, the simulations of these equations show that the velocity
convergence follows our analytical prediction (\ref{5051}). An example
of our results is shown in Fig.\ 12 in a plot as in Figs.\ 8($a$) and 11,
where we track various level curves of the electron density $\sigma$.
The dimensionless time is $T=1/f_{str}(E^+)=e^1=2.718$ for $E^+=-1$. 
We plot our data for times $40 \le t \le 200$, and again find our 
predictions to hold.

\end{multicols}

\begin{figure} \label{fig13}
\vspace{0.5cm}
\begin{center}
\epsfig{figure=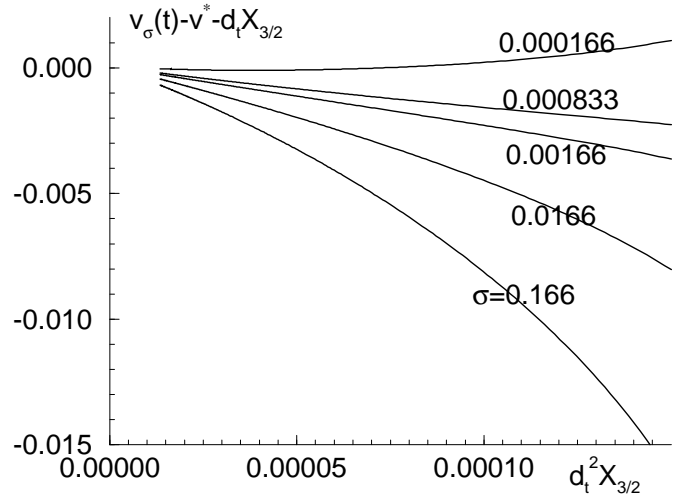,width=0.4\linewidth}
\end{center}
\vspace{0.5cm}
\caption{Velocity relaxation in the streamer equations (\ref{508})
for $E^+=-1$ and $D=0.1$, plotted as in Figs.\ 8($a$) and 11
for times $40 \le t \le 200$.
Initial condition: Gaussian electron density $\sigma(x,0)=0.9\;
e^{-x^2}$ (thus $\lambda_{init}=\infty$), $E(x,0)=-1$.
System size $L=400$, front position shifted back to $x_0=100$,
after it is reached.
Grid sizes: $\Delta x = 0.01$, $\Delta t = 0.0025$.}
\end{figure}

\begin{multicols}{2}

\subsubsection{A difference-differential equation}
\label{S564}

We now summarize some key elements of our analysis \cite{evsp} of
the difference-differential equation 
\begin{equation}
\label{ramses2}
\partial_tC_j(t)=-C_j+C_{j-1}^2~~,~C_0(t)=0~,~C_{j\gg1}(t)=1~,
\end{equation}
with $j$ integer. This equation originates from kinetic theory
\cite{ramses}. If we transform with $\phi_j(t)=1-C_j(t)$ to
\begin{eqnarray} \label{509}
\partial_t \phi_j(t)&  = & -\phi_j+2\phi_{j-1}-\phi_{j-1}^2~,~\\
\phi_0(t)& = & 1
~~,~\phi_{j\gg1}(t)=0~,\nonumber
\end{eqnarray}
we have our usual notation with the state $\phi_j=0$ being
unstable and the state $\phi_j=1$ stable. As usual  we consider fronts
between these states starting from sufficiently steep initial
conditions. It is easy to see that such initial conditions 
will create a pulled front \cite{evsp}.

Equation (\ref{509}) provides the first illustration of our argument from
Sect.\ \ref{S55} that our analysis applies to difference equations 
as well --- with the only difference that the spatial Fourier modes 
$k$ now extend over a finite interval or ``Brillouin zone'' 
$0\le k < 2\pi$ only. Substitution of the Fourier ansatz 
$\phi_j \sim e^{-i\omega t +ik j}$ into the equation of motion
linearized about the unstable state $\phi_j=0$
\begin{equation}
\partial_t \phi_j = -\phi_j +2 \phi_{j-1}~\label{Cjeq}
\end{equation}
yields the dispersion relation
\begin{equation}
-i\omega(k) =2 e^{-ik } -1
~~\Longleftrightarrow~~s(\lambda)=2e^\lambda-1~.
\end{equation}
As discussed before, the long time asymptote of the leading edge
is again determined by the saddle point which obeys (\ref{5a19}),
(\ref{5a24}) and (\ref{5a24a}). This results in
\begin{equation}
v^*=2 e^{\lambda^*} = \frac{2 e^{\lambda^*}-1}{\lambda^*}
\end{equation}
When we choose the solution with $v^*>0$, the saddle point equations  
are solved by
\begin{eqnarray}
-ik^*&=&\lambda^*>0~~\mbox{real}\nonumber\\
\label{v*diff}
v^*&=&2e^{\lambda^*}=4.31107~,\nonumber\\
\lambda^* &=&  (2 e^{\lambda^*} -1)/(2 e^{\lambda^*})=0.768039~,
\\
D&=&D_2=v^*/2~~,~~D_n=(-)^n v^*/n!~.\nonumber
\end{eqnarray}
The $D_n$ are determined from (\ref{diffdef}). 
We now perform the leading edge transformation 
\begin{equation}
\phi_j(t) = e^{-\lambda^* \xi}\; \psi(\xi,t) ~, ~~~\xi=j-v^* t~.
\end{equation}
The large-time, small-gradient expansion in the leading edge 
now results in the {\em p.d.e.}\
\begin{equation}
\partial_t \psi = D \partial^2_\xi \psi + D_3 \partial^3_\xi \psi +
\dots~. \label{psidiffeq}
\end{equation}
The velocity convergence is again given by (\ref{5051}), 
with $v^*$, $\lambda^*$ and $D$ given by (\ref{v*diff}). We do find
indeed that the fronts in this equation are pulled, and that the
velocity convergence follows (\ref{5051}). This is illustrated
in Fig.\ 13, where we plot $(v(t)-v^*+3/(2\lambda^*t))/t^{-3/2}$
as a function of $1/\sqrt{t}$. $v(t)=\dot{x}(t)$ is the velocity
of the front defined as $x(t)=\sum_{j=0}^\infty\phi_j(t)$.
The curve in Fig.\ 13 should extrapolate to 
$3/(2\lambda^*)\cdot\sqrt{\pi/({\lambda^*}^2D)}=3.0699$ as $1/\sqrt{t}\to 0$.
This predicted asymptote is marked by the cross on the axis.
Indeed, the data of $(v(t)-v^*+3/(2\lambda^*t))/t^{-3/2}$ for
$40\le t\le 4000$ extrapolate very well to the predicted asymptote, especially
 in view of the fact, that $t^{3/2}\approx 2\cdot10^5$ at the latest times.
The slight offset at the end might be due either to finite system size
$L$ or to finite numerical discretization $\Delta t$.

\end{multicols}

\begin{figure} \label{fig14}
\vspace{0.5cm}
\begin{center}
\epsfig{figure=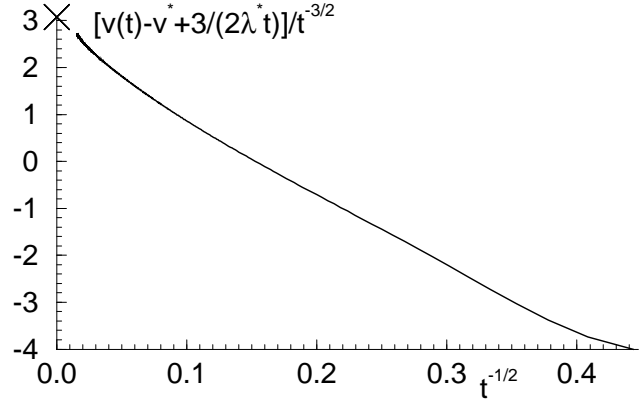,width=0.4\linewidth}
\end{center}
\vspace{0.5cm}
\caption[]{Velocity relaxation for the difference-differential equation
(\ref{ramses2}), where $v(t)=\dot{x}(t)$, and 
$x(t)=\sum_{j=0}^\infty\phi_j(t)$, see Eq.\ (\ref{509}).
Plotted is here $\big(v(t)-v^*+3/(2\lambda^*t)\big)/t^{-3/2}$
as a function of $1/\sqrt{t}$ for times $40\le t\le 4000$. 
The curve is predicted to extrapolate to $c_{3/2}$ as $1/\sqrt{t}\to0$. 
The predicted value of $c_{3/2}$ is marked by the cross on the axis.
Initial condition $\phi_j(0)=e^{-j^2}$. System size $N=4000$ grid points.
Front shifted back to $n_0=75$, after it has been reached.
Temporal grid size $\Delta t = 0.0005$.}
\end{figure}

\begin{multicols}{2}

\subsubsection{Diffusion equation with second order time derivative}
\label{S562}

Quite recently, it was shown \cite{gallay} that, not surprisingly,
fronts in a second order extension of the F-KPP equation,
\begin{equation}
\tau_2 \frac{\partial^2 \phi}{\partial t^2} + \frac{\partial
  \phi}{\partial t} = \frac{\partial^2 \phi}{\partial x^2} +\phi -
\phi^3 ~,
\end{equation}
are also pulled. One interesting aspect of this equation is that while 
the diffusive spreading in a first order diffusion equation is, in a
sense, infinitely fast, the second order term gives a finite speed of
propagation of the disturbances. 

As discussed in Section \ref{S54}, our results immediately apply to
this equation, so the velocity and front relaxation is then given by
Eqs. (\ref{M15}) and (\ref{M16}), with
\begin{eqnarray}
\nonumber
v^* & = & \frac{2}{\sqrt{1+4\tau_2}}~,\\
\lambda^* & = & \sqrt{1+4 \tau_2}~,\\
D & = & \frac{1}{(1+4\tau_2)^2}~.\nonumber
\end{eqnarray}
The expression for $D$ nicely illustrates the effictive
renormalization of the diffusion coefficient due to the second order
time derivative.

\subsubsection{An extension of the F-KPP equation with a memory
  kernel}\label{S565}
  
As an example of an equation with a memory kernel,
consider the extension of the F-KPP equation
\begin{eqnarray}
\label{memory}\nonumber
\partial_t \phi(x,t)& =& \partial^2_x \phi(x,t) +  \int_0^t dt' K(t-t')
\phi(x,t') \\ & & ~~~- \phi^k(x,t)~, ~~~(k>1)~.
\end{eqnarray}
Upon Fourier--Laplace transformation as in (\ref{5a17b}), 
this equation is a scalar
version of (\ref{gen1}) with $S(k,\omega) =i \omega -k^2 + \tilde{K}(
\omega)$, and so according to our discussion of Section \ref{S55},
our analysis directly applies. If we take for instance 
\begin{equation}
\label{memory2}
K(t-t') = \frac{1}{\sqrt{\pi} \tau_3} e^{-(t-t')^2/4\tau^2_3}~,
\end{equation}
the equation reduces to the F-KPP equation in the limit $\tau_3 \to
0$, and the characteristic equation becomes
\begin{equation}
\lambda^2-s+ e^{\tau_3^2 s^2} \mbox{erfc}(\tau_3 s) =0~,\label{erf}
\end{equation}
where we follow the notation of Section \ref{S532} in writing
$s=\mbox{Im}\;\omega$, $\lambda=\mbox{Im}\;k$, and where erfc
is the complementary error function. The 
results for $v^*$, $\lambda^*$ and $D$, obtained by solving (\ref{erf})
together with the saddle point condition $\partial s /\partial \lambda
=s/\lambda  |_{\lambda^*}$ numerically, are shown in Fig.\ 14.

Other examples of equations with memory kernels can, e.g., be found in
\cite{odo,pik}.

\end{multicols}

\begin{figure} \label{fig15}
\vspace{0.5cm}
\begin{center}
\epsfig{figure=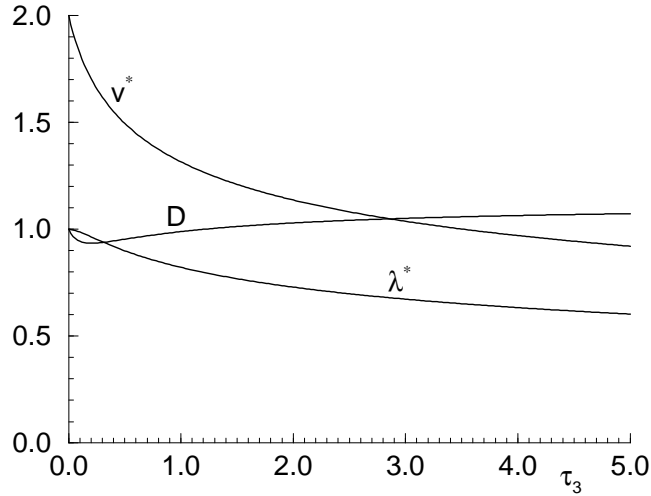,width=0.4\linewidth}
\end{center}
\vspace{0.5cm}
\caption{Plot of $v^*$, $\lambda^*$, and $D$ as a function of $\tau_3$ 
for the extension (\ref{memory}) of the F-KPP equation with 
a memory kernel (\ref{memory2}).}
\end{figure}

\begin{multicols}{2}

\subsubsection{Exact results for numerical finite difference schemes}
\label{S566}

The fact that our results also apply to finite difference equations
has the important implication that if we study a {\em p.d.e.}\
with pulled fronts numerically using a finite difference approximation
with gridsize $\Delta x$ and timestep $\Delta t$, we can calculate
$v^*(\Delta x, \Delta t) $ as well as $v(t; \Delta x, \Delta t)$ {\em
  exactly}. This allows us to estimate analytically the intrinsic
discretization error in these quantities, and hence to decide
{\em beforehand} which grid and step size are needed to obtain a given
accuracy. 

As a first illustration, suppose,  that one integrates the F-KPP 
equation (\ref{101}) numerically with an explicit Euler scheme. 
This amounts to approximating the {\em p.d.e.}\ by
\begin{eqnarray}
& & \frac{ u_j(t+\Delta t) - u_j(t)}{\Delta t} = \nonumber \\ 
& & ~~~~ \frac{u_{j+1}(t) -2
  u_j(t) +u_{j-1}(t)}{(\Delta x)^2}  + u_j(t) - u^k_j(t)~.
\end{eqnarray}
Upon substitution of $u_j(t) \sim e^{st-\lambda x}$, $x=j\;\Delta x$ 
into the linearized equation (we again
follow the notation of Section \ref{S532} by writing
$s = \mbox{Im} \; \omega$, $\lambda = \mbox{Im}\; k$), we obtain 
\begin{equation}
\frac{e^{\:s\Delta t}-1}{\Delta t}=1+
\left(\frac{\sinh \lambda \Delta x/2}{\Delta x/2}\right)^2~,
\end{equation}
which is straight forward to solve for $s(\lambda; \Delta x, \Delta t)$. 
As we emphasized
above, by solving the saddle point condition $\partial s / \partial
\lambda = s / \lambda |_{\lambda^*}=v^*$, we can obtain the exact values
of $v^*$, $\lambda^*$ and $D$ for any step and grid size, 
and in this way determine the accuracy of the
numerical scheme. In general, these equations have to be solved by a
simple numerical iteration routine, but for small $\Delta x$ and
$\Delta t$, the result can easily be calculated analytically:
Expanding in $\Delta x$ and $\Delta t$, we find the dispersion relation 
\begin{equation}
\label{euler}
s (\lambda; \Delta x, \Delta t) = 1+\lambda^2 
+\frac{\lambda^4(\Delta x)^2}{12}-\frac{(1+\lambda^2)^2\Delta t}{2}+\ldots~.
\end{equation}
For $\Delta t \to 0$, $\Delta x \to 0$, this reduces to the continuum
result $s(\lambda) = 1+\lambda^2$, as it should. For the saddle point 
parameters, we find
\begin{eqnarray}
v^* & = & 2 - 2 \Delta t + \frac{1}{12} (\Delta x)^2 + \cdots
\nonumber\\
\mbox{Euler:}~ \lambda^* & = &  1+ \Delta t - \frac{1}{8} (\Delta x)^2
+\cdots\\
D &=& 1-  4 \Delta t + \frac{1}{2}(\Delta x)^2 + \cdots ~,\nonumber
\end{eqnarray}

In practice, the Euler scheme is not used very often, because it is
numerically very unstable and not very accurate. We have
done all our simulations in Section \ref{S4} and in the present section 
with a more stable and accurate semi-implicit method
\cite{potter}, which for the F-KPP equation amounts to the
discretization
\begin{eqnarray} \label{5098}
& & \frac{ u_j(t+\Delta t) - u_j(t)}{\Delta t} = \nonumber \\ & & ~~~~
\frac{1}{2} \left[ \frac{u_{j+1}(t) -2
  u_j(t) +u_{j-1}(t)}{(\Delta x)^2}  \right] + \nonumber \\ & & ~~~~ 
+ \frac{1}{2} \left[ \frac{u_{j+1}(t+\Delta t) -2
  u_j(t+\Delta t) +u_{j-1}(t+\Delta t)}{(\Delta x)^2}  \right]
+\nonumber \\
& & ~~~~ +\frac{1}{2} \left[  u_j(t) + u_j(t+\Delta t) \right] - \nonumber\\
 & &  ~~~~ -\frac{1}{2} \left[ 2 u^k_j(t) + k u^{k-1}_j (t) (u_j(t+\Delta
 t)-u_j(t)) \right] ~.
\end{eqnarray}
The term on the last line is obtained by expanding $u^k_j(t+\Delta t)$
about $u_j^k(t)$ to first order in $u_j(t+\Delta t)-u_j(t)$, so that
one obtains a linear equation for the $u_j(t+\Delta t)$. This
expansion makes what would otherwise have been an implicit method,
into a semi-implicit method. This difference, however, does not matter
for the leading edge dynamics evaluated below.

The dispersion relation is now given by 
\begin{equation}
\frac{\tanh s\Delta t/2}{\Delta t/2}=1+
\left(\frac{\sinh \lambda \Delta x/2}{\Delta x/2}\right)^2~,
\end{equation}
which immediately yields $s(\lambda; \Delta x,\Delta t)$.
For small $\Delta x$ and $\Delta t$ the result is
\begin{equation}
\label{semi-im}
s (\lambda; \Delta x, \Delta t) = 1+\lambda^2 
+\frac{\lambda^4(\Delta x)^2}{12}+
\frac{(1+\lambda^2)^3(\Delta t)^2}{12}+\ldots~.
\end{equation}
For this integration scheme, it is now straight forward to find
\begin{eqnarray}
\label{semi-im2}
v^* & = & 2 + \frac{2}{3}( \Delta t)^2 + \frac{1}{12} (\Delta x)^2 + \cdots
\nonumber\\
\mbox{Semi-implicit:}~ \lambda^* & = &  1-\frac{2}{3} (\Delta t)^2 - 
\frac{1}{8} (\Delta x)^2
+\cdots\\
D &=& 1+  3 ( \Delta t)^2 + \frac{1}{2}(\Delta x)^2 + \cdots ~\nonumber
\end{eqnarray}
We stress that these are the {\em exact} expressions for the application 
of this numerical scheme to the nonlinear diffusion equation, scaled to
the normal form (\ref{205}), (\ref{206}). They are therefore the ``ideal'' 
finite difference correction terms in the absence of numerical 
instabilities, round-off errors et cetera. The correctness and
accuracy of the prediction (\ref{semi-im2}) for $v^*$ is demonstrated
in Section \ref{S4} in Fig.\ 8($b$).

We finally note that an early example of pulled front relaxation 
observed in a finite difference
equation in space {\em and} time was seen in a mean-field model of ballistic
growth \cite{krug}. In this paper, the prefactor of the $1/t$ term, obtained
by plotting $v$ versus $1/t$, was found to be about 9 percent too {\em small}.
Presumably, this discrepancy is due to the corrections from the $1/t^{3/2}$ 
term: According to (\ref{M15}), the term $(1- \sqrt{\pi /({\lambda^*}^2D t)})$ 
generally gives rise to a {\em lowering} of the effective slope in a $v$ 
versus $1/t$ plot, as Fig.\ 7($b$) clearly demonstrates.

\end{multicols}

\newpage

\begin{multicols}{2}

\section{Summary and outlook} \label{S6} 

\subsection{Summary of the main results}

The essential result of this paper is that for front propagation into
unstable states, starting from steep initial conditions, the
convergence of front velocity and shape is given  in the pulled regime
by the {\em universal} expressions 
\begin{eqnarray} \label{6a1}
v(t) & = & v^* + \dot{X}(t)~,\\ \label{6a2}
\dot{X}(t) & = &  -\frac{3}{2\lambda^* t}
\left(1-\sqrt{\frac{\pi}{(\lambda^*)^2Dt}}\;\right) + O(t^{-2})~,\\
\label{6a3}
  \phi &=& \Phi_{v(t)} (\xi_X) + O\left(\frac{1}{t^2}\right)~,
  ~~~\xi_X \lesssim \sqrt{t} ~,\\
\xi_X&  =& x-v^*t -X(t) ~, 
\end{eqnarray}
provided the asymptotic front profile is uniformly translating. All
terms in the expression for $v(t)$, $\lambda^*$, $v^*$ and $D$, are
given explicitly  in terms of the dispersion relation of dynamical
equation, linearized about the unstable state [see Eqs.\ (\ref{5a24}),
(\ref{5a24a}) and (\ref{5a25}) or Table V]. These results are also
summarized in Table II. The dependence on pushing or pulling and on the
initial conditions is sketched in Table IV.

With {\em universal} we mean that not only the asymptotic profile is
unique, but also the relaxation towards it, provided we start with
sufficiently steep initial conditions which decay exponentially faster than $e^{-\lambda^*x}$ for $x\rightarrow \infty$. Moreover, the relaxation is
universal in that it is independent of the precise nonlinearities in
the equation, and of the precise form of the equation: It holds for
{\em p.d.e.}'s, sets
of {\em p.d.e.}'s, difference--differential equations, equations with
memory kernels, etc., provided fronts are pulled and that the
asymptotic front solution is 
uniformly translating, and provided that we are not at the bifurcation point
from the pulled to the pushed regime, or at the bifurcation point
$D=0$ towards pattern forming fronts (e.g., at $\gamma=1/12$ 
in the EFK-equation). The fact that the results also
apply to finite difference equations has a nice practical consequence:
If a {\em p.d.e.}\ is studied numerically using a finite difference
approximation scheme, both $v^*$ and the prefactors of the algebraic 
relaxation terms
can also be calculated {\em exactly} for the numerical scheme. This allows
one to estimate in advance how big step and grid sizes need to be, in
order to achieve a particular numerical accuracy (see Section
\ref{S566}).

The remarkable relaxation properties are reminiscent of the universal
corrections to scaling in critical phenomena, if we think of the
relaxation as the approach to a unique fixed point in function space
along a unique trajectory. An alternatively way to express this in
more  mathematical terms is to say  that we
have constructed the {\em center manifold} for front relaxation in the 
pulled regime. 

The above expressions contain {\em all} universal terms: those of
order $t^{-2}$ depend on the precise initial conditions and on the
nonlinearities in the equations. The order of the limits is important
here: Our results  are the exact
expressions in a $1/t$ expansion, i.e., when we take  the large time
limit while  tracking the velocity of a particular fixed value of $\phi$.
To order $1/t^2$, this is equivalent to keeping  
$\xi_X$ fixed. When we interchange the limits by taking $\xi_X$ large
at fixed time, 
there is a cross-over to a different intermediate asymptotic regime
for $ \xi_X \gtrsim \sqrt{t}$. The different dynamical regions of a 
pulled front are sketched in Fig.\ 2.

The slow algebraic convergence of pulled fronts to the asymptotic 
velocity has important consequences, as it prohibits the derivation 
of a standard moving boundary approximation for patterns in more than 
one dimension that consist of propagating pulled fronts whose width 
is much smaller than their radius of curvature \cite{ebertmba}. 

While we have limited the analysis in this paper to equations that
admit uniformly
translating front solutions, it turns out that most elements of our
analysis can be extended to pattern forming fronts for which
$\mbox{Re~}k^* \neq 0$ and $\mbox{Re~}\omega^* \neq 0$. In this
case, the  expression (\ref{6a2}) with $1/\sqrt{D}$ replaced by
$\mbox{Re~}( 1/\sqrt{D})$ applies \cite{spruijt,kees}.

In addition to our derivation of the above expressions for the
convergence of pulled fronts, we have reformulated and extended
the connection 
between front selection and the stability properties of fronts. 
This leads to an essentially complete picture also of 
front relaxation in the pushed regime
and in the case of leading edge dominated dynamics resulting from 
flat initial conditions. For an interpretation of these 
results, again a consideration of the different dynamical regions of
a front as in Fig.\ 2 is helpful. The relaxation behavior in the
pulled regime with sufficiently steep initial conditions can not be
obtained simply from the properties   
of the stability operator of the pulled front solution, and therefore 
had to be obtained along a different route, which is summarized below.

\subsection{Summary of the main conceptual steps of the analysis}
\label{S61}

The derivation of our central result on pulled front relaxation 
is based on the following steps:

{\em 1.} From the dispersion relation $\omega(k)$ or 
from the characteristic function $S(k,\omega)$, we obtain $v^*$, 
$\lambda^*$ and $D$ (see Table V).

{\em 2.} The double root condition which determines $v^*$ and
$\lambda^*$ implies that the asymptotic large $\xi$ behavior of
uniformly translating front solutions is as $\Phi^*(\xi)=(\alpha \xi +
\beta) e^{-\lambda^* \xi}$, where generically $\alpha \neq 0$.

{\em 3.} The double root condition which determines $v^*$ and
$\lambda^*$ also implies that the lowest order spatial derivative term
in the dynamical equation for the leading
edge representation $\psi= e^{\lambda^* \xi} \phi(\xi,t)$ is of the
diffusion type, $D \partial^2 \psi / \partial \xi^2$ (see Sections
\ref{S533}, \ref{S54}, and \ref{S55}).

{\em 4.} The diffusion type dynamics implied by {\em 3.} shows that in
the co-moving frame $\xi=x-v^*t$, the front profile shifts back with
the collective coordinate $X(t)$ which grows logarithmically in
time. Linearization about the asymptotic front solution $\Phi^*(\xi)$
in the $\xi $ frame is therefore impossible (see Sections \ref{S312}
and \ref{S313}). Instead, we introduce the
frame $\xi_X= x-v^*t -X(t)$ with the expansion $\dot{X}(t)=c_1/t +
c_{3/2} / t^{3/2} +\cdots$ and the corresponding leading edge
transformation $\psi(\xi_X,t) = e^{\lambda^* \xi_X} \phi(x,t)$. 

{\em 5.} In the {\em front
interior}, the long time expansion for $\dot{X}$ generates an expansion
for the corrections to the front profile in inverse powers of $t$. To
order $t^{-2}$ temporal derivatives of the front corrections do not
come in, so that to this order the equations for the profile shape
reduce to those for $\Phi_{v(t)}$. This immediately leads to (\ref{6a3})
for the time dependence of the front 
profile.  

{\em 6.} In the {\em leading edge}, where nonlinearities can be
neglected, we use an asymptotic expansion for $\psi(\xi_X,t)$, 
linearized about $\psi = 0$, in terms of
functions of the similarity variable $z=\xi^2_X/(4Dt)$ of the
diffusion equation. Now for small values of $z$ the
expansion has to match the boundary condition $\psi\approx
e^{-\lambda^*\xi_X} \Phi^*(\xi_X) \approx \alpha \xi_X +\beta$ (implied
by observation {\em 2.}), and for large $\xi_X$ the terms in the
(intermediate) asymptotic expansion have to decay as a Gaussian $e^{-z}=
e^{-\xi^2_X /(4Dt)}$ times a polynomial in the similarity variable $z$.
These two requirements fix the constants $c_1, c_{3/2}, \ldots$ in
the expansion of $\dot{X}$, and hence (\ref{6a2}).

\subsection{Open problems}

What one considers as remaining open problems concerning pulled front
propagation, will depend largely on one's background and standards
regarding the desired mathematical rigour. While our results are exact
and yield an almost complete understanding of the general mechanism of
pulled front propagation, they have, of course, not been derived
rigorously. In physics, such a situation is often not just quite
acceptable but even quite gratifying, but more mathematically inclined
readers may wish to take up the challenge to provide a more rigorous
justification. More work could also be done on enlarging the classes
of equations for which the assumptions underlying our approach can be
shown to hold, i.e., for which one can show that fronts are pulled and
that there exists a family of uniformly translating front solutions.

Within the realm of our approach, one can consider slight  extensions
of our method to two nongeneric  
special cases. First of all, we have focussed on the case of
sufficiently  steep initial conditions
such that the steepness $\lambda = - \lim_{x\to \infty} \ln \phi(x,0)$
is larger than $\lambda^*$. As we discussed at the end of Section
\ref{S3}, the intermediate case in which for large $x$ $\phi(x,0) \simeq
x^{-\nu} 
e^{-\lambda^* x}$ with $\nu <2$, does give a $\nu$-dependent result for 
the coefficient of the $1/t$ term. According to Bramson
\cite{bramson}, the next 
order correction  is of order $1/(t \ln t)$. This
suggests that for this special case logarithmic terms will have to be
included in the expansion.   Second, at the bifurcation
point from uniformly translating solutions to pattern forming fronts,
which in the EFK equation (\ref{502}) happens at $\gamma= 1/12$, the
diffusion coefficient $D$ vanishes [see Eq. (\ref{v*efk})]. At this
bifurcation point, the equation for the leading edge representation
$\psi$ is not of the diffusion type, so our asymptotic expansion
breaks down right at this point. We have not investigated what happens
then.

As mentioned before, we will elsewhere address what we consider the
most interesting remaining challenges, the extension of (part of)
these results to pattern forming and chaotic fronts
\cite{spruijt,kees} and the question whether weakly curved fronts can
be analyzed with a moving boundary approximation \cite{ebertmba}, an
issue which is of central 
importance for understanding fronts in two and three dimensions like
streamers \cite{ebert}.

\subsection{The multiplicity of front solutions and of solutions of the saddle
  point equations}\label{S62}

As we discussed in Section \ref{S52}, our general discussion of the
convergence of pulled fronts to their  asymptotic velocity and shape
is based on 
the assumption that a uniformly translating front solution
$\Phi^*(\xi)$ exists (see (\ref{5a25}) for a definition), and that 
it is a member of a one-parameter family
of front solutions. What happens if this family of front solutions
does not exist has, to
our knowledge, not been investigated systematically for real
equations. However, experience with various pattern generating fronts ---
especially with a similar case in which no generalized uniformly translating
solutions exist  in the quintic complex Ginzburg-Landau
equation \cite{vs3}, even though the dynamics is pulled ---
yields the scenario that the leading edge just
spreads according to the linearized equations, and that the front
interior ``just follows'', in the sense that if there are uniformly
translating fronts solutions, the front interior and the region behind 
it relax smoothly, while 
if there are none, it is forced to follow the spreading
in some other way. This leads one to 
conjecture that if there is no family of uniformly translating front solutions,
the velocity relaxation will still be described by Eqs.\
(\ref{6a1}) and (\ref{6a2}) in the leading edge, but that  in the
interior front region the dynamics will inherently time-dependent,
e.g., incoherent \cite{kees}.  

This can occur in particular in the following situation:  
As mentioned in Section \ref{S531}, it can happen that the dispersion 
relation is such that there is more than one allowed nontrivial 
solution for the equations for $v^*$ and $\lambda^*$. According to the
linearized  
equation,  arbitrary sufficiently steep initial conditions will spread
out asymptotically with the largest of the speeds $v^*$. Hence the
asymptotic spreading speed of 
pulled fronts emerging from steep initial conditions is simply
the largest velocity $v^*$. Now,  according to a counting argument 
for the multiplicity of
uniformly translating front solutions, the multiplicity of front
solutions associated with different solutions of the saddle point
equations for $v^*$  will differ: If there are two solutions
$v_1^*$ and $v^*_2$ with $\lambda^*_1 < \lambda^*_2$, the multiplicity
of front solutions with velocity near $v^*_2$ and an asymptotic
spatial decay rate 
near $\lambda^*_2$ will be smaller than that of
those with velocity near $v^*_1$ and a spatial decay rate near
$\lambda^*_1$. Investigations of the issue of the competition between
various solutions $v^*$ will therefore also bear on
the issue raised in the beginning of this section, the question what
happens when there is no uniformly translating solution $\Phi^*$. In
particular, the dynamics in an equation that has  a family of
uniformly translating  
fronts associated with the solution $v^*_1$, should show a transition
from smoothly relaxing interior dynamics for $v^*_1>v^*_2$ to
incoherent interior dynamics for $v^*_1<v^*_2$.

\subsection{A step by step guideline for applying these results}
\label{S63}

If one just wants to apply our results to a given dynamical
equation with a given initial condition without worrying 
about the derivation and justification,
one can simply follow the following guidelines:

{\em (i)} Linearize the dynamical equation about the unstable state,
and determine the characteristic equation $S(k,\omega)=0$ for modes
$e^{-i\omega t+ikx}$ in the linearized equation.

{\em (ii)} Solve the double root or saddle point conditions
from Table V to determine $v^*$, $k^* $ and $D$.

{\em (iii)} Check whether the leading edge of the initial conditions
is steeper than $e^{-\lambda^*x}$ with Im~$k^*=\lambda^*$. Only then 
the front is a candidate for pulling with an asymptotic velocity $v^*$.

{\em (iv)} Check whether the conditions (\ref{5a25}) under which fronts are
expected to be uniformly translating, $\mbox{Re}\;k^*=
\mbox{Re}\;\omega^*=0$, $\mbox{Im}\;D=0$ are satisfied. If not, the
fronts will be pattern generating rather than uniformly translating
(see Section \ref{S62} above).

{\em (v)} Assuming the conditions under {\em (iv)} are obeyed, so
that the asymptotic front is expected to be uniformly translating,
investigate by a counting argument or otherwise whether there is a
one-parameter 
family of uniformly translating front profiles $\Phi_v(\xi)$ that
includes $\Phi^*(\xi)$.

{\em (vi)} Determine,  by using bounds, comparison theorems or physical
arguments, whether the fronts will be pushed or pulled. This determines,
which particular regime from Table IV applies.

{\em (vii)} If according to points {\em (iv)}--{\em (vi)} 
there is a family of front solutions that includes
$\Phi^*$, and if the dynamics is pulled, then our predictions
(\ref{6a1})--(\ref{6a3}) or Table II apply. If the conditions under 
{\em (iv)} are satisfied but there is no family of uniformly translating
solutions according to {\em (v)}, then our formula (\ref{6a2}) should
apply but one then expects  intrinsic nontrivial
dynamics in the front interior to remain, so that (\ref{6a3}) does not
apply. If {\em (iv)} is not satisfied (as for the EFK equation
(\ref{502}) for
$\gamma> 1/12$), one expects pattern generating fronts with a similar
algebraic convergence \cite{spruijt,kees}.

\subsection{The subtle role of the nonlinearities: an alternative
  intuitive explanation} \label{S64}

As we have seen in (\ref{2055}) and (\ref{5a26}), the convergence of the 
{\em linear} spreading
velocity to the asymptotic value $v^*$ is as $v(t)=v^*
-1/(2\lambda^*t)+\cdots $, while the convergence of nonlinear fronts
is as $v(t)=v^*-3/(2\lambda^*t)+\cdots$. The prefactor of the $1/t$ in
the latter case is just three times larger than for the linear
spreading velocity. What is this subtle difference due to?

In this paper, we have attributed the difference to the presence of
the term $\alpha \xi$ in the large-$\xi$ asymptotics
$(\alpha\xi + \beta)e^{-\lambda^*\xi}$ of $\Phi^*(\xi)$. 
We used an argument closely related to the one presented below, 
to prove in Section \ref{S252}, that $\alpha\ne0$. The functional
form of $\Phi^*$ leads to
the requirement that the  leading term in the expansion in similarity
solutions in the leading edge  is $(\xi/t^{3/2}) e^{-\xi^2/(4Dt)}$,
{\em not}  $(1/t^{1/2}) e^{-\xi^2/(4Dt)}$ (see Section
\ref{S311}). Nevertheless, one may want to have a better intuitive
understanding of why the asymptotics of the linear spreading velocity
is not correct for  the nonlinear front relaxation --- after all, one
might at first sight think that the linear spreading results should be
correct sufficiently far into the leading edge where the
nonlinearities can be neglected. The following picture allows us to
understand why this is wrong, and why the same type of algebraic
convergence also applies to pattern forming and chaotic fronts
\cite{spruijt,kees}.

Consider for simplicity the F-KPP equation (\ref{101}). As discussed
in the introduction and Section \ref{S252}, the dynamical equation for
the leading edge representation $\psi(\xi,t)$ of $\phi$ is
\begin{equation}
\partial_t \psi(\xi,t) = \partial^2_\xi \psi(\xi,t) -  
\psi^3(\xi,t) e^{-2\lambda^*
  \xi}~.\label{sinkeq}
\end{equation}
We can think of the nonlinear $\psi^3 e^{-2\lambda^*\xi}$ term as a
localized sink term in the diffusion equation for $\psi$: the term
vanishes for positive $\xi$ due to the exponential term, and for large
negative $\xi$ since $\psi$ vanishes exponentially in the region to
the left where $\phi$ saturates, see (\ref{2061}). 
Thus, if we think of $\psi$ as
representing the density of diffusing particles, then in the region
where this term is nonvanishing it describes the 
annihilation of particles. For the half space to the right of it,
where the particles freely diffuse, this term therefore acts like an absorbing 
boundary on the left. This is actually all that remains of the
nonlinearities in the equation! Whenever the
integrated sink strength $\alpha$ [the spatial integral of the
nonlinear term, in agreement with (\ref{2061})]  is
nonzero, the problem in the leading edge reduces to that of the
buildup of a diffusion field in the presence of an absorbing boundary
(and at the same time,  as (\ref{2061}) shows, $\alpha \neq 0$). In
this language, the  pulled to pushed transition occurs precisely when
the absorption strength $\alpha$ vanishes, and indeed precisely at
this point the velocity convergence is as  $v(t)=v^*
-1/(2\lambda^*t)+\cdots $ [see Eq.\ (\ref{3067})].

There is one complication: unlike the usual problems of diffusion in
the presence of a given absorbing boundary, the ``sink'' in
(\ref{sinkeq}) depends on the relaxing field $\psi$ itself. In fact, as
we discussed extensively in the paper, the diffusive dynamics of
$\psi$ leads to a logarithmic shift of the sink in time, in the frame
$\xi$. That is why in this interpretation we have to go, for  
selfconsistency,  to the frame
$\xi_X = \xi-X(t)$. In this frame, the ``sink'' or ``absorbing wall''
remains essentially fixed in time, and so the dynamics of $\psi$ is,
in leading order, that of a diffusion field in the presence of a fixed
absorbing wall. As is well known, in such a case a linear gradient
$\psi \propto \xi_X$ will build up in front of the wall, to balance
the constant annihilation of particles in the wall region.

Clearly, even if the ``sink'' strength is not stationary in time, the
buildup of the linear diffusion gradient far ahead of it will not be
affected. The present interpretation therefore yields a natural
starting point for analyzing the velocity relaxation of non-uniformly
translating fronts. This will be explored elsewhere
\cite{evsp,spruijt,kees}.  

We end this paper by stressing that while we have shown  that
nonlinear fronts relax according to the ``3/2 law''
$v(t)=v^*-3/(2\lambda^*t) +\cdots$, one can not apply this result
completely with closed eyes. An amusing illustration of this warning
is the following. It has been noted, that
the spreading velocity in the equation
\begin{equation}\label{carpeq}
\frac{\partial \phi}{\partial t} = \frac{\partial^2 \phi}{\partial x^2} +\phi
+ e \left( \frac{\partial \phi}{\partial x}\right)^2 
\end{equation}
follows the ``1/2 law'' $v(t)=v^* -1/(2\lambda^*t) \cdots =
2-1/(2t)+ \cdots $\cite{carpentier2}. At first sight,
this equation therefore might appear to yield a counterexample to our
assertions. In fact, it does not. Our results only hold for equations
where the growth of the dynamical field saturates behind the front,
{\em not} in the case in which the growth is unbounded. If the growth is
 unbounded, our arguments for why $\alpha \neq 0$, and hence for the
 ``3/2 law'', break down. The above
equation is precisely an example in which the growth does not
saturate: For $e>0$ and positive $\phi$, the nonlinear term only increases 
the growth. Hence there is no saturation and the spreading velocity 
$v_{nl}(t) $ in the
presence of the nonlinearities is larger than the one of the linear
equation: $v_{nl}(t) \geq v^* - 1/(2t)+ \cdots$. Apparently, in
practice the equality is obeyed asymptotically. Of course, if we add a
saturation term of the type $-\phi^k$ with $k>1$ to the right hand
side of (\ref{carpeq}), we obtain regular fronts and our usual
expression for $v(t)$ is recovered. 
~\\
~\\
{\bf Acknowledgement:} The work described in this paper started 
from discussions with C.\ Caroli. The work of U.E.\
was supported by the Dutch Science Foundation (NWO) 
and the EU-TMR-network ``Patterns, Noise, and Chaos''.

\end{multicols}

\newpage

\begin{multicols}{2}

\begin{appendix}

\section{An upper bound for $v_c$ in the nonlinear diffusion equation}
\label{bound}

With a generalization of the leading edge transformation
introduced in Sect.\ \ref{S26},
it is straight forward to prove the well-known
upper bound $v_c\le v_{sup}$, where
\begin{equation}
\label{vsup}
v_{sup}=2\sup_{0\le\phi\le1} \sqrt{\frac{f(\phi)}{\phi}}~,
\end{equation}
for the selected front velocity in the nonlinear diffusion equation,
if the initial conditions have steepness $\lambda>v_{sup}/2$.
The steepness $\lambda$ of a front is defined in (\ref{lambdadef}).
To prove this bound, transform (\ref{205}) to a frame $\xi=x-vt$, and write 
\begin{equation}
\label{l.e.rep}
\psi(\xi,t)=e^{v\xi/2}\phi(x,t)~.
\end{equation} 
The equation of motion is now
\begin{equation}
\partial_t \psi=\partial_\xi^2 \psi 
-\frac{v^2}{4}\;\psi+e^{v\xi/2}\;f\left(\psi\;e^{-v\xi/2}\right)~.
\end{equation}
If the initial steepness is $\lambda>v/2$, then
\begin{equation}
\lim_{\xi\to\pm\infty}\psi(\xi,t)=0~~\mbox{for all }0\le t<\infty~,
\end{equation}
since the steepness of the leading edge ($\xi\to\infty$) is conserved 
for all finite times, cf.\ the discussion in Sect.\ \ref{S25};
and since convergence at $\xi\to-\infty$ is garantueed by $\phi\to1$
behind the front together with the transformation (\ref{l.e.rep}). 
Thus the decay of $\psi$ at $\xi\to\pm\infty$ is exponential in $\xi$
for $t<\infty$. Hence, the whole equation can be multiplied by $\psi$ 
and integrated over $\xi$. This yields
\begin{equation}
\partial_t\int_\xi\frac{\psi^2}{2}=
-\int_\xi\left\{(\partial_\xi \psi)^2+\psi^2\;\left[\frac{v^2}{4}-
\frac{f\left(\psi\;e^{-v\xi/2}\right)}{\psi\;e^{-v\xi/2}}\right]\right\} ~,
\end{equation}
where all integrals are finite.
The r.h.s.\ of this equation is strictly negative, if
$v> v_{sup}$ (\ref{vsup}). Therefore, in a frame moving with
velocity $v>v_{sup}$, the integral $\int_\xi\psi^2$ decays in time.
This means, that the frame is propagating too rapidly, so that
the front shrinks away in the leading edge representation 
$\psi$ (\ref{l.e.rep}). Only a frame moving with velocity 
$v\le v_{sup}$ can propagate along with the speed of the front.
$v_{sup}$ is therefore an upper bound for the asymptotic velocity
of any initial condition with $\lambda>v_{sup}/2$.

For nonlinearity $f_{KPP}=\phi-\phi^k$, we have $v_{sup}=2$.
But on the other hand, we know (see Sect.\ \ref{S2}), that 
$v_{sup}\ge v_c\ge v^*=2$. Hence, these fronts are pulled
with $v_c=v^*=2$.
For nonlinearity $f_\epsilon=\epsilon\phi+\phi^{n+1}-\phi^{2n+1}$,
we have $v_{sup}=\sqrt{1+4\epsilon}>2\sqrt{\epsilon}=v^*$.

This version of the argument for $v<v_{sup}$ \cite{willem1}
can be generalized to equations with higher spatial derivatives,
forming both uniformly translating fronts or pattern forming fronts
\cite{spruijt}.

\section{The generalized nonlinear diffusion equation}
\label{genKPP} \label{A3}

Analyse a general equation with first temporal and second spatial derivative:
\begin{equation} \label{c01}
F\left(\phi,\;\partial_x\phi,\;\partial_x^2\phi,\;\partial_t\phi\right)=0~.
\end{equation} 
A front translating uniformly with velocity $v$ solves
\begin{equation} \label{c02}
F\left(\Phi_v,\;d_\xi\Phi_v\;,d_\xi^2\Phi_v,\;
-vd_\xi\Phi_v\right)=0~~,~~\xi=x-vt~.
\end{equation} 
The stability analysis of such a solution and the further treatment of
convergence is identical with what we did for the nonlinear diffusion
equation (\ref{101}) in Sections \ref{S2} and \ref{S3}.  We only need
 to transform the linear operators as discussed below. Our analysis is
 directly relevant for   
the equation studied in \cite{wang}.

We use the definition of functional derivatives as in (\ref{5038}) --
(\ref{5041}).  A linear perturbation $\eta(\xi,t)$ (\ref{2023}) about
a uniformly translating state $\Phi_v$ then solves the linear equation
$\partial_t\eta={\cal L}_v\eta$ (\ref{2024}) resp.\ (\ref{5036}) with
the linear operator being now
\begin{equation} \label{c03}
{\cal L}_v=f_2(\xi)\;\partial_\xi^2 +f_{1,v}(\xi)\;\partial_\xi 
           + f_0(\xi)~~,~~f_{1,v}=v+f_1~.
\end{equation}
For transforming to a Schr\"odinger problem $\partial_t\psi = {\cal
  H}_v\psi+o(\psi^2 e^{-\alpha})$, ${\cal H}_v=-\partial_y^2+V_v(y)$,
we now have to make the coefficient of the first order derivative
$\partial_\xi$ vanish, and the coefficient of the second order
derivative $\partial_\xi^2$ constant. This can be achieved through a
transformation similar to (\ref{2026}) and (\ref{2027}), combined with
a nonlinear transformation $y(\xi)$ of the length scale $\xi$:
\begin{eqnarray} \label{c04}
\psi &=& e^\alpha\;\eta ~~~,~~~ d \alpha(\xi) = 
\frac{2 f_{1,v}-\partial_{\xi}f_2}{4 f_2}\;d\xi~,
\\ \label{c05}
{\cal H}_v(y) &=& 
-\;e^{\alpha(\xi)}\;{\cal L}_v\;e^{-\alpha(\xi)}
=-\;\partial_y^2+V_v(y)~,
\\ \label{c06}
d y(\xi) &=& \frac{d\xi}{\sqrt{f_2(\xi)}}~~
\left(~\Leftrightarrow~\partial_y=\sqrt{f_2(\xi)}\;\partial_\xi~\right)~,
\\ \label{c07}
V_v(y(\xi)) &=& \frac{f_{1,v}^2-4f_0f_2}{4f_2} 
+\frac{f_2\;d_\xi f_{1,v}-f_{1,v}\;d_\xi f_2}{2f_2}
\nonumber\\
&& +\;\frac{3\;(d_\xi f_2)^2 - 4f_2\;d_\xi^2f_2}{16 f_2}~.
\end{eqnarray}
We use again the convention $\lim_{\xi\to\infty}\Phi_v(\xi)=0$.  By
construction the pulled velocity $v^*$ is the velocity, where
$V_{v^*}(\infty)= 0$. Accordingly now
\begin{equation} \label{c08}
v^*=2\sqrt{f_0(\infty)\;f_2(\infty)}-f_1(\infty) ~.
\end{equation}
The steepness of the leading edge is
\begin{equation} \label{c09}
\lambda^*=\partial_\xi\alpha \Big|_{\xi\to\infty,\;v=v^*} =
\sqrt{\frac{f_0(\infty)}{f_2(\infty)}}~.
\end{equation}
(In the convention of Section \ref{S52}: $f_n(\infty)=c_n$.)  In the
leading edge region, the relation between $y$ and $\xi$ is linear:
$y=\xi/\sqrt{f_2(\infty)}$.

If $V_{v^*}(y)\ge 0$ for all $y$, there are no destablizing linear
modes within the Hilbert space of (\ref{c05}). Then the front
propagating with $v^*$ is stable. The remaining analysis translates
from Sections \ref{S2} and \ref{S3} step by step with only the
explicit form of the linear operators ${\cal L}_v$ and ${\cal
  H}_v$ and the transformation operator $e^\alpha$ being  more
involved.

If there is a range of $y$ such that $V_{v^*}(y)$ becomes negative,
there might be a destabilizing mode in the spectrum of linear
perturbations.  In this case, there must be a pushed front solution
with some velocity $v^\dag>v^*$ with steepness
$\lambda=\lambda_+(v^\dag)>\lambda_0(v^\dag)=v^\dag/2$. Such a pushed
front might even be integrated analytically, if one can find an
analytic solution $\psi(\phi)$ of
\begin{equation} \label{c010}
F\left(\phi,\;\psi,\;\psi\;\frac{d \psi}{d \phi},\;-v\psi\right)=0~,
\end{equation}
equivalent to (\ref{a03}). $\xi(\phi)$ can then be integrated as in
(\ref{a05}). (Again, a closed form for $\psi(\phi)$ cannot be found
for pulled fronts, except possibly for equations at the pushed/pulled
transition.)

\section{Analytical solutions for pushed nonlinear diffusion fronts
and transition to pulling} \label{analKPP} \label{A1}

We here discuss, how to find analytical solutions for uniformly
translating fronts $\phi(\xi)$ in the equation
\begin{equation} \label{a01}
\partial_\xi^2\phi+v\partial_\xi\phi+f(\phi)=0~.
\end{equation}
$\bullet$ We rephrase and straighten the method from \cite{vs2} (see
also \cite{kaliappan,otwinowski})
how to find analytical front solutions.\\
$\bullet$ We recall, that analytical solutions can be found only for 
pushed fronts (propagating either into a meta- or into an unstable state,
Cases I and II from Table IV),
but not for pulled fronts (Case IV).\\
$\bullet$ We recall, that only a strongly heteroclinic orbit, i.e., a
front approaching $\phi=0$ with $\lambda>\lambda_0(v)$, is a candidate
for a pushed front. This allows us to calculate the critical $\epsilon$ 
for the pushed/pulled transition in the case of the nonlinearity (\ref{103}).

Write the equation as a flow in phase space as in (\ref{2022})
\begin{equation} \label{a02}
\partial_\xi {\psi\choose\phi}={-v\psi-f(\phi)\choose\psi}~,
\end{equation} 
where $\xi$ parametrizes the flow. If $\phi$ is monotonic in $\xi$, 
$\psi$ can be parametrized by $\phi$ instead of by $\xi$. 
This substitution yields for $\psi(\phi)$
\begin{equation} \label{a03}
\psi\;\frac{\partial\psi}{\partial\phi} + v\psi+f(\phi)=0~.
\end{equation}
This is the differential equation for the trajectory in phase space, 
where now the translational degree of freedom is removed together
with the parametrization $\xi$ of the flow. The resulting differential 
equation is one order lower than the original differential equation
(\ref{a01}). According to (\ref{2015}), the initial condition for the 
integration at $\phi\approx 1$ is 
\begin{eqnarray} \label{a04}
\psi(\phi=1-\delta)&=&-\tilde{\lambda}_-\delta+o(\delta^2)~,
\\
\tilde{\lambda}_-&=&v/2-\sqrt{v^2/4-f'(1)} ~,
\nonumber
\end{eqnarray}
so the front trajectory is unique and can be integrated. In some cases,
the integration can be done analytically, if one is lucky enough to find
an analytical solution $\psi(\phi)$ of Eq.\ (\ref{a03}) for a given $f(\phi)$.
If we have a solution $\psi(\phi)$, then the function $\xi(\phi)$
can be integrated as
\begin{equation} \label{a05}
\xi=\int_{\phi(0)}^{\phi(\xi)} \frac{d\phi}{\psi(\phi)}~.
\end{equation}
The final step consists in finding the inverse function 
$\phi=\phi(\xi)$, if this is possible.

Note now, that solutions $\psi(\phi)$ can be found analytically only,
if $\phi$ approaches $\phi=0$ with a single exponential
$\phi\propto e^{-\lambda\xi}$, since only then $\psi(\phi)$ has
the simple analytic form $\psi(\phi)=-\lambda\phi+o(\phi^2)$. 
Any other form of the approach to $\phi=0$, cf.\ (\ref{2018}),
would not be expressible in a simple analytic expression for
$\psi(\phi)$. In particular, a generic $\Phi^*$ front with
$\Phi^*\propto(\alpha\xi+\beta)e^{-\lambda^*\xi}$ in the leading edge
does not have a simple analytical expression for $\psi(\phi)$
since $\psi(\phi)=-\lambda^*\phi+\alpha\phi/(\alpha\xi+\beta)
+o(\phi^2)$, so a pulled front generically cannot be integrated 
analytically. Again $\alpha\ne0$ spoils the conventional tools 
of analysis!

Given an analytical front solution with velocity $v$ and decay
rate $\lambda$, one has to check the nature of the front.
A pushed front is a strongly heteroclinic front, i.e., it has
leading edge steepness $\lambda=\lambda_+(v)>\lambda_0(v)$. (For the notation 
of $\lambda$'s, compare Eq.\ (\ref{2018a}).) If $\lambda=\lambda_0(v)=
\lambda_\pm(v)$, we have found a front at the transition point from 
pushed to pulled with leading edge behavior $\phi\propto e^{-\lambda^*\xi}$. 
This is the only pulled front, we can integrate. 
If $\lambda=\lambda_-(v)<\lambda_0(v)$, we have a particular flat front, 
that has evolved from an initial condition with the same flatness in the 
leading edge.

Finding analytical solutions for pushed fronts can even be turned
into a machinery, if we don't fix $f$ and look for a $\psi$, but if we
define $\psi(\phi)$ and then calculate $f(\phi)$. For
\begin{equation} \label{a06}
\psi=-\lambda\phi(1-\phi^n)
\end{equation}
we calculate, e.g.,
\begin{eqnarray} \label{a07}
f(\phi)&=&\lambda(v-\lambda)\phi + \lambda(\lambda(n+2)-v)\phi^{n+1} 
- \lambda^2(n+1)\phi^{2n+1} 
\nonumber\\
&=&\bar{\epsilon}\phi+\phi^{n+1}-(1+\bar{\epsilon})\phi^{2n+1}~,
\end{eqnarray}
where we have to identify $v=(n+2)\lambda-1/\lambda$, and 
$\bar{\epsilon}=\lambda(v-\lambda)$. The analytic front solution for 
(\ref{a06}) can be calculated from (\ref{a05}) and inverted to yield
\begin{equation} \label{a08}
\phi(\xi)=\left[ 1+\left(\phi(0)^{-n}-1\right) e^{\lambda n\xi}\right]^{-1/n}~.
\end{equation}
This solution is a pushed front, if $\lambda\ge\lambda_0(v)=v/2$, 
which implies $\bar{\epsilon}\le1/n$. For such $\bar{\epsilon}$, 
we find pushed fronts with decay rate 
$\lambda=\sqrt{(\bar{\epsilon}+1)/(n+1)}$, velocity 
$v=(1+\bar{\epsilon}(n+2))/\sqrt{(n+1)(1+\bar{\epsilon})}$, 
and analytical form (\ref{a08}). For $\bar{\epsilon}=1/n$, the solution is 
a front on the transition point from pushed to pulled fronts with asymptotic
decay $\phi\propto e^{-\lambda^*\xi}+o(\phi^2)$.
For $\bar{\epsilon}>1/n$, the solution (\ref{a08})
is a flat front evolving from flat initial conditions. Fronts evolving
from sufficiently steep initial conditions are then pulled, propagate 
with velocity $2\sqrt{\bar{\epsilon}}$, have decay rate
$\lambda^*=\sqrt{\bar{\epsilon}}$ and no analytic form of the front solution 
can be found. 

Eq.\ (\ref{101}) with nonlinearity (\ref{a07}) can also be rescaled
to bring the equation to the more familiar form
\begin{eqnarray} \label{a09}
\partial_\tau\varphi &=& \partial_y^2\varphi
+\epsilon\varphi+\varphi^{n+1}-\varphi^{2n+1}
\\ \label{a010}
\epsilon&=&\bar{\epsilon}(1+\bar{\epsilon})~~,~~
(1+\bar{\epsilon})=\frac{t}{\tau}=\left(\frac{x}{y}\right)^2
= \left(\frac{\varphi}{\phi}\right)^n~.
\end{eqnarray}
This reproduces precisely the form of the nonlinearity (\ref{103})
with the stable state now at 
$\varphi_s=\left[\left(1+\sqrt{1+4\epsilon}\right)/2\right]^{1/n}$.
Accordingly the critical $\epsilon$ is now $\epsilon_c=(n+1)/n^2$.
Fronts propagate for $\epsilon<\epsilon_c$ with the pushed
velocity $v^\dag=\big[(n+2)\sqrt{1+4\epsilon}-n\big]
/\big[2\sqrt{n+1}\big]$ and decay rate 
$\lambda_+(v^\dag)=\big[1+\sqrt{1+4\epsilon}\big]
/\big[2\sqrt{n+1}\big]$. For $\epsilon>\epsilon_c$, they propagate 
with the pulled velocity $v^*=2\sqrt{\epsilon}$ and decay rate
$\lambda^*=\sqrt{\epsilon}$.

\section{Linear stability analysis of moving front solutions} 
\label{stabilityapp}

In this appendix we study the linear perturbations $\eta$ of
a uniformly translating front $\Phi_v(\xi)$ in the nonlinear
diffusion equation. The problem is defined in Eqs.\ (\ref{2023})
-- (\ref{2025}) and can be transformed to a Schr\"odinger problem
(\ref{2026}) -- (\ref{2028}) with linear operator ${\cal H}_v$.
Since ${\cal H}_v$ is self-adjoint, we
can decompose functions that lie in the Hilbert space of ${\cal
  H}_v$, into the orthonormal set of eigenfunctions of ${\cal H}_v$.
Eigenfunctions in this Hilbert space form a complete set. However, it
is obvious that not all linear perturbations with $|\eta|\ll1$ are in
this space: Only perturbations with
\begin{equation}
\label{Hilberteta}
\lim_{\xi\to\infty} |\eta|\;e^{\:\lambda_0(v)\;\xi}<\infty ~~\mbox{with}~
\lambda_0(v)=\frac{v}{2}
\end{equation}
can lie in the Hilbert space (which consists of square integrable
functions and of solutions proportional to plane waves $e^{ik\xi}$ as
$\xi\to\pm\infty$).

\subsection{Schr\"odinger stability analysis} \label{A231}

The general properties of the spectrum and eigenfunctions of ${\cal
  H}_v$ within the Hilbert space can be immediately obtained from a
few well known results which to physicists are known from quantum
mechanics (see, e.g., \cite{messiah}), since ${\cal H}_v$ is the
Hamiltonian operator for a (quantum) wave in a potential in one
dimension. The potential is asymptotically lower on the right than on
the left, since
\begin{equation} \label{2030}
V(\infty)={\textstyle\frac{v^2}{4}}-f'(0)\;\;<\;\;
{\textstyle\frac{v^2}{4}}-f'(1)=V(-\infty)~,
\end{equation}
according to (\ref{206}). If we write the temporal behavior of an
eigenfunction as $\tilde{\psi}_\sigma(\xi)\; e^{-\sigma t}$, one finds
that the spectrum of
\begin{equation} \label{2031}
{\cal H}_v \tilde{\psi}_\sigma=\sigma\tilde{\psi}_\sigma
\end{equation} 
is continuous for $\sigma\ge V(\infty)$, and that the eigenfunctions
are distributions, i.e., essentially plane waves $e^{ik\xi}$ with
$k=\pm\sqrt{\sigma-V(\infty)}$ as $\xi\to\infty$. One immediately
concludes that a front $\Phi_v$ with velocity $v<v^*=2\sqrt{f'(0)}$
will be unstable against the continuous spectrum of linear
perturbations with ``energies'' $V(\infty)<\sigma<0$.

For a front $\Phi_v$ with velocity $v\ge v^*=2\sqrt{f'(0)}$, there
still might be a point spectrum of bound and square integrable states
with $\sigma<0$.  Bound states have a finite number of nodes, and
there is a one-to-one correspondence between the number of nodes and
the eigenvalue of the bound state ``wavefunction'' $\psi_\sigma$: the
eigenfunction with the lowest eigenvalue $\sigma$ is nodeless (if it
exists), the eigenfunction corresponding to the next largest bound
state eigenvalue has one node, etc.  Therefore, the point spectrum is
bounded from below by the ``energy'' $\sigma$ of the nodeless
eigenfunction, if it exists. Now, one eigenfunction is known: the
translation mode $\tilde{\psi}_0$ clearly has $\sigma=0$. It
 can be generated by an infinitesimal translation of
$\Phi_v$:
\begin{equation} \label{2032}
\tilde{\psi}_0 = e^{\lambda_0(v)\xi}\;\partial_\xi\Phi_v~~,~~
{\cal H}_v\tilde{\psi}_0 = 0~.
\end{equation}
If $\Phi_v$ is monotonic, $\tilde{\psi}_0$ will be nodeless.  If
$\Phi_v$ is non-monotonic, $\tilde{\psi}_0$ will have nodes.

From this one might be tempted to immediately draw conclusions on the
stability of monotonic or non-monotonic front solutions.  However,
this is only possible if $\tilde{\psi}_0$ is in the Hilbert space!
Comparison with (\ref{2018}) shows, that this is the case, if either
$v=v^*$ and $\alpha=0$, or if $v>v^*$ and $A_v=0$, i.e., for one of
the strongly heteroclinic orbits.

If  a front $\Phi_v$  obeys one of these conditions and if it is
monotonic, then 
$\tilde{\psi}_0$ is the 
eigenfunction in the Hilbert space with the lowest ``energy''
$\sigma=0$.  Therefore all other eigenfunctions will have $\sigma>0$
and will decay in time as $e^{-\sigma t}$. An arbitrary linear
perturbation in the Hilbert space can be decomposed into the complete
set of eigenfunctions, and therefore it will decay too (apart of
course from  the nondecaying translation
mode $\tilde{\psi}_0$).

If such a front $\Phi_v$ is non-monotonic it will have $n$ extrema,
with $n>0$ some integer.  The translation mode then has $n$ nodes, and
hence there are then $n$ bound eigenfunctions $\tilde{\psi}_\sigma$
with negative $\sigma$.  The front profile is then linearly unstable
with respect to these modes.  Since any generic initial condition will
have a nonvanishing contribution from these destabilizing modes, a
non-monotonic $\Phi_v$ will generically not be approached for long
times. Such a $\Phi_v$ is called dynamically unstable.

The analysis of the spectrum and eigenfunctions of the Schr\"odinger
operator in the Hilbert space therefore yields the following results:

1) A front $\Phi_v$ with velocity $v<v^*$ is intrinsically unstable
against a continuous band of linear perturbations from the Hilbert
space.  Such a front generically will not be approached under the
dynamics.

2) A front $\Phi^*$ with velocity $v=v^*$ and $\alpha=0$ is unstable
against perturbations from the Hilbert space, if it is non-monotonic,
and it is stable, if it is monotonic. There is a continuous band of
linear perturbations with $\sigma\ge0$, that continuously extends down
to $\sigma=0$. Accordingly, there is no gap in the excitation
spectrum, which already hints at the non-exponential convergence
towards a monotonic $\Phi^*$.

3) A strongly heteroclinic orbit $\Phi_v$ with $v>v^*$ and $A_v=0$, if
it exists, is unstable against perturbations from the Hilbert space,
if it is non-monotonic, and it is stable, if it is monotonic.  If
strongly heteroclinic orbits exist, by construction (see Section
\ref{S22}) only the one with the largest velocity $v=v^\dag$ is
monotonic, and the front $\Phi^*$ with velocity $v=v^*<v^\dag$ is
non-monotonic and thus unstable. So only for $\Phi^\dag$, the spectrum
of linear perturbations is purely positive: $\sigma\ge0$.  For
$\Phi^\dag$ there is at best a discrete spectrum of linear
perturbations in the Hilbert space in the range
$0<\sigma<V(\infty)=({v^\dag}^2-{v^*}^2)/4$, and the continuous
spectrum begins at $\sigma\ge V(\infty)$. Convergence of all
perturbations in the Hilbert space will thus be exponential in time
like $e^{-\sigma t}$, with $\sigma$ the smallest positive eigenvalue.

Note the restrictions of this analysis:

$(a)$ Up to now, we have no predictions for fronts with velocity $v\ge
v^*$, whose translation mode $\tilde{\psi}_0$ (\ref{2032}) is outside
the Hilbert space. We will see that the equivalence of stability and
monotonicity extends beyond the Hilbert space analysis.

$(b)$ The analysis of general initial conditions might require linear
perturbations, that lie outside the Hilbert space, even if
$\tilde{\psi}_0$ is in the Hilbert space.

\subsection{Linear perturbations outside the Hilbert space} \label{A232}

The mapping to the Schr\"odinger problem is a powerful method for
perturbations $\eta$ about a front $\Phi_v$, that lie within the
Hilbert space, because we then can work with a complete set of
orthogonal functions. In general, however, this space of perturbations
needs to be 
completed by functions from outside the Hilbert space.

To see this, consider for simplicity an initial condition, that is
close to some $\Phi_v$ with $v\ge v^*$, but steeper than this
asymptotic front: $\lim_{x\to\infty}\phi(x,0)/\Phi_v(x) = 0$.  Then
the steepness in the leading edge of $\eta=\phi-\Phi_v$ will be
dominated by $\Phi_v$, and
\begin{equation} \label{2035}
\psi=\eta\;e^{\lambda_0(v)\xi}\;
\stackrel{\xi\to\infty}{\sim}\left\{\begin{array}{ll}
  \alpha\xi+\beta ~&\mbox{ for } v=v^*\\ 
  e^{-\mu(v)\xi}~&\mbox{ for } v=v^\dag \mbox{ or generally}\\
  &~\mbox{ for }v>v^*\mbox{ and }A_v=0\\ 
  e^{\:\mu(v)\xi}~&\mbox{ for }v>v^*\mbox{ and }A_v\ne0\\ 
\end{array}\right.
\end{equation}
with $\mu(v)=\sqrt{v^2/4 -1}>0$ from (\ref{2018}).  Accordingly, only
for a pushed front propagating with velocity $v=v^\dag$ (or more
generally for a strongly heteroclinic orbit with $v>v^*$ and $A_v=0$)
or for a pulled front with velocity $v^*$ and $\alpha=0$, the linear
perturbation $\eta\;e^{\lambda_0(v)\xi}$ is in the Hilbert space of
${\cal H}_v$.  The decay of the zero mode $\tilde{\psi}_0$
(\ref{2032}) is asymptotically the same as that of $\psi$ in
(\ref{2035}).  So a treatment of linear perturbations outside the
Hilbert space is clearly called for.

In general, we want to decompose perturbations $\eta$ that obey
\begin{equation}
\label{decay}
\lim_{\xi\to\pm\infty}|\eta(\xi,t)|\ll 1~.
\end{equation} 
This is required for the linearization of $\phi$ about $\Phi_v$ in
(\ref{2024}).  We aim at a decomposition of $\eta(\xi,t)$ into
eigenfunctions $\eta_\sigma(\xi)\:e^{-\sigma t}$. We therefore return
to the eigenvalue equation for such an eigenmode, which according to
(\ref{2024}) and (\ref{2025}) is given by
\begin{equation} \label{etamode1} 
  \left[ \partial_\xi^2 + v\partial_\xi
    +f'\Big(\Phi_v(\xi)\Big)+\sigma \right] \eta_\sigma = 0~.
\end{equation} 
Our previous analysis in the Hilbert space already has identified many
of these eigenmodes, in fact all those, which obey (\ref{Hilberteta}).
This criterium on $\eta_\sigma$ is too strict at $\xi\to\infty$, so we
now need to additionally analyze perturbations with
$e^{-\lambda_0(v)\xi}<|\eta_\sigma(\xi)|<1$ as $\xi\to\infty$, which
lie outside the Hilbert space. On the other hand, for $\xi\to-\infty$,
Eq.\ (\ref{Hilberteta}) is less restrictive than (\ref{decay}).  This
gives us the freedom to impose only $|\eta_\sigma(\xi)|\lesssim
e^{\lambda_0(v)|\xi|}$ as $\xi\to-\infty$, since such a divergence can
be compensated for by perturbations from inside the Hilbert space,
where we make use of its completeness.  We therefore now impose the
boundary conditions
\begin{equation}
\label{etaeta}
\label{etasigma2}
\lim_{\xi\to\infty}|\eta_\sigma(\xi)|<\infty
~~,~~\lim_{\xi\to-\infty}e^{\lambda_0(v)\xi}|\eta_\sigma(\xi)|<\infty~,
\end{equation}
where perturbations that additionally obey
$e^{-\lambda_0(v)\xi}|\eta_\sigma(\xi)|<\infty$ as $\xi\to\infty$,
are in the Hilbert space of ${\cal H}_v$.

First of all, we note that the translation mode
$\eta_0(\xi)=\partial_\xi\Phi_v(\xi)$ (\ref{2032}) now is always
included in the larger space (\ref{etaeta}) of perturbations.

Second, solve (\ref{etamode1}) for $\xi\to\infty$ and find in analogy
to (\ref{2018}) that
\begin{equation} \label{etasigma}
  \eta_\sigma (\xi) = A_\sigma e^{-\Lambda_-\xi} + B_\sigma
  e^{-\Lambda_+\xi}~,
\end{equation}
with
\begin{equation} \label{sigmaeq}
  \Lambda_\pm(\sigma,v) = \frac{v}{2} \pm \sqrt{ \frac{v^2}{4} -f'(0)
    -\sigma}~.
\end{equation}
For brevity of notation, we here allowed $\Lambda_\pm(\sigma,v)$ to be
complex. In Fig.\ 15 we plot $\Lambda_\pm$ versus $\sigma$, both for
the case of a front propagating into an unstable state ($f'(0)>0$),
and for the case of a front between a stable and a metastable state
($f'(0)<0$), and for $f'(0)>0$, we furthermore distinguish between
$v>v^*$ and $v=v^*$.
The leading edge solution (\ref{etasigma}), of course,
precisely coincides with the leading edge behavior of the Hilbert
space functions, except that one case was excluded from the Hilbert
space: A leading edge with $A_\sigma\ne0$ and $\sigma\le
V(\infty)=v^2/4-f'(0)$ does not obey the boundary condition
(\ref{Hilberteta}).  It does obey the boundary condition
(\ref{decay}), if $\sigma\ge-f'(0)$.  Let us therefore now focus on
the additional perturbations with
\begin{equation}
\label{sigma}
-f'(0)<\sigma\le V(\infty)=\frac{v^2}{4}-f'(0)~.
\end{equation}
If $A_\sigma\ne0$, such perturbations are outside the Hilbert space,
but they do obey (\ref{etaeta}).

Are there such perturbations for a given $\sigma$, and how many?  For
answering this question we need to analyze $\eta_\sigma$ globally, in
close analogy to the global analysis of the $\Phi_v$ as a function of
$v$ in Section \ref{S22}.  Solving (\ref{etamode1}) at $\xi\to-\infty$
yields two exponents
\begin{eqnarray}
\label{kl}
\tilde{\Lambda}_\pm(\sigma,v)&=&\frac{v}{2}\pm
\sqrt{\frac{v^2}{4}-f'(1)-\sigma}
\\
&=&\lambda_0(v)\pm\sqrt{V(-\infty)-\sigma}~,
\nonumber
\end{eqnarray}
in analogy with (\ref{2014}) and (\ref{2015}).  Since
$V(-\infty)>V(\infty)$ (\ref{2031}), for $\sigma\le V(\infty)$ we
certainly have $V(-\infty)-\sigma>0$.  The coefficient of
$e^{-\tilde{\Lambda}_+(\sigma,v)\xi}$ therefore needs to vanish for
$\eta_\sigma$ to obey (\ref{etaeta}).  Behind the front for
$\xi\to-\infty$, we therefore find that
\begin{equation}
\label{mininf}
\eta_\sigma(\xi)=\pm e^{-\tilde{\Lambda}_-(\xi-\xi_0)}
+o\left(e^{-2\tilde{\Lambda}_-\xi}\right)~,
\end{equation}
for an $\eta_\sigma$ obeying (\ref{etaeta}) and (\ref{sigma}).  Eq.\ 
(\ref{mininf}) determines $\eta_\sigma$ uniquely because the
arbitrary constant coefficient $\pm e^{\tilde{\Lambda}_-\xi_0}$ can be
scaled out of a linear equation like (\ref{etamode1}). Such a linear
equation can always be integrated towards $\xi\to\infty$, where it
uniquely determines the coefficients $A_\sigma$ and $B_\sigma$ in
(\ref{etasigma}). Accordingly, $A_\sigma$ and $B_\sigma$ generically
are non-vanishing, in complete analogy to the argument for $A_v$ and
$B_v$ in (\ref{2018}) to be generically non-vanishing in
$\Phi_v$.

What do we gain with these extra solutions?  The eigenfunctions in the
Hilbert space had a continuous spectrum for $\sigma\ge
V(\infty)=(v^2-{v^*}^2)/4\ge0$ and at best a discrete spectrum defined
by $A_\sigma=0$ for $\sigma<V(\infty)$. Adding the solutions, that
obey (\ref{etaeta}), we extend the continuous spectrum down to
$\sigma\ge-f'(0)=V(\infty)-v^2/4<0$ and find at best a discrete
spectrum defined by $A_\sigma=0$ for $\sigma<-f'(0)$. These discrete
solutions for $\sigma<-f'(0)$ all lie in the Hilbert space.

Let us now look at the steepness in the leading edge of the solutions
outside the Hilbert space. They have a $\sigma$ from the interval
(\ref{sigma}), and $A_\sigma\ne0$. For these we observe (cf.\ Fig.\ 15) that
\begin{eqnarray} 
  \Lambda_-(\sigma,v) & > & \lambda_-(v) ~~\mbox{for } ~~ \sigma >0
  ~~~~\mbox{(decaying)}~, \nonumber \\ \Lambda_- (\sigma,v) & < &
  \lambda_-(v) ~~\mbox{for } ~~ \sigma <0 ~~~~\mbox{(destabilizing)}~,
  \label{hoiute} \\ 
  \Lambda_- (0,v) & = & \lambda_-(v) ~~\mbox{for } ~~ \sigma =0
  \nonumber ~~~~\mbox{(marginal)}~,
\end{eqnarray}
with $\lambda_-(v)$ from (\ref{2015}). This means, that these linear
eigenmodes $\eta_\sigma$ of $\Phi_v$ will decay ($\sigma>0$), if they
are steeper than $e^{-\lambda_-(v)\:\xi}$, and that they will
destabilize a front $\Phi_v$, if they are flatter. Note that the
spectrum of decaying modes is continuous down to zero, as $\sigma
\downarrow 0 $ as $\Lambda_- \downarrow \lambda_-$.

It is tempting to conclude here immediately, that a front
$\Phi_v(\xi)$ with velocity $v\ge v^*$ will be stable against all
perturbations, which are steeper in the leading edge than
$e^{-\lambda_-(v)\xi}$.  However, the possible existence of the
discrete set of solutions with $A_\sigma=0$ and $\sigma<0$ requires
special attention, since these perturbations are steeper than
$e^{-\lambda_0(v)\xi}$, but destabilizing ($\sigma<0$). Now, if
$\Phi_v$ is strongly heteroclinic ($A_v=0$), we
already found  in Section \ref{A231}, that such destabilizing
perturbations exist, if and only if $\Phi_v$ is non-monotonic.  
We now need to show that this argument also holds for fronts $\Phi_v$ 
with $v>v^*$ and $A_v\ne0$ or for fronts $\Phi^*$ with velocity $v^*$
and $\alpha\ne0$.  The following five steps {\em (i)--(v)} prove this:
$(i)$ Impose (\ref{mininf}) at $\xi\to-\infty$. This defines a unique
solution of equation (\ref{etamode1}) for $\eta_\sigma$ for every
$\sigma<V(-\infty)$. In fact, we only need to analyze
$\sigma<V(\infty)$, since we know the spectrum for larger $\sigma$.
$(ii)$ Integrate (\ref{etamode1}) forward towards $\xi\to\infty$ for a
very large negative $\sigma$. The variation of $f'(\Phi_v(\xi))$ in
space then can be almost neglected.  Therefore at $\xi\to\infty$, we
will find (\ref{etasigma}) with $|A_\sigma/B_\sigma|\gg1$. For our
further construction it is crucial to observe, that such a
perturbation for sufficiently large negative $\sigma$ will be
nodeless.  It does not matter, on the other hand, that this solution
typically will not obey our bound (\ref{etasigma2}), since we only use
it as a means for constructing the solutions with $A_\sigma=0$, which
will not only obey (\ref{etasigma2}), but even lie inside the Hilbert
space.  $(iii)$ Upon increasing $\sigma$ continuously, at discrete
values of $\sigma<V(\infty)$, $\eta_\sigma$ will gain an extra node.
Since the generation of every new node is associated with a change of
sign of the perturbation at $\xi\to\infty$, if the sign at
$\xi\to-\infty$ is kept fixed, the appearance of an additional mode
can only occur at a $\sigma$, where the sign of $A_\sigma$ changes.
$(iv)$ We know the number of nodes of the zero mode $\eta_0$. It is
identical to the number of extrema of $\Phi_v$. We therefore know the
number of particular perturbations with $A_\sigma=0$ and $\sigma<0$.
$(v)$ From this it follows that if $\Phi_v$ is monotonic, there are 
no particular perturbations with $A_\sigma=0$ and $\sigma<0$.  
If $\Phi_v$ is non-monotonic, there are such perturbations.

In Section \ref{S22}, we have counted the multiplicity of front
solutions $\Phi_v$ as a function of $v$. Here we have counted the
multiplicity of perturbations $\eta_\sigma$ of a front $\Phi_v$ as a
function of $\sigma$. This counting was based on the proper
asymptotics of the solutions at $\xi\to\pm\infty$, which is of the
same structure for both $\Phi_v$ and $\eta_\sigma$, so the counting argument
follows exactly the same lines in both cases.

The conclusions from this appendix are summarized in Section \ref{S23}.

\section{Stability analysis, selection and 
  rate of convergence} \label{consequencesapp}

In this appendix, we analyze the implications  of the results of the 
stability analysis of Appendix \ref{stabilityapp} for understanding
the selection of fronts and for the rate of convergence towards the
asymptotic front solution. For pushed fronts, the stability analysis
implies that the relaxation towards the pushed front solution is
exponentially fast, while for pulled fronts the spectrum is gapless,
and the convergence can not be obtained from the stability spectrum.

\subsection{Pushed regime: $v_c=v^\dag$}

\label{A241}

We first consider equations with the nonlinearity $f(\phi)$ such that
the slowest stable front is a strongly heteroclinic orbit in phase
space with $A_{v^\dagger}=0$ in (\ref{2018}). We have denoted this
asymptotic front with $\Phi^\dag$ and its velocity with $v^\dag$.  Its
steepness is $\lambda_+(v^\dag)=\lambda_0(v^\dag)+\mu(v^\dag)$, cf.\ 
(\ref{2021}). There is a continuous family of stable front solutions
$\Phi_v$ with velocity $v>v^\dag$ which are all flatter than
$\lambda_-(v^\dag)=\lambda_0(v^\dag)-\mu(v^\dag)$.  Their steepness
$\lambda=\lambda_-(v^\dagger)$ is related to their velocity $v$
through
\begin{equation}
\label{20044}
v(\lambda)=\lambda+\frac{1}{\lambda}~,
\end{equation}
as can be obtained by inverting (\ref{lambda1}).

CASE I:$~$ Consider an initial condition with steepness
$\lambda_{init}>\lambda_0(v^\dag)$.  We let $\phi$ evolve some time,
and then linearize it about $\Phi^\dag$.  According to (\ref{lameta})
the perturbation $\eta$ will have steepness
$\lambda_{\eta}>\lambda_0(v^\dag)$. It then is in the Hilbert space
analyzed in Section \ref{A231}.  We can decompose the perturbation
into the known eigenperturbations.  The spectrum of decay
rates has no negative eigenvalues, one eigenvalue zero and then a gap 
above zero. A contribution from the zero mode can be made vanishing
by adjusting the position of the subtracted asymptotic front
$\Phi_v$, by making use of the translational freedom
of $\Phi_v$. The perturbation then can be decomposed into Hilbert
space functions $\eta_\sigma$ with $\sigma$ all positive and bounded 
away from zero. Thus, for large times the perturbation 
will decay exponentially. This means that an initial condition with
$\lambda_{init}>\lambda_0(v^\dag)$ will converge to $\Phi^\dag$
exponentially in time, generically with $e^{-\sigma_1 t}$,
where $\sigma_1$ is the smallest positive eigenvalue.

CASE II:$~$ If the initial steepness is
$\lambda_-(v^\dag)<\lambda_{init}\le\lambda_0(v^\dag)$, the
perturbation of $\phi$ about $\Phi^\dag$ will not be in the Hilbert
space. However, we do know from the results illustrated in Fig.\ 15 
that there is an eigenmode $\eta_\sigma$ of the linear stability
operator of $\Phi^\dag$ with the proper steepness $\lambda_{init}=
\lambda_\eta$, that will decay exponentially in time, see Section
\ref{A232}.  The remaining linear perturbation $\eta-\eta_\sigma$
might lie in the Hilbert space, in which case we are back to Case I. 
If it does not, we have to identify the subleading $\lambda$, its 
corresponding eigenmode $\eta_\sigma$ etc.  The iteration of this 
construction leads
us to conclude that the perturbation indeed will decay exponentially
in time.  (Examples of exponential convergence towards pushed fronts
which is dominated by such modes can be found in Fig.\ 19 of
\cite{vs2}.) Another way of putting the argument is that only
perturbations with $\lambda< \lambda_-(v^\dag)$ can grow in time, but
these cannot be involved in the decomposition of a perturbation with
$\lambda_\eta>\lambda_-(v^\dag)$. A more elegant way of analyzing
this case and the following ones  is discussed in Section \ref{S25}.

CASE III:$~$ If the initial steepness is
$\lambda_{init}<\lambda_-(v^\dag)$, and we linearize $\phi$ about
$\Phi^\dag$, there is a perturbation $\eta_\sigma$ with steepness
$\lambda_\eta=\lambda_{init}$ that is growing in time ($\sigma<0$).  
So such an initial condition cannot approach $\Phi^\dag$ or any other 
asymptotic front $\Phi_v$ with steepness $\lambda_{asympt}>\lambda_{init}$.  
If we linearize $\phi$ about the asymptotic front $\Phi_v$ with the same
steepness $\lambda_{init}=\lambda_{asympt}$, the remaining
perturbation will be steeper, so contributions from the zero mode
are excluded by construction, and the perturbation can be decomposed 
into eigenperturbations of $\Phi_v$, which all decay in time. 

In summary: All initial conditions with
$\lambda_{init}>\lambda_-(v^\dag)$ converge exponentially in time to
the ``selected'' front with velocity $v_{sel}=v^\dag$ and steepness
$\lambda_{sel}=\lambda_+(v^\dag)$.  Initial conditions with
$\lambda_{init}<\lambda_-(v^\dag)$ will converge to a quicker
asymptotic front with steepness $\lambda_{asympt}=\lambda_{init}$ and
velocity $v(\lambda_{init})$ given by (\ref{20044}).

In Section \ref{S21}, we have termed an initial condition sufficiently
steep ($\lambda_{init}>\lambda_{steep}$), if it approached the
``selected'' front for large times. We have denoted the steepness of
the selected front with $\lambda_{sel}$.  In the pushed regime, one
can thus identify these parameters with
\begin{eqnarray}
\label{20045}
\lambda_{steep}&=&\lambda_-(v^\dag)=\frac{v^\dag}{2}-\mu(v^\dag)~,\\
\lambda_{sel}&=&\lambda_+(v^\dag)=\frac{v^\dag}{2}+\mu(v^\dag)~,~~
\mu(v^\dag)=\sqrt{\frac{v^{\dag\:2}-4}{4}}~, \nonumber\\
v_{sel}&=&v^\dag~. \nonumber
\end{eqnarray}

\subsection{Fronts into metastable states}

\label{A242}

The only difference between a pushed front propagating into an unstable 
state, i.e., with a nonlinearity $f$ such that $f'(0)>0$ and
$v_c=v^\dag>v^*$, and a front propagating into a metastable state, 
i.e., with $f'(0)<0$, is the sign of $\lambda_{-}(v)$:
For a front into a metastable state, we have
\begin{equation}
\label{2046}
\mu(v)=\sqrt{\frac{v^2-4f'(0)}{4}} > \frac{v}{2}~~\mbox{for}~~
f'(0)<0~,
\end{equation}
so  $\lambda_-(v)<0$ and $\lambda_+(v)>0$ for all $v>0$ (the sign of
$\lambda_0(v)$ is the same as the sign of $v$).
Suppose, that the selected front still travels with positive speed 
$v_{sel}=v^\dag$ (otherwise reverse $x$).
Because now $\lambda_-(v)<0$,
\begin{equation}
\label{2047}
\lambda_{steep}=0~,
\end{equation} 
so all initial conditions are sufficiently
steep and converge to $\Phi^\dag$. The continuous spectrum of asymptotic
solutions $\Phi_v$ with $\lambda_{asympt}<\lambda_{steep}$ ceases
to exist, and the asymptotic front $\Phi^\dag$ therefore now is unique.

For the convergence of an initial condition $\phi$ towards 
$\Phi^\dag$ we still need to distinguish whether $\lambda_{init}$
is larger or smaller than $\lambda_0(v^\dag)=v^\dag/2$.
If $\lambda_{init}>\lambda_0(v^\dag)$ the perturbation about
$\Phi^\dag$ lies in the Hilbert space, while for 
$\lambda_{init}<\lambda_0(v^\dag)$, it does not. 
This corresponds to the Cases I and II for $v_c=v^\dag$ above,
which apply literally. In both cases the initial conditions converge  
to $\Phi^\dag$ exponentially in time. Case III does not occur
for fronts into metastable states.

\subsection{Pulled regime: $v_c=v^*$}

\label{A243}

At the transition from fronts propagating into metastable
towards fronts into unstable states, $f'(0)$ changes sign, 
and so does $\lambda_{-}(v)$. At this point a continuum of possible
attractors $\Phi_v$ of the dynamics comes into existence, but the
convergence behavior of sufficiently steep initial conditions
is completely unchanged. In other words: Cases I and II are completely
unchanged and only Case III needs to be considered additionally
for initial conditions with $\lambda_{init}<\lambda_{steep}$.

A qualitative change  in the convergence behavior of sufficiently steep
initial conditions $\lambda_{init}>\lambda_{steep}$ only
takes place at the transition from the pushed to the pulled regime. 
This happens for $f$ changing such that $v^\dag$ approaches $v^*$.
Then
\begin{equation}
\label{2048}
\lambda_{steep}=\lambda_0(v^*)=\lambda_{sel}~.
\end{equation} 
This transition leaves the multiplicity of possible attractors unchanged,
but the resulting changes in the spectrum have  deep consequences for
the convergence behavior of sufficiently steep initial conditions.

We now need to distinguish but two Cases for the initial condition,
namely $\lambda\ge\lambda^*$ and $\lambda<\lambda^*$, where
we use the short hand notation 
$\lambda^*=\lambda_0(v^*)=\lambda_\pm(v^*)=v^*/2$.

For flat initial conditions $\lambda_{init}<\lambda^*$, the arguments
from Case III above apply literally. Such an initial condition
will approach a front $\Phi_v$ with velocity $v(\lambda_{init})>v^*$
given by (\ref{20044}) and with steepness $\lambda_{asympt}=\lambda_{init}$.
Sufficiently steep initial conditions, however, exhibit a new behavior:

CASE IV:$~$ Consider a sufficiently steep initial condition with
$\lambda>\lambda^*$. As before we linearize the profile $\phi(x,t)$
after a sufficient evolution time about the selected front $\Phi^*$.
The corresponding  perturbation $\eta=\phi-\Phi^*$ then decays like
$\Phi^*$ (\ref{2019}), because the steepness of $\phi(x,t)$ remains larger
than that of $\Phi^*$ at any finite time $t$, cf.\ Eq.\ (\ref{lameta}). 
As a result, $\eta$ is just outside the Hilbert space in the generic
case of $\alpha\ne0$ (\ref{2018}), just like the zero mode (\ref{2032}). 
The Hilbert space
has a continuous spectrum for all decay rates $\sigma>0$, and there are
no growing perturbations with $\sigma<0$. The perturbation $\eta$ can be written as 
a multiple of the zero mode $\eta_0$ plus a remainder inside the Hilbert space.
From this we might be tempted to argue that the perturbation will decay,
and that we only can not tell how quickly --- probably non-exponential,
because the spectrum is gapless.
However, in contrast to Cases I -- III, there is no way to get rid
of the zero mode, because no matter at which position $\xi_0$ one
places  the subtracted
$\Phi^*(\xi-\xi_0)$, $\Phi^*$ will always  dominate the large $\xi$ behavior, 
and therefore the coefficient of the zero mode in the decomposition
of the perturbation will always be non-vanishing. A convergence argument 
based on simply neglecting the contribution from the zero mode is bound
to be wrong: In the very same way we could argue, that a steep
initial condition converges to $\Phi_v$ with just any $v\ge v^*$. 
Strictly speaking, the linear stability analysis does
not even allow us to conclude, that sufficiently steep initial conditions
approach $\Phi^*$ at all. We only can reason that there is no steeper
attractor than $\Phi^*$, and that one therefore expects that the
pulled front solution $\Phi^*$ is selected from  steep initial
conditions. The different analytical tools that are  developed in
Section \ref{S25} to analyze the convergence behavior, confirm this.

\section{General integration of $g_{n/2}^{sp}(z)$} \label{g0app}

We here show how to find special solutions $g_{n/2}^{sp}(z)$
of inhomogeneous equations like (\ref{3045}) or (\ref{3046}) 
in general. The general form of such an equation is
\begin{equation}
\label{d11}
\hat{T}_n[z,d_z]\;g(z)=i_n(z)~,
\end{equation}
with $i_n(z)$ the inhomogeneity and $\hat{T}_n[z,d_z]$ the operator
\begin{equation}
\label{d12}
\hat{T}_n[z,d_z]=
z\;\frac{d^2}{dz^2}+\left(\frac{1}{2}-z\right)\frac{d}{dz}+\frac{n}{2}~.
\end{equation}
We search for a particular solution $g(z)$ of Eq.\ (\ref{d11}).
A particular solution of the homogeneous equation ($i_n(z)=0$)
can be expressed by Hermite polynomials:
\begin{eqnarray}
\label{d13}
\hat{T}_n[z,d_z]\;h_n(z)&=&0\\
h_0(z)=1~~,~~h_1(z)&=&\sqrt{2z}~~,~~h_2(z)=1-2z~~\mbox{etc. }\nonumber
\end{eqnarray}
The ansatz $g(z)=h_n(z)\;u_n(z)$ reduces (\ref{d11}) to an
equation for $d_zu_n(z)$ of first order:
\begin{eqnarray}
\label{d14}
\hat{T}_n\;g&=&z\;h_n(z)\;\left(\frac{d}{dz}
+\frac{d\;\ln h_n(z)}{dz}+\frac{1-2z}{2z}\right)\frac{du_n(z)}{dz}
\nonumber\\
&=&z\;h_n(z)\;\frac{d_z\left(M_n(z)\;d_zu_n(z)\right)}{M_n(z)}~,
\end{eqnarray}
where in the last line we introduced the integrating factor
\begin{equation}
\label{d15}
M_n(z)=h_n^2(z)\;\sqrt{z}\;e^{-z}~.
\end{equation}
Identify now $\hat{T}_n\;g_n=i_n$, integrate twice, and substitute 
$M_n$ by the full expression. A special solution of (\ref{d11})
then reads
\begin{equation}
\label{d16}
g(z)=h_n(z)\int_a^z dx\;
\frac{\int_b^x dy\;i_n(y)\;h_n(y)\;e^{-y}/\sqrt{y}}
{h_n^2(x)\;\sqrt{x}\;e^{-x}}~,
\end{equation}
where the integration constants $a$ and $b$ are free.
If we in particular choose $b=\infty$, the integrated exponential
$e^{x-y}$ cannot exceed unity, and $g_n^{sp}(z)$ can at most diverge
algebraically, if the integrated inhomogeneity $i_n(z)$ is algebraic.

Integrating Eq.\ (\ref{3056}) for $g_0(z)$ as in (\ref{d16}) with $b=\infty$,
we find for the algebraic divergence of $g_0(z)$ for large $z$:
\begin{equation}
\label{d17}
g_0(z)\sim 3\alpha\;z\;\ln z~~\mbox{as }z\to\infty~,
\end{equation}
while the solution of the homogeneous equation diverges 
only as $h_2(z)\sim z$.
For determining the small $z$ expansion of (\ref{d16}),
it must be noted that the factor $h_n(x)^{-2}$ is singular
at the zeroes of $h_n(x)$. Hence, (\ref{d16}) needs 
to be evaluated separately in each interval between the zeroes of 
$h_n(x)$. This can be done by a proper choice of $a$.
It can be shown, that the results in each interval join smoothly.

\section{Algebraic convergence at the pushed/pulled transition}
\label{pu/pu} \label{A2}

In Section \ref{S3} we have analyzed equations, that are within
the pulled regime. We here analyze equations, that are at the 
pushed/pulled transition. Leading edges of fronts within the pulled regime have 
the form $\Phi^*=(\alpha\xi+\beta)\;e^{-\xi}\propto\xi\; e^{-\xi}$ 
($\xi\gg1$), cf.\ (\ref{2018}). Leading edges of fronts within the pushed 
regime are given by $\Phi^\dag\propto e^{-\lambda_+(v^\dag)\;\xi}$, cf.
(\ref{2021}). Leading edges of fronts at the pushed/pulled transition 
accordingly behave as
\begin{equation} \label{b01} 
\Phi^*=\beta\; e^{-\xi} \mbox{ for } \xi\gg 1~~,~~\lambda_+(v^*)=\lambda^*=1~.
\end{equation} 
For our example nonlinearity (\ref{103}), fronts are within the
pulled regime for $\epsilon>(n+1)/n^2$ and at the pushed/pulled transition  
for $\epsilon=(n+1)/n^2$. The analysis below can again be extended to
more general equations along the lines of Section \ref{S5}. We will
come back to this at the end of this Appendix.

At the pushed/pulled transition, the spectrum of linear perturbations
is still gapless, and convergence therefore is algebraic.
On the other hand, the form of the leading edge played a crucial role 
in determining the velocity corrections $\dot{X}$. Compare our
qualitative discussion in Section \ref{S311}. The leading edge behavior 
(\ref{b01}) immediately lets us expect, that now $v(t)=2-1/(2t)+\ldots$,
in contrast to (\ref{305}) and (\ref{3067}) for fronts within the
pulled regime, and in agreement with (\ref{2055}) for the spreading
of perturbations under the linearized equation. Intuitively, we can argue,
that the slower convergence of fronts within the pulled regime is due to 
the leading edge having to pull the interior part of the front along. This also
makes the leading edge flatter. The quicker convergence of fronts at the 
pushed/pulled transition and in the linearized equation then 
resembles the fact, that the leading edge and the interior part of the front 
``impose the same speed''.

Let us now do the explicit convergence analysis for fronts developing
from initial conditions steeper than $e^{-x}$ for $x\gg1$ and aproaching
(\ref{b01}) for large times. The analysis of the interior is identical
with Section \ref{S32}, where constants $c_{\frac{n}{2}}$ are yet 
undetermined. When expanding the interior shape towards the leading edge as in
Section \ref{S33}, the inhomogeneities created by $\Phi^*$ (\ref{b01})
are different, because now $\alpha=0$. The differential equations for the $\psi_{\frac{n}{2}}$ result from (\ref{3035}) with $\alpha=0$, $\gamma=\beta$
and start with 
\begin{eqnarray} \label{b02}
\partial_\xi^2 \psi_1     &=& c_1 \beta~, \qquad \qquad
\partial_\xi^2 \psi_{\frac{3}{2}} = c_{\frac{3}{2}}\beta ~,
\\
\partial_\xi^2 \psi_2     &=& [-1+c_1(1-\partial_\xi)] \psi_1 
+ c_2\beta +o(e^{-\xi_X})~ \mbox{ etc.}
\nonumber
\end{eqnarray}
Integrating and resumming, we now find for $\xi\gg1$
\begin{eqnarray} \label{b03}
\psi\qquad\quad=\qquad\quad\beta&+&
\\
+\frac{c_1\beta\;\xi_X^2}{2!\;t}&+&\frac{c_1\delta\;\xi_X}{t}
+O\left(\frac{1}{t}\right)
\nonumber\\
&+&\frac{c_{\frac{3}{2}}\beta\;\xi_X^2}{2!\;t^{3/2}}
+O\left(\frac{\xi_X}{t^{3/2}}\right)
\nonumber\\
+\frac{c_1(c_1-1)\beta\;\xi_X^4}{4!\;t^2}&+&
\frac{c_1(c_1\delta-\delta-c_1\beta)\;\xi_X^3}{3!\;t^2}
+O\left(\frac{\xi_X^2}{t^2}\right)
\nonumber\\
&+& \frac{c_{\frac{3}{2}}(2c_1-\frac{3}{2})\beta\;\xi_X^4}{4!\;t^{5/2}} 
+ O\left(\frac{\xi_X^3}{t^{5/2}}\right)
\nonumber\\
+ \ldots &+& \ldots~.
\nonumber
\end{eqnarray}
Here $\delta$ is an unknown integration constant fixed by condition
(\ref{309}). We will see below, that it is not involved in fixing
the velocity, just as also the subleading $\beta$ for the leading edge (\ref{3033})
within the pulled regime is not involved in fixing the velocity,
cf.\ calculation till (\ref{3066}).

Again, for $\xi_X\gg \sqrt{t}$ we have to reorder the expansion in powers
of $\sqrt{z}=\sqrt{\xi_X^2/(4t)}$ and $1/\sqrt{t}$, and find
\begin{eqnarray} \label{b04}
\psi &=& \beta\;\left(1+\frac{c_1(4z)}{2!}+\frac{c_1(c_1-1)(4z)^2}{4!}
 + O(z^3)\right)
\nonumber\\
&&+\;\frac{1}{\sqrt{t}}\;\left(c_1\delta\;(4z)^{1/2}+
\frac{c_{\frac{3}{2}}\beta\;(4z)}{2!}+\right.
\nonumber\\
&&~~+\left.\frac{c_1(c_1\delta-\delta-c_1\beta)\;(4z)^{3/2}}{3!}+
\frac{c_{\frac{3}{2}}(2c_1-\frac{3}{2})\beta\;(4z)^2}{4!}\right)
\nonumber\\
&&+\;O\left(\frac{1}{t}\right)~.
\end{eqnarray}
The structure of the expansion is the same as in (\ref{3043}),
except that now the leading order term is of order $t^0$:
\begin{equation} \label{b05} 
G(z,t) = e^z \psi = g_0(z)
+\frac{g_{\frac{1}{2}}(z)}{\sqrt{t}} + \ldots ~.
\end{equation} 
The equations of motion for the leading and subleading term are 
derived from (\ref{3044}) -- (\ref{3046}) through putting
$g_{\frac{-1}{2}}=0$. For $g_0$ we find now the homogenous equation
\begin{equation} \label{b06}
\left[z\partial_z^2+\left(\frac{1}{2}-z\right)\partial_z 
-\frac{1}{2}-c_1\right]\;g_0 = 0~.
\end{equation}
Just like (\ref{3044}) was solved by (\ref{3053}), we now 
solve (\ref{b06}) with
\begin{equation} \label{b07} 
c_1=\frac{-1}{2} ~~,~~ g_0(z)=\beta~.
\end{equation} 
The equation for $g_{\frac{1}{2}}$ is now, cf.\ (\ref{3046}) and (\ref{b07}),
\begin{equation} \label{b08} 
\left[z\partial_z^2+\left(\frac{1}{2}-z\right)\partial_z 
+\frac{1}{2}\right]\;g_{\frac{1}{2}} 
= \beta\left[c_{\frac{3}{2}} -\frac{\sqrt{z}}{2}\right]~.
\end{equation} 
Again a special solution of the inhomogeneous equation can be found,
and the general solution contains the constants of integration 
$k_{\frac{1}{2}}$ and $l_{\frac{1}{2}}$:
\begin{eqnarray} \label{b09}
g_{\frac{1}{2}}&=&\beta\left[2c_{\frac{3}{2}}-\frac{\sqrt{z}}{2}
\sum_{n=1}^\infty\frac{(1)_{n-1}\;z^n}{\left(\frac{3}{2}\right)_n n!}\right]
\nonumber\\
&&+\;k_{\frac{1}{2}}\;M\left(\frac{-1}{2},\frac{1}{2},z\right)
+l_{\frac{1}{2}}\;\sqrt{z}
\\ \label{b010}
&\stackrel{z\ll1}{=}& 2\beta c_{\frac{3}{2}}+k_{\frac{1}{2}}
+l_{\frac{1}{2}}\sqrt{z}+O(z)
\\ \label{b011}
&\stackrel{z\to\infty}{\sim}& 
\frac{-\beta\sqrt{\pi}}{4z}\;e^z - \frac{k_{\frac{1}{2}}}{2z}\;e^z~.
\end{eqnarray} 
Comparing (\ref{b010}) to the order $1/\sqrt{t}$ in (\ref{b04}) and 
imposing proper convergence of (\ref{b011}) for $z\to\infty$, we find
\begin{equation} \label{b012} 
2\beta c_{\frac{3}{2}} + k_{\frac{1}{2}} = 0 ~~,~~
l_{\frac{1}{2}}=-\delta ~~,~~ 
\beta\sqrt{\pi} + 2 k_{\frac{1}{2}}=0~.
\end{equation}
With these constants, the velocity correction $c_{\frac{3}{2}}$ is
\begin{equation} \label{b013}
c_{\frac{3}{2}}=\frac{\sqrt{\pi}}{4}~,
\end{equation}
and for $g_{\frac{1}{2}}$ we find
\begin{eqnarray} \label{b014}
g_{\frac{1}{2}}&=&\frac{\beta\sqrt{\pi}}{2}\;
\left[1-M\left(\frac{-1}{2},\frac{1}{2},z\right)
-\sqrt{\frac{z}{\pi}}
\sum_{n=1}^\infty\frac{(1)_{n-1}\;z^n}{\left(\frac{3}{2}\right)_n n!}\right]
\nonumber\\
&&-\;\delta\sqrt{z}~.
\end{eqnarray}

In summary, we find for the convergence to a front at the pushed/pulled
transition, whose leading edge accordingly takes the form (\ref{b01}), that
the velocity correction is given by
\begin{equation} \label{b015} 
\dot{X}=-\:\frac{1}{2t}\left(1-\frac{1}{2}\;\sqrt{\frac{\pi}{t}}\:\right)
+O\left(\frac{1}{t^2}\right)~.
\end{equation} 
In the interior, i.e., for $\xi_X\ll\sqrt{t}$, the front is given 
by (\ref{3031}) just like a front within the pulled regime.
In the leading edge, where $\xi_X\gg\sqrt{t}$, the front is given by
\begin{equation} \label{b016} \phi(\xi_X,t)=e^{-\xi_X-\xi_X^2/(4t)}\;G\left(\frac{\xi_X^2}{4t},t\right)~,
\end{equation} 
where
\begin{equation} \label{b017} 
G(z,t)=\beta+\frac{g_{\frac{1}{2}}(z)}{\sqrt{t}}+O\left(\frac{1}{t}\right)~.
\end{equation} 

The extension along the lines of Section \ref{S5} to more general
equations is straightforward. The general expression for $\dot{X}(t)$ is
\begin{equation} \label{b015b} 
\dot{X}=-\:\frac{1}{2\lambda^* t}\left(1-\frac{1}{2\lambda^* }\;\sqrt{\frac{\pi}{Dt}}\:\right)
+O\left(\frac{1}{t^2}\right)~,
\end{equation} 
but the subleading function $g_{\frac{1}{2}} (z) $ will depend on
the additional terms in the expansion, just like the subleading
$g_0(z)$ in Section \ref{S53}.

\section{Multiplicity of fronts and linear 
eigenmodes for reflection symmetric equations of first order in time}
\label{A6}

The generical multiplicity of uniformly translating fronts $\Phi_v$ 
can be determined by counting arguments analogous to those
performed in Section \ref{S22}. Uniformly translating solutions
$\Phi_v(\xi)$ of (\ref{5a2}) can be 
understood as a heteroclinic orbit in $N$-dimensional phase space
between fixed points characterized by $\Phi_v=1$ at $\xi\to-\infty$
and $\Phi_v=0$ at $\xi\to\infty$. For a linear perturbation
$\delta=1-\Phi_v$ about the fixed point $\phi=1$ from (\ref{5a5}), 
we get the equation
\begin{equation}
\label{5a11}
{\cal L}_v(-\infty)\;\delta +O(\delta^2)=0~,
\end{equation}
which is a linear ordinary differential equation with constant 
coefficients with the linear operator ${\cal L}$ being defined in (\ref{5a6}). 
The same is true for a linear perturbation
$\Phi_v=0+\delta$ of the fixed point $\phi=0$, which solves
\begin{equation}
\label{5a12}
{\cal L}_v(\infty)\;\delta +O(\delta^2)=0~.
\end{equation}
In linear order of $\delta$, each of these equations has $N$
solutions $e^{-\lambda_n(v)\;\xi}$, $n=1,\ldots,N$.

Let us restrict the analysis to real equations which are isotropic
in space, i.e., where (\ref{506}) is invariant under $x\to-x$.
Such equations are even in $\partial_x$, so $N$ needs to be even. 
According to arguments presented in Appendix A of \cite{vs2}, Eqs.\
(\ref{5a11}) and (\ref{5a12}) for $v>0$ will have $N/2+1$ eigenvalues 
$\lambda_n$ with positive real part and $N/2-1$ ones with a negative real 
part, if the state, about which we linearize, is linearly unstable against 
a range of Fourier modes. If it is stable, we will have $N/2$ 
eigenvalues with positive real part and $N/2$ ones with a negative real part. 
We assume $\phi=1$ to be stable, so at $\xi\to-\infty$ there are
$N/2$ directions in phase space with negative real part of $\lambda$,
that need to be excluded. If $\phi=0$ is unstable, we have only 
$N/2-1$ bad eigendirections at $\xi\to\infty$. We then generically 
have a front connecting these fixed points for arbitrary values of $v$.
If, however, the state $\phi=0$ is metastable, there are $N/2$ bad 
eigendirections at $\xi\to\infty$. Then also $v$ needs to be tuned
to find a solution. So for fronts propagating into unstable
states, we generically have a front solution $\Phi_v$ for a continuum
of velocities, while for fronts into metastable states,
there are solutions $\Phi_v$ only for discrete values of $v$,
in generalization of the arguments from Section \ref{S22}.

The multiplicity of linear perturbations is determined along the 
same lines. We again decompose the linear perturbations $\eta$
(\ref{5036}) into $\eta(\xi,t)=\eta_\sigma(\xi)\;e^{-\sigma t}$
by separation of variables. The $\eta_\sigma$ then solve the 
{\em o.d.e.}\
\begin{equation}
\label{5a13}
\left[ {\cal L}_v(\xi)+\sigma \right] \;\eta_\sigma(\xi)=0~.
\end{equation}
For counting the generic multiplicity of solutions, we need to
linearize the equations about $\xi\to\pm\infty$, which amounts to 
a problem equivalent to (\ref{5a11}) and (\ref{5a12}), except for a shift
of the constant contribution of ${\cal L}_v(\xi)$ by $\sigma$.
For fronts propagating into unstable states, we in general expect
a continuous spectrum $\sigma$ of linear perturbations at least in
some finite interval of $\sigma$, in generalization of Section
\ref{S23}.

\section{Strongly heteroclinic orbits and change of stability at $v^\dag$}
\label{A7}

According to the counting argument from Appendix \ref{A6},
the front $\Phi^*(\xi)$ propagating uniformly with velocity $v^*$
does exist. The question is now, whether it is stable and 
whether it will be approached by steep initial conditions.
In particular, we want to analyze initial conditions $\phi(x,0)$, 
that are steeper than $e^{-\lambda^*x}$ in the leading edge.

This amounts to the question, whether in the spectral decomposition
$\eta_\sigma$ (\ref{5a13}) of a generic $\phi(x,0)-\Phi^*(x)$, 
there are destablizing modes with $\sigma<0$. As in Section
\ref{S23}, the contributing modes in general will all decay at least
as quick as $\Phi^*$ in the leading edge.
The leading edge properties of the $\eta_\sigma$ in general
will depend smoothly on $\sigma$, just as in (\ref{hoiute}),
so generically $\Phi^*$ will still be stable against all
perturbations, that in the leading edge decay quicker than $\Phi^*$.

An exemption is again the generalization of $A_\sigma=0$ from 
(\ref{etasigma}). For an equation of order $N$ with $\phi=0$ unstable, 
there are $N/2+1$ exponents $\Lambda_n(\sigma,v)>0$. The leading edge 
will be a superposition of all the exponentials 
\begin{equation}
\label{5a31}
\eta_\sigma(\xi) = \sum_{n=1}^{N/2+1} A_\sigma^{(n)}\;e^{-\Lambda_n \xi}
~~\mbox{ as }~\xi\gg1~.
\end{equation}
The condition $A_\sigma^{(1)}=0$, where $\Lambda_1$ is the smallest one
of the positive $\Lambda_n$, fixes a discrete set (which can be either 
empty or not empty) of negative $\sigma$'s whose eigenfunctions
$\eta_\sigma$ have a steepness in the leading edge
larger than $\Phi^*$. The stability of the pulled front $\Phi^*$ thus
again depends on the ``strongly heteroclinic'' perturbations.

If there are strongly heteroclinic perturbations, that destabilize
the pulled front propagating with velocity $v^*$, then there
will be a steeper and quicker front $\Phi^\dag$, which can be constructed
as a strongly heteroclinic orbit of (\ref{5a1}). The zero mode
$\partial_\xi\Phi^\dag$ then again is a strongly heteroclinic
perturbation, and as discussed in  Appendix \ref{stabilityapp} on the
stability of front solutions, we can conclude
that the quickest of all strongly heteroclinic orbits cannot be
destabilized, so it will attract all sufficiently steep initial conditions.

{\em We conclude, that Table IV generalizes to higher order equations,
which form uniformly translating fronts,
if we only appropriately adjust the explicit definitions
of the velocities $v$ and steepnesses $\lambda$.}

\section{Relation between the generalized diffusion
  constants $D_n$ and the dispersion relation} \label{A8}
  
If we use the expansion (\ref{5a15})  for the dispersion relation
$\omega(k)$, we get
\begin{eqnarray}
{\cal D} & = & e^{\:\lambda^*\xi}\;\left( \sum_{m=0}^N a_m\partial_\xi^m
-v^*\partial_\xi \right)  \;e^{-\lambda^*\xi}~,\nonumber \\
 & = & \sum_{m=0}^Na_m(\partial_\xi -\lambda^*)^m -v^*(\partial_\xi
 -\lambda^*) ~,\nonumber \\
 & = & \sum_{m=0}^N \sum_{n=0}^m a_m \frac{m!}{n!(m-n)!}
 (-\lambda^*)^{(m-n)} \partial_\xi^n - \nonumber \\ & & \hspace{2.5cm}
 - v^*(\partial_\xi -\lambda^*)
 ~,\nonumber \\
 & = & \sum_{n=0}^N \left( \frac{\partial^n}{\partial (-\lambda^*)^n} 
 \sum_{m=n}^N a_m (-\lambda^*)^m \right) \frac{1}{n!} \partial_\xi^n
 \nonumber \\ & & \hspace*{2.5cm} -v^*(\partial_\xi -\lambda^*) ~.
\end{eqnarray}
This immediately yields the expansion (\ref{5a29}) with the identification 
(\ref{5a30}).

\section{Edge analysis of uniformly translating pulled fronts with $M=1$}
\label{hot}

We analyze the leading edge representation (\ref{5042}) 
for a uniformly translating front whose equation of motion (\ref{5a1})
is of arbitrary order $N$ in space and of first order in time $M=1$:
\begin{equation} \label{e01}
\partial_\tau\psi=
\left(\partial_\zeta^2+\sum_{n=3}^Nd_n\partial_\zeta^n\right)\psi
  +\dot{Y}\left(\partial_\zeta-1\right)\psi~.
\end{equation}
We generalize the leading edge analysis from Sections \ref{S33} and \ref{S34}.

With the notions and ansatz
\begin{eqnarray}
\label{e02}
{\cal D}&=&\partial_\zeta^2+\sum_{n=3}^Nd_n\partial_\zeta^n
~~,~~\dot{Y}=\sum_{n=2}^\infty \frac{C_{\frac{n}{2}}}{\tau^{n/2}}~,\\
\label{e03}
\psi(\zeta_Y,\tau)&=&\alpha\;\zeta_Y+\beta+\frac{\psi_{\frac{1}{2}}}{\tau^{1/2}}
+\frac{\psi_1}{\tau}+\frac{\psi_{\frac{3}{2}}}{\tau^{3/2}}+\ldots~,
\end{eqnarray}
the expansion of the interior in the region of $\zeta_Y\gg1$
at the crossover towards the leading edge reads 
\begin{eqnarray} \label{e04}
{\cal D}\psi_{\frac{1}{2}}
&=& 0~,
\nonumber\\
{\cal D}\psi_1   
&=& C_1 (\alpha\;\zeta_Y+\gamma) 
~~~,~~~\gamma=\beta-\alpha~,
\nonumber\\
{\cal D}\psi_{\frac{3}{2}}  
&=& C_{\frac{3}{2}} (\alpha\zeta_Y+\gamma) ~,~~\ldots~~,
\end{eqnarray} 
in generalization of (\ref{3035}). These equations can be integrated 
explicitely. The result can be written in leading edge variables 
$z=\zeta_Y^2/(4\tau)$ as
\begin{eqnarray} \label{e05}
\psi&=&
\sqrt{\tau}\;
\alpha\left((4z)^{1/2}+\frac{C_1(4z)^{3/2}}{3!}
+\ldots\right)
\nonumber\\
&&+\;\tau^0\;\left(\beta+\frac{C_1(\beta-\alpha(1+d_3))(4z)}{2!}+\ldots\right)
\nonumber\\
&&+\;O(1/\sqrt{\tau})~.
\end{eqnarray}
This generalizes the results of Section \ref{S33} and supplies us with
the small $z$ expansion of the leading edge function
\begin{equation} \label{e06}
\psi(\zeta,\tau)=e^{-z}\;G(z,\tau)~,~~~~~z=\frac{\zeta^2}{4\tau}~.
\end{equation}
$G$ solves [compare Eq.\ (\ref{3042})]
\begin{eqnarray} \label{e07}
\lefteqn{ \left[ z\partial_z^2 + \left( \frac{1}{2}-z \right)\partial_z
                 -\frac{1}{2}-\tau \partial_\tau -C_1 \right]\;G\;=}
\nonumber\\
&=&\frac{1}{\sqrt{\tau}} \left[C_{\frac{3}{2}}+C_1\sqrt{z}(1-\partial_z)
\right]\;G
\nonumber\\
&&~-\;\frac{d_3\sqrt{z}}{\sqrt{\tau }}
\left[\frac{3}{2}\left(\partial_z-1\right)^2
+z\left(\partial_z-1\right)^3\right]\;G
\nonumber\\
&&+\;O\left(\frac{1}{\tau }\right)~,
\end{eqnarray}
where we wrote all operators of order $\tau^0$ on the l.h.s.\ of the
equation and the operators of order $\tau^{-1/2}$ on the r.h.s.

With the ansatz 
\begin{equation} \label{e08}
G(z,\tau)=\sqrt{\tau}\;g_{\frac{-1}{2}}(z)+g_0(z)
+\frac{g_{\frac{1}{2}}(z)}{\sqrt{\tau}}+\ldots
\end{equation}
as in (\ref{3043}), we find that $g_{\frac{-1}{2}}(z)$
solves again (\ref{3044}), so we copy from Section \ref{S34}, that
\begin{equation} \label{e09}
C_1=\frac{-3}{2}~~,~~g_{\frac{-1}{2}}(z)=2\alpha\sqrt{z}~.
\end{equation}
For $g_0(z)$ we then find instead of (\ref{3056}):
\begin{eqnarray} \label{e010}
\lefteqn{
\left[z\partial_z^2+\left(\frac{1}{2}-z\right)\partial_z 
+1\right]\;g_0 =}
\\
&=& 2\alpha\;\left[\frac{3\;(1+d_3)}{4}+c_{\frac{3}{2}} \sqrt{z}
-\frac{3}{2}\;z+d_3(z^2-3z)\right]~.
\nonumber
\end{eqnarray}
A special solution of the inhomogeneous equation is now instead of 
(\ref{3057}):
\begin{equation} \label{e011}
g_0^{sp}(z) = 2\alpha\;\left(\frac{3\;(1+d_3)}{4}+2c_{\frac{3}{2}} \sqrt{z}
-\frac{3}{4}\;F_2(z)-d_3z^2\right)~,
\end{equation}
with $F_2(z)$ from (\ref{3058}).
The general solution is
\begin{eqnarray} \label{e012} 
g_0(z)&=&g_0^{sp}(z) + k_0 \;(1-2z) + 
l_0\;\sqrt{z}\;M\left(\frac{-1}{2},\frac{3}{2},z\right)
\nonumber\\
&\stackrel{z\ll1}{=}& \left(\frac{3\alpha}{2}\;(1+d_3)+k_0\right)
+\left(4\alpha c_{\frac{3}{2}}+l_0\right) \sqrt{z}+O(z)
\nonumber\\ 
&\stackrel{z\to\infty}{\sim}&
-\left(\frac{3}{2}\;\alpha \;\sqrt{\pi}+ \frac{l_0}{4}\right)\; 
z^{-3/2}\;e^z ~,
\end{eqnarray} 
Note, that $d_3\ne0$ does not cause any divergences at $z\to\infty$.
It only shifts the constant contribution at $z\to0$.

Suppressing the divergence at $z\to\infty$ in (\ref{e012}),
and comparing its small $z$ expansion to (\ref{e05}) yields again
\begin{equation} \label{e013}
C_{\frac{3}{2}}=\frac{3\sqrt{\pi}}{2}~,
\end{equation} and
\begin{eqnarray} \label{e014} 
g_0(z)&=&\beta\;(1-2z) +3\alpha(1+d_3)z-2\alpha d_3z^2
\nonumber\\
&&~~-\;\frac{3\alpha}{2}\;F_2(z)+6\alpha\;\sqrt{\pi\:z}\;
\left(1-M\left(\frac{-1}{2},\frac{3}{2},z\right)\right)~.
\nonumber
\end{eqnarray}

\section{Leading edge projections for coupled equations: an example}
\label{wim} \label{Adoublekpp}

As a simple illustration of the various questions related to the
projection discussed in Section \ref{S553}, we consider two coupled
F-KPP equations,
\begin{eqnarray}
\partial_t \phi_1 & = & \partial^2_x \phi_1 + \phi_1-
\phi^3_1~,\label{A91}\\
\partial_t \phi_2 & = & D \partial^2_x \phi_2 + \phi_2-
\phi^3_2 + K \phi_1~.\nonumber
\end{eqnarray}
The dynamics of this set of equations for fronts propagating into the
state $\phi_1=\phi_2=0$ with steep initial conditions, is of course
immediately obvious: when $K=0$, the two equations are uncoupled, and
fronts in the first equation propagate with speed $v_1^*=2$, while
those in the second equation propagate with speed $v^*_2=2
\sqrt{D}$. The dynamics of $\phi_1$ is always independent of that of
$\phi_2$, even for $K\neq 0$, so for $K>0$ and $D<1$, the dynamics of
the coupled equations amounts to a normal F-KPP $\phi_1$ front, with
relaxation given by our usual expressions. This front  entrains a front with 
speed $v=v^*_1=2$ in $\phi_2$. For $D>1$, the $\phi_1$ and $\phi_2$
fronts keep on propagating with different speeds. We consider the case 
$D<1$ and make a leading edge transformation
$\phi_1=e^{-\xi} \psi_1$, $\phi_2=e^{-\xi} \psi_2$ (where $\lambda^*=1$)
to the frame moving with velocity $\xi=x-v^*_1t$. The 
linearized equations then become
\begin{eqnarray}
\partial_t \psi_1 & = & \partial^2_\xi \psi_1 ~, \label{A92}\\
\partial_t \psi_2 & = & D \partial^2_\xi \psi_2 +2(D-1)\partial_\xi
\psi_2 + (D-1) \psi_2 + K \psi_1~.\nonumber
\end{eqnarray}
The matrix $\underline{\underline{S}}^*(q,\Omega) $ of the linearized
equations is in this case
\begin{equation}
 \underline{\underline{S}}^*(q,\Omega)=
 \left(\begin{array}{cc}
 i\Omega -q^2  &  0 \\
K &  i\Omega - q^2 + J(q)
\end{array}\right)~,
\end{equation}
where $J(q)=(D-1)(1+2iq-q^2)$.
Since the element $S_{12}^*(q,\Omega)=0$, the eigenvalues $u_1^* $ and 
$u_2^*$ are simply the diagonal element of 
 $\underline{\underline{S}}^*(q,\Omega) $, $u^*_1(q,\Omega)=i\Omega
 -q^2$ and $u^*_2(q,\Omega) =  i\Omega - q^2 +J(q)$. 
However, the eigenvectors are {\em not} both along the $\psi_1$ 
 and $\psi_2$ axis. Indeed, we have in the notation of \ref{S55} 
\begin{eqnarray}
\underline{U}_1^*(q) = \left(\begin{array}{c} 1 \\ -K /J(q)
  \end{array} \right)  &  \underline{U}_1^{*\dagger} = \left(
  1, 0 \right) ~,\\
\underline{U}_2^* = \left(\begin{array}{c} 0 \\ 1
  \end{array} \right)~~~~~~~~~~~  & ~~~~ \underline{U}_2^{*\dagger}(q) = \left(
  K/J(q) , 1 \right) ~,
\end{eqnarray}
 The appropriate saddle point is
$\Omega=q=0$, and since $J(0)=(D-1)$, we have 
\begin{equation}
\underline{U}_1^*(0) = \left(\begin{array}{c} 1 \\ -K/(D-1)
  \end{array} \right)~.
\end{equation}
The fact that the second component is nonzero just expresses the fact
that the variable $\psi_2$ is entrained by the leading edge in
$\psi_1$. We can now illustrate our assertion that different choices of
projection lead to different dynamical equations for the projected
leading edge variable $\psi^p$, but that the universal results from
Table II are independent of the particular choice of projection. 
Clearly, one obvious intuitively
appealing choice is to take $\psi^p=\psi_1$, since the $\psi_1$
dynamics is independent of that of $\psi_2$. In this case, the
dynamical equation for $\psi^p$ is nothing but the single F-KPP
equation, and all the results for this equation carry over in
detail. Likewise, the  choice $\psi^p=\pi_1(q,\Omega)$ 
(\ref{funnychoice}) leads to the linearized F-KPP equation 
for $\psi^p$ since $u^*_1(q,\Omega(q))=0$ gives the
dispersion relation of the F-KPP equation. However, this choice is
more formal than practical, since the direction in the vector
space $(\psi_1,\psi_2)$ is 
not fixed, but depends on the variable $q$ which influences the
dynamics. A more practical choice for the coupled variables would be
to take $\psi^p$ as the component along  $\underline{U}^*_1(0)$, as
this corresponds to a fixed ratio of $\psi_1 $ and $\psi_2$. Since 
$\underline{U}_2^{*\dagger} \cdot \underline{U}_1^{*}=$ $
K(J(q)-J(0)=$ $ -2Kiq/(D-1)+O(q^2)$, the projected equation for in this
case picks up a third order derivative term $D_3 \partial_\xi^3 \psi^p$,
amoungh other ones.

 {\em Thus, we observe in this particular example,
that indeed the universal results from Table II on velocity and
shape relaxation are independent of the choice of projection,
while the subleading contribution $g_0(z)$ in the leading edge
is universal in the sense, that it is independent of the precise
initial conditions, but it does depend on the direction of projection.}

\section{Pinch point versus saddle point analysis}\label{utesfavorite}

In this appendix, we briefly discuss the major differences
and similarities between the saddle point and the pinch point
approach for evaluating the integral
\begin{equation}
\label{J01}
{\cal I}_m=\int_{i\gamma-\infty}^{i\gamma+\infty}
\frac{d\omega}{2\pi}
\int_{-\infty}^\infty \frac{dk}{2\pi}\;e^{ik\xi-i(\omega-vk) t}\;
\frac{\underline{\underline{\hat{M}}}_{\:m}(k,\omega)}{u_m(k,\omega)}
\end{equation}
from Eq.\ (\ref{green}) on a given branch $m$. Here $\gamma>0$ needs to
be large enough, that the integrand is analytic along and above the
path of $\omega$ integration in the complex $\omega$ plane.
We introduced the abbreviation
$\underline{\underline{\hat{M}}}_{\:m}(k,\omega)=
\underline{\hat{U}}_{\:m}(k,\omega) \times
\underline{\hat{U}}_{\:m}^\dag(k,\omega)$.
In the moving frame $\xi$, it obviously is convenient to 
transform to the variable $\Omega=\omega-vk$, and to introduce
\begin{equation}
\label{umv}
u_m^v(k,\Omega)=u_m(k,\Omega+vk)=u_m(k,\omega)~~,~~\Omega=\omega-vk~.
\end{equation}
The characteristic equation 
\begin{equation}
u_m(k,\omega_m(k))=0~~\Longleftrightarrow~~u_m^v(k,\Omega_m(k))=0
\end{equation}
defines the dispersion relation $\omega_m(k)$ or $\Omega_m(k)$.
The integrals are now of the form
\begin{equation}
\label{J04}
{\cal I}_m=\int_{i\gamma-\infty}^{i\gamma+\infty}
\frac{d\Omega}{2\pi}
\int_{-\infty}^\infty \frac{dk}{2\pi}\;e^{ik\xi-i\Omega t}\;
\frac{\underline{\underline{\hat{M}}}_{\:m}(k,\Omega+vk)}{u_m^v(k,\Omega)}~.
\end{equation}
The ``saddle point'' type approach, that we follow in Sects.\ 
\ref{S53} -- \ref{S55} of this paper, is based on first evaluating
the $\Omega$ integral by closing the $\Omega$ contour in the lower
half plane for $t>0$ around the simple pole $\propto (\Omega-\Omega_m(k))$. 
The integral then yields
\begin{equation}
\label{J05}
{\cal I}_m=\int_{-\infty}^\infty \frac{dk}{2\pi}
\;e^{ik\xi-i\Omega_m(k)t}\;
\frac{\underline{\underline{\hat{M}}}_{\:m}(k,\Omega_m(k)+vk)}
{i\partial_\Omega u_m^v(k,\Omega_m(k))}~,
\end{equation}
where $\gamma$ needs to be larger than $\max_{k~{\rm real}}~($Im~$\Omega_m(k))$.
From here on, the saddle point analysis proceeds essentially as in
Sect.\ \ref{S54}: the $k$ contour is deformed continuously
such that it passes through a saddle point of $\Omega_m(k)$
that allows for a steepest descent evaluation of the $k$ integral.
A saddle point is a double root in $k$ of $u_m^v(k,\Omega)$, so that
\begin{eqnarray}
\label{pinch1}
\left.u_m^v(k,\Omega)\right|_{sp}=0 
~~&\Longleftrightarrow&~~\omega_{sp}=\omega_m(k_{sp})
\\
~~&\Longleftrightarrow&~~\Omega_{sp}=\omega_m(k_{sp})-vk_{sp}~,
\nonumber
\end{eqnarray}
and
\begin{eqnarray}
\label{pinch2}
\left.\partial_k u_m^v(k,\Omega)\right|_{sp}=0 
~~&\Longleftrightarrow&~~
\left.\left(\partial_k+v\partial_\omega\right) u_m(k,\omega)\right|_{sp}=0 
\nonumber\\
~~&\Longleftrightarrow&~~v=-\;\frac{\left.\partial_k u_m(k,\omega)\right|_{sp}}
{\left.\partial_\omega u_m(k,\omega)\right|_{sp}}~.
\end{eqnarray}
By expanding about such a saddle point, we  get for large $t$ 
to leading order
\begin{equation}
{\cal I}_m=\left.\frac{\underline{\underline{\hat{M}}}_{\:m}(k,\omega)}
{i\partial_\omega u_m(k,\omega)}\right|_{sp}\;
e^{ik_{sp}\xi-i\Omega_{sp} t}\;
\int_q\;e^{iq\xi-D_{sp} q^2 t} + \ldots~,
\end{equation}
with the diffusion constant
\begin{equation}
\label{J09}
D_{sp}=\frac{-i\;\left.\left(\partial_k+v\partial_\omega\right)^2 
u_m\right|_{sp}}
{2\;\left.\partial_\omega u_m\right|_{sp}}
=\frac{-i\;\left.\partial_k^2 u_m^v\right|_{sp}}
{2\;\left.\partial_\omega u_m^v\right|_{sp}}~.
\end{equation}
The remaining integral over real $q=k-k_{sp}$ is a simple Gaussian
integral of the form discussed previously in Sect.\ \ref{S531}. 
As before, we are in the comoving frame, if 
\begin{equation}
\label{pinch3}
\mbox{Im }\Omega_{sp}=0 ~~\Longleftrightarrow~~
v=\frac{\mbox{Im }\omega_m(k_{sp})}{\mbox{Im }k_{sp}}~.
\end{equation}
Differentiating the dispersion relation $u(k,\omega_m(k))=0$
with respect to $k$: $\partial_k u(k,\omega_m(k))=0$, and comparing
to (\ref{pinch2}), we can immediately identify
\begin{equation}
\label{pinch4}
v=\left.\frac{\partial\omega_m(k)}{\partial k}\right|_{sp}~.
\end{equation}
From $\partial_k^2 u(k,\omega_m(k))=0$ and (\ref{J09}), we get
\begin{equation}
\label{pinch5}
D=\left.\frac{i\partial^2\omega_m(k)}{2\partial k^2}\right|_{sp}~.
\end{equation}
Choosing in (\ref{green}) the branch $m$ with the largest velocity 
$v_{sp}=v^*$, Eq.\ (\ref{GVD2}) immediately results.

If the denominator of an integral like (\ref{J04}) 
contains a product of characteristic functions 
$\prod_{m=1}^M u_m^v(k,\Omega)$,
then each factor $u_m^v(k,\Omega)$ will contribute with its
pole and yield an integral as in (\ref{J05}), so that the
total integral amounts to a sum of $M$ integrals of the form 
(\ref{J05}). Again the dominating contribution for $\xi$ fixed
and $t\gg1$ will be the one with the largest velocity $v_{sp}$ through 
which the contour of $k$ integration can be deformed.

The pinch point analysis \cite{bers} is based on evaluating (\ref{J01})
by a different order of the integrations, i.e., by first closing
the $k$ contour to get $k=k(\Omega)$ and then evaluating the
remaining $\Omega$ integral. (For $\xi>0$, the $k$ contour must be closed
in the upper half plane.) As discussed most clearly by Bers \cite{bers},
this is done as follows. $\gamma$ in (\ref{J04}) has to be large enough
to lie above the maxima of the dispersion relation $\Omega_m(k)$
for real $k$. When $\Omega$ varies along the integration path, 
the poles in the $k$ plane move. Now when $\gamma$ is lowered 
sufficiently, that it approaches the maximum of the line $\Omega_m(k)$
traced out by the real $k$ values, a pole in the $k$ plane will
approach the real $k$ axis. When that happens, the $k$ contour
can be continuously deformed to avoid this pole. This in turn
allows one to lower the value of $\gamma$. This process can continue
until two poles in the $k$ plane approach the $k$ contour 
from opposite sides, and ``pinch off'' the $k$ contour at a 
particular value of $\Omega^*$. Clearly, that point corresponds to
a double root, since for that given value of $\Omega$ the two
$k$ roots coincide. When the $k$ contour is closed,
this point generates a branch-cut in the $\Omega$ plane,
since near $\Omega^*$ we have $k-k^*=\pm\sqrt{(\Omega-\Omega^*)/D}$.
When the $\Omega$ contour is subsequently closed in the lower half
$\Omega$ plane, these branch points then generate the usual
leading asymptotic behavior (\ref{5a20}), (\ref{5a22}).

In both approaches, there are conditions for a saddle or pinch
point to be dynamically relevant; these arise from the global
properties of the dispersion relation $\omega(k)$.
In the saddle point approach, only saddle points that will dominate
the $k$ integral along the deformed contour of integration,
are relevant for the dynamics. Pictorially, a saddle point that obeys
this conditions is located between ``valleys'' of Im~$\omega(k)$
in the direction of real $k$ that are not completely separated by
``ridges'' from the real $k$ axis. In this formulation, the condition
Re~$D>0$ naturally comes out. If there is more than one such saddle
point, the one with the highest velocity $v^*$ determines the
asymptotic spreading velocity. In the pinch point formulation,
the condition usually mentioned is that the poles in the $k$ plane
``pinch off'' the $k$ contour, while the condition Re~$D>0$ 
is usually not mentioned, but it is actually hidden in the formulation
as well: it just expresses that the pinch point is associated with
a point of the dispersion relation, where the growth rate is maximal.
In fact, the examples discussed on pages 466, 467 in \cite{bers}
for solutions of the saddle point equations which are no pinch points,
are just cases where Re~$D<0$, i.e., solutions which are
excluded by a saddle point formulation as well. In the pinch point
formulation, the improper solutions of the saddle point equations
correspond to solutions where two poles in the $k$-plane do not
``pinch off'' the deformed $k$-contour, but instead just merge by themselves.

\end{appendix}
\end{multicols}

\begin{figure} \label{fig4}
\vspace{0.5cm}
\epsfig{figure=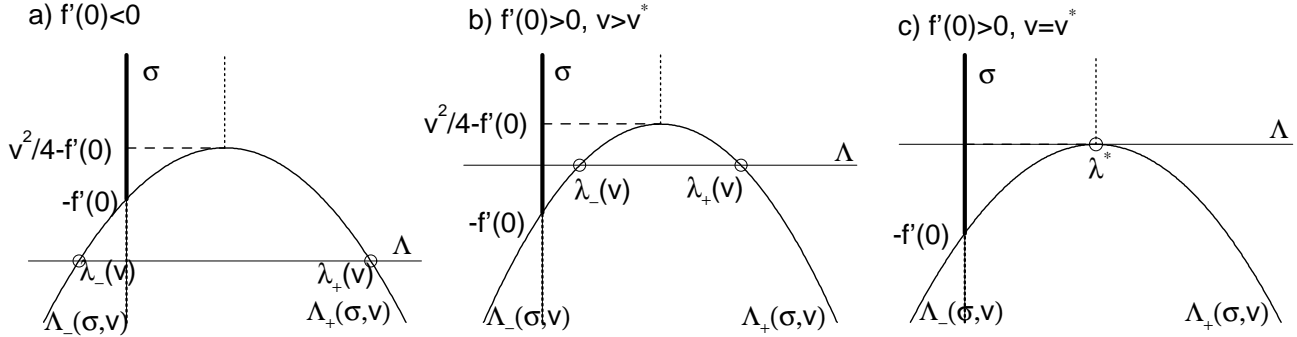,width=0.9\linewidth}
\vspace{0.5cm}
\caption{Steepness $\Lambda(\sigma,v)$ (\ref{sigmaeq}) versus decay 
rate $\sigma$ of linear perturbations $\eta_\sigma$ (\ref{etamode1}),
(\ref{etaeta}) of a given front $\Phi_v$ with velocity $v\ge v^*$.
The solid curve denotes real $\Lambda$, the dotted curve the real part 
of complex $\Lambda$. 
$\lambda_\pm(v)$ and $\lambda^*$ are the steepnesses of $\Phi_v$
and of the zero mode $\eta_0=\partial_\xi\Phi_v$. 
They are marked by circles on the $\Lambda$ axis. The generic
steepness of a front $\Phi_v$ with $v>v^*$ is $\lambda_-(v)$,
while in the particular case of $A_v=0$, it is $\lambda_+(v)$.
The continuous spectrum of $\sigma$ is denoted by a fat solid line
on the $\sigma$-axis, the interval in which there may be discrete 
eigenvalues $\sigma$ by the fat dotted line. 
The continuous spectrum within the Hilbert space
of ${\cal H}_v$ exists only at $v^2/4-f'(0)\le\sigma$.
The continuous spectrum for $-f'(0)<\sigma<v^2/4-f'(0)$
is on the $\Lambda_-$-branch. There might be discrete
solutions characterized by $A_\sigma=0$. They lie on the $\Lambda_+$-branch,
might exist for all $\sigma< v^2/4-f'(0)$, and need to be constructed.
$a)$ The front $\Phi_v$ propagates into a metastable state ($f'(0)<0$).
Its steepness is $\lambda_+(v)$. It is stable against all linear 
perturbations with $\Lambda<\lambda_+(v)$. The discrete spectrum 
of steep perturbations with $\Lambda>\lambda_+(v)$ needs to be investigated.
$b)$ The front propagates into an unstable state ($f'(0)>0$)
with velocity $v>v^*$. It is stable against all linear 
perturbations with $\lambda_-(v)<\Lambda<\lambda_+(v)$, it is unstable
against the continuous spectrum of very flat perturbations with 
$0<\Lambda<\lambda_-(v)$, which might be excluded by the initial 
conditions. The discrete spectrum of steep perturbations with 
$\Lambda>\lambda_+(v)$ needs to be investigated.
$c)$ The front propagates into an unstable state ($f'(0)>0$)
with velocity $v=v^*$. The discussion is as for $(b)$ after identifying
$\lambda_\pm(v^*)=\lambda^*$.}
\end{figure}

\newpage

\begin{multicols}{2}



\end{multicols}


\begin{references}
\bibitem{collet2} For a discussion of the few mathamatically precise
  results that are available for more complicated or higher order
  equations, we refer to the book by Collet and Eckmann,
  Ref. \onlinecite{collet}.
\bibitem{spruijt} A new and simple proof that fronts in the
  Swift-Hohenberg equations are pulled, and a new mode expansion that
  leads to a generalization of (\ref{107}) for pattern forming fronts
  which asympotically are periodic in the comoving frame, such as
  those arising in the Swif-Hohenberg equation, will be given in a
  future publication by W. Spruijt, U. Ebert, and W. van Saarloos.
\bibitem{note1} A brief overview of comparison type arguments can be
  found in the Appendix of Ref. \onlinecite{bj}.
\bibitem{note2} We stress that we claim no originality here. 
 In the physics literature, this type of analysis has
  appeared in various places, quite often in relation to
  Ginzburg-Landau or mean-field type approaches. See, e.g.,
  Refs. \onlinecite{gunton,bray,langer}.
\end{references}
\end{document}